\documentclass[10pt]{article}
\usepackage{graphicx,color}
\usepackage{hyperref}
\usepackage{footmisc}
\usepackage{amsmath}
\usepackage{tikzsymbols}
\usepackage{amsfonts, dsfont}
\usepackage{amssymb}
\usepackage{bbm}
\usepackage{epstopdf}
\usepackage{color}
\usepackage{cleveref}
\usepackage{dirtytalk}
\usepackage{titlesec}
\usepackage{comment}
\usepackage{mathtools}

\usepackage[utf8]{inputenc}
\graphicspath{{images/}}
\usepackage[a4paper,margin=1in]{geometry}
\usepackage{tikz}
\usetikzlibrary{arrows.meta}
\usepackage{amsthm}
\usepackage{float}

\DeclareMathOperator{\sgn}{sgn}

\definecolor{dimerred}{RGB}{237,34,39}
\definecolor{dimerorange}{RGB}{246,135,18}
\definecolor{edgeblack}{RGB}{35,31,32}
\definecolor{medgeblue}{RGB}{31,104,196}
\definecolor{dimorange}{RGB}{246,135,18}
\definecolor{dimred}{RGB}{237,34,39}
\definecolor{mblue}{RGB}{31,104,196}

\numberwithin{equation}{section}

\usepackage{graphicx}

\newtheorem{theorem}{Theorem}

\newtheorem{remark}{Remark}

\newtheorem{lemma}{Lemma}

\newtheorem{proposition}[theorem]{Proposition}

\newcommand{\un}[1]{\underline{#1}}

\titleformat{\subsection}[runin]
{\normalfont\normalsize\bfseries}{\thesubsection}{1em}{}

\titleformat{\subsubsection}[runin]
{\normalfont\normalsize\bfseries}{\thesubsubsection}{1em}{}

scaled\magstep1 
 
scaled\magstep1

\let\a=\alpha \let\b=\beta  \let\g=\gamma  \let\d=\delta \let\eps=\epsilon  
\let\e=\varepsilon
\let\z=\zeta      \let\k=\kappa \let\l=\lambda
    \let\n=\nu             
\let\s=\sigma \let\t=\tau   \let\f=\varphi 
   \let\o=\omega
   \let\L=\Lambda

\def\be#1\ee{\begin{equation}#1\end{equation}}
\def\bsp#1\esp{\begin{split}#1\end{split}}
\def\bal#1\eal{\begin{align}#1\end{align}}
\def\bald#1\eald{\begin{aligned}#1\end{aligned}}
\def\ba#1#2\ea{\begin{array}{#1}#2\end{array}}

\def\PP{{\cal P}}\def\EE{{\cal E}}\def\VV{{\cal V}}
\def\WW{{\cal W}}
\def\TT{{\cal T}}\def\BB{{\cal B}}
\def\RR{{\cal R}}\def\LL{{\cal L}}\def\JJ{{\cal J}}
\def\DD{{\cal D}}\def\AA{{\cal A}}\def\SS{{\cal S}}

\def\AAA{{\cal A}}

\def\xx{{\bf x}}
 
\def\nn{{\bf n}}

\def\\{\hfill\break}
\def\={:=}
\let\io=\infty
\let\0=\noindent

\def\media#1{{\langle#1\rangle}}

\def\tende#1{\,\vtop{\ialign{##\crcr\rightarrowfill\crcr\noalign{\kern-1pt
    \nointerlineskip} \hskip3.pt${\scriptstyle #1}$\hskip3.pt\crcr}}\,}
\def\otto{\,{\kern-1.truept\leftarrow\kern-5.truept\to\kern-1.truept}\,}

\def\to{\rightarrow}

\def\qed{\hfill\raise1pt\hbox{\vrule height5pt width5pt depth0pt}}

\def\un#1{{\underline#1}}

\def\be{\begin{equation}}
\def\ee{\end{equation}}
\def\bea{\begin{eqnarray}}
\def\eea{\end{eqnarray}}
\def\nn{\nonumber}

\def\hpsp{\hat \psi^+}
\def\hpsm{\hat \psi^-}

\let\a=\alpha \let\b=\beta  \let\g=\gamma  \let\d=\delta \let\eps=\epsilon  
\let\e=\varepsilon
\let\z=\zeta      \let\k=\kappa \let\l=\lambda
    \let\n=\nu             
\let\s=\sigma \let\t=\tau   \let\f=\varphi 
   \let\o=\omega
   \let\L=\Lambda

\def\PP{{\mathcal P}}

\def\nn{\nonumber}

\numberwithin{equation}{section}

\theoremstyle{definition}

\theoremstyle{definition}

\theoremstyle{definition}

\title{Rough-to-frozen transition in non-integrable dimer models}

\author{Alessandro Giuliani, Vieri Mastropietro,  Bruno Renzi,   Fabio Toninelli}
\date{\today}

\begin{document}

\maketitle

\begin{abstract}
The transition from rough to frozen phases in models of random surfaces is expected to be governed by the Pokrovsky-Talapov law, which predicts a  $\varepsilon^{3/2}$-type singularity of the macroscopic height profile (limit shape), as the distance  $\varepsilon$ from the frozen phase (facet) is sent to zero. This has been proved so far only in exactly solvable cases, notably the classical, determinantal, 2D dimer model. In this paper we prove the validity of the Pokrovsky-Talapov law for a non-integrable 2D dimer model on the hexagonal lattice, whose Gibbs weight depends upon a weak plaquette interaction among dimers. Our proof is based on a multiscale analysis (constructive fermionic Renormalization Group), which allows us to identify two different regimes with distinct scaling properties and to explain the universality of the Pokrovsky-Talapov exponent $3/2$, in contrast with the interaction-dependent critical exponents characterizing the rough phase. The first regime, corresponding to smaller distances, is characterized by an effective quadratic dispersion relation, while the second comes with an effective linear dispersion relation, typical of  Luttinger liquid behavior. As a byproduct of our construction, we control the scaling limit of the interacting dimer model uniformly in the distance from the frozen phase. 
\end{abstract}

\tableofcontents

\section{Introduction}

The study of random surfaces revealed the emergence of deep universality phenomena in many different models, ranging from dimer coverings of planar graphs to interlaced particle systems and stochastic partial differential equations, with unexpected connections with random matrix theory. Universal fluctuation properties, characterized by a Gaussian Free Field (GFF) behavior or by a Tracy-Widom-type distribution, have been observed in seemingly disparate systems \cite{Spohn2006,Corwin,Sheffield2007}. Among these, the dimer model on planar bipartite graphs plays a prominent role due to its exact solvability: correlation functions can be computed explicitly via the Kasteleyn formalism, starting from which its scaling limit can be rigorously analyzed \cite{Kenyon_notes,Gorin}. As such, the dimer model provides an ideal laboratory for investigating the universal properties of random surfaces.

Dimer models on planar, periodic, bipartite graphs exhibit a rich phase diagram with three distinct phases, each characterized by specific properties of the height function distribution and of the large distance behavior of the dimer correlations \cite{Kenyon_notes,KOS}.
In the \textit{frozen phase}, the random surface becomes flat on a microscopic scale: the height function is deterministic and dimer configurations are rigidly ordered.
In the \textit{gaseous or smooth phase}, the variance of height fluctuations is bounded uniformly in the system size but, differently from the frozen phase, the system has positive configurational entropy. Truncated dimer-dimer correlations do not vanish, but they decay to zero exponentially fast at large distances, so that the height fluctuations at faraway points are essentially independent. 
The \textit{rough or liquid phase} exhibits the richest structure: the height function converges at large scales to a profile with non-trivial curvature, whose form depends upon the shape of the domain and the boundary conditions, and can be computed as the solution to a variational problem involving the surface tension (or free energy) of the model; moreover, at large distances fluctuations around the average profile converge to a GFF, after appropriate rescaling. At the microscopic level, dimer-dimer correlations decay polynomially to zero at large distances, signaling a critical behavior analogous to the one displayed by statistical mechanics systems at a second order phase transition. In this phase, the scaling limit of height fluctuations displays conformal invariance properties \cite{Kenyon2001}.

At the transition between these phases, in particular at the \emph{rough–frozen boundary} and at the \emph{rough–smooth boundary}, the macroscopic height profile 
displays a singular behavior characterized by additional universal features. More specifically, the height profile is governed by the \textit{Pokrovsky-Talapov (PT) law} \cite{PT,RW}: as one moves from the rough to the frozen (or to the smooth) phase, the average height profile (which is not affine within the rough phase) approaches an affine one, with deviations of the order $\varepsilon^{3/2}$ as the distance $\varepsilon$ from the frozen (or smooth) boundary is sent to zero. The scaling exponent $3/2$ has been observed in a variety of exactly solvable models \cite{ADPZ}, and it is intimately connected with the \textit{Kardar-Parisi-Zhang (KPZ)} universality class and the Airy line ensemble, which describes, among others,
the fluctuations of the level lines of the height function near a frozen-rough boundary \cite{Corwin,Johansson,AggarwalH} (see also the recent \cite{AI}, that announced a proof in the case of the the six-vertex model at the ice point).

These phenomena have been rigorously established for integrable dimer models,
and recently \cite{M-T} also for a non-interacting dimer model 
with {\it random} layered edge weights. On the other hand, as soon as interactions 
among dimers are included in the picture, Kasteleyn's theory breaks down and most of 
the results mentioned above do not hold anymore, even though the general expectation
is that the same qualitative features, characterizing the different phases and the transitions from one to the other, should be valid irrespective of the
integrability or non-integrability properties of the microscopic model. 

In a series 
of previous papers \cite{GMT17,GMT17b, GMT20,GRT} we considered a class of non-integrable dimer models, for which we proved that, well-inside the rough phase, the scaling limit of height fluctuations around the average height profile is described by a GFF, like in the case of integrable dimer models; on the other hand, the critical exponents characterizing the large distance decay properties of dimer correlations change, in general: in particular, in \cite{GMT17b, GMT20,GRT} we prove that 
the critical exponent associated with the oscillatory part of the two-point dimer correlations is a non-trivial analytic function of the interaction strength, equal to the `stiffness' coefficient of the GFF. The identity among these two a-priori unrelated quantities is the content of a deep universality relation, first predicted by Kadanoff on the basis of non-rigorous bosonization methods. The results of 
\cite{GMT17,GMT17b, GMT20,GRT} hold in the weak coupling regime, i.e., they require 
that the interaction strength $\lambda$, which measures the distance from a reference integrable model, is small enough, a priori non-uniformly in the distance $\varepsilon$ from the rough-frozen (or rough-smooth) boundary. 

These results motivate several additional questions on the behavior of weakly interacting dimer models: is there a well defined separation line, possibly 
deformed by the interaction, separating the rough from the frozen (or smooth) phase? 
If so, does the GFF behavior of height fluctuations, as well as the Kadanoff universality relation between the stiffness of the GFF and the dimer-dimer critical exponent, hold in the whole rough phase, up to the transition line? Does the exponent 3/2, characterizing the speed at which the height profile within the rough phase approaches an affine profile, remain unchanged, or is it modified by the interaction? That is, is the PT critical behavior robust under the addition of perturbations breaking the integrability of the model or not? 

In this paper we investigate the rough-to-frozen transition for a class of non-integrable interacting dimer models on the hexagonal lattice. Motivated by the questions above, we analyze the critical behavior of the free energy as we approach the frozen phase, and prove the stability of the 3/2 exponent characterizing the 
PT critical behavior. A key technical aspect of our proof is a generalization 
of the multiscale analysis developed in previous papers, which converges 
for small enough interaction strength, {\it{uniformly}} in the distance from the 
frozen boundary. More specifically, we identify two different scaling regimes, 
characterized by different scaling properties, and we employ two different multiscale 
procedures in order to analyze them; the ideas and method we apply are analogous to 
those introduced in \cite{BM36} to study the transition from a metallic to an insulating phase in 1D fermionic chains. 

\subsection{Model and results}

\begin{figure}[ht]
\centering
\begin{tikzpicture}[x=0.64cm,y=0.64cm]
\begin{scope}
\clip (-1.0000,0.0000)--(5.0000,3.4641)--(11.0000,0.0000)--(5.0000,-3.4641)--cycle;
\draw[black,line width=0.45pt] (0.0000,0.0000)--(1.0000,0.0000) (1.0000,0.0000)--(1.5000,-0.8660) (1.5000,-0.8660)--(1.0000,0.0000) (1.0000,0.0000)--(1.5000,0.8660) (1.5000,0.8660)--(1.0000,0.0000) (1.5000,0.8660)--(2.5000,0.8660) (2.5000,0.8660)--(3.0000,0.0000) (3.0000,0.0000)--(2.5000,0.8660) (2.5000,0.8660)--(3.0000,1.7321) (3.0000,1.7321)--(2.5000,0.8660) (3.0000,1.7321)--(4.0000,1.7321) (4.0000,1.7321)--(4.5000,0.8660) (4.5000,0.8660)--(4.0000,1.7321) (4.0000,1.7321)--(4.5000,2.5981) (4.5000,2.5981)--(4.0000,1.7321) (4.5000,2.5981)--(5.5000,2.5981) (5.5000,2.5981)--(6.0000,1.7321) (6.0000,1.7321)--(5.5000,2.5981) (5.5000,2.5981)--(6.0000,3.4641) (0.0000,0.0000)--(-0.5000,-0.8660) (1.5000,-0.8660)--(2.5000,-0.8660) (2.5000,-0.8660)--(3.0000,-1.7321) (3.0000,-1.7321)--(2.5000,-0.8660) (2.5000,-0.8660)--(3.0000,0.0000) (3.0000,0.0000)--(2.5000,-0.8660) (3.0000,0.0000)--(4.0000,0.0000) (4.0000,0.0000)--(4.5000,-0.8660) (4.5000,-0.8660)--(4.0000,0.0000) (4.0000,0.0000)--(4.5000,0.8660) (4.5000,0.8660)--(4.0000,0.0000) (4.5000,0.8660)--(5.5000,0.8660) (5.5000,0.8660)--(6.0000,0.0000) (6.0000,0.0000)--(5.5000,0.8660) (5.5000,0.8660)--(6.0000,1.7321) (6.0000,1.7321)--(5.5000,0.8660) (6.0000,1.7321)--(7.0000,1.7321) (7.0000,1.7321)--(7.5000,0.8660) (7.5000,0.8660)--(7.0000,1.7321) (7.0000,1.7321)--(7.5000,2.5981) (1.5000,-0.8660)--(1.0000,-1.7321) (3.0000,-1.7321)--(4.0000,-1.7321) (4.0000,-1.7321)--(4.5000,-2.5981) (4.5000,-2.5981)--(4.0000,-1.7321) (4.0000,-1.7321)--(4.5000,-0.8660) (4.5000,-0.8660)--(4.0000,-1.7321) (4.5000,-0.8660)--(5.5000,-0.8660) (5.5000,-0.8660)--(6.0000,-1.7321) (6.0000,-1.7321)--(5.5000,-0.8660) (5.5000,-0.8660)--(6.0000,0.0000) (6.0000,0.0000)--(5.5000,-0.8660) (6.0000,0.0000)--(7.0000,0.0000) (7.0000,0.0000)--(7.5000,-0.8660) (7.5000,-0.8660)--(7.0000,0.0000) (7.0000,0.0000)--(7.5000,0.8660) (7.5000,0.8660)--(7.0000,0.0000) (7.5000,0.8660)--(8.5000,0.8660) (8.5000,0.8660)--(9.0000,0.0000) (9.0000,0.0000)--(8.5000,0.8660) (8.5000,0.8660)--(9.0000,1.7321) (3.0000,-1.7321)--(2.5000,-2.5981) (4.5000,-2.5981)--(5.5000,-2.5981) (5.5000,-2.5981)--(6.0000,-3.4641) (0.0000,0.0000)--(-0.5000,0.8660) (5.5000,-2.5981)--(6.0000,-1.7321) (6.0000,-1.7321)--(5.5000,-2.5981) (6.0000,-1.7321)--(7.0000,-1.7321) (7.0000,-1.7321)--(7.5000,-2.5981) (1.5000,0.8660)--(1.0000,1.7321) (7.0000,-1.7321)--(7.5000,-0.8660) (7.5000,-0.8660)--(7.0000,-1.7321) (7.5000,-0.8660)--(8.5000,-0.8660) (8.5000,-0.8660)--(9.0000,-1.7321) (3.0000,1.7321)--(2.5000,2.5981) (8.5000,-0.8660)--(9.0000,0.0000) (9.0000,0.0000)--(8.5000,-0.8660) (9.0000,0.0000)--(10.0000,0.0000) (10.0000,0.0000)--(10.5000,-0.8660) (4.5000,2.5981)--(4.0000,3.4641) (10.0000,0.0000)--(10.5000,0.8660) (4.5000,-2.5981)--(4.0000,-3.4641);
\draw[dimorange,line width=2.2pt] (1.5000,0.8660)--(2.5000,0.8660) (7.5000,0.8660)--(8.5000,0.8660) (5.5000,2.5981)--(6.0000,1.7321) (6.0000,1.7321)--(5.5000,2.5981) (5.5000,0.8660)--(6.0000,0.0000) (6.0000,0.0000)--(5.5000,0.8660) (7.0000,0.0000)--(7.5000,-0.8660) (7.5000,-0.8660)--(7.0000,0.0000) (8.5000,-0.8660)--(9.0000,-1.7321) (3.0000,1.7321)--(2.5000,2.5981) (4.0000,1.7321)--(4.5000,2.5981) (4.5000,2.5981)--(4.0000,1.7321) (4.0000,0.0000)--(4.5000,0.8660) (4.5000,0.8660)--(4.0000,0.0000) (2.5000,-0.8660)--(3.0000,0.0000) (3.0000,0.0000)--(2.5000,-0.8660) (7.0000,1.7321)--(7.5000,2.5981) (1.5000,-0.8660)--(1.0000,-1.7321);
\draw[dimred,line width=2.2pt] (0.0000,0.0000)--(1.0000,0.0000) (4.5000,-0.8660)--(5.5000,-0.8660) (3.0000,-1.7321)--(4.0000,-1.7321) (6.0000,-1.7321)--(7.0000,-1.7321) (4.5000,-2.5981)--(5.5000,-2.5981) (9.0000,0.0000)--(10.0000,0.0000);
\end{scope}
\draw[dash pattern=on 5pt off 3.5pt,line width=0.45pt] (-1.0000,0.0000)--(5.0000,3.4641) (5.0000,-3.4641)--(11.0000,0.0000);
\draw[dotted,line width=0.6pt] (-1.0000,0.0000)--(5.0000,-3.4641) (5.0000,3.4641)--(11.0000,0.0000);
\fill[black!88] (0.0000,0.0000) circle (0.12);
\filldraw[fill=white,draw=black,line width=0.45pt] (1.0000,0.0000) circle (0.12);
\fill[black!88] (1.5000,0.8660) circle (0.12);
\filldraw[fill=white,draw=black,line width=0.45pt] (2.5000,0.8660) circle (0.12);
\fill[black!88] (3.0000,1.7321) circle (0.12);
\filldraw[fill=white,draw=black,line width=0.45pt] (4.0000,1.7321) circle (0.12);
\fill[black!88] (4.5000,2.5981) circle (0.12);
\filldraw[fill=white,draw=black,line width=0.45pt] (5.5000,2.5981) circle (0.12);
\fill[black!88] (1.5000,-0.8660) circle (0.12);
\filldraw[fill=white,draw=black,line width=0.45pt] (2.5000,-0.8660) circle (0.12);
\fill[black!88] (3.0000,0.0000) circle (0.12);
\filldraw[fill=white,draw=black,line width=0.45pt] (4.0000,0.0000) circle (0.12);
\fill[black!88] (4.5000,0.8660) circle (0.12);
\filldraw[fill=white,draw=black,line width=0.45pt] (5.5000,0.8660) circle (0.12);
\fill[black!88] (6.0000,1.7321) circle (0.12);
\filldraw[fill=white,draw=black,line width=0.45pt] (7.0000,1.7321) circle (0.12);
\fill[black!88] (3.0000,-1.7321) circle (0.12);
\filldraw[fill=white,draw=black,line width=0.45pt] (4.0000,-1.7321) circle (0.12);
\fill[black!88] (4.5000,-0.8660) circle (0.12);
\filldraw[fill=white,draw=black,line width=0.45pt] (5.5000,-0.8660) circle (0.12);
\fill[black!88] (6.0000,0.0000) circle (0.12);
\filldraw[fill=white,draw=black,line width=0.45pt] (7.0000,0.0000) circle (0.12);
\fill[black!88] (7.5000,0.8660) circle (0.12);
\filldraw[fill=white,draw=black,line width=0.45pt] (8.5000,0.8660) circle (0.12);
\fill[black!88] (4.5000,-2.5981) circle (0.12);
\filldraw[fill=white,draw=black,line width=0.45pt] (5.5000,-2.5981) circle (0.12);
\fill[black!88] (6.0000,-1.7321) circle (0.12);
\filldraw[fill=white,draw=black,line width=0.45pt] (7.0000,-1.7321) circle (0.12);
\fill[black!88] (7.5000,-0.8660) circle (0.12);
\filldraw[fill=white,draw=black,line width=0.45pt] (8.5000,-0.8660) circle (0.12);
\fill[black!88] (9.0000,0.0000) circle (0.12);
\filldraw[fill=white,draw=black,line width=0.45pt] (10.0000,0.0000) circle (0.12);
\draw[dashed,line width=0.4pt] (5.0000,-0.8660) ellipse (0.78 and 0.35);
\draw[dashed,line width=0.4pt] (6.5000,-1.7321) ellipse (0.78 and 0.35);
\draw[dashed,line width=0.4pt] (6.5000,0.0000) ellipse (0.78 and 0.35);
\draw[-stealth,line width=0.5pt] (4.5000,-0.8660)--(5.6700,-1.5415);
\node[font=\footnotesize] at (4.9500,-1.6660) {$\vec e_1$};
\draw[-stealth,line width=0.5pt] (4.5000,-0.8660)--(5.6700,-0.1905);
\node[font=\footnotesize] at (4.9700,-0.1060) {$\vec e_2$};
\node[font=\footnotesize] at (6.5000,0.1700) {$A$};
\node[font=\footnotesize] at (7.4100,0.4200) {$C$};
\node[font=\footnotesize] at (7.0000,-0.6000) {$B$};
\begin{scope}
\clip (11.6000,0.0000)--(17.6000,3.4641)--(23.6000,0.0000)--(17.6000,-3.4641)--cycle;
\draw[mblue,line width=0.45pt] (11.1000,-3.5641)--(11.1000,3.5641) (12.6000,-3.5641)--(12.6000,3.5641) (14.1000,-3.5641)--(14.1000,3.5641) (15.6000,-3.5641)--(15.6000,3.5641) (17.1000,-3.5641)--(17.1000,3.5641) (18.6000,-3.5641)--(18.6000,3.5641) (20.1000,-3.5641)--(20.1000,3.5641) (21.6000,-3.5641)--(21.6000,3.5641) (23.1000,-3.5641)--(23.1000,3.5641) (24.6000,-3.5641)--(24.6000,3.5641) (12.1000,-3.5641)--(12.1000,3.5641) (13.6000,-3.5641)--(13.6000,3.5641) (15.1000,-3.5641)--(15.1000,3.5641) (16.6000,-3.5641)--(16.6000,3.5641) (18.1000,-3.5641)--(18.1000,3.5641) (19.6000,-3.5641)--(19.6000,3.5641) (21.1000,-3.5641)--(21.1000,3.5641) (22.6000,-3.5641)--(22.6000,3.5641) (24.1000,-3.5641)--(24.1000,3.5641) (25.6000,-3.5641)--(25.6000,3.5641);
\draw[black,line width=0.45pt] (12.6000,0.0000)--(13.6000,0.0000) (13.6000,0.0000)--(14.1000,-0.8660) (14.1000,-0.8660)--(13.6000,0.0000) (13.6000,0.0000)--(14.1000,0.8660) (14.1000,0.8660)--(13.6000,0.0000) (14.1000,0.8660)--(15.1000,0.8660) (15.1000,0.8660)--(15.6000,0.0000) (15.6000,0.0000)--(15.1000,0.8660) (15.1000,0.8660)--(15.6000,1.7321) (15.6000,1.7321)--(15.1000,0.8660) (15.6000,1.7321)--(16.6000,1.7321) (16.6000,1.7321)--(17.1000,0.8660) (17.1000,0.8660)--(16.6000,1.7321) (16.6000,1.7321)--(17.1000,2.5981) (17.1000,2.5981)--(16.6000,1.7321) (17.1000,2.5981)--(18.1000,2.5981) (18.1000,2.5981)--(18.6000,1.7321) (18.6000,1.7321)--(18.1000,2.5981) (18.1000,2.5981)--(18.6000,3.4641) (12.6000,0.0000)--(12.1000,-0.8660) (14.1000,-0.8660)--(15.1000,-0.8660) (15.1000,-0.8660)--(15.6000,-1.7321) (15.6000,-1.7321)--(15.1000,-0.8660) (15.1000,-0.8660)--(15.6000,0.0000) (15.6000,0.0000)--(15.1000,-0.8660) (15.6000,0.0000)--(16.6000,0.0000) (16.6000,0.0000)--(17.1000,-0.8660) (17.1000,-0.8660)--(16.6000,0.0000) (16.6000,0.0000)--(17.1000,0.8660) (17.1000,0.8660)--(16.6000,0.0000) (17.1000,0.8660)--(18.1000,0.8660) (18.1000,0.8660)--(18.6000,0.0000) (18.6000,0.0000)--(18.1000,0.8660) (18.1000,0.8660)--(18.6000,1.7321) (18.6000,1.7321)--(18.1000,0.8660) (18.6000,1.7321)--(19.6000,1.7321) (19.6000,1.7321)--(20.1000,0.8660) (20.1000,0.8660)--(19.6000,1.7321) (19.6000,1.7321)--(20.1000,2.5981) (14.1000,-0.8660)--(13.6000,-1.7321) (15.6000,-1.7321)--(16.6000,-1.7321) (16.6000,-1.7321)--(17.1000,-2.5981) (17.1000,-2.5981)--(16.6000,-1.7321) (16.6000,-1.7321)--(17.1000,-0.8660) (17.1000,-0.8660)--(16.6000,-1.7321) (17.1000,-0.8660)--(18.1000,-0.8660) (18.1000,-0.8660)--(18.6000,-1.7321) (18.6000,-1.7321)--(18.1000,-0.8660) (18.1000,-0.8660)--(18.6000,0.0000) (18.6000,0.0000)--(18.1000,-0.8660) (18.6000,0.0000)--(19.6000,0.0000) (19.6000,0.0000)--(20.1000,-0.8660) (20.1000,-0.8660)--(19.6000,0.0000) (19.6000,0.0000)--(20.1000,0.8660) (20.1000,0.8660)--(19.6000,0.0000) (20.1000,0.8660)--(21.1000,0.8660) (21.1000,0.8660)--(21.6000,0.0000) (21.6000,0.0000)--(21.1000,0.8660) (21.1000,0.8660)--(21.6000,1.7321) (15.6000,-1.7321)--(15.1000,-2.5981) (17.1000,-2.5981)--(18.1000,-2.5981) (18.1000,-2.5981)--(18.6000,-3.4641) (12.6000,0.0000)--(12.1000,0.8660) (18.1000,-2.5981)--(18.6000,-1.7321) (18.6000,-1.7321)--(18.1000,-2.5981) (18.6000,-1.7321)--(19.6000,-1.7321) (19.6000,-1.7321)--(20.1000,-2.5981) (14.1000,0.8660)--(13.6000,1.7321) (19.6000,-1.7321)--(20.1000,-0.8660) (20.1000,-0.8660)--(19.6000,-1.7321) (20.1000,-0.8660)--(21.1000,-0.8660) (21.1000,-0.8660)--(21.6000,-1.7321) (15.6000,1.7321)--(15.1000,2.5981) (21.1000,-0.8660)--(21.6000,0.0000) (21.6000,0.0000)--(21.1000,-0.8660) (21.6000,0.0000)--(22.6000,0.0000) (22.6000,0.0000)--(23.1000,-0.8660) (17.1000,2.5981)--(16.6000,3.4641) (22.6000,0.0000)--(23.1000,0.8660) (17.1000,-2.5981)--(16.6000,-3.4641);
\draw[dimorange,line width=2.2pt] (20.1000,-0.8660)--(21.1000,-0.8660) (14.1000,-0.8660)--(15.1000,-0.8660) (18.1000,-2.5981)--(18.6000,-3.4641) (12.6000,0.0000)--(12.1000,0.8660) (18.1000,2.5981)--(18.6000,1.7321) (18.6000,1.7321)--(18.1000,2.5981) (19.6000,1.7321)--(20.1000,0.8660) (20.1000,0.8660)--(19.6000,1.7321) (21.1000,0.8660)--(21.6000,0.0000) (21.6000,0.0000)--(21.1000,0.8660) (22.6000,0.0000)--(23.1000,0.8660) (17.1000,-2.5981)--(16.6000,-3.4641) (16.6000,1.7321)--(17.1000,2.5981) (17.1000,2.5981)--(16.6000,1.7321) (15.1000,0.8660)--(15.6000,1.7321) (15.6000,1.7321)--(15.1000,0.8660) (13.6000,0.0000)--(14.1000,0.8660) (14.1000,0.8660)--(13.6000,0.0000) (17.1000,-0.8660)--(17.1000,0.8660) (17.1000,0.8660)--(17.1000,-0.8660) (18.1000,-0.8660)--(18.1000,0.8660) (18.1000,0.8660)--(18.1000,-0.8660);
\draw[dimred,line width=2.2pt] (18.6000,-1.7321)--(19.6000,-1.7321) (15.6000,0.0000)--(16.6000,0.0000) (18.6000,0.0000)--(19.6000,0.0000) (15.6000,-1.7321)--(16.6000,-1.7321);
\end{scope}
\draw[dash pattern=on 5pt off 3.5pt,line width=0.45pt] (11.6000,0.0000)--(17.6000,3.4641) (17.6000,-3.4641)--(23.6000,0.0000);
\draw[dotted,line width=0.6pt] (11.6000,0.0000)--(17.6000,-3.4641) (17.6000,3.4641)--(23.6000,0.0000);
\fill[black!88] (12.6000,0.0000) circle (0.12);
\filldraw[fill=white,draw=black,line width=0.45pt] (13.6000,0.0000) circle (0.12);
\fill[black!88] (14.1000,0.8660) circle (0.12);
\filldraw[fill=white,draw=black,line width=0.45pt] (15.1000,0.8660) circle (0.12);
\fill[black!88] (15.6000,1.7321) circle (0.12);
\filldraw[fill=white,draw=black,line width=0.45pt] (16.6000,1.7321) circle (0.12);
\fill[black!88] (17.1000,2.5981) circle (0.12);
\filldraw[fill=white,draw=black,line width=0.45pt] (18.1000,2.5981) circle (0.12);
\fill[black!88] (14.1000,-0.8660) circle (0.12);
\filldraw[fill=white,draw=black,line width=0.45pt] (15.1000,-0.8660) circle (0.12);
\fill[black!88] (15.6000,0.0000) circle (0.12);
\filldraw[fill=white,draw=black,line width=0.45pt] (16.6000,0.0000) circle (0.12);
\fill[black!88] (17.1000,0.8660) circle (0.12);
\filldraw[fill=white,draw=black,line width=0.45pt] (18.1000,0.8660) circle (0.12);
\fill[black!88] (18.6000,1.7321) circle (0.12);
\filldraw[fill=white,draw=black,line width=0.45pt] (19.6000,1.7321) circle (0.12);
\fill[black!88] (15.6000,-1.7321) circle (0.12);
\filldraw[fill=white,draw=black,line width=0.45pt] (16.6000,-1.7321) circle (0.12);
\fill[black!88] (17.1000,-0.8660) circle (0.12);
\filldraw[fill=white,draw=black,line width=0.45pt] (18.1000,-0.8660) circle (0.12);
\fill[black!88] (18.6000,0.0000) circle (0.12);
\filldraw[fill=white,draw=black,line width=0.45pt] (19.6000,0.0000) circle (0.12);
\fill[black!88] (20.1000,0.8660) circle (0.12);
\filldraw[fill=white,draw=black,line width=0.45pt] (21.1000,0.8660) circle (0.12);
\fill[black!88] (17.1000,-2.5981) circle (0.12);
\filldraw[fill=white,draw=black,line width=0.45pt] (18.1000,-2.5981) circle (0.12);
\fill[black!88] (18.6000,-1.7321) circle (0.12);
\filldraw[fill=white,draw=black,line width=0.45pt] (19.6000,-1.7321) circle (0.12);
\fill[black!88] (20.1000,-0.8660) circle (0.12);
\filldraw[fill=white,draw=black,line width=0.45pt] (21.1000,-0.8660) circle (0.12);
\fill[black!88] (21.6000,0.0000) circle (0.12);
\filldraw[fill=white,draw=black,line width=0.45pt] (22.6000,0.0000) circle (0.12);
\draw[mblue,line width=0.45pt] (7.1000,-3.3840)--(7.1000,-5.1160) (8.1000,-3.3840)--(8.1000,-5.1160);
\draw[black,line width=0.45pt] (8.1000,-3.3840)--(7.1000,-3.3840) (7.1000,-3.3840)--(6.6000,-4.2500) (6.6000,-4.2500)--(7.1000,-5.1160) (7.1000,-5.1160)--(8.1000,-5.1160) (8.1000,-5.1160)--(8.6000,-4.2500) (8.6000,-4.2500)--(8.1000,-3.3840);
\draw[dimred,line width=2.2pt] (8.1000,-3.3840)--(7.1000,-3.3840) (7.1000,-5.1160)--(8.1000,-5.1160);
\filldraw[fill=white,draw=black,line width=0.45pt] (8.1000,-3.3840) circle (0.12);
\fill[black!88] (7.1000,-3.3840) circle (0.12);
\filldraw[fill=white,draw=black,line width=0.45pt] (6.6000,-4.2500) circle (0.12);
\fill[black!88] (7.1000,-5.1160) circle (0.12);
\filldraw[fill=white,draw=black,line width=0.45pt] (8.1000,-5.1160) circle (0.12);
\fill[black!88] (8.6000,-4.2500) circle (0.12);
\node[font=\scriptsize] at (7.6000,-3.0300) {1};
\node[font=\scriptsize] at (6.5434,-3.6400) {2};
\node[font=\scriptsize] at (6.5434,-4.8600) {3};
\node[font=\scriptsize] at (7.6000,-5.4700) {4};
\node[font=\scriptsize] at (8.6566,-4.8600) {5};
\node[font=\scriptsize] at (8.6566,-3.6400) {6};
\node[font=\footnotesize] at (7.6000,-4.2500) {$f$};
\draw[mblue,line width=0.45pt] (10.8000,-3.3840)--(10.8000,-5.1160) (11.8000,-3.3840)--(11.8000,-5.1160);
\draw[black,line width=0.45pt] (11.8000,-3.3840)--(10.8000,-3.3840) (10.8000,-3.3840)--(10.3000,-4.2500) (10.3000,-4.2500)--(10.8000,-5.1160) (10.8000,-5.1160)--(11.8000,-5.1160) (11.8000,-5.1160)--(12.3000,-4.2500) (12.3000,-4.2500)--(11.8000,-3.3840);
\draw[dimred,line width=2.2pt] (10.3000,-4.2500)--(10.8000,-5.1160) (12.3000,-4.2500)--(11.8000,-3.3840);
\filldraw[fill=white,draw=black,line width=0.45pt] (11.8000,-3.3840) circle (0.12);
\fill[black!88] (10.8000,-3.3840) circle (0.12);
\filldraw[fill=white,draw=black,line width=0.45pt] (10.3000,-4.2500) circle (0.12);
\fill[black!88] (10.8000,-5.1160) circle (0.12);
\filldraw[fill=white,draw=black,line width=0.45pt] (11.8000,-5.1160) circle (0.12);
\fill[black!88] (12.3000,-4.2500) circle (0.12);
\draw[mblue,line width=0.45pt] (14.5000,-3.3840)--(14.5000,-5.1160) (15.5000,-3.3840)--(15.5000,-5.1160);
\draw[black,line width=0.45pt] (15.5000,-3.3840)--(14.5000,-3.3840) (14.5000,-3.3840)--(14.0000,-4.2500) (14.0000,-4.2500)--(14.5000,-5.1160) (14.5000,-5.1160)--(15.5000,-5.1160) (15.5000,-5.1160)--(16.0000,-4.2500) (16.0000,-4.2500)--(15.5000,-3.3840);
\draw[dimred,line width=2.2pt] (14.5000,-3.3840)--(14.0000,-4.2500) (15.5000,-5.1160)--(16.0000,-4.2500);
\filldraw[fill=white,draw=black,line width=0.45pt] (15.5000,-3.3840) circle (0.12);
\fill[black!88] (14.5000,-3.3840) circle (0.12);
\filldraw[fill=white,draw=black,line width=0.45pt] (14.0000,-4.2500) circle (0.12);
\fill[black!88] (14.5000,-5.1160) circle (0.12);
\filldraw[fill=white,draw=black,line width=0.45pt] (15.5000,-5.1160) circle (0.12);
\fill[black!88] (16.0000,-4.2500) circle (0.12);
\end{tikzpicture}
\caption{Left: the hexagonal graph $G_L$ for $L=4$; the fundamental domain, containing one black and one white vertex sharing the same coordinate $x$, is encircled.  Right: the graph $G_{L,m}$ can be thought of as $G_L$ (whose edges are colored black) with additional edges of type $m$ (which are colored blue); 
Bottom: three local dimer configurations contributing a factor $e^\lambda$ to the Gibbs weight. 
Every edge can be determined by specifying a face $f$ of $G_L$ and a type label $j=1,\dots,6$, running anti-clockwise from the top horizontal edge of $f$; the two edges of type $m$ inside $f$ join its two black vertices on the left and its two white vertices on the right. The dimers colored orange or red form a close-packed dimer configuration (containing no dimer of type $m$); the red dimers are those giving a non-zero contribution to the interaction $V$.
The weight of horizontal dimers (i.e. edges of types 1 or 4) is equal to $A$, that of northwest-to-southeast, i.e. type 3 or 6 is $B$, that of  northeast-to-southwest, types 2 or 5 dimers is $C$, and that of dimers of type $m$ is $m$.}
\label{fig:toroLesagono}
\end{figure}
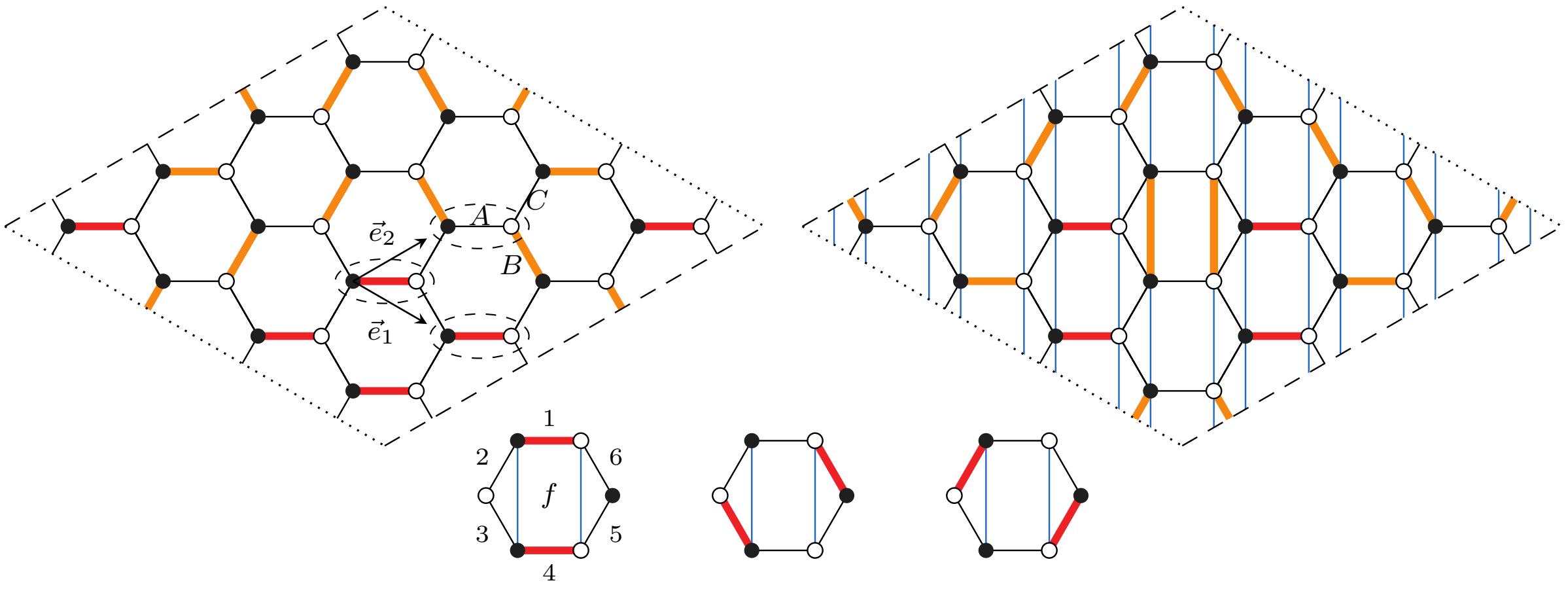

In order to study the problem we choose the simplest possible setting; possible generalizations are discussed after the main theorem.
Let $G_L:=\mathcal{H}/(L \mathbb{Z}\times L \mathbb{Z})$, be the hexagonal graph $\mathcal{H}$ periodized in directions $\vec e_1,\vec e_2$ with period $L\in 2\mathbb N$, (see Figure \ref{fig:toroLesagono}) and let $\Omega_L$ be the set of close-packed dimer configurations or perfect matchings of $G_L$. Alongside with $G_L$, we consider a modified, non-bipartite graph $G_{L,m}$ obtained from $G_L$ by adding in each hexagonal face $f \in G^*_L$ the two vertical edges joining the upper-left to the lower-left vertex of $f$ (both black) and the upper-right to the lower-right vertex of $f$ (both white), see Figure \ref{fig:toroLesagono}. We call these additional edges \emph{of type $m$} while the other edges are of type $A,B$ or $C$ according to their orientation, see Fig. \ref{fig:toroLesagono}; we denote $\Omega_{L,m}$ the set of perfect matchings of $G_{L,m}.$

The interacting dimer model is defined by assigning to each $M\in\Omega_{L,m}$ the probability
\begin{equation}\label{defintermodel}
\begin{split}
&\mathbb{P}_{\lambda,L,m}(M):=\frac{A^{N_A} B^{N_B} C^{N_C} m^{N_m}e^{\lambda V}}{Z_{\lambda,L,m}}, \qquad Z_{\lambda,L,m}=\sum_{M \in \Omega_{L,m}}A^{N_A} B^{N_B} C^{N_C} m^{N_m}e^{\lambda V} \\ &  V=\sum_{\substack{f \in G_L^*}} \sum_{\ell=1}^3\mathbbm{1}_{f;\ell}\mathbbm{1}_{f;\ell+3}
\end{split}
\end{equation}
where $N_s=N_s(M)$ is the total number of dimers of type $s\in\{A,B,C,m\}$, see Fig. \ref{fig:toroLesagono}; moreover, $\l\in\mathbb R$, $m>0$ and, given a face $f$ of $G_L$, $\mathbbm{1}_{f;j}(M)$ is the indicator function of the event that the edge of $f$ with label $j=1,\dots,6$ belongs to $M$, see Fig. \ref{fig:toroLesagono}. The non interacting model corresponds to $\l=0$. 

\begin{remark}
The specific form of the interaction $V$ in \eqref{defintermodel} is chosen just for simplicity and for illustrative purposes: our results and methods of proof can be straightforwardly adapted to 
the case of a general translationally invariant local interaction, at the cost of a few technical and notational complications, as discussed in \cite{GMT20}, see also Section \ref{subsec:generalizations}. Choosing 
the interaction in a different and special way, namely equal to $W=\sum_{f \in G^*_L} (\mathds 1_{f;2}\mathds 1_{f;6}+\mathds 1_{f;3}\mathds 1_{f;5})$, leads to an integrable model: as discussed in Section \ref{subsec:vertexmodels}, if we replace $V$ by $W$ in \eqref{defintermodel} the resulting model can be mapped exactly into the five-vertex (5V) model, which is integrable by Bethe ansatz. However, a generic choice of the interaction, including the simple one in \eqref{defintermodel}, corresponds to a non-integrable model. 
\end{remark}

\begin{remark}
The weight $m$ plays the role of an infrared regulator: as discussed below, it produces a `mass' which makes the dimer correlations exponentially decaying at large distances. We compute 
all the observables of interest in the massless limit $m\to 0^+$, taken \emph{after} the thermodynamic limit  $L\to\infty$. For technical reasons, this is easier than setting $m=0$ and taking $L\to\infty$ afterwards.
\end{remark}

The observables of interest are the \emph{free energy} per unit volume
and the \emph{density of dimers} of type $s\in\{A,B,C\}$:
\be 
F_{\l,L,m}:=\frac{1}{L^2}\log Z_{\l,L,m}, \qquad \rho_{s;\l,L,m}:=\mathbb{E}_{\l,L,m}\big(L^{-2} N_s\big) 
\label{eq:deffandrho}
\ee where $\mathbb{E}_{\l,L,m}$ is the expectation under $\mathbb{P}_{\l,L,m}$ (cf. \eqref{defintermodel}). Their $L\to\infty$ limits 
followed by the limit $m\to 0^+$ are denoted by $F_{\l}$ and $\rho_{s;\l}$, respectively.

\medskip

{\it The non-interacting case.} If $\l=0$, the model is integrable via the Kasteleyn's solution, and exhibits a transition between a rough phase (for values of $A,B,C$ satisfying the strict triangular inequality) and a frozen phase (for values of $A,B,C$ such that one of the three weights is larger than the sum of the other two). In the rough region, the free energy $F_0$ is a strictly convex function of $(\log B,\log C)$, while in the frozen region it is affine. From the microscopic point of view, the transition corresponds to the following: in the rough (or massless) region all three types of edges have a positive density
and correlations decay polynomially with the distance; in the frozen regions edges of one type (those with largest weight) have density 1, while the others have density 0.

We are interested in the rough-frozen transition.  Without loss of generality, we consider the case $A=1$ and $(B,C)\to(t,1-t)$ for some $t\in (0,1)$. The case where $t=0$ or $t=1$ is somewhat degenerate  and will not be considered here. 

\begin{figure}[ht]
    \centering
    \begin{minipage}{0.45\textwidth}
        \centering
        \includegraphics[width=0.8\textwidth]{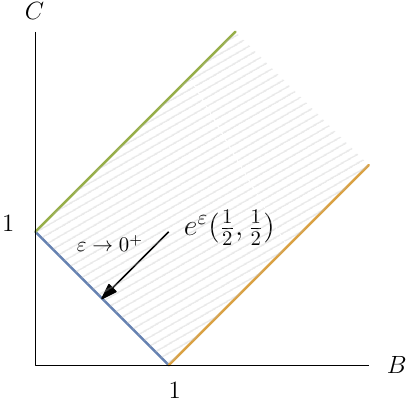}
    \end{minipage}\hfill
    \begin{minipage}{0.45\textwidth}
        \centering        \includegraphics[scale=0.3]{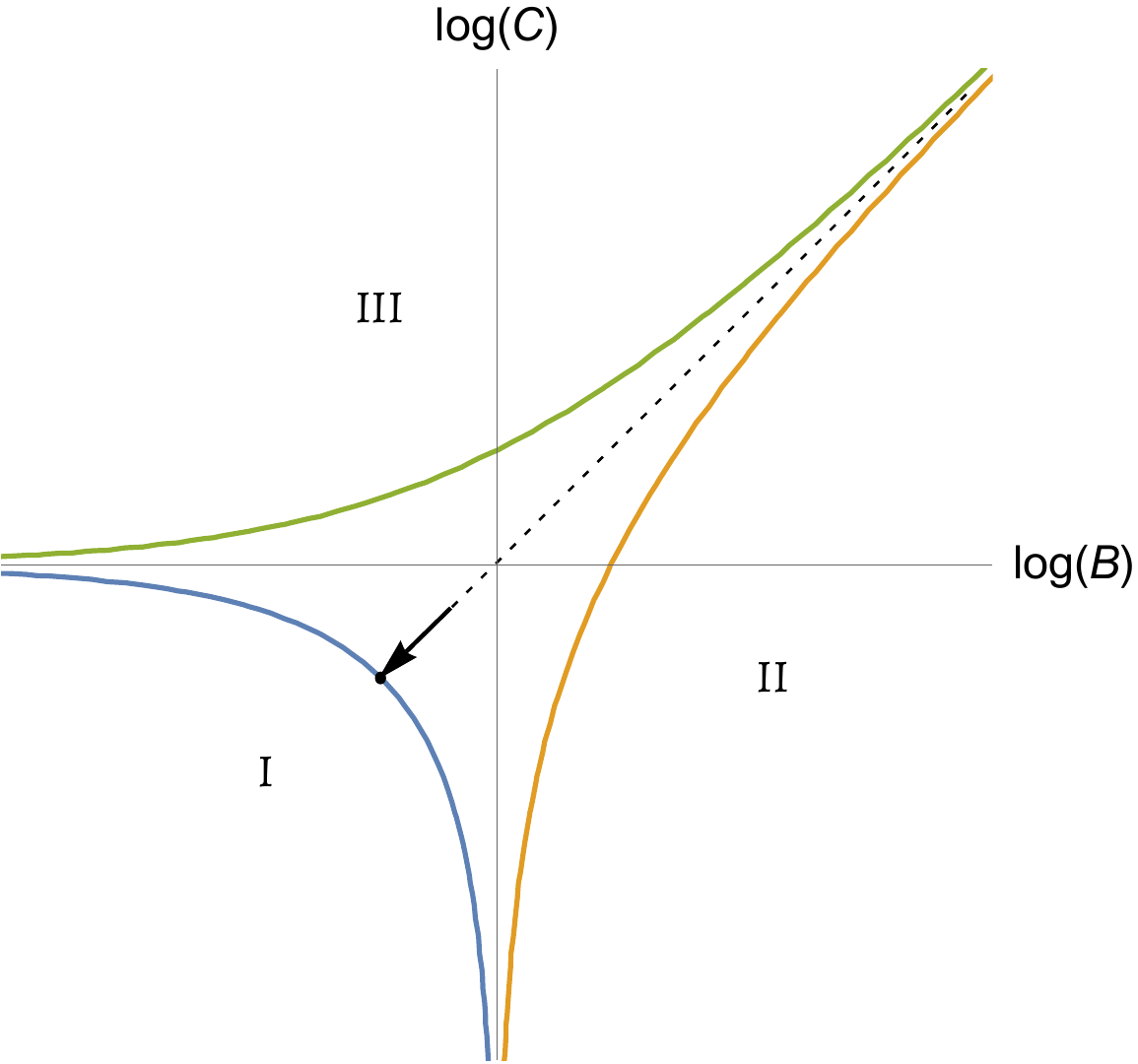} 
    \end{minipage}
    \caption{Left: in the shaded region the weights $A=1,B,C$ satisfy the strict triangular inequality. The blue segment corresponds to $(B,C)=(t,1-t)$, 
with $t\in[0,1]$. The midpoint can be approached for instance by choosing $(B,C)=(\frac{e^\varepsilon}{2},\frac{e^\varepsilon}{2})$, with $\varepsilon\to 0$. Right: the free energy $F_0$ is a strictly convex function of $\log B,\log C$ in the infinite deformed triangular region and affine in the three connected component of the complement: it is equal to zero in region I, to $\log B$ in region II and to $\log C$ in region III.
    \label{fig:03}}
\end{figure} 

In order to approach the point $(B,C)=(t,1-t)$, we parameterize the weights as \be \label{eq:parametr}(B,C)=(B(\varepsilon),C(\varepsilon))={e^\varepsilon}(t,1-t):\ee  $\varepsilon>0$ (resp. $\varepsilon<0$) corresponds to the rough (resp. frozen) region. The densities $\rho_{s;0}=\rho_{s;0}(B,C)$ can be explicitly computed via Kasteleyn theory, and the result is (see e.g. \cite{Kenyon_notes}) \be \rho_{s;0}(B,C)=s\int\limits_{[-\pi,\pi]^2} \frac{dk}{(2\pi)^2} \frac{e^{i k \cdot \hat e_s}}{1+Be^{ik_1}+Ce^{ik_2}}=\frac{\theta_s}{\pi},\label{eq:1.3bis}
\ee
where $k$ stands for $k=(k_1,k_2)$,  $\hat e_A=(0,0),\hat e_B=(1,0),\hat e_C=(0,1)$  and $\theta_s$ is the angle opposite to the side $s=A,B,C$ in a triangle of side-lengths $A,B,C$.   For the choice \eqref{eq:parametr}
with $\varepsilon>0$, 
 elementary trigonometry gives \begin{equation}
 \begin{split}
 &\theta_B=\sqrt{2\varepsilon\frac{t}{1-t}}+O(\varepsilon^{3/2}), \quad  \theta_C=\sqrt{
2\varepsilon\frac{1-t}t
}+O(\varepsilon^{3/2}). \label{eq:A.2}
\end{split}
\end{equation}
Using the fact that $\partial_\varepsilon F_0(B(\varepsilon),C(\varepsilon))=\rho_{B;0}(B(\varepsilon),C(\varepsilon))+\rho_{C;0}(B(\varepsilon),C(\varepsilon))$ and that $F_0(B(0),C(0))=\log A=0$, one derives by integration 
\begin{equation}\label{eq:2.1.1}
F_0\left(t e^\varepsilon, {(1-t)e^\varepsilon}\right)=\mathbbm{1}_{\varepsilon \geq 0} \left (\frac{2\sqrt{2}}{3\pi \sqrt{t(1-t)}}\varepsilon^{3/2}+O(\varepsilon^{5/2})\right )
\end{equation}
which is the Pokrovsky-Talapov (PT) law in the non-interacting, $\lambda=0$, case. 

{\it The interacting case.} 
In order to study the rough-frozen transition in the interacting, $\l\neq0$, model, 
we choose the edge weights in \eqref{defintermodel} to be $A=1$ and, for simplicity, \be (B,C)=e^\varepsilon\left(\frac12,\frac12\right),
\label{eq:parame}
\ee 
with $\varepsilon$ a parameter to be thought of as being small. As we shall see, 
the transition between the rough and frozen phases is located at $\varepsilon=\varepsilon_0(\lambda)$, with $\varepsilon_0(\lambda)$ an analytic function of $\lambda$, vanishing linearly as $\lambda\to 0$. By tuning $\varepsilon$ we can induce a transition
from the rough phase (corresponding to $\varepsilon>\varepsilon_0(\lambda)$) 
to the frozen one (corresponding to $\varepsilon<\varepsilon_0(\lambda)$). The special choice \eqref{eq:parame} of the weights guarantees that we are on a symmetry line of the phase diagram (symmetry under exchange of $B$ with $C$, see Fig. \ref{fig:03}), which comes with a few simplifications 
in the technical analysis; these are not crucial, but they allow us to relate quite directly the multiscale computations of the free energy, discussed below, with the analogous computations worked out in \cite{BM36} for a 
quantum spin chain in the vicinity of the metallic to insulating phase transition. 
The connection between these two seemingly unrelated problems is based on their common formulation in terms of Grassmann functional integrals, which is closer in the presence 
of appropriate symmetries, thus allowing us to import part of the results and methods of 
\cite{BM36} from theirs to our setting. A generalization of the method would allow us to 
investigate the rough-to-frozen transition at general points of the rough-frozen boundaries, see Section \ref{subsec:generalizations} for additional comments. 

From now on, we focus on the case that $\varepsilon$ is smaller than a positive small constant, since the complementary case has already been studied in \cite{GMT20}. By making the dependence upon $\varepsilon$ explicit, we denote by $F_{\varepsilon,\lambda,L,m}$ the free energy per unit volume in the first equation of \eqref{eq:deffandrho} computed at $B=C=e^{\varepsilon}/2$, and by 
$\rho_{\varepsilon,\lambda,L,m}$ the common dimer densities of the dimers of type $B$ or $C$. We let
\begin{equation}\label{eq:orderedlimit}
F_{\varepsilon,\lambda}:=\lim_{m\to0^+}\lim_{L\to\infty}F_{\varepsilon,\lambda,L,m},\qquad \rho_{\varepsilon,\lambda}:=\lim_{m\to0^+}\lim_{L\to\infty}\rho_{\varepsilon,\lambda,L,m}
\end{equation}
be the massless limits of the thermodynamic limits of the free energy per unit volume and of the dimer density of dimers of type $B$ or $C$, whenever the combined limits exist.
 Our main result is the following.
\begin{theorem}\label{th:5} 
There exists $\bar{\epsilon},\l_0>0$, and real-analytic functions $\varepsilon_0,A :(-\lambda_0,\lambda_0)\mapsto \mathbb R$, satisfying $\varepsilon_0(\l)=2\l+O(\l^2)$,  $A(\lambda)=1+O(\l^2)$, 
  such that the functions $(-\lambda_0,\lambda_0) \ni \l \mapsto F_{\epsilon+\varepsilon_0(\lambda),\lambda},$ and $(-\lambda_0,\lambda_0) \ni \l \mapsto\rho_{\eps+\varepsilon_0(\l),\l}$ are  
  infinitely differentiable in $\lambda$,  uniformly in $\eps\le \bar\eps$. Moreover, the functions $\epsilon \mapsto \rho_{\epsilon+\varepsilon_0(\lambda),\lambda}$ and 
  $\epsilon \mapsto   F_{\epsilon+\varepsilon_0(\lambda),\lambda}$ are $C^\infty$ in $\eps$ for $\eps\in (-\infty,0)\cup(0,\bar{\epsilon})$, and
\begin{eqnarray}
&& \rho_{\epsilon+\varepsilon_0(\lambda),\lambda}=\mathbbm{1}_{\{\epsilon \geq 0\}}\frac{\sqrt{2\eps}}{\pi}A(\lambda)(1+O(\epsilon^{\theta})), \label{eq:mainthrm}
\\
&& F_{\epsilon+\varepsilon_0(\lambda),\lambda}=\l+\mathbbm{1}_{\{\epsilon \geq 0\}} \frac{4\sqrt{2} \epsilon^{3/2}}{3\pi}A(\l)(1+O(\epsilon^{\theta})), \label{eq:PTbehavior}
\end{eqnarray}
where $\theta$ can be chosen in $(0,1/2)$ (the remainders $O(\epsilon^\theta)$ are not uniform in $\theta$ as $\theta\to (1/2)^-$). 
\end{theorem}

\begin{remark}
    A few comments are in order:
    \begin{enumerate}
        \item The exponent $3/2$ in \eqref{eq:PTbehavior} is the hallmark of the Pokrovsky-Talapov law. Its remarkable persistence in the presence of interactions provides a strong universality result: while interactions are known to modify the critical exponents of the model within the rough phase ($\epsilon>0$), the singular behavior of the free energy at the rough/frozen transition remains governed by the same exponent $3/2$ as in the non-interacting model. In the rough phase, universality is expected only in the weaker Kadanoff sense, namely via exact relations among interaction-dependent critical exponents. For the interacting dimer model, this phenomenon was established in \cite{GMT17,GMT20} for the critical exponent governing the oscillating part of the dimer-dimer correlation, which is related to the stiffness coefficient of the height field fluctuations at large distances and to the critical exponent of the electric correlators. The fact that the exponent $3/2$ is not renormalized can ultimately be traced back to the dependence of the rough-phase critical exponents on the dressed value of the interaction strength, which vanishes as $\epsilon\to0$, see also the discussion below \eqref{eq:1.9}.
        \item The theorem also identifies $\varepsilon_0(\lambda)$ as the rough/frozen transition point of the interacting model. Indeed, \eqref{eq:mainthrm} shows that the density $\rho_{\epsilon+\varepsilon_0(\lambda),\lambda}$ of $B/C$ dimers vanishes for $\epsilon<0$ and is strictly positive for $\epsilon>0$, thereby separating the frozen and rough regions of the phase diagram.
\item The amplitude $A(\lambda)$ is, a priori, a non-trivial real-analytic function of the interaction strength, admitting a perturbative expansion whose coefficients depend on the full renormalization-group flow. However, one of the outcomes of our analysis is an explicit representation of $A(\lambda)$ in terms of the dressed parameters of the theory at $\epsilon=0$
(such as the location of the poles of the interacting Green's function in momentum space, its residues at the poles, etc.), see \eqref{eq:2.81}. The fact that $A(\lambda)-1=O(\lambda^2)$ is a special feature of the interaction considered here and follows from the symmetry under the exchange $B\leftrightarrow C$. We refer to Sec. \ref{subsec:generalizations} for a discussion of the generic situation.
\item Eq. \eqref{eq:PTbehavior} follows by integrating \eqref{eq:mainthrm} with respect to $\epsilon$. For this reason, the derivation of \eqref{eq:mainthrm} will be the main goal of the paper. 
\item The smallness condition on $\lambda$ is uniform in the distance $\epsilon$ from the rough/frozen transition line. As explained in the introduction, this is a crucial novelty with respect to \cite{GMT20}, whose analysis allowed one to compute thermodynamic quantities in the thermodynamic limit only for $|\lambda|<\lambda_0(\epsilon)$, with $\lambda_0(\epsilon)\to0$ as $\epsilon\to0$.

\item We do not think that the stated regularity in $\lambda$ at $\eps>0$ is optimal: we expect that 
the limiting free energy and density are real-analytic in $\lambda$ uniformly in $\eps$, for all
$\eps\le \bar\eps$, but our proof falls short of proving this for $\eps>0$, due to an application of the implicit function theorem in the class of $C^\infty$, rather than analytic, functions.
Such an application is related to the shift at $\eps>0$ of 
the poles of the interacting Green 
function as compared to the non-interacting one 
(i.e., the shift of the `Fermi points' induced by 
the interaction of strength $\lambda$). In order to take this into account, using a QFT jargon, we 
first add a `counterterm', i.e., an auxiliary additional term to the model Hamiltonian, proportional to $\nu=\nu(\bar p_F,\lambda)$, in such a way that 
the Fermi points in the theory with counterterm $\nu$ and interaction strength $\lambda$ are located at $\pm \bar p_F$; next, we invert the counterterm and show that for the original Hamiltonian (the one with $\nu=0$ and interaction strength $\lambda$) the Fermi points are located at $\pm\bar p_F(\lambda)$, for a suitable function $\bar p_F(\lambda)$ obtained via an application of the implicit function theorem. For technical reasons, our counterterm $\nu(\bar p_F,\lambda)$ is $C^\infty$ in $\bar p_F$ and analytic in $\lambda$; after inversion, this implies that $\bar p_F(\lambda)$ is `just' $C^\infty$ in $\lambda$, rather than analytic. We think that this is just a by-product of a few sub-optimal estimates\footnote{We take the occasion to point out that a similar problem is present in \cite{GMT20}, where our proof falls short of proving the analyticity of 
the functions $\nu$, $\bar K_{\omega,r}$, $\bar H_{\omega,r}$, $\bar\alpha$, $\bar\beta$, $\bar p^\omega$ stated in \cite[Theorem 1]{GMT20};
strictly speaking, the proof in \cite{GMT20}
only provides $C^\infty$ regularity of those functions. We plan to come back to this issue 
in an erratum or auxiliary paper soon. This issue does not affect the results on \cite{GMT17}: due to the symmetries of the dimer model in that case, the Fermi points are not modified by the interaction and the functions $K$, $\tilde K$, $\kappa$ in \cite[Th. 1 and 2]{GMT17} are indeed analytic as stated.}
    \end{enumerate}
\end{remark}
\subsubsection{Origin of the Pokrovsky-Talapov law}

We would like to emphasize that the derivation of the PT law \eqref{eq:2.1.1} for the non-interacting model does not really require the exact form of the integrand in 
\eqref{eq:1.3bis}, but just the knowledge of the asymptotic form of its Fourier transform in the vicinity of its singularities. In fact, using the parametrization \eqref{eq:parametr},
and letting $k=(\pi,\pi)+k'$, we can rewrite $\rho_{s;0}$ in \eqref{eq:1.3bis} for $s\in\{B,C\}$ as follows: 
 \be \rho_{s;0}= \int\limits_{[-\pi,\pi]^2}\frac{dk'}{(2\pi)^2}f_s(\varepsilon,k'), \qquad f_s(\varepsilon,k):= -\tau_s \frac{e^{i k\cdot\hat e_s}}{e^{-\varepsilon}-te^{ik_1}-(1-t)e^{ik_2}},
\label{eq:1.3bistris}
\ee
where $\tau_B:=t$ and $\tau_C:=1-t$. Note that $\lim_{\varepsilon\to 0}\rho_{s;0}=0$, both for $s=B$ and $s=C$ (see \eqref{eq:6.12},\eqref{eq:A.16}): therefore $\rho_{s;0}=\lim_{\delta\to 0}\int_{[-\pi,\pi]^2}\big(f_s(\varepsilon,k)-f_s(\delta,k)\big)\tfrac{dk}{(2\pi)^2}$. We further rewrite this expression by adding and subtracting the integral of the function obtained by expanding 
$f_s(\varepsilon,k)$ in Taylor series around $\varepsilon=|k|=0$ and by retaining only the dominant terms, in the following way: the denominator in $f_s(\varepsilon,k)$ equals
$-\varepsilon-i(tk_1+(1-t)k_2)+\tfrac12[t(k_1)^2+(1-t)(k_2)^2]+O(\varepsilon^2)+O(|k|^3)$; by performing the change of variables $q_1=tk_1+(1-t)k_2$, $q_2=k_1-k_2$
(which has Jacobian determinant equal to 1, and will be denoted $q(k)$, with inverse $k(q)$), we get $$f_s(\varepsilon,k(q))=-\tau_s\frac{1+O(q)}{-\varepsilon-iq_1+t(1-t)q_2^2/2+O(\varepsilon^2,q_1^2,|q_2|^3)}.$$ We then 
rewrite, for $s\in\{B,C\}$: 
\begin{equation} \rho_{s;0}=-\tau_s\lim_{\delta\to 0}
\int_{\mathbb R^2} \Big(\frac1{-\varepsilon-iq_1+t(1-t)q_2^2/2}-\frac1{-\delta-iq_1+t(1-t)q_2^2/2}\Big)\frac{dq}{(2\pi)^2}+R_s(\varepsilon,t),\end{equation}
where $R_s$ is a remainder term, which can be (patiently but straightforwardly) bounded from above proportionally to $\varepsilon$, for $\varepsilon$ small and positive (details left to the reader). On the other hand, rescaling $q_1\to \varepsilon q_1$ and $q_2\to\sqrt{\varepsilon}q_2$, one can rewrite
\begin{equation}\begin{split}
& \lim_{\delta\to 0}
\int_{\mathbb R^2} \Big(\frac1{-\varepsilon-iq_1+t(1-t)q_2^2/2}-\frac1{-\delta-iq_1+t(1-t)q_2^2/2}\Big)\frac{dq}{(2\pi)^2}\\
&=\sqrt{\varepsilon}
\lim_{\delta'\to 0}
\int_{\mathbb R^2} \Big(\frac1{-1-iq_1+t(1-t)q_2^2/2}-\frac1{-\delta'-iq_1+t(1-t)q_2^2/2}\Big)\frac{dq}{(2\pi)^2}=-\frac{\sqrt{\varepsilon}}\pi\sqrt{\frac{2}{t(1-t)}},\label{eq:1.9}\end{split}\end{equation}
where the last identity follows from an explicit computation of the integral, via the Residue Theorem. 
Therefore, for $\varepsilon>0$, $\rho_{B;0}+\rho_{C;0}=\frac1\pi\sqrt{\frac{2\varepsilon}{t(1-t)}}+O(\varepsilon)$ and, by integrating in $\varepsilon$ (recall the equation for $F_0$ in the line following \eqref{eq:A.2}) we obtain the PT law \eqref{eq:2.1.1}, even if with a sub-optimal bound on the sub-leading term, $O(\varepsilon^2)$, rather than $O(\varepsilon^{5/2})$. In this sense, the PT law only depends upon the asymptotic form of the dimer density 
in momentum space (i.e., the integrand in \eqref{eq:1.3bis}), in the vicinity of the singularities.  In the interacting case, a multiscale analysis will allow us to compute the fine asymptotics of the Fourier transform of the dimer densities in the vicinity of the singularities, and this will allow us to prove the analogue of \eqref{eq:2.1.1} at $\lambda\neq0$, at the same precision in $\varepsilon$, 
notwithstanding the absence of an exact solution. This is the content of Proposition \ref{prop:7} below. We remark that the extraction of the fine asymptotics of the Fourier transform of the dimer densities requires to exhibit subtle and non-trivial cancellations in the infrared multiscale analysis of the dimer density. This is due to the fact that the dimer interaction is marginal, rather than irrelevant, in the Renormalization Group jargon, with asymptotically vanishing beta function (i.e., the effective dimer interaction does not scale to zero at large distances, rather it tends to a non-zero
value $\lambda_{-\infty}=\lambda_{-\infty}(\lambda,\epsilon)$, of the same order as the interaction strength $\lambda$ in \eqref{defintermodel} as $\lambda\to 0$). If the effective dimer interaction were irrelevant, the invariance of the critical exponent would be a simpler consequence of the multiscale analysis; instead, in our case, it follows from the fact that the `infrared fixed point' value $\lambda_{-\infty}(\lambda,\epsilon)$, which measures the effective interaction at large distances, tends to zero as the distance $\epsilon$ from the rough-frozen boundary tends to zero, in a way that we can quantitatively control as a function of $\epsilon$. 

\subsubsection{Heuristic connection with KPZ universality}\label{sec:kpz}

\begin{figure}[ht]
    \centering
    \begin{minipage}{0.42\textwidth}
         \centering
    \includegraphics[width=0.9\textwidth]{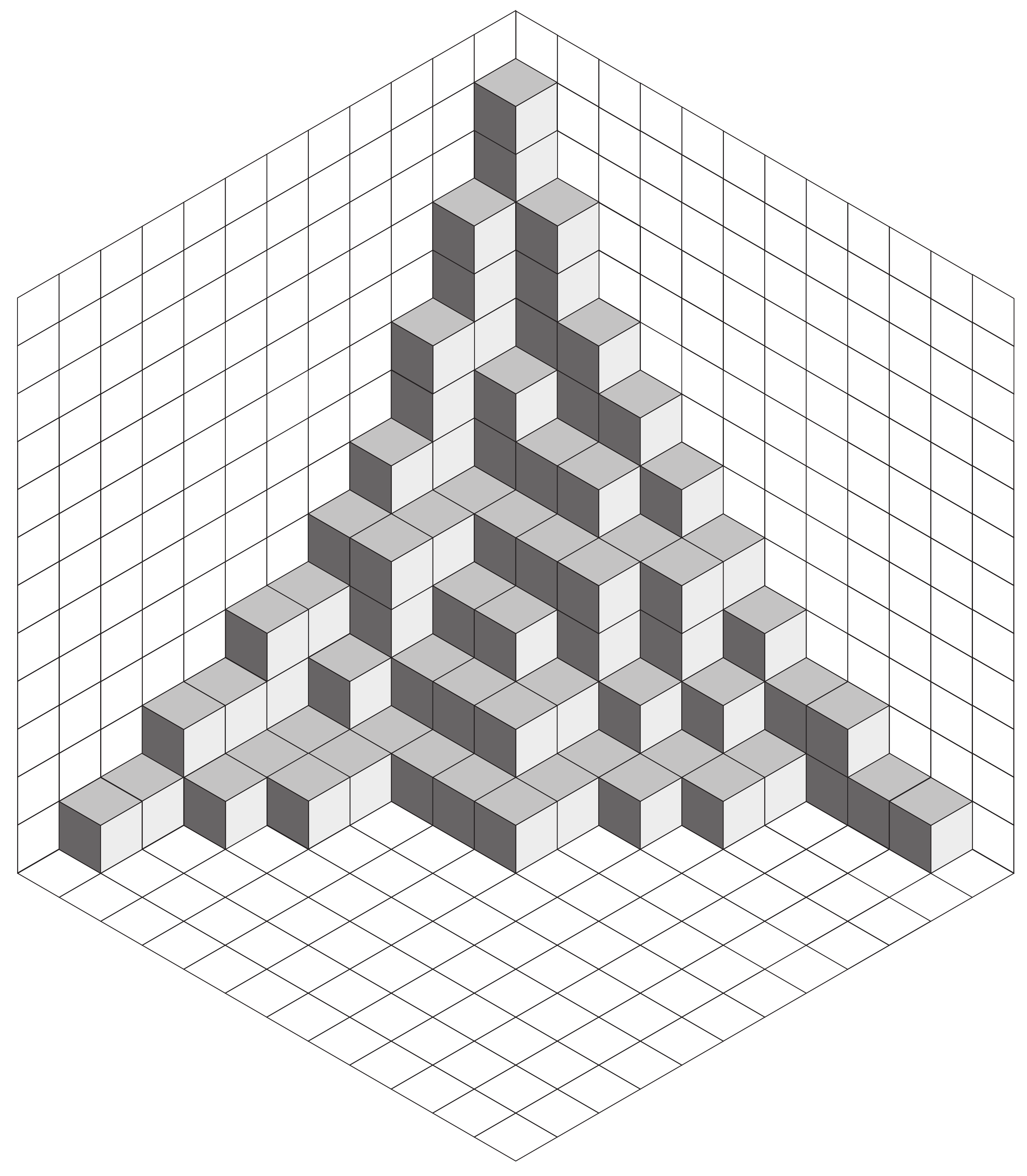}
    \end{minipage}\hfill
    \begin{minipage}{0.45\textwidth}
        \centering
        \includegraphics[width=0.87\textwidth]{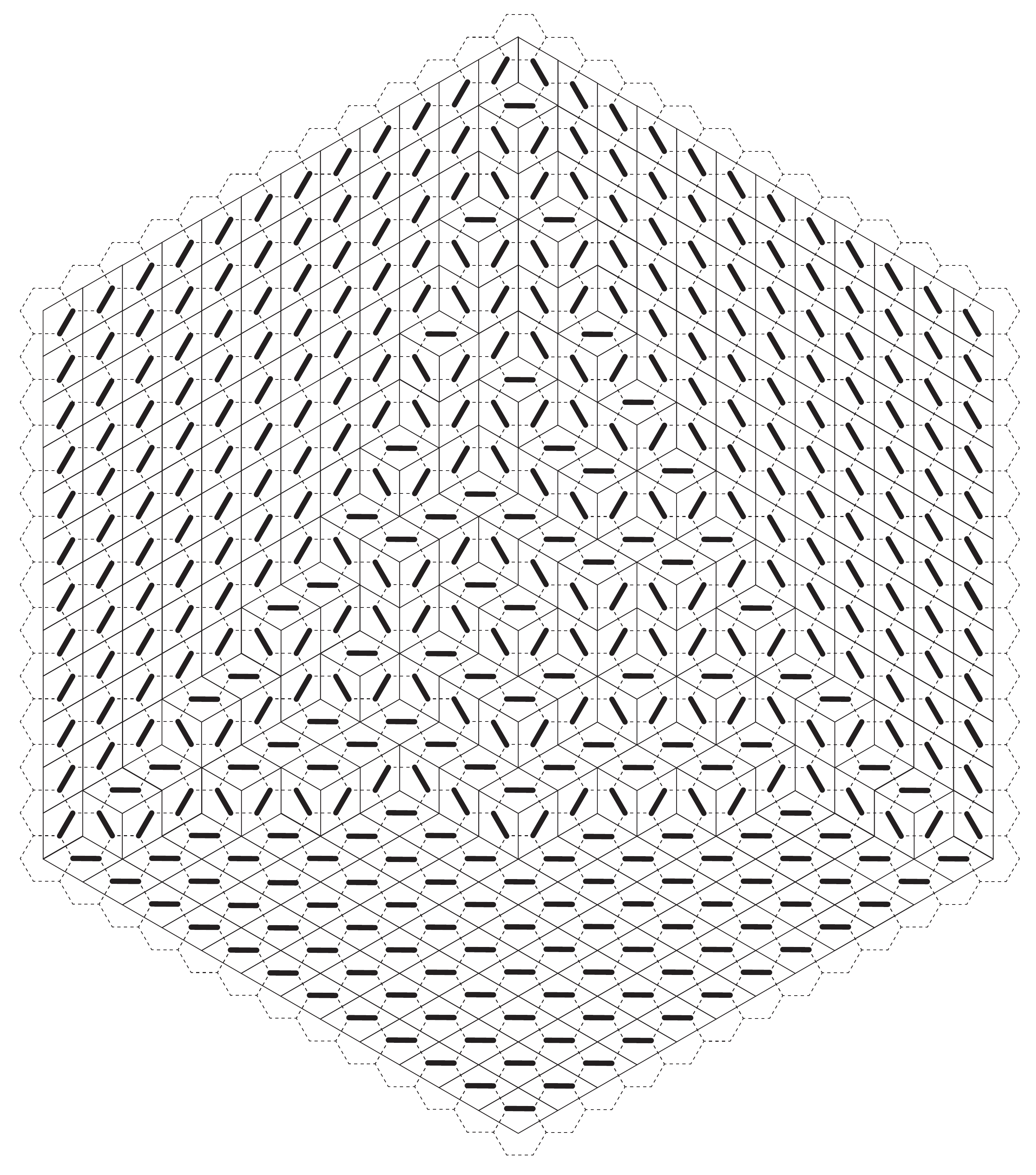}
    \end{minipage}
    \caption{Left: a discrete monotone surface $S$ or equivalently a planar rhombus tiling. The volume $\mathrm{Vol}(S)$ of the configuration is the number of cubic boxes, including those hidden from view. Right: the dimer configuration $M=M(S)$ corresponding to $S$, under the `usual' bijection between monotone surfaces and close-packed dimer configurations, see \cite{Kenyon_notes}. For $M=M(S)$, we shall write $\mathrm{Vol}(M):=\mathrm{Vol}(S)$.}
    \label{fig:motivationkpz}
\end{figure}

The Pokrovsky-Talapov law \cite{PT} heuristically implies $1/3$ fluctuations of the height function’s level lines near frozen boundaries, a hallmark of KPZ-class line ensembles \cite{Corwin}. Specifically, near frozen regions of the limit shape, level lines are expected to fluctuate on a scale $L^{1/3}$, where $L$ is the local radius of curvature, and converge under rescaling (by $L^{-1/3}$ transversally and $L^{-2/3}$ longitudinally) to a universal limit. Such frozen regions arise, e.g., in polygonal domains on the hexagonal lattice as $L\to\infty$, or on the infinite graph with a volume-tilted measure fixing the volume under the discrete surface to be proportional to $L^3$ (see Fig. \ref{fig:motivationkpz}). These predictions have been proven in several integrable cases \cite{AggarwalH,Johansson}, while the connection is to date unproven in the interacting, non-integrable, case. 

In order to illustrate the connection between the $3/2$ free energy exponent and $1/3$ fluctuation exponent, let us first discuss the non-interacting case: consider the infinite hexagonal graph with a dimer configuration $M$ sampled proportionally to $\exp(-\alpha\, \mathrm{Vol}(M)/L)$, where $\alpha>0$ and $L>0$ (for the definition of Vol$(M)$, see caption of Fig. \ref{fig:motivationkpz}). Rescaling the graph to mesh size $1/L$ and letting $L\to\infty$, the configuration concentrates on a deterministic limit shape \cite{Cerf-Kenyon}, whose height function\footnote{We recall that, microscopically, 
the height $h(f)$ at a face $f$ associated to a dimer configuration $M$ can be defined by fixing zero height at a reference face, and by setting $h(f')-h(f)=\sum_{e\in C_{f\to f'}}\sigma_e(\mathds 1_e(M)-\mathds 1_e(M_0))$, where $M_0$ is a reference dimer configuration, say the one consisting of all horizontal dimers while $\sigma_e=+1$ ($\sigma_e=-1$) if $e$ is crossed with the white site on the right (left). By using this definition, and by choosing 
the reference face $f_0$ at which $h(f_0)=0$ to be the one at the `center of the hexagon' in the right part of Fig. \ref{fig:motivationkpz}, we see that the height at faces in the bottom frozen region is identically zero, while the one at faces $f=(x,y)$ in the right (resp. left) upper frozen region is equal to $y$ (resp. $-x$), where $(x,y)$ are the coordinates of the center
of the face with respect to the origin $f_0$, in the basis $\vec e_1,\vec e_2$ of Fig. \ref{fig:toroLesagono}.\label{footnote.1}}
$H$ is the free energy $F_0(B,C)$, up to a coordinate change.
To see this, one can use the fact that the height $H$ minimizes the functional (see, e.g., \cite{CKP} in the case of tilings of a bounded domain and $\alpha=0$)
$$-\int [\sigma(\nabla H(x,y))+\alpha H(x,y)]dx dy,$$ 
where the convex function $\sigma$ is the surface tension, that is the Legendre transform of the convex function $(s,t)\mapsto G_0(s,t):=F_0(e^t,e^{-s})$. It is immediate to check that the Euler-Lagrange equation corresponding to this functional is the 
PDE
\begin{equation}
\mathrm{div}(\nabla \sigma \circ \nabla H) = \alpha \label{Legendre}.
\end{equation}
Up to a linear scaling of coordinates we can assume that $\alpha=2$, which is convenient for the rest of this argument.
By definition of  Legendre transform, $G_0$  satisfies $\nabla \sigma \circ \nabla G_0 = \mathrm{Id}$ (with $\mathrm{Id}$ the identity map on $\mathbb R^2$), so $G_0$ solves \eqref{Legendre} (with $\alpha=2$). Moreover, recalling the boundary conditions satisfied by $F_0(B,C)$, equal respectively to $0, \log B, \log C$ in regions I,II,III (see Fig. \ref{fig:03}), 
it follows that $G_0(x,y)\equiv F_0(e^y,e^{-x})$ satisfies the same boundary conditions as $H(x,y)$ in the three infinite frozen regions occupying the bottom, right-upper and left-upper 
parts of the plane (recall that $H(x,y)$ is the rescaling of the microscopic height function defined in footnote \ref{footnote.1}, whose boundary conditions are described there). Therefore, $G_0=H$, which is the announced relation between macroscopic shape and free energy.

In particular, $H$ is strictly convex in the rough phase, i.e., in the region $R:=\{(x,y)\in\mathbb R^2: \text{$1, e^{-x}, e^y$ are the sides of a non-degenerate triangle}\}$, 
and affine in the frozen regions, i.e., in the three connected components of $R^c$. The connection between $H$ and $G_0(x,y)\equiv F_0(e^y,e^{-x})$ that we just described, in combination with the PT law \eqref{eq:2.1.1}, implies that, if we approach the point $(-\log(1-t), \log t)$ on the bottom frozen boundary (with $t\in(0,1)$) from the interior of the rough phase, along the curve $(-\log(1-t)-\varepsilon,\log t +\varepsilon)$ with $\varepsilon\in [0,\varepsilon_0)$, the height $H$ at a distance of order $\varepsilon$ if of the order $\varepsilon^{\gamma}$, with $\gamma=3/2$. This implies that, before rescaling, the typical microscopic height function at a distance $L\varepsilon$ from the frozen boundary 
is of the order $L\varepsilon^\gamma$, 
implying that in a region of width $L\varepsilon$ there are $O(L\varepsilon^\gamma)$ level lines. Taking heuristically $\varepsilon \sim L^{-1/\gamma}$ suggests that the 
$n^{th}$ level line (with $n$ fixed in the limit $L\to\infty$ and/or $\varepsilon\to 0$) typically has fluctuations of the order $L\varepsilon \sim L^{\nu}$ with $\nu = 1 - 1/\gamma = 1/3$. 

This heuristics, spelled out above in the case $\lambda=0$, is expected to hold beyond the integrable case: in fact, while the exact limit shape changes in general 
in the presence of interactions, the link between surface tension, free energy, and height is purely thermodynamical, and is expected to hold quite generally (for results in this direction, see e.g. \cite{LT24, MT20,Sh05}), so that the scaling relation $\nu = 1 - 1/\gamma$ should be valid in the non-integrable case, as well. 
Therefore, our proof of the PT law (that is, of the fact that $\gamma=3/2$), suggests that $\nu=1/3$, i.e., that the KPZ scaling is the correct one to use in order to obtain a non-trivial scaling limit for the level lines fluctuations in the interacting, non-integrable, case, as well.

\subsubsection{Connection with vertex models: the PT law from the Bethe ansatz}\label{subsec:vertexmodels}
As mentioned above, replacing the interaction $V$ in \eqref{defintermodel} by $W=\sum_{f\in G^*_L} (\mathds 1_{f;2}\mathds 1_{f;6}+\mathds 1_{f;3}\mathds 1_{f;5})$ leads to an integrable model, equivalent to the 5V model. The equivalence can be proved via a one-to-one mapping of the dimer coverings of the hexagonal lattice into 5V configurations of the square lattice, first discussed by F. Wu in \cite{Wu68}. Note that this re-interpretation of an interacting dimer model as a vertex model solvable via Bethe ansatz is not the only possible one, see e.g. \cite[Section 2.3]{GMT20} for the review of a different correspondence between an interacting dimer model on the square lattice and the six vertex (6V) model on the square lattice. 

\begin{figure}[ht]
    \centering
    \includegraphics[width=0.7\textwidth]{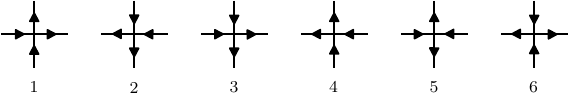}
    \caption{The six allowed vertex types in the 6V model.}
    \label{fig:6V}
\end{figure}

In order to make the correspondence with the vertex model explicit, let us recall that a 6V configuration $\o$ on the square lattice is an assignment of arrow orientations to the edges of the square lattice satisfying at each vertex the ice rule, that is two arrows are entering the vertex and two are exiting, cf. Figure \ref{fig:6V}. We denote by $\Omega^{6V}_L$ the set of allowed 6V configurations on a square lattice of side $L$ with periodic boundary conditions. 
The partition function reads:
\[
Z^{6V}_L(\un a) =\sum_{\omega \in \Omega^{6V}_L} \prod_{i=1}^6 a_i^{n_i(\omega)},
\]
where $\un a:=(a_1,a_2,a_3,a_4,a_5,a_6)$; here $a_i$ are the energy weights associated with vertices of type $i$, with $i=1,\dots,6$ as in Fig \ref{fig:6V}, 
and $n_i(\omega)$ is the number of vertices of type $i$. 

Setting $a_1=0$, we get the partition function of the 5V model: $Z^{5V}_L(a_2,a_3,a_4,a_5,a_6):=Z^{6V}_L(\un a)\big|_{a_1=0}$. Wu's mapping  \cite{Wu68} implies that (see \cite[Section 2]{HWKK95})
\begin{equation}
\sum_{M \in \Omega_L}A^{N_A} B^{N_B} C^{N_C}e^{\lambda W}=Z^{5V}_L(0,A,B,C,e^\lambda\sqrt{BC},e^\lambda\sqrt{BC}), \label{eq:id_dimer5v}
\end{equation}
where the notations in the left hand side are the same as in \eqref{defintermodel}; in fact, the sum in the left hand side defines the partition 
function of an interacting dimer model, which is essentially the same as $Z_{\lambda,L}$ in \eqref{defintermodel}, modulo the replacement of $V$ by $W$. 

The Bethe Ansatz solution of the 5V model provides an explicit formula for the free energy
$f_\lambda(B,C)=\lim_{L\to\infty}L^{-2}\log 
Z^{5V}_L(0,1,C,B,e^\lambda\sqrt{BC},e^\lambda\sqrt{BC})$, from which the phase boundaries can be determined exactly, see \cite[Fig.9]{HWKK95}, and the PT law derived explicitly, see \cite[Eqs.(8.18) and (8.20)]{HWKK95}. See also \cite{dGKW21} for the connection between the free energy of the 5V model and the limit shape, and for an explicit solution of the variational equation for the limit shape in domains of specific form and/or with specific boundary conditions. 

Our main result, summarized in Theorem \ref{th:5}, can be interpreted as an extension of the proof of the PT law beyond the integrable case. Notably, the methods behind the proof of our main result are completely different from those underlying the computation of \eqref{eq:id_dimer5v} via Bethe ansatz, and are robust under additional weak, local, translationally invariant perturbations: as already mentioned, our choice of the interaction $V$ in \eqref{defintermodel}
is made just for simplicity and definiteness: it is possibly the simplest one leading to a non-integrable model, and our analysis can be straightforwardly  extended to generic weak finite range interactions, as discussed in detail in \cite{GMT20}, see also next subsection. Note that while our approach is robust with respect to 
the specific choice of the interaction, it requires the strength $\lambda$ to be small enough, contrary to the Bethe ansatz solution, which suggests that the PT behavior should remain valid even beyond the weak coupling regime, up to a well defined critical line separating a rough from a frozen phase. 

\subsubsection{Generalizations and perspectives}\label{subsec:generalizations}
As mentioned above, the Pokrovsky--Talapov law, characterized by the critical exponent $3/2$ in \eqref{eq:PTbehavior}, is expected to remain valid for a wide class of weak and finite-range interactions, independently of their specific form, at least in the same generality as \cite[Sec. 2.3]{GMT20}. Moreover, 
it should remain valid at generic points along the rough-frozen boundary, as well as at generic points of the boundary separating the rough from the gaseous phase, in settings where both phases are present; similarly, the specific direction along which we approach the boundary point should be unimportant for the 
validity of the PT law, as long as it is transversal to the boundary. Using the Grassmann representation described in next section, see also \cite[Section 3]{GMT20}, the analysis of the critical behavior of the free energy in the vicinity of generic points of the rough-frozen boundary, or of the rough-gaseous boundary, in the presence of generic finite range, translationally invariant, dimer interactions, can be reduced to the analysis of a Grassmann functional integral similar to \eqref{eq:Zinteract} below, displaying, however, fewer symmetries than those spelled in subsection \ref{ssec:symmetri} below. 
The lack of certain discrete symmetries will moderately complicate the infrared multiscale analysis of the Grassmann integral, in particular it 
will induce the presence of additional running coupling constants (in the Renormalization Group jargon) such as those in \cite{GMT20}, see e.g. \cite[Eq.(6.49)]{GMT20}. We expect that, by combining the method of proof discussed in this paper with that in \cite{GMT20}, we can prove that the free energy 
of a general interacting dimer model with interaction strength $\lambda$, at a distance $\epsilon$ from a boundary point $Q$ along a prescribed curve transversal to the boundary (with the convention that $\eps>0$ corresponds to the rough phase, while $\eps<0$ to the frozen or gaseous phase), will be of the form: 
\begin{equation}
    F_{\l}(\epsilon)=a(\l,Q)+\eps \, b(\l,Q)+\mathbbm{1}_{\{\eps \geq 0\}}c(\l,Q)\eps^{3/2}(1+O(\eps^\theta)),
    \label{eq:1.4.14}
\end{equation}
where $a,b,c$ are analytic functions of the interaction parameter $\l$, depending both on the boundary point $Q$ and on the specific dimer interaction; generically, both the coefficients of order 0 and 1 in their Taylor expansions in $\lambda$ are different from zero. 
Moreover, the critical line separating the rough from the frozen or from the gaseous region admits itself a parametrization that is analytic in $\lambda$. 
Therefore, the interaction is expected to analytically deform the reference affine function $a+\eps b$, as well as the pre-factor in front of $\eps^{3/2}$; however, the critical exponent $3/2$ is expected to be unchanged, in line with the main result of this work and with the Bethe ansatz computation mentioned in the previous subsection. 

While we expect that the proof of \eqref{eq:1.4.14} is feasible on the basis of straightforward generalizations of currently available methods, a proof 
of the convergence of the fluctuations of the level lines to a non-trivial limiting stochastic process under KPZ scaling (see subsection \ref{sec:kpz}) seems to 
be harder, and beyond our current technical abilities. A related problem is to understand whether there is a way to rescale the microscopic height function 
in a way inspired by the KPZ scaling so that the corresponding characteristic function converges to a non-trivial limit as the scaling parameter tends to zero, in 
a way analogous to the convergence of the height function to the Gaussian Free Field studied and proved in \cite[Theorem 3]{GMT17}.

\subsection{Organization of the work}
\label{sec:organization} 
The proof of Theorem \ref{th:5} is organized as follows.

We begin by rewriting the partition function and the generating function of dimer correlations as Grassmann integrals; see Section \ref{sec:grassmann}. This representation provides the natural fermionic framework in which to set up the multiscale analysis. In the same section, we also collect the symmetry properties of the bare Grassmann theory; see Subsection \ref{ssec:symmetri}.

Section \ref{sec:multiscale} introduces the multiscale analysis. For $\eps \neq 0$, the free energy and the density can be computed at $m=0$ in terms of a single Grassmann integral, see \eqref{eq:mlimit}, which is the object of the multiscale analysis.  After introducing Grassmann variables adapted to the analysis of the infrared problem and the counterterm fixing the location of the interacting critical point, we first integrate out the ultraviolet degrees of freedom; see Subsection \ref{ssec:uv}. The remaining infrared integration is then performed in two different regimes, in the same spirit as \cite{BM36}. The first regime, described in Subsection \ref{sec:firstregime}, corresponds to the anisotropic quadratic scaling close to the merging point of the two singularities. The second regime, described in Subsection \ref{sec:secondregime}, appears for $\eps>0$, after reaching a scale allowing us to resolve two distinct singularities, and is characterized by an effective linear dispersion relation.

Section \ref{sec:2.3} collects the output of the Renormalization Group analysis. In particular, Proposition \ref{prop:7} provides a convergent expansion for the dimer density $\rho_{\eps+\varepsilon_0(\lambda),\lambda}$, uniformly in the distance from the rough-frozen transition. This expansion is then used in Subsection \ref{sec:2.3.1} to extract the sharp $\eps$-asymptotics in \eqref{eq:mainthrm}. The Pokrovsky–Talapov law for the free energy, \eqref{eq:PTbehavior}, follows by integration in $\eps$. Finally, Subsection \ref{sec:2.3.2} contains the proof of \eqref{eq:mlimit}, based on Proposition \ref{prop:rgmassa}.

The appendices collect additional technical aspects of the proof. Appendix \ref{app:symmetries} proves the symmetry properties of the effective theory at all steps of the multiscale flow. Appendix \ref{app:A} recalls the Gallavotti–Nicolò tree expansion and uses it to control the flow in the first regime, to determine the counterterm $\varepsilon_0(\lambda)$, and to prove Proposition \ref{prop:7}. Additionally: in Appendix \ref{app:scalaheps} we discuss the estimates on the running constants at the transition scale $h_\eps$; in Appendix  \ref{app:regularitydensityeps} we prove the statement of Theorem \ref{th:5} on the regularity in $\epsilon$ of the free energy.  Appendix \ref{app:firstorder} contains the first-order computations leading to $\varepsilon_0(\lambda)=2\lambda+O(\lambda^2)$ and $A(\lambda)=1+O(\lambda^2)$. Appendix \ref{app:proofprop7} contains the proof of Proposition \ref{prop:rgmassa}. Finally, Appendix \ref{A:6.3} collects further technical details on the frozen phase of the non-interacting model and on the flow of the effective parameters in the second regime. 

\section{Proof of Theorem \ref{th:5}}

\subsection{Grassmann representation}\label{sec:grassmann} 
We start by rewriting the free energy and the density in   \eqref{eq:deffandrho} and, more generally, the `truncated', or `connected', dimer correlations 
\begin{equation} \label{eq:dimercorr}\mathbb E_{\lambda,L,m}(\mathds 1_{e_1};\cdots;\mathds 1_{e_n}):=\frac{\partial^n}{\partial A_{e_1}\cdots\partial A_{e_n}} \mathcal{W}_{\varepsilon,\l,L,m}(\mathcal A)\Big|_{A_e\equiv 0},\end{equation}
where $e_1,\ldots,e_n\in E_{L,m}$, with $E_{L,m}$ the edge set of $G_{L,m}$, in terms of a Grassmann integral. In \eqref{eq:dimercorr}, 
$\mathbbm{1}_e$ is the indicator function of the event that the edge $e$ belongs to the dimer configuration, i.e., if $M\in\Omega_{L,m}$, $\mathds 1_e(M)=1$ if $e\in M$, and $\mathds 1_e(M)=0$ otherwise. Moreover, letting $\mathcal{A}=\{A_e\}_{e \in E_{L,m}}$, with $A_e\in\mathbb R$, 
\begin{equation}
    \mathcal{W}_{\varepsilon,\l,L,m}(\mathcal A):=\log\sum_{M \in \Omega_{L,m}} w_{\varepsilon,\l,m}(M)e^{\sum_{e \in E_{L,m}} \mathbbm{1}_e A_e},   \quad w_{\varepsilon,\l,m}(M):=\Big(\frac{e^{\varepsilon}}{2}\Big)^{N_B+N_C}m^{N_m}e^{\lambda V},
\label{eq:3.2.1}
\end{equation}
where $w_{\varepsilon,\l,m}(M)$ is nothing but the rewriting of the probability weight in \eqref{defintermodel} with $A=1$ and $B,C$ as in \eqref{eq:parame}. 

In analogy with \cite{GMT17,GMT20}, the generating function in \eqref{eq:3.2.1} can be written as a perturbation of a Gaussian Grassmann integral, defined via the following conventions.
 
 We color the vertices of $G_L$ black or white, as in Fig. \ref{fig:toroLesagono}, and we assign the same coordinate $x=(x_1,x_2)\in\Lambda:=\{1,\dots,L\}^2$ to each pair of white/black endpoints of the same horizontal edge\footnote{Given two black vertices of coordinates $x,y\in\Lambda$, $y_1-x_1$ and $y_2-x_2$ denote the components of the vector joining the first to the second vertex in the basis $\vec e_1,\vec e_2$, to be identified modulo $L$.}; we denote by $\mathrm{b}_x,\mathrm{w}_x$ the black and the white vertex of coordinate $x \in \Lambda,$ respectively. Letting $\hat e_A=(0,0),\hat e_B=(1,0), \hat e_C=(0,1)$ and $d:=\hat e_C-\hat e_B=(-1,1)$, the edges of type $s \in \{A,B,C\}$ are $\{\mathrm{w}_x,\mathrm b_{x+\hat e_s}\}$, and those of type $m$ are $\{\mathrm b_x, \mathrm b_{x+d}\}$ and $\{\mathrm w_x, \mathrm w_{x+d}\}$, $x \in \Lambda$.
 \par\emph{Kasteleyn orientation.} We orient each edge of $G_L$ from its white to its black endpoint. This orientation extends uniquely to a Kasteleyn orientation of $G_{L,m}$, i.e. such that every face of $G_{L,m}$ has an odd number of edges oriented clockwise.
 \par \emph{Kastelyn matrices.} We assign a weight $K_e$ to each edge $e \in G_{L,m}$, such that $K_e:=K_s$ if $e$ is of type $s \in \{A,B,C\}$, with $K_A:=1,K_B=K_C=e^{\varepsilon}/2$, and $K_e:=m$ if $e$ is of type $m$. For $\vartheta=(\vartheta_1,\vartheta_2) \in \{0,1\}^2$, the Kasteleyn matrix $\mathcal{K}_{\vartheta,m}$ is the antisymmetric matrix, indexed by the vertices of $G_{L,m}$, whose only non vanishing elements are those indexed by an ordered pair $(u,v)$ 
 whose vertices are the endpoints of an edge in $G_{L,m}$. If $(u,v)$
corresponds to an edge $e=\{u,v\}\in G_{L,m}$ oriented in agreement with the Kasteleyn orientation defined above, then 
  \begin{equation}
     \left (\mathcal{K}_{\vartheta,m}\right )_{u,v}=(-1)^{\vartheta_1 n_1(e)+\vartheta_2 n_2(e)}K_e,
 \end{equation}
where $n_i(e)=1$ if $e$ joins a vertex of coordinate $x$ with $x_i=L$ to a vertex of coordinate $x'$ with $x_i'=1$ and $n_i(e)=0$ otherwise.\par \emph{Grassmann integration.} We assign Grassmann variables $\psi_x^-,\psi^+_x$ to $\mathrm w_x, \mathrm b_x$, respectively; when needed we also use the alternative notation 
$\psi_u$ for the Grassmann variable associated with the vertex $u$. The Grassmann integration $\int D\psi$ is the linear functional on the Grassmann algebra generated by $\{\psi^+_x,\psi^-_x\}_{x \in \Lambda}$ vanishing on the monomials of degree smaller than $2L^2$ and such that 
$\int D\psi \prod_{x \in \Lambda} \psi^-_x \psi^+_x=1.$ With this normalization we have $\int D\psi\, e^{\frac{1}{2}\sum_{u,v} \psi_u ( \mathcal{K}_{\vartheta,m})_{u,v} \psi_v}={\rm Pf} \mathcal{K}_{\vartheta,m}$ where the sums over $u$ and $v$ at exponent run over the vertex set of $G_L$, and ${\rm Pf}$ is the Pfaffian \cite[(2.1)-(2.3)]{GMT17} (in order to fix the sign of the Pfaffian, we assume that the vertices of $G_L$ are ordered as $\mathrm w_{x_1}, \mathrm b_{x_1}, \dots,  \mathrm w_{x_{L^2}}, \mathrm b_{x_{L^2}}$, for an arbitrary enumeration of $\Lambda$).

Given these conventions and proceeding as in \cite[Proposition 4]{GMT17}, we find:
\begin{equation}
\label{eq:Zinteract}
\begin{split}
e^{\mathcal{W}_{\varepsilon,\l,L,m}(\AAA)}=\sum_{\vartheta \in \{0,1\}^2} \frac{c_\vartheta}{2}\int D\psi\, e^{S_{\vartheta,m}(\psi)+ V_\vartheta(\psi,\AAA)}, 
\end{split}
\end{equation}
where  $(c(0,0),c(1,0),c(0,1),c(1,1))=(-1,1,1,1)$, and
\begin{equation}\label{eq:Scoordinates}\begin{split}
    & S_{\vartheta,m}(\psi)=\frac12\sum_{u,v} \psi_u (\mathcal{K}_{\vartheta,m})_{u,v}\psi_v\equiv-\sum_{e\in E_{L,m}}E_{\vartheta,e}(\psi) \\ &  
     V_\vartheta(\psi,\AAA)=\sum_{e \in E_{L,m}}(1-e^{A_e})E_{\vartheta,e}(\psi)+\sum_{\gamma} c(\gamma) \prod_{e \in \gamma}  e^{A_e}E_{\vartheta,e}(\psi),
\end{split}
\end{equation} 
where: if $e$ is an edge of type $s\in\{A,B,C\}$ with 
white endpoint of coordinate $x\in\Lambda$,
then 
$E_{\vartheta,e}(\psi)=K_s\psi_{x+\hat e_s}^+\psi_x^-$; 
if $e=\{\mathrm b_x,\mathrm b_{x+d}\}$, then $E_{\vartheta,e}(\psi)=m\psi^+_{x+d}\psi^+_x$; if $e=\{\mathrm w_x,\mathrm w_{x+d}\}$, then $E_{\vartheta,e}(\psi)=m\psi^-_{x}\psi^-_{x+d}$, with the understanding that 
\begin{equation} \label{eq:boundarycond}
\psi_{(L+1,x_2)}^\pm:=(-1)^{\vartheta_1}\psi_{(1,x_2)}^\pm, \qquad \psi_{(x_1,L+1)}^\pm:=(-1)^{\vartheta_2}\psi_{(x_1,1)}^\pm\end{equation}
(note that the dependence of $E_{\vartheta,e}(\psi)$ upon $\vartheta$ is hidden in the boundary conditions \eqref{eq:boundarycond}).
Moreover, the sum over $\gamma$ in the r.h.s. of the second definition in \eqref{eq:Scoordinates} runs over collections $\gamma \subset E_L$, $|\g| \geq 2$ of parallel edges at distance $1$, that is,  $\g=\{e_1,\dots,e_{|\g|}\}$ with $e_i$ and $e_{i+1}$ parallel and belonging to the same face of $G_L$, for $i=1,\cdots,|\g|-1$; finally,   $c(\gamma)=(e^\lambda-1)^{|\gamma|-1} (-1)^{|\gamma|}$.

We can also represent the Grassmann variables in 
Fourier space, as follows. For any choice of $\vartheta\in\{0,1\}^2$, we rewrite the Grassmann variables satisfying the
$\vartheta$-boundary conditions in \eqref{eq:boundarycond} as:
\begin{equation}\label{fourierfieldsk} \psi_x^\pm=L^{-2}\sum_{k \in \mathcal{D}_\vartheta} e^{\pm ikx}\hat \psi_k^\pm \quad \Leftrightarrow \quad \hat\psi^\pm_k=\sum_{x\in\Lambda}e^{\mp ikx}\psi_x^\pm
\end{equation}
where $\mathcal{D}_\vartheta:=\left \{(k_1,k_2): k_i=\frac{2\pi}{L}\left (n_i+\frac{\vartheta_i}{2}\right ), \, n_i=0,\dots,L-1\right \}$. In particular, we can rewrite: 
\begin{equation}
\label{Steta}
\begin{split}
    & S_{\vartheta,m}(\psi)=-L^{-2}\sum_{k \in \mathcal{D}_\vartheta}\left ( \mu(k) \hat{\psi}^+_k\hat{\psi}^-_k+im \sin(k \cdot d)(\hat \psi^+_k \hat \psi^+_{-k}+\hat \psi^-_{k}\hat \psi^-_{-k})\right ) 
\end{split}
\end{equation} 
where the function $\mu:\mathbb T^2\mapsto \mathbb C$ on $\mathbb T^2:=[0,2\pi)^2$, given by $\mu(k)=1+\frac{1}{2}e^{\varepsilon}(e^{ik_1}+e^{ik_2})$, satisfies $\mu(-k)=\overline{\mu(k)}$, and $\mu(\sigma_1 k)=\mu(k)$, 
with $\sigma_1=\begin{pmatrix}
    0&1\\1&0
\end{pmatrix}$  the first Pauli matrix; moreover, $\sin((\sigma_1 k)\cdot d)=-\sin (k \cdot d).$ Note also that $\DD_\vartheta=-\DD_{\vartheta}$ mod $(2\pi,2\pi)$ and $\sigma_1 \DD_{\vartheta}=\DD_{\sigma_1\vartheta}.$ 
\subsubsection{Symmetries}
\label{ssec:symmetri}

We introduce the following mappings of the even part of the Grassmann algebra $\mathfrak G$ generated by $\psi:=\{\psi^+_x,\psi^-_x\}_{x\in\Lambda}$
into itself: if 
$$P(\psi)=\sum_{\substack{n\ge 0\\ n \ \text{even}}}\sum_{\underline x\in\Lambda^n}\sum_{\un \a \in \{+,-\}^n}
c_n(\underline x,\un \a)
\psi^{\a_1}_{x_1}\psi^{\a_2}_{x_2}\cdots\psi^{\a_n}_{x_{n}}$$
 is an even element of the algebra, with $c_n(\underline x, \un \a)$ complex coefficients and $\underline x=(x_1,\ldots,x_n), \un \a=(\a_1,\dots,\a_n)$, 
 we define  
${\mathcal{S}}_1$ (\textit{conjugation} map) and $\mathcal{S}_2$ (\textit{reflection} map) as follows: 
\begin{equation}
\begin{split}
 &{\mathcal{S}}_1(P(\psi))=
\sum_{\substack{n\ge 0\\ n \ \text{even}}}\sum_{\underline x\in\Lambda^n}\sum_{\un \a \in \{+,-\}^n}\overline{c_n(\underline x,\un \a)}
\psi^{\a_1}_{x_1}\psi^{\a_2}_{x_2}\cdots\psi^{\a_n}_{x_{n}}, \\
 &{\mathcal{S}}_2(P(\psi))=
\sum_{\substack{n\ge 0\\ n \ \text{even}}}\sum_{\underline x\in\Lambda^n}\sum_{\un \a \in \{+,-\}^n}i^{Q(\un \a)}{c_n(\underline x,\un \a)}
\psi^{\a_1}_{\sigma_1x_1}\psi^{\a_2}_{\sigma_1 x_2}\cdots\psi^{\a_n}_{\sigma_1x_{n}},
\end{split}\label{eq:S1S2'}
\end{equation}
where $\overline{z}$ denotes the complex conjugate of $z\in\mathbb C$, and $Q(\un \a):= \sum_{j=1}^n \a_j$, so that $i^{Q(\un \a)}\in \{-1,1\}$. In particular, on monomials with the same number of $\psi^+$ and $\psi^-$, $\SS_2$ simply acts as the reflection $x \mapsto \sigma_1 x$.

In Fourier space, these mappings act as follows: given a generic even element $P(\hat\psi)$ of the Grassmann algebra $\mathfrak G_\vartheta$ generated by $\hat\psi_\vartheta:=\{\hat\psi^+_k,\hat\psi^-_k\}_{k\in\mathcal D_\vartheta}$ with complex coefficients, its image under $\mathcal S_1$ is the 
element of $\mathfrak G_\vartheta$ obtained from $P(\hat \psi)$
by changing every $\hat\psi^\pm_k$ into $\hat\psi^\pm_{-k}$ (with $-k$ identified with an element of $\mathcal D_\vartheta$ modulo $(2\pi,2\pi)$) and every complex coefficient into its complex conjugate; similarly, its image under $\mathcal S_2$ is the element of $\mathfrak G_{\sigma_1\vartheta}$ obtained from 
$P(\hat\psi)$ by changing every $\hat\psi^\pm_k$ into 
$\pm i\hat\psi^\pm_{\sigma_1k}$. Informally, we write: 
\begin{equation}\label{eq:S1S2} \begin{split} &\mathcal S_1:\hat\psi^\pm_k\to \hat\psi^\pm_{-k},\hskip1.2truecm c\to\bar c,\\
&\mathcal S_2:\hat\psi^\pm_k\to \pm i\hat\psi^\pm_{\sigma_1 k},\qquad \vartheta\to\sigma_1\vartheta,\end{split}
\end{equation}
Note that   $\mathcal S_2$ changes the boundary conditions $\vartheta$ into $\sigma_1\vartheta$.

Under the actions of $\mathcal S_1$ and $\mathcal S_2$, the 
terms in the Grassmann action in the r.h.s. of \eqref{eq:Zinteract} change as follows: 
\begin{equation} \begin{split}
& \mathcal S_1(S_{\vartheta,m}(\psi))=S_{\vartheta,m}(\psi),\hskip.9truecm 
\mathcal S_2(S_{\vartheta,m}(\psi))=S_{\sigma_1\vartheta,m}(\psi),\\
& \mathcal S_1(V_\vartheta(\psi,\mathcal A))=V_\vartheta(\psi,\mathcal A),\qquad 
\mathcal S_2(V_\vartheta(\psi,\mathcal A))=V_{\sigma_1\vartheta}(\psi,\mathcal S_2\mathcal A),\\
\end{split}\label{eq:S1S2psiA}\end{equation}
where $(\mathcal S_2\mathcal A)_e=A_{\mathcal S_2e}$ and
$\mathcal S_2e$ is the reflection of the edge $e$, i.e., denoting by $(x,s)$ an edge of type $s\in\{A,B,C\}$ with white vertex of 
coordinate $x$, $\mathcal S_2(x,A)=(\sigma_1x,A)$,
$\mathcal S_2(x,B)=(\sigma_1x,C)$ and $\mathcal S_2(x,C)=(\sigma_1x,B)$, while, for the edges of type $m$, $\SS_2\{\mathrm b_x,\mathrm b_{x+d}\}=\{\mathrm b_{\sigma_1 x},\mathrm b_{\sigma_1 x-d}\}$ and $\SS_2\{\mathrm w_x,\mathrm w_{x+d}\}=\{\mathrm w_{\sigma_1 x},\mathrm w_{\sigma_1 x-d}\}$. Note that the factor $i^{Q(\un \a)}$ in the definition of $\SS_2$ is due to the presence of edges of type $m$: the reflection $k \mapsto \sigma_1 k$ maps $d$ to $-d$  and therefore changes sign to the last term in \eqref{Steta}, which is compensated by $(\pm i)^2=-1,$ and a similar compensation occurs in the source terms in $V_\vartheta$ associated with edges of type $m$. In particular, 
$$\mathcal W_{\varepsilon,\lambda,L,m}(\mathcal A)=
\mathcal W_{\varepsilon,\lambda,L,m}(\mathcal S_2\mathcal A);$$
the latter can also be seen directly from \eqref{eq:3.2.1}, without introducing fermions.

\subsection{Setting up the multiscale analysis}
\label{sec:multiscale}

A possible naive approach to the computation of the r.h.s. 
of \eqref{eq:Zinteract} would be to expand  
the exponential $e^{V_\vartheta(\psi,\mathcal A)}$ in powers 
of $V_\vartheta$ and then to integrate term by term with 
respect to the Gaussian integration $\int 
D\psi\,e^{S_{\vartheta,m}(\psi)}(\cdot)$, via the fermionic Wick rule, see 
\cite[Eq.(4.20)]{GM01}. We assume that $\varepsilon\neq 0$, which guarantees that the Gaussian integration $\int 
D\psi\,e^{S_{\vartheta,m}(\psi)}(\cdot)$ is normalizable for all $\vartheta$. In fact, letting $\mathcal N_{\vartheta,m}:=
\int D\psi\,e^{S_{\vartheta,m}(\psi)}$ and $\EE_m(k):=|\mu(k)|^2+4m^2\sin^2(k\cdot d)$, we find that, for all $\vartheta\neq(0,0)$, $\mathcal N_{\vartheta,m}=
[\prod_{k\in\mathcal D_{\vartheta}}\mathcal E_m(k)]^{1/2}$, while, 
for $\vartheta=(0,0)$, $\mathcal N_{(0,0),m}=-\text{sign}\,\varepsilon
[\prod_{k\in\mathcal D_{\vartheta}}\mathcal E_m(k)]^{1/2}$, 
to be interpreted as $0$ at $\varepsilon=0$; it can be checked that 
$\mathcal N_{\vartheta,m}\neq0$ for all $\vartheta$, provided $m>0$ and $\varepsilon\neq0$. In this case, we let $P_{\vartheta,m}(D\psi):=\mathcal N_{\vartheta,m}^{-1}(D\psi)e^{S_{\vartheta,m}(\psi)}$ and define the {\it propagator} with $\vartheta$ boundary conditions to be $\mathfrak g^{\alpha,\beta}_{L,m}(x-x'):=\int P_{\vartheta,m}(D\psi)\,\psi^\alpha_x\psi^\beta_{x'}$. 
A computation shows that
{\begin{equation}\label{eq:proppsi}
\mathfrak g^{-+}_{L,m}(x)=
L^{-2}\sum_{k\in\mathcal D_\vartheta}e^{-ik\cdot x}\,\frac{\overline{\mu(k)}}{\mathcal E_m(k)}, \qquad 
\mathfrak g^{++}_{L,m}(x)=
L^{-2}\sum_{k\in\mathcal D_\vartheta}e^{ik\cdot x}\,\frac{2im\sin(k\cdot d)}{\mathcal E_m(k)},
\end{equation}
while $\mathfrak g^{--}_{L,m}(x)=-\mathfrak g^{++}_{L,m}(x)$.} Note that $\mathfrak{g}^{\a\b}_m(x)$ is $\vartheta$-periodic, in the sense that $\mathfrak{g}^{\a\b}_m(x+(L,0))=(-1)^{\vartheta_1}\mathfrak{g}^{\a\b}_m(x)$ and $\mathfrak{g}^{\a\b}_m(x+(0,L))=(-1)^{\vartheta_2}\mathfrak{g}^{\a\b}_m(x)$. 
In the limit $L \to \io$, at fixed $x\in\mathbb Z^2$, the propagators in \eqref{eq:proppsi} tend to 
\begin{equation}
    \mathfrak{g}_m^{-+}(x)=\int_{\mathbb{T}^2}\frac{dk}{(2\pi)^2}e^{-ikx}\frac{\overline{\mu(k)}}{\EE_m(k)}, \qquad \mathfrak{g}_m^{++}(x)=\int_{\mathbb{T}^2}\frac{dk}{(2\pi)^2}e^{-ikx}\frac{2mi \sin (k \cdot d)}{\EE_m(k)}\label{eq:2comptsprop}
\end{equation} irrespective of the choice of $\vartheta.$ Note that, taking $m\to0^+$ in \eqref{eq:2comptsprop}, the $++$ component tends to 0, while the $-+$ component tends to 
$\mathfrak g(x)=\int_{\mathbb T^2}\frac{dk}{(2\pi)^2}e^{-ikx}(\mu(k))^{-1}$. 
Convergence (or divergence) of perturbation theory in the limit $L\to \infty$ followed by $m\to 0^+$ critically depends upon the large-distance decay properties of \eqref{eq:proppsi}, of their infinite volume limits, and of the corresponding $m\to 0^+$ limit theoreof. 

Direct inspection shows that, for $\varepsilon<0$, $\mathcal E_m(k)\ge |\mu(k)|^2\ge (1-e^{\varepsilon})^2>0$; 
correspondingly, one can check that the components of the propagator decay exponentially to zero at large distances, uniformly in $L$ at large $L$, and uniformly in $m$ for $m$ small and positive (note, however, that the rate of the exponential decay vanishes as $\varepsilon\to 0^-$). As a consequence, for $\varepsilon<0$ naive perturbation theory can be shown to converge uniformly in $L$ for $L$ large and in $m$ for $m$ small and positive, but non-uniformly in $\varepsilon$ as $\varepsilon\to0^-$. 

For $\varepsilon>0$ and $m>0$ sufficiently small, let's say much smaller than $\sqrt\varepsilon$, direct inspection shows that
$\mathcal E_m(k)\ge c m^2\varepsilon$ for some 
$c$ independent of $m,\varepsilon$; 
correspondingly, one can check that the components of the propagator decay exponentially to zero at large distances, uniformly in $L$ for $L$ large, but 
non-uniformly in $m$ and $\varepsilon$. 
As a consequence, for $\varepsilon>0$ naive perturbation theory can be shown to converge uniformly in $L$ for $L$ large, but non-uniformly in $m$ as $m\to 0^+$. For these reasons, in order to be able to compute thermodynamic and correlation functions for small values of $\lambda$, independent of $\varepsilon$, in the limit $L\to\infty$ followed by $m\to0$, we need to appeal to a more sophisticated multiscale procedure, described in the following.

\medskip

In our construction, the behavior of $\mu(k)\big|_{\varepsilon=0}$ in the vicinity of its zero $(\pi,\pi)$ is particularly relevant, and it reads as follows: 
\begin{equation} \mu(\pi+k'_1,\pi+k_2')\big|_{\varepsilon=0}=
-i\frac{k_1'+k_2'}2+\frac{|k'|^2}4+O(|k'|^3).\end{equation}
This asymptotics motivates the following change of variables: we let 
\begin{equation}
k=Rq, \qquad R=\begin{pmatrix} 1 & -1 \\ 1 & 1 \end{pmatrix} \label{eq:transf}
\end{equation} so that 
\begin{equation}\mu((\pi,\pi)+Rq)=-iq_1+\frac{q_2^2}2+O(q_1^2,q_2^4,q_1q_2^2).\end{equation}
For later reference, for any $\vartheta\in\{0,1\}^2$, we define $\mathcal{D}_{\vartheta,R}:=R^{-1} \mathcal{D}_\vartheta=\{q: Rq \in \mathcal{D}_\vartheta\}$; moreover, we let $\tilde \Lambda:=R^T \Lambda=\{(y_1,y_2): y=R^T x,\,  x \in \Lambda\}$. In the following, we will typically use the symbol $y$ for coordinates belonging to $\tilde\Lambda$ (rather than $x$, which is 
used for coordinates in $\Lambda$). The change of variables from $k$ to $q$ and correspondingly from $x$ to $y$ induces the following change of Grassmann variables: 
\begin{equation}\label{eq:3.29}
\varphi_y^\pm=L^{-2}\sum_{q \in \mathcal{D}_{\vartheta,R}} e^{\pm iqy} \hat{\varphi}^\pm_q, \qquad 
\hat \varphi^\pm_q:=\hat \psi^\pm_{Rq}, \qquad   \varphi^\pm_y:=\psi^\pm_{(R^T)^{-1}y},
\end{equation}
in terms of which \eqref{Steta} becomes \begin{equation}S_{\vartheta,m}(\psi(\varphi))=-L^{-2}\sum_{q \in \mathcal{D}_{\vartheta,R}}(\mu(Rq)\hat \varphi^+_q \hat \varphi^-_q+im \sin (2q_2) (\hat \varphi^+_q\hat \varphi^+_{-q}+\hat \varphi^-_q\hat\varphi^-_{-q})),\nonumber \end{equation}
with $\mu(Rq)=1+e^{\varepsilon+iq_1}\cos q_2$.

We now rewrite the generating function of dimer correlations in terms of the $ \varphi$ field. In addition to this, we shift $\varepsilon$, in order to take into account the possible shift of the critical point induced by the presence of the interaction, see the discussion after \eqref{eq:parame} and after Theorem \ref{th:5}. 
So we set $\varepsilon=\eps+\varepsilon_0$ with $\varepsilon_0$ a free parameter, to be fixed later on as a function of $\lambda$, in such a way that $\eps=0$ is the critical point separating a regime $\epsilon>0$ in which dimer correlations decay polynomially, from a frozen regime at $\epsilon<0$. We
rewrite \eqref{eq:Zinteract}  as
$e^{\mathcal{W}_{\varepsilon,\l,L,m}(\AAA)}=\sum_{\vartheta \in \{0,1\}^2} \frac{c_\vartheta}{2} e^{\mathcal{W}^{(\vartheta)}_{\varepsilon,\l,L,m}(\AAA)}$ where for a fixed $\vartheta \in \{0,1\}^2$
\begin{equation}\label{eq:5.71}
\begin{split}
& e^{\mathcal{W}^{(\vartheta)}_{\epsilon+\varepsilon_0,\l,L,m}(\AAA)}:=e^{L^2(\epsilon+\varepsilon_0)}\int D\varphi\, e^{\tilde{S}_{\epsilon,\vartheta,m}(\varphi)+\nu N_\vartheta(\varphi) +\tilde V_{\eps,\vartheta}(\varphi,\AAA)}, \\ & \tilde{S}_{\epsilon,\vartheta,m}(\varphi):=-L^{-2}\hspace{-3pt} \sum_{q \in \mathcal{D}_{\vartheta,R}} (\hat{\varphi}^+_{q}  (1+r+e^{iq_1}\cos q_2 ) \hat{\varphi}^-_{q}+ime^{-\eps-\varepsilon_0}\sin(2q_2) (\hat \varphi^+_q\hat\varphi^+_{-q}+\hat\varphi^-_q\hat\varphi^-_{-q})),  \\ & N_\vartheta(\varphi):=L^{-2} \sum_{q \in \mathcal{D}_{\vartheta,R}} \hat{\varphi}^+_{q} \hat{\varphi}^-_{q}, \qquad r:=(e^{-\epsilon}-1)e^{-\varepsilon_0}, \qquad \nu:=1-e^{-\varepsilon_0},
\end{split}
\end{equation}
    with $\tilde V_{\epsilon,\vartheta}$ the rewriting of $V_\vartheta$ in  \eqref{eq:Scoordinates} in terms of the composition of the transformations $\psi \to \sqrt{e^{-\varepsilon_0-\eps}}\psi$ and $\psi \to \psi(\varphi)$ (cf. \eqref{eq:3.29}).
The \emph{counter-term}  $\nu$  will be fixed appropriately (as a function of $\l$) at the end of the multiscale procedure, and its choice will provide the function $\varepsilon_0(\l)$ via the last equation in the third line of \eqref{eq:5.71}. Note that \eqref{eq:5.71} is covariant under the (image of the) transformations in \eqref{eq:S1S2} that, in the new variables, read:
\begin{equation}\label{eq:S1S2bis} \begin{split} &\mathcal S_1:\hat\varphi^\pm_q\to \hat\varphi^\pm_{-q},\hskip.8truecm c\to\bar c,\\
&\mathcal S_2:\hat\varphi^\pm_q\to \pm i\hat\varphi^\pm_{\sigma_3 q},\qquad \vartheta\to\sigma_1\vartheta\end{split}
\end{equation}
where $\sigma_3=\begin{pmatrix}
    1&0\\0&-1
\end{pmatrix}$ is the third Pauli matrix.

{Recall that we are interested in the free energy and dimer densities in the limit $L\to \infty$ followed by $m\to 0^+$: \begin{equation}\label{eq:interested} F_{\eps+\varepsilon_0,\l}=\lim_{m\to 0^+}\lim_{L\to \infty}L^{-2}\WW_{\eps+\varepsilon_0,\lambda,L,m}(0),\qquad \rho_{\epsilon+\varepsilon_0,\lambda}=\lim_{m\to 0^+}\lim_{L\to\infty}\partial_{A_{(0,B)}}\mathcal W_{\eps+\varepsilon_0,\lambda,L,m}(0),\end{equation}
with $
\WW_{\eps+\varepsilon_0,\lambda,L,m}(\mathcal A)$ defined as in \eqref{eq:5.71} and preceding line (we recall that $A_{(0,B)}$ denotes the auxiliary field $A_e$ for $e$ the edge of type $B$ whose white site has coordinate $0$, see the definition of edge label $(x,s)$ after \eqref{eq:S1S2psiA}). As discussed in more detail in Sect.\ref{sec:2.3.2} below, for any $\eps\neq0$ and $m>0$, the exponential decay of the propagator mentioned after \eqref{eq:2comptsprop} implies that the effect of boundary conditions vanishes exponentially as $L\to\infty$; as a consequence, for the purpose of computing $F_{\eps+\varepsilon_0,\l}$ and $\rho_{\epsilon+\varepsilon_0,\lambda}$, we can actually replace $\mathcal W_{\eps+\varepsilon_0,\lambda,L,m}$ by 
$\mathcal W_{\eps+\varepsilon_0,\lambda,L,m}^{(\vartheta_0)}$ under the limit sign, for any boundary condition $\vartheta_0$; even more, for fixed $\vartheta_0\neq(0,0),(1,1)$, we can safely exchange the order of limits $L\to\infty$ and $m\to 0^+$, see again Sect.\ref{sec:2.3.2}. Additionally, by convexity of the limiting free energy, the limiting 
density $\rho_{\eps+\varepsilon_0(\lambda),\lambda}$ is 
a monotone increasing function of $\eps$. Thanks to this
fact,
in order to prove Theorem \ref{th:5}, it is not restrictive to restrict to small values of $\eps$
in absolute value: in fact, since we prove in Sect.\ref{sec:2.3.1} below that 
$\rho_{\eps+\varepsilon_0(\lambda),\lambda}$, thought of as a function of $\eps$, vanishes identically in a left neighborhood of $\eps=0$, by monotonicity we find that 
$\rho_{\eps+\varepsilon_0(\lambda),\lambda}=0$ for all $\eps<0$, and this also shows that the limiting free energy is affine in $\eps$ for all $\eps<0$. 
In conclusion:}

\medskip

{\it For any $\eps\neq0$, for the purpose of computing the free energy and the dimer density, we can choose $\vartheta_0$ equal either to $(1,0)$ or to $(0,1)$, and $m=0$; unless otherwise stated, we shall do so from now on. More precisely, we will use the following identities\footnote{In the proof of \eqref{eq:mlimit}, 
one needs to distinguish the cases $\eps\neq0$ and $\eps=0$. In first place, we show that the order of limits $L\to\infty$ and $m\to0^+$ can be safely exchanged 
for $\vartheta_0=(1,0),(0,1)$ and $\eps\neq0$; a priori, 
this would leave the possibility that at $\eps=0$ 
the two limits in \eqref{eq:interested} are different from $\lim_{L\to\infty}W^{(\vartheta_0)}_{\varepsilon_0(\lambda),\lambda,L,0}(0)$ and $\lim_{L\to\infty}\rho^{(\vartheta_0)}_{\varepsilon_0(\lambda),\lambda,L,0}(0)$, respectively. On the other hand, 
we prove that $\rho^{(\vartheta_0)}_{\eps+\varepsilon_0(\lambda),\lambda,L,0}(0)$ is uniformly continuous in $\eps$ for large $L$; therefore its $L\to\infty$ limit is continuous in $\eps$ as well. In combination with the fact that 
$\rho_{\eps+\varepsilon_0(\lambda),\lambda}$ is 
non-decreasing in $\eps$, this a posteriori implies the validity of the identities in \eqref{eq:mlimit} at 
$\eps=0$, as well. For more details, see Sect.\ref{sec:2.3.2} below.}, proved in Sect.\ref{sec:2.3.2}: 
\begin{equation}\label{eq:mlimit}
F_{\epsilon+\varepsilon_0(\lambda),\l}=\lim_{L \to \io} L^{-2}\mathcal W^{(\vartheta_0)}_{\epsilon+\varepsilon_0(\lambda),\lambda,L,0}(0), \qquad \rho_{\epsilon+\varepsilon_0(\lambda),\l}=
\lim_{L\to\infty}\rho^{(\vartheta_0)}_{\epsilon+\varepsilon_0(\lambda),\lambda,L,0}, 
\qquad \vartheta_0=(1,0), (0,1),
\end{equation}
where $\rho^{(\vartheta_0)}_{\epsilon+\varepsilon_0(\lambda),\lambda,L,0}:=\partial_{A_{(0,B)}}W^{(\vartheta_0)}_{\epsilon+\varepsilon_0(\lambda),\lambda,L,0}(0)$.
Moreover, from now on we restrict to $|\eps|\le \bar\eps$, for a suitable $\bar\eps$ positive and sufficiently small.}

\medskip

In view of the previous discussion and of \eqref{eq:mlimit}, we focus on the Grassmann integral in the first line of \eqref{eq:5.71}, with $m=0$ and boundary condition $\vartheta_0$ equal either to $(1,0)$ or $(0,1)$, which is computed via a multiscale procedure to be described next (the procedure is identical for the two cases $\vartheta_0=(1,0),(0,1)$). The starting point is the normalization of the integral, which depends on the zeros of $(q_1,q_2)\mapsto 1+r+e^{iq_1}\cos q_2$: if $\epsilon<0$ then $r>0$ and it does not vanish; if $\epsilon\ge0$ its zeros are
\begin{equation}\label{eeps}
\pm\mathbf q^0_\epsilon=(\pi,\pm q_\epsilon),\qquad q_\epsilon=\arccos(1+r),
\end{equation}
with $q_\epsilon^2=2|r|(1+O(\epsilon))\asymp\epsilon$ for $\epsilon>0$; the two zeros coalesce at $(\pi,0)$ when $\epsilon=0$, and they are always at distance at least $\pi/(2L)$ from $\DD_{\vartheta_0,R}$. Thus, the Gaussian integration can be normalized immediately and gives (dropping for simplicity the labels $\vartheta_0$ in the right hand side)
\begin{equation}\label{eq:2.2bis}
e^{\mathcal W_{\epsilon+\varepsilon_0,\lambda,L}(\mathcal A)}=e^{L^2t^{(0)}_\epsilon}\int P(D\varphi)\,e^{\nu N(\varphi)+\tilde V_\epsilon(\varphi,\mathcal A)},
\end{equation}
where $t^{(0)}_\epsilon=\epsilon+\varepsilon_0+L^{-2}\sum_{q\in\mathcal D'_{\vartheta_0,R}}\log(1+r-e^{iq_1}\cos q_2)$, and $P(D\varphi)$ is the Grassmann Gaussian integration \cite[Sect. 4]{GM01} with covariance, or propagator,
\begin{equation}\label{eq:2.3}
g_\epsilon(y-y'):=\int P(D\varphi)\,\varphi^-_y\varphi^+_{y'}=\frac{(-1)^{y_1-y_1'}}{L^2}\sum_{q\in\mathcal D'_{\vartheta_0,R}}\frac{e^{-iq(y-y')}}{1+r-e^{iq_1}\cos q_2},
\end{equation}
with $\DD'_{\vartheta_0,R}:=\DD_{\vartheta_0,R}-q_F$ with $q_F=(\pi,0).$
Note that in terms of the coordinates in $\tilde\Lambda$, the $\vartheta_0$-periodicity of $g_\epsilon$ reads $g_\epsilon(y+(L,(-1)^iL))=(-1)^{i}g_\epsilon(y)$, $i=1,2$.

 The idea now is to decompose $g_\eps$ as a sum of single-scale propagators, $g_\eps=\sum_{h\le 1}g^{(h)}$, each decaying stretched-exponentially at infinity, with a rate independent of $L$ and $\eps$; the integer $h$ labelling the single-scale propagators takes values $h=1,0,-1,-2,\ldots$, and the specific  range of scales which it needs to be summed over depends in general upon $\eps$; the specific scaling properties of $g^{(h)}$ will also depend on the choice of the scale;
roughly speaking, $g^{(h)}$ corresponds to the restriction of the original propagator to a shell in momentum space for which 
$1+r-e^{iq_1}\cos q_2$ is of the order $2^h$. 
Thanks to the addition principle for Gaussian integrals \cite[eq.(4.21)]{GM01}, the Grassmann Gaussian integration $P(D\varphi)$ is 
decomposed in a product of Gaussian integrations associated with the single-scale propagators, and the computation of the Grassmann integral of interest performed iteratively, starting 
with the largest scale index, $h=1$, then moving to $h=0$, $h=-1$, etc.

\begin{figure}[ht]
           \centering
           \includegraphics[width=0.87\linewidth]{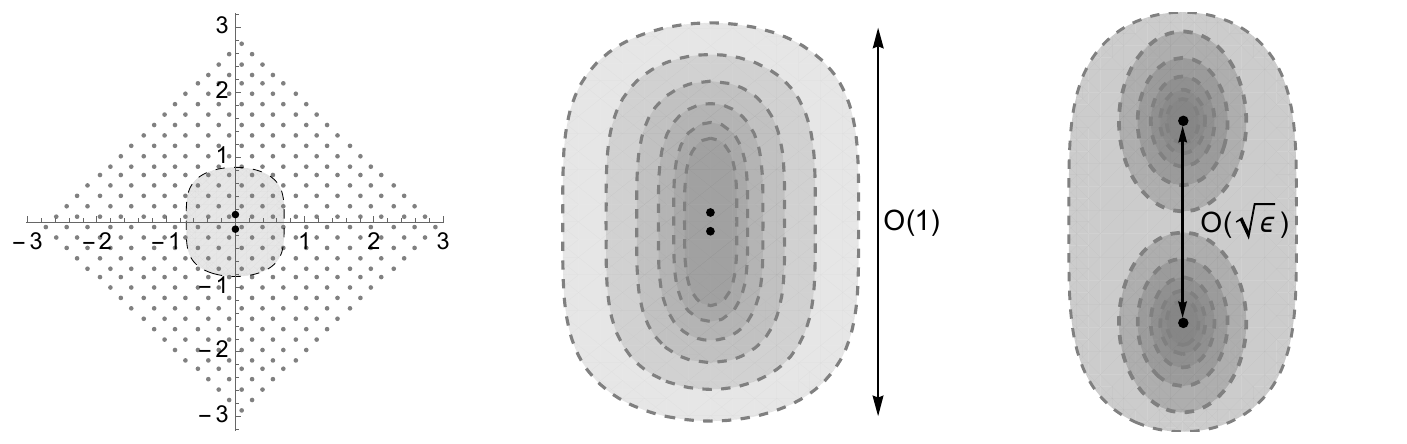}
           \caption{On the left, the allowed discrete set of momenta  
           $\DD'_{\vartheta,R}$, depicted in gray for $\vartheta=(0,1)$  and $L=25$; the zeros of $1+r-e^{iq_1}\cos q_2$ are depicted in black. In the center, the shells corresponding to scale labels larger than $\log_2\eps$. On the right, the shells corresponding to scale labels smaller than $\log_2\eps$, and larger than $\log_2\frac\pi{L}$.}
           \label{fig:enter-label}
       \end{figure}

If $\eps>0$, the decomposition of $\mathcal D'_{\vartheta_0,R}$ 
in momentum shells inducing the above mentioned rewriting of the propagator is pictorially described in Fig. \ref{fig:enter-label}. Recall the definition \eqref{eeps} of $q_\epsilon$; since $r$ is approximately equal to $-\eps$, up to corrections of relative order $\eps$ or $\varepsilon_0$ (where, as we will see, $\varepsilon_0(\lambda)$ is of order $\lambda$), the two zeros are at a mutual distance of the order $\sqrt{\eps}$. {Recall also that $1+r-e^{iq_1}\cos q_2=-\eps-iq_1+q_2^2/2+O(q_1^2,q_1q_2^2,q_2^4,\eps^2)$}. Therefore, 
for $h\gtrsim\log_2\eps$, the $h$-th shell, corresponding to momenta s.t. $2^{h-1}\lesssim\sqrt{q_1^2+q_2^4}\lesssim 2^h$, 
is connected, with the elongated shape shown in the central part of Fig. \ref{fig:enter-label}; {at scales of the order of $\log_2\eps$ one has $\eps \asymp \sqrt{q_1^2+q_2^4}$ and so,} for $h\lesssim\log_2\eps$, the $h$-th shell, corresponding to momenta s.t. $2^{h-1}\lesssim\sqrt{q_1^2+(q_2\pm q_\eps)^2}\lesssim 2^h$, 
consists of two disconnected ellipsoidal regions as shown in the right part of Fig. \ref{fig:enter-label}. The iterative procedure goes on until we reach a scale s.t. $2^h$ is comparable with the lattice mesh $2\pi/L$ of the allowed discrete set of momenta, $\mathcal D'_{\vartheta_0,R}$. This procedure is well defined if $L$ is much larger that $\eps^{-1}$, which clearly is in the regime of validity in which we intend to prove the main Theorem \ref{th:5}.
       
If $\eps=0$, the two zeros of $1+r-e^{iq_1}\cos q_2$ merge, and the momentum shells, which we decompose $\mathcal D'_{\vartheta_0,R}$ into, all have the shape depicted in the central part of Fig. \ref{fig:enter-label}; the scale label $h$ ranges from $1$ to $\log_2\frac\pi{L}$. 

If $\eps<0$, the combination $1+r-e^{iq_1}\cos q_2$ has no zeros. Still, in order to obtain bounds that are uniform in $L$ and $\eps$, we decompose $\DD'_{\vartheta_0,R}$ in shells, of the same shape as depicted in the central part of Fig. \ref{fig:enter-label}, but only for scale labels ranging from $1$ to $\log_2|\eps|$. For $\sqrt{q_1^2+q_2^4}\lesssim |\eps|$ the combination $1+r-e^{iq_1}\cos q_2$ is larger than a constant times $|\eps|$, uniformly in $q$; correspondingly, once we reach scale $\log_2|\eps|$, we integrate the Grassmann degrees of freedom we are left with all at once. 

In the next subsections, we will first describe the integration of the degrees of freedom corresponding to the first momentum shell, $h=1$. Then we will describe the integration procedure associated with scales `in the first regime', i.e., those for which the momentum shells have the shape depicted in the central part of Fig. \ref{fig:enter-label}; this exhausts the integration procedure both for $\eps=0$ and $\eps<0$. Next, if $\eps>0$, we describe the integration procedure associated with scales `in the second regime', i.e., those for which the momentum shells have the shape depicted in the right part of Fig. \ref{fig:enter-label}.

\subsubsection{Ultraviolet integration}
\label{ssec:uv} Let $\chi: \mathbb{R}^+ \to [0,1]$ be a smooth, compactly supported function in the Gevrey-$2$ class\footnote{We refer to \cite[Appendix C]{GMT17} and \cite[Appendix A]{GMR} for the definition of these functions, and for an explicit example of $\chi$ with the required properties.} such that
$\chi(z)= 1$ for $z \leq 1$ and $\chi(z)=0$ for $z \geq 2$. Letting $\|q\|_{(1)}^2:=q_1^2+q^4_2$, define $\chi^{(1)}_{0}(q):=\chi(\|q\|_{(1)})$ and $f^{(1)}_1(q):=1-\chi^{(1)}_{0} (q)$. We introduce a smooth partition of $\mathcal{D}_{\vartheta_0,R}'$ via $1=\chi^{(1)}_{0}(q)+f^{(1)}_1(q)$, which induces  $g_\eps(y,0)=(-1)^{y_1}(g_\eps^{(\leq 0)}(y,0)+g_\eps^{(1)}(y,0))$, with
\begin{equation}\label{eq:5.39}
g^{(\leq 0)}_\eps(y,0)=\int dq\frac{e^{-iqy} \chi^{(1)}_{0}(q)}{1+r-e^{iq_1}\cos q_2}, \qquad 
g^{(1)}_\eps(y,0)=\int dq\frac{e^{-iqy} f^{(1)}_{1}(q)}{1+r-e^{iq_1}\cos q_2},
\end{equation}
where $\int dq$ stands for $L^{-2}\sum_{q \in \mathcal{D}'_{\vartheta_0,R}}$.
Correspondingly we rewrite $\varphi^\pm_y=(-1)^{y_1}(\varphi^{\pm, (\leq 0)}_y+\varphi^{\pm,(1)}_{y})$ and, using the Addition Principle  (see \cite{GM01} Eq. (4.21)), 
we find: 
\begin{equation}\label{eq:4.20}
\begin{split}
e^{\mathcal{W}^{(\vartheta_0)}_{\eps+\varepsilon_0,\l,L}(\AAA)} &=e^{L^2 t^{(0)}_\eps} \int P_\epsilon^{(\leq 0)}(D\varphi^{(\leq 0)}) \int P^{(1)}_\epsilon(D\varphi^{(1)}) e^{\nu N(\varphi^{(\leq 0)}+\varphi^{(1)})+\tilde V_\eps(\varphi^{(\leq 0)}+\varphi^{(1)},\mathcal A)}\\ & =e^{L^2( F^{(0)}_\eps+t^{(0)}_\eps)+S^{(0)}_\eps(J)} \int P^{(\leq 0)}_\epsilon(D\varphi^{(\leq 0)}) e^{ V_\eps^{(0)}(\varphi^{(\leq 0)},J)}
\end{split}
\end{equation}
where in the last line $J=\{J_{z,s}\}_{z\in\tilde \Lambda}^{s\in\{A,B,C\}}$ and, if $e$ is an edge of type $s\in\{A,B,C\}$ with coordinate of the white site $z\in\tilde \Lambda$, we denoted $J_{z,s}:=e^{A_e}-1$; moreover, $F_\eps^{(0)}$, $S_\eps^{(0)}$ and $V_\eps^{(0)}$ are defined by 
\begin{equation}\label{effectivepot}
\begin{split}
& L^2 F_\eps^{(0)}+ S_\eps^{(0)}(J)+V_\eps^{(0)}(\varphi,J) =\sum_{n \geq 1} \frac{1}{n!} \mathcal{E}^T_{1}(\nu N(\varphi+\cdot)+\tilde V_\eps(\varphi+\cdot,J);n),
\end{split}
\end{equation}
where $\mathcal{E}^T_{1}$ is the truncated expectation of order $n$ with respect to the propagator $g^{(1)}_\eps$ (see \cite[Sec. 4; (4.13)-(4.18)]{GM01}) and $S_\eps^{(0)}$ and $V_\eps^{(0)}$
are fixed uniquely by the conditions $S_\eps^{(0)}(0)=V_\eps^{(0)}(0,J)=0$.
Note that, since the cutoff function $f^{(1)}_{1}$ in \eqref{eq:5.39} is a Gevrey-2 function, the propagator $g^{(1)}$ has
stretched-exponential decay at large distances (cf. \cite[Eq. (6.21)]{GMT20}). From \eqref{effectivepot} it follows that $V^{(0)}_\eps$ is a polynomial in $(\varphi,J)$ of the form 
$V^{(0)}_\eps(\varphi,J)=\sum_{n \geq 1, m \geq 0} V^{(0)}_{2n,m;\eps}(\varphi,J)$, where 
\begin{equation}\label{eFPotent(0)}
\begin{split}
V^{(0)}_{2n,m;\eps}(\varphi,J) &= \int d\un s d\un y d \un z  \, W^{(0)}_{2n,m;\eps,\un s}(\underline{y},\underline{z}) \varphi_{\un y}J_{\un z, \un s}=   \int d \un s d \un q d \un p \,  \widehat{W}_{2n,m;\eps,\un s}^{(0)}(\un q,\un p)\hat\varphi_{\un{q}+\un{q_F}}\hat J_{\un p,\un s}\delta({\underline{q}},\un p).
\end{split}
\end{equation}
In this formula, $\int d \un s d \un y d\un z$ is a shorthand notation for  $\sum_{\un s, \un y, \un z}$ with $\underline{y}=(y_1,\dots,y_{2n}) \in \tilde \Lambda^{2n}$, $\un z=(z_1,\dots,z_m) \in \tilde \Lambda^m$, $\un s=(s_1,\dots,s_m) \in \{A,B,C\}^m$, and $\varphi_{\un y}:=\varphi^+_{y_1}\varphi^{-}_{y_2}\dots \varphi^{+}_{y_{2n}-1}\varphi^-_{y_{2n}}$, $J_{\un z,\un s}=J_{z_1,s_1}\dots J_{z_m,s_m}$. Note that we included the oscillating factor $(-1)^{\sum_{i=1}^{2n}y_{i1}}$ multiplying the Grassmann variables $\varphi_{\un y}$ (see below \eqref{eq:5.39}) in the definition of the kernels: $\widetilde{W}_{2n,m;\eps,\un s}^{(0)}(\un y,\un z):=(-1)^{\sum_{i=1}^{2n}y_{i1}} {W}_{2n,m;\eps,\un s}^{(0)}(\un y,\un z)$ is translation invariant and $\vartheta_0$-periodic in each variable, in the same sense as below \eqref{eq:2.3}. 
Similarly, $\int d \un s d \un q d\un p$ is a compact notation for $L^{-2(n+m)}\sum_{\un s, \un q, \un p}$  with $\un q=(q_1,\dots,q_{2n}) \in (\mathcal{D}_{\vartheta_0,R}')^{2n}, \un p=(p_1,\dots,p_m) \in (\mathcal{D}_{(0,0),R})^m$; 
$\widehat{W}_{2n,m;\eps,\un s}^{(0)}(\un q,\un p)$  denotes the Fourier transform of $\widetilde{W}_{2n,m;\eps,\un s}^{(0)}(\un y,\un z)$ computed at momenta $(q_1+q_F,q_2+q_F,\dots, q_{2n}+q_F,p_1,\dots,p_m)$; 
   $\hat\varphi_{\un q+\un q_F}$ denotes a product of $\hat\varphi^\pm_{q_i+q_F}$ (we recall that $q_F=(\pi,0)$), see \eqref{eq:3.29}, and $\hat J_{\un p,\un s}$ is a product of 
$\hat J_{p,s}:=\sum_{z\in \tilde\Lambda}e^{i p z}J_{z,s}$; finally, the momentum conservation due to translation invariance is guaranteed by 
$\d(\un q, \un p)=L^2$ if $\sum_{i=1}^{2n} (-1)^{i-1} q_i=\sum_{i=1}^m p_i$ and $\d(\un q, \un p)=0$ otherwise: with some abuse of notation we will write $\widehat W^{(0)}_{2n,m;\eps,\un s}(q_2,\dots,q_{2n},\un p)$ in place of $\widehat W^{(0)}_{2n,m;\eps,\un s}(\un q,\un p)$ whenever $q_1$ is fixed via the relation given by $\d(\un q,\un p)\neq 0.$ 

A similar expression holds for $S^{(0)}(J)$ with $n=0, m>0.$ The kernels $W^{(0)}_{2n,m;\eps,\un s}(\un y,\un z)$ are antisymmetric under the exchange of any pair of $y_i,y_j$ with $i+j$ even, symmetric under the exchange of any pair $(z_i,s_i),(z_j,s_j)$.  

Using the so called {\it Battle-Brydges-Federbush-Kennedy} determinant formula, see \cite[eq.(4.43)]{GM01} and the Gram-Hadamard bound for determinants, see \cite[App.A.3]{GM01}, applied to the truncated expectation in \eqref{effectivepot}, we find that the series representation for $F^{(0)}-F^{(1)}$ and $W^{(0)}_{2n,m;\eps,\un s}(\underline{y},\un z)$ following from \eqref{effectivepot} is absolutely convergent  and, therefore, they are real analytic function of $(\lambda,\nu)$, provided that $\max\{|\nu|,|\lambda|\}\leq \bar \delta$, with $\bar \delta >0$ small enough. For details, see e.g. \cite[Sect.4.2]{GM01}. From analyticity and an evaluation of the lowest non-trivial order contributing to $F^{(0)}$ and $W^{(0)}_{2n,m;\eps,\un s}(\underline{y},\un z)$, 
it readily follows that, letting $
\lVert W^{(0)}_{2n,m;\eps, \un s}\rVert:=L^{-2} \sum_{\underline{y} \in \tilde{\Lambda}^{2n}}\sum_{\un z \in \tilde \Lambda^m}|W^{(0)}_{2n,m;\eps, \un s}(\underline{y}, \un z)|$,
\begin{equation}\label{kernestim}
\begin{split}
& |F^{(0)}| \leq c \bar{\delta}; \qquad \lVert W^{(0)}_{2n,m;\un s} \| \leq C^{n+m} \bar{\delta}^{d_{n,m}}, \quad (n,m) \neq (0,0)
\end{split}
\end{equation}
for positive constants $c,C$ independent of $L,\eps$, where  $d_{n,0}=\max\{1,cn\}$, $d_{n,m}=\max\{0,c(n-1)\}$ if $m\geq 1$, and $m_A=\sum_{i=1}^m \d_{s_i,A}$, with $\d_{\cdot,\cdot}$ the Kronecker's delta. From \eqref{effectivepot} and the definitions of $N$ and $\tilde V_\epsilon$, it turns out that the coefficients of the series expansion for $F^{(0)}$ and $W^{(0)}_{2n,m;\eps,\un s}$, which are convolutions of the kernels of 
$N$ and $\tilde V_\epsilon$ with propagators $g^{(1)}_\eps$, admit a well-defined 
$L\to\infty$ limit, consisting in the same kind of convolutions, with propagators replaced by their infinite volume limits
\begin{equation}\label{inftvolprop}
g^{(1)}_{\eps,\io}(y,0)=\frac{1}{2\pi^2}\int_{R^{-1}[0,2\pi]^2} dq e^{-iqy}f^{(1)}_1(q)\hat g(q), \qquad \hat g(q) =\frac{1}{1+r-e^{iq_1}\cos q_2},
\end{equation}
and sums over coordinates extended to $R^T\mathbb{Z}^2$ rather than restricted to $\tilde\Lambda$. This fact, combined with the absolute, uniform-in-$L$, convergence of the series, implies that $F^{(0)}$ and $W^{(0)}_{2n,m;\eps,\un s}$ converge to well-defined limits as $L\to\infty$, independent of  $\vartheta_0$; these are denoted by $F^{(0);\infty}$ and $W^{(0);\infty}_{2n,m;\eps,\un s}$, respectively, and they are analytic in the same domain as their finite-volume approximations. 

\medskip

A couple of additional remarks are in order: 
\begin{enumerate}
\item {\bf Comparison with \cite{BM36}.} The Fourier transform of the propagator, 
$\hat g(q)$, as well as its single-scale counterpart, $f^{(1)}_1(q)\hat g(q)$, satisfies the following symmetries: $\hat{g}(q)=\overline{\hat g(-q)}=\hat g(\sigma_3 q)$, which are the same displayed by the propagator $\hat g_Q(q)$ of the quantum model in \cite[(28), Sec. 1.5]{BM36}. Even more, the inverse of our propagator, $(\hat g(q))^{-1}=1+r-e^{iq_1}\cos q_2=1+r-iq_1-\cos q_2+O(q_1^2)+O(q_1q_2^2)$, has the same form as the propagator in \cite{BM36}, $(\hat g_Q(q))^{-1}=-iq_1+\cos q_2+h$, up to subleading corrections for $q_1,q_2$ small. In addition to 
this common structure of the propagator, both our model and the one in \cite{BM36} 
have an interaction that is invariant under the same conjugation and reflection symmetries, which make the multiscale analysis used in both cases to study the partition function and correlations very similar. In particular, the structure and 
flow of the running coupling constants, defined below, are qualitatively the same in the two models and they can be studied in the same way, as anticipated in the discussion after \eqref{eq:parame}. A minor difference in our case as compared to 
\cite{BM36} is the presence of a non-zero shift $\varepsilon_0(\lambda)$ in the location of the critical line, see \eqref{eq:5.71} and preceding lines. We will show 
(cf. Lemma \ref{lem:firstordercalc}, where $\nu^*(\lambda)=1-e^{-\varepsilon_0(\lambda)}$), that $\varepsilon_0(\l)=2T \l+O(\l^2)$ where $T=\int_{R^{-1}[0,2\pi]^2} dq \hat g(q)=1$; on the contrary, the analogue of this shift is zero in \cite{BM36}: in particular, $T_Q=\int_{\mathbb{R}\times [0,2\pi]}d q \hat g_Q(q)=0$ from  residue calculus, and a similar cancellation holds, in fact, at all orders, i.e., $\varepsilon_0^Q(\l)=0$ cf. \cite[Eq. (99) and below]{BM36}.  
\item {\bf Dependence on $\eps$.} \label{item2after} The dependence on $\eps$ of the kernels $W^{(0)}_{2n,m;\epsilon,\un s}$ can be completely re-expressed in terms of $r$, see \eqref{effectivepot} and \eqref{eq:5.71}. In fact, since $\nu$ is independent of $\eps$, the dependence on $\eps$ is induced either by a propagator $g^{(1)}_\eps$ \eqref{inftvolprop} or by the coefficients of the monomials in $\tilde V_\epsilon$ (cf. \eqref{eq:5.71} and following line), and can be obtained by inverting $e^{-\eps}=1+e^{\varepsilon_0}r$ as $\eps=\epsilon(r)$. When looking at the kernels as functions of $r$, we will use the notation $W^{(0)}_{2n,m;\un s}(\un y, \un z; r)$ (and similarly for their Fourier transform).
\end{enumerate}

We now have a representation for  the generating function (cf. \eqref{eq:4.20}) in terms of an effective interaction $S^{(0)}(J)+V^{(0)}(\varphi^{(\leq 0)},J)$ for the infrared field $\varphi^{(\leq 0)}$, with bounds on its kernels, see \eqref{kernestim}. As already anticipated,  one could naively iterate the integration by decomposing $\varphi^{(\leq 0)}=\sum_{h= h_\eps-1}^{0}\varphi^{(h)}+\varphi^{(\leq h_\eps)}$, integrating $\varphi^{(h)}$  down to a scale $h_\eps$ such that infrared cutoff $\chi^{(1)}_{h_\eps-1}$ no longer contains the singularities $\pm \boldsymbol{q}_\eps$ (see the discussion below \eqref{eq:2.3}).  However, this naive iteration fails to yield $L$-uniform bounds due to divergences arising from a finite number of kernels with bad scaling behavior (see also the discussion \cite[below (5.8)]{GMT17}). To identify the dangerous kernels, one studies the scaling of the propagator $G^{(h)}$, obtained from $g^{(1)}$ in \eqref{eq:5.39} by replacing $f^{(1)}_1$ with $f^{(1)}_h=\chi^{(1)}_h-\chi^{(1)}_{h-1}$ and $\hat g^{-1}(q)$ with its dominant Taylor approximation at $q=0$, which scales as  \be G^{(h)}(y_1,y_2)=2^{\frac h2}G^{(0)}(2^h y_1,2^{\frac h2}y_2).\label{eq:propcontinuo}\ee 
From this, repeating the same discussion of \cite[Sec. 5.2.1-5.2.2]{GMT17} one finds that a kernel $W_{2n,m}^{(h)}$ at scale $h$ can be bounded as $\rVert W_{2n,m}^{(h)}\rVert \leq C^n \l^{d_{n,m}} 2^{h D_{sc}}$, where $d_{n,m}$ was defined right below \eqref{kernestim}, the \emph{scaling dimension} $D_{sc}$ is given by \be D_{sc}(2n,m):= 3/2-2n/4-m, \label{eq:scdim} \ee and $C=C(L)>0$ is a priori not uniformly bounded in $L$.  As explained in \cite[Sec. 5.2.2]{GMT17}, the dangerous diagrams potentially leading 
to non-uniform-in-$L$ bounds on $C$ 
are those associated to a non-negative scaling dimension, namely $
D_{sc} \geq 0$, which correspond to $ (2n,m)=(2,0),(4,0),(6,0),(2,1)$. These terms are called relevant, if $D_{sc}>0$, or marginal, if $D_{sc}=0$, and they require a renormalization procedure, while those with $D_{sc}<0$ are called irrelevant.

\begin{remark}
Despite the fact that the scaling dimension \eqref{eq:scdim} allows \emph{a priori} for several potentially divergent terms (i.e., marginal and relevant), more than those in previously studied models featuring linear dispersion relations (see, e.g., \cite[p. 133]{GMT17}), the actual emergence of dangerous divergences is prevented by the anticommutativity of the Grassmann fields:  indeed, many of the local contributions to marginal or relevant kernels vanish identically. This fact is related crucially to the fact that the Grassmann field has just two components per site, $\varphi^+_y$ and $\varphi^-_y$. 

When the Grassmann variables carry additional internal degrees of freedom (such as spin in quantum systems or flavor indices in classical models such as the dimer model on a decorated lattice) such cancellations no longer occur automatically. In these cases, one in general observes qualitatively different critical behavior as $\epsilon \to 0$, and a different renormalization strategy is needed \cite{BM19}. 
\end{remark}

\subsubsection{Infrared integration: first regime}
\label{sec:firstregime}
 In this section we mainly follow \cite{BM36}, implementing when needed a few technical modifications; we will emphasize the dependence on $\eps$ in all the involved quantities. If $\eps>0$, recall that $L$ is much larger than $\eps^{-1}$ (see also the discussion below \eqref{eq:2.3}). 
 
 We start from the outcome of the ultraviolet integration \eqref{eq:4.20} and, in view of the previous discussion, we split the effective interaction $V^{(0)}_\eps$ in two parts: the first, denoted the `local' part, includes the relevant and marginal contributions, while the second, called the `renormalized' part, includes the irrelevant ones. In 
 formulae, we write: $V_\epsilon^{(0)}=\mathcal{L}_{1,\epsilon}V_\epsilon^{(0)}+\mathcal{R}_{1,\epsilon}V_\epsilon^{(0)}$, with the operator $\mathcal{R}_{1,\eps}=1-\mathcal{L}_{1,\eps}$, and $\LL_{1,\eps}$ acting on the effective potential \eqref{eFPotent(0)} as follows:  $\mathcal{L}_{1,\eps}V^{(0)}_{2n,m;\eps}(\varphi,J)=0$ if $n>4$ or $m>1$, while 
\begin{equation}
\begin{split}
& \hspace{-6pt} \mathcal{L}_{1,\epsilon}V_{2,0;\eps}^{(0)}(\varphi,J)= \\ 
& \hspace{-6pt} \int dq \,  \hat \Phi_{q} \Big [ \hat{W}^{(0);\io}_{2,0;0}(0)+q \cdot \partial \hat W^{(0);\io}_{2,0;0}(0)+\frac{q_2^2}{2}\partial^2_2 \hat{W}^{(0);\io}_{2,0;0}(0) + \mathcal T \left (\hat W_{2,0;\epsilon}^{(0);\io}(0)-\hat W_{2,0;0}^{(0);\io}(0) \right ) \Big ],  \\ & \hspace{-6pt} \mathcal{L}_{1,\epsilon}V_{2,1;\epsilon}^{(0)}(\varphi,J)=\sum_{s=A,B,C}  \int dy \, \varphi^+_{y}\varphi^-_{y}J_{y,s} \hat W^{(0);\io}_{2,1;\epsilon,s}(0,0)
\end{split}\label{eq:loc1reg}
\end{equation}
where  we used the Fourier representation \eqref{eFPotent(0)} with $\hat \Phi_q:=\hat \varphi^+_{q+q_F}\hat \varphi^-_{q+q_F}$. In the second line, $\mathcal{T}$ is an operator that extracts the linear part in $r$ (cf. \eqref{eq:5.71}) of  $\hat W_{2,0;\epsilon}^{(0);\io}$ namely, \be \mathcal{T} ( \hat W_{2,0;\epsilon}^{(0);\io}(0)-\hat W_{2,0;0}^{(0);\io}(0)):= r \partial_{{r}}\hat W^{(0);\io}_{2,0;0}(0).\label{eq:operatorT} \ee 

\begin{remark}
    As pointed out in item 2 after \eqref{inftvolprop}, since the dependence on $\eps$ of the kernels is all encoded in $r$, and $r=0$ iff $\eps=0$ as it follows from \eqref{eq:5.71},   $\TT$ is nothing but the leading term in the Taylor expansion around $r=0$ of $\hat W^{(0);\io}_{2,0}(0;r):=\hat W^{(0);\io}_{2,0;\eps}(0)$ (here the notation is the same as the one used in item 2 after \eqref{inftvolprop}). 
\end{remark}

Note that $\mathcal{L}_{1,\eps}$ acts non trivially only on quadratic terms  since the contributions $V_{4,0;\eps}^{(0)},V_{6,0,\eps}^{(0)}$ are automatically renormalized (i.e., their local part is zero), thanks to the anticommutativity of the Grassmann variables, i.e. $(\varphi^\s_y)^2=0$. In fact we can rewrite
\begin{equation}
V^{(0)}_{4,0;\eps}(\varphi,0)\equiv \mathcal{R}_{1,\eps} V^{(0)}_{4,0;\eps}(\varphi,0)=\int d\un y\,  \varphi^+_{y_1}\varphi^-_{y_2}(\varphi^+_{y_3}-\varphi^+_{y_1})(\varphi^-_{y_4}-\varphi^-_{y_2}) W^{(0)}_{4,0;\eps}(\underline{y})\label{eq:3.3.30}
\end{equation}
which is the analogue of \cite[Eq. (45)]{BM36}; a similar rewriting holds for the sextic Grassmann term as well. The right hand side of \eqref{eq:3.3.30} is the Riemann sum approximation on the lattice 
$\tilde \Lambda$ to the integral $\int_{\mathbb R^4}d\underline y  \varphi^+_{y_1}\varphi^-_{y_2}(\varphi^+_{y_3}-\varphi^+_{y_1})(\varphi^-_{y_4}-\varphi^-_{y_2}) W^{(0)}_{4,0;\eps}(\underline{y})$ that, by proceeding as 
in \cite[Sect.5.2.1, 5.2.2, App.C]{GMR}, can be rewritten in {\it interpolated form} as:
\begin{equation}
\begin{split}& \sum_{\ell,\ell'=1,2}\int_{\mathbb R^4}d\underline y\int_0^1 ds\int_0^1 ds' \varphi^+_{y_1}\varphi^-_{y_2}\partial_{\ell}\varphi^+_{y_1+s(y_3-y_1)}
\partial_{\ell'}\varphi^-_{y_1+s'(y_4-y_1)} (y_3-y_1)_{\ell}(y_4-y_1)_{\ell'}W^{(0)}_{4,0;\eps}(\underline y)\\
&\equiv \sum_{\ell,\ell'=1,2}\int_{\mathbb R^4}dy_1\, dy_2\, dz_3\, dz_4 \,
\varphi^+_{y_1}\varphi^-_{y_2}\partial_{\ell}\varphi^+_{z_3} \partial_{\ell'}\varphi^-_{z_4}\,[\mathcal I W^{(0)}_{4,0;\eps}]_{\ell,\ell'}(y_1,y_2,z_3,z_4),  \end{split}
\end{equation}
where in passing from the first to the second line we performed the change of variables $y_1+s(y_3-y_1)\equiv z_3$, $y_1+s'(y_4-y_1)\equiv z_4$ 
and implicitly defined the interpolated potential $[\mathcal I W^{(0)}_{4,0;\eps}]_{\ell,\ell'}$. In the rewriting above we neglected lattice and finite volume 
effects; if desired, they can be accomodated for, at the cost of slightly more involved expressions, see \cite{AGG} for a thorough 
description of the interpolation procedure in the presence of a finite lattice and a finite box. In conclusion, \eqref{eq:3.3.30} can be rewritten as
\begin{equation}
V^{(0)}_{4,0;\eps}(\varphi,0)=\sum_{\ell,\ell'=1,2}\int d\un y\,  \varphi^+_{y_1}\varphi^-_{y_2} \bar \partial_{\ell}\varphi^+_{y_3}\bar\partial_{\ell'}\varphi^-_{y_4}
[\mathcal I W^{(0)}_{4,0;\eps}]_{\ell,\ell'}(\underline{y})+({\text{finite size corrections}}),\label{eq:3.3.30bis}
\end{equation}
where $\partial_\ell$, $\ell=1,2$, are the components of the discrete gradient defined as\footnote{There is some freedom in the specific choice of the discrete gradient to be used; the choice made here is convenient 
for symmetry reasons, it guarantees that its two components have the same parity under the natural symmetries of the model. Note that the elementary steps $\pm(2,0),\pm(0,2)$ on the rotated lattice $\tilde \Lambda$, involved in the definition of $\bar\partial_\ell$, correspond to $\pm(1,1), \pm(-1,1)$, respectively, in the original lattice $\Lambda$.} 
$\bar \partial_1 \varphi^\sigma_y=\tfrac14(\varphi^\sigma_{y+(2,0)}-\varphi^\sigma_{y-(2,0)})$, $\bar \partial_2 \varphi^\sigma_y=\tfrac14(\varphi^\sigma_{y+(0,2)}-\varphi^\sigma_{y-(0,2)})$, and $[\mathcal I W^{(0)}_{4,0;\eps}]_{\ell,\ell'}$ is a suitable interpolated potential, which is dimensionally irrelevant with scaling dimension 
equal to\footnote{Due to the different scaling properties of the propagator in the directions $y_1$ and $y_2$, 
a discrete derivative in direction $y_1$ comes with a decrease of $-1$ in the scaling dimension of the corresponding kernel, 
while a discrete derivative in direction $y_2$ comes with a decrease of $-1/2$ in the scaling dimension of the corresponding kernel.} $D_{sc}(4,0)-1/\ell-1/\ell'\le -1/2$. The finite size corrections are also dimensionally irrelevant, see \cite[App.B]{GMT20} for details. 

By proceeding similarly, in the case $(2n,m)=(2,0)$, using \eqref{eq:loc1reg}, we can rewrite $\RR_{1,\eps}V^{(0)}_{2,0;\eps}=\widetilde \RR_{1,\eps}V^{(0)}_{2,0;\eps}+\RR_{1,\eps}^{\mathcal{T}}V^{(0)}_{2,0;\eps}$, where 
\begin{equation} 
\mathcal{R}_{1,\eps}^{\mathcal{T}} V^{(0)}_{2,0;\eps}(\varphi,0)= 
\int dq \, \hat \Phi_q \left ( \hat W^{(0);\io}_{2,0;\eps}(q)-\hat W^{(0);\io}_{2,0;0}(q)-\mathcal{T}(\hat W^{(0);\io}_{2,0;\eps}(0)-\hat W^{(0);\io}_{2,0;0}(0))   \right ) \label{eq:2.33biss}
\end{equation}
and 
\begin{equation}
\begin{split}
 \widetilde \RR_{1,\eps}V^{(0)}_{2,0;\eps}(\varphi,0)&=
 \sum_{\substack{\ell,\ell'=1,2\\ (\ell,\ell')\neq(2,2)}} \int d\underline y \bar\partial_\ell\varphi^+_{y_1} \bar \partial_{\ell'} \varphi^-_{y_2}[\mathcal{I} W^{(0)}_{2,0;0}]_{\ell,\ell'}(\underline y) +\sum_{\ell=1,2} \int d\underline y \bar\partial_\ell\varphi^+_{y_1} \bar\partial^2_{2} \varphi^-_{y_2}[\mathcal{I}W^{(0)}_{2,0;0}]_{\ell,(2,2)}(\underline y)\\
 & +({\text{finite size corrections}}),\label{eq:interpopsi2}
\end{split}
\end{equation}
for suitable interpolated potentials $[\mathcal{I} W^{(0)}_{2,0;0}]_{\ell,\ell'}$ and $[\mathcal{I}W^{(0)}_{2,0;0}]_{\ell,(2,2)}$, which are dimensionally irrelevant:
more precisely, for any $(\ell,\ell')\neq(2,2)$, the scaling dimension of the former is $D_{sc}(2,0)-1/\ell-1/\ell'\le -1/2$, while that of the latter is 
$D_{sc}(2,0)-1/\ell-1\le -1/2$. Once again, the finite size corrections are irrelevant, as well, cf. with \cite[app.B]{GMT20}. On the other hand, the dimensional bounds on $\mathcal{R}_{1,\eps}^{\mathcal{T}} V^{(0)}_{2,0;\eps}$ are discussed below, after Lemma \ref{prop:an}. Analogous rewritings are valid for the other terms requiring non-trivial renormalization, i.e., those with $(2n,m)=(6,0),(2,1)$. We will not belabor the details of these rewritings here. We refer the 
interested reader to the analogous discussions in, e.g., \cite{AGG,GMT17,GMT20,GMR}. 

\begin{remark}
\label{rem:2}
Since the bare potential $V$ and the Gaussian integration $P^{(1)}(D\varphi)$ are both invariant under the symmetries $\mathcal{S}_1,\mathcal{S}_2$ (cf. Sec. \ref{ssec:symmetri}), then $V^{(0)}$ is invariant as well and thus its kernels satisfy  
 $\hat{W}^{(0);\io}_2(0),i \partial_1\hat{W}^{(0);\io}_{2,0;0}(0), \partial^2_2 W^{(0);\io}_{2,0;0}(0) \in \mathbb{R}$ and $\partial_2\hat{W}^{(0);\io}_{2,0;0}(0)=0,$ see Appendix \ref{app:symmetries} for more details.
\end{remark}

\begin{remark}
\label{rem:2bis}
 By construction the operator  $\mathcal{L}_{1,\eps}$ is idempotent, i.e.  $\mathcal{L}_{1,\eps}^2=\mathcal{L}_{1,\eps}$, which implies that $\mathcal{L}_{1,\eps}\mathcal{R}_{1,\eps}=0$.  
\end{remark}

 Now letting \be 2^{-1}\nu_{-1}:=\hat W^{(0);\io}_{2,0;0}(0),\label{eq:numeno1} \ee we move all the terms appearing in the first of \eqref{eq:loc1reg} except $\nu_{-1}$ in the Grassmann integration \eqref{eq:4.20} in such a way to rewrite
\begin{equation}
\eqref{eq:4.20}=e^{L^2(F_\eps^{(0)}+t_\eps^{(-1)})+S^{(0)}(J)}\int \tilde{P}_\eps^{(\leq 0)}(D\varphi)e^{2^{-1}\nu_{-1} N(\varphi)+\sum_{s=A,B,C}Y_{-1,s;\epsilon}F_{Y,s}(\varphi,J)+\mathcal{R}_{1,\eps}V_\eps^{(0)}(\varphi,J)}\label{eq:4.62}
\end{equation}
where we defined $N(\varphi)=\int dq\,  \hat \Phi_q$ (cf. \eqref{eq:loc1reg}) and
\be F_{Y,s}(\varphi,J)= \int dy \,  \varphi^+_{y}\varphi^-_{y} J_{y,s}, \qquad  
 Y_{-1,s;\epsilon}:=W^{(0);\io}_{2,1;\epsilon,s}(0,0). \label{eq:Yfirstreg}\ee  The new integration $\tilde{P}^{(\leq 0)}_\eps$ has covariance $\tilde{g}_\eps^{(\leq 0)}(y,y'):=\int P^{(\leq 0)}(D\varphi) \varphi^-_y \varphi^+_{y'}$ given by
\begin{equation}\label{eq:4.64p}
 \tilde{g}_\eps^{(\leq 0)}(y,y')=\int dq \frac{e^{-iq(y-y')}\chi^{(1)}_{0}(q)}{\boldsymbol{r}_{-1}-iq_1(1+ {\boldsymbol \z}_{-1}) +\frac{q_2^2}{2}(1+{\boldsymbol \a}_{-1})+\rho(q)}
\end{equation}
where, defining $\z_0:=0$, $ \a_0:=0$, \be \begin{split} & {\boldsymbol r_{-1}(q)}=r-\chi^{(1)}_{0}(q)r \partial_{{r}} \hat W^{(0);\io}_{2,0;0}(0), \qquad {\boldsymbol \z}_{-1}(q)=\z_0-i \chi^{(1)}_{0}(q) \partial_{1} \hat W^{(0);\io}_{2,0;0}(0), \\ & {\boldsymbol \a}_{-1}(q)=\a_0-
\chi^{(1)}_{0}(q) \partial^2_{2} \hat W^{(0);\io}_{2,0;0}(0), \qquad \rho(q)=1-e^{iq_1}\cos q_2+iq_1-q_2^2/2, \end{split}\label{eq:2.26biss}\ee  while $t_{\eps}^{(-1)}-t^{(0)}_\eps$ is obtained from the change of normalization of the Grassmann integration.  Remark \ref{rem:2} implies that $ {\boldsymbol r_{-1}(q)}, {\boldsymbol \z}_{-1}(q), {\boldsymbol \a}_{-1}(q) \in \mathbb{R}$.

Next we decompose $\chi^{(1)}_{0}=f^{(1)}_{0}+\chi^{(1)}_{-1}$, with $\chi^{(1)}_{-1}(q):=\chi^{(1)}_{0}(2q)$ which implies that $
\tilde{g}_{\eps}^{(\leq 0)}=\tilde{g}_{\eps}^{(0)}+g_{\eps}^{(\leq -1)}$   where $\tilde{g}_{\eps}^{(0)}$ is obtained from \eqref{eq:4.64p} by replacing $\chi^{(1)}_{0}$ with $f^{(1)}_{0}$. 
Then using the addition principle of Grassmann integration \cite{GM01} and letting   $\varphi=\varphi^{(0)}+\varphi'$, $\varphi':=\varphi^{(\leq -1)}$ we rewrite \eqref{eq:4.62} as
\begin{equation}
    \eqref{eq:4.62}=e^{L^{2}(F_\eps^{(-1)}+t_\eps^{(-1)})+S_\eps^{(-1)}(J)}\int P_\eps^{(\leq -1)}(D \varphi')e^{V_\eps^{(-1)}(\varphi',J)}\label{eq:2.24bis}
\end{equation}
where $P_\eps^{(\leq -1)}$ has covariance $g^{(\leq -1)}_\eps$ obtained from \eqref{eq:4.64p} by replacing $\chi^{(1)}_{0}$ with $\chi^{(1)}_{-1}$ (note that ${\boldsymbol r}_{-1}(q)={\boldsymbol r}_{-1}(0)=:r_{-1}$ in the support of $\chi^{(1)}_{-1}$, and similarly for ${\boldsymbol \z, \boldsymbol \a}$) while 
\begin{equation}
\hspace{-8pt}
     L^{2}\widetilde F^{(-1)}+\widetilde S_\eps^{(-1)}(J)+V_\eps^{(-1)}(\varphi',J) = \sum_{n \geq 1} \frac{1}{n!}\mathcal{E}^T_{0} \left (\mathcal{V}_\eps^{(0)}(\varphi'+\cdot,J); n \right ) \label{eq:4.68bis}
\end{equation}
with $\widetilde F_\eps^{(-1)}:=F_\eps^{(-1)}-F_\eps^{(0)}$, $\widetilde S_\eps^{(-1)}(J):=S_\eps^{(-1)}(J)-S_\eps^{(0)}(J)$,  $\mathcal{V}_\eps^{(0)}(\varphi,J):=$ $2^{-1}\nu_{-1} N(\varphi) +$ $\sum_{s}Y_{-1,s;\eps}F_{Y,s}(\varphi,J)$ $+\mathcal{R}_{1,\eps}V_\eps^{(0)}(\varphi)$
 and  $\mathcal{E}^T_{0}$ is the truncated expectation associated with the Gaussian integration $\tilde{P}_\eps^{(0)}$ with propagator $\tilde g^{(0)}_\eps$;
 moreover, $\widetilde S_\eps^{(-1)}(J)$ and $\mathcal{V}_\eps^{(-1)}(\varphi,J)$ are normalized in such a way that ${V}_\eps^{(-1)}(0,J)=\widetilde S_\eps^{(-1)}(0)=0$ that, 
 together with \eqref{eq:4.68bis}, fixes $\widetilde F_\eps^{(-1)}$, $\widetilde S_\eps^{(-1)}(J)$ and $V_\eps^{(-1)}(\varphi,J)$ uniquely.
 
 As a consequence of the localization procedure \eqref{eq:loc1reg} and the rewriting \eqref{eq:3.3.30}, the effective potential can be written as $\widetilde S_\eps^{(-1)}(J)+V_\eps^{(-1)}(\varphi,J)=\sum_{\substack{n,m \in \mathbb{N}, n+m \geq 1}} V^{(-1)}_{2n,m;\eps}(\varphi,J)$ with \begin{equation}\label{eq:3.3.49}
    \begin{split}
         & V_{2n,m;\eps}^{(-1)}(\varphi,J):=\int d\un y d\un z d \un s d \un D\,  \bar \partial^{\un D} \varphi_{\un y} J_{\un z, \un s}W^{(-1)}_{2n,m;\eps, \un s,\un D}(\un y,\un z) 
        \end{split}
        \end{equation}
        where $\bar \partial^{\un D}\varphi_{\un y}:=\prod_{j=1}^{2n}\bar \partial^{D_{j}}\varphi^{\tilde \sigma_j}_{y_{j}}$ with $\tilde \sigma_j=+$ for $j$ odd while $\tilde \sigma_j=-$ otherwise and  $\bar \partial^{D_j} \varphi^\pm_y:=\bar \partial^{D_{j1}}_1\bar \partial^{D_{j2}}_2 \varphi^\pm_y$; $\int d \un D$ stands for $\sum_{\un D}$ where $\un D=\{D_1,\dots,D_{2n}\}$, with $D_j=(D_{j1},D_{j2})\in \{0,1\}\times\{0,1,2\}$, which 
        keeps track of the action of the derivatives arising  from the interpolation procedure, see \eqref{eq:3.3.30bis} and \eqref{eq:interpopsi2}; $J_{\un z,\un s}$ and the other conventions are defined as in \eqref{eFPotent(0)} and following lines. Note as in \eqref{eFPotent(0)} that the kernels $W^{(-1)}_{2n,m;\eps,\un s,\un D}$ include the oscillating factors, i.e. $W^{(-1)}_{2n,m;\eps,\un s,\un D}=e^{\sum_{j=1}^{2n} i \tilde \sigma_j q_F y_{j1}} \widetilde W^{(-1)}_{2n,m;\eps,\un s,\un D},$ with $\widetilde W^{(-1)}_{2n,m;\eps,\un s,\un D}$ being translation invariant and $\vartheta_0$-periodic in each variable, see the discussion after  \eqref{eFPotent(0)}.
        The representation \eqref{eq:3.3.49} is a priori not unique (e.g., in the terms not involving the $J$ field, the discrete derivatives can be moved by integration by parts from one $\varphi$ field to another) but if needed it can be made so by, e.g., symmetrizing over all possible ways of choosing the derivative indices; however, this won't be needed in the following. If desired, the effective potential can be rewritten in Fourier space:
        \begin{equation}
             V^{(-1)}_{2n,m;\eps}(\varphi,J)=\int d \un s d \un q d \un p\,  \widehat{W}_{2n,m;\eps,\un s}^{(-1)}(\un q, \un p)\hat\varphi_{\un{q}+\un{q_F}}\hat J_{\un p,\un s}\delta({\underline{q}},\un p),\label{eq:2.27biss}
        \end{equation}
        where $\hat W^{(-1)}_{2n,m;\eps,\un s}$ is a combination of the Fourier transforms of the kernels $\widetilde W^{(-1)}_{2n,m;\eps, \un s, \un D}$ and includes the factors coming from the effect of derivatives; all the other conventions are again as in \eqref{eFPotent(0)} and following lines. 
 
 \paragraph{The inductive statement} After the integration of the scales $h' \geq h+1$, with $h\ge h_{\eps}$ and $h_\epsilon$ a cutoff scaled defined below (see \eqref{eq:4.69}), we can rewrite \eqref{eq:4.20}  as 
 \be 
e^{\mathcal{W}_{\eps+\varepsilon_0,\l,L}(\AAA)}=e^{L^2( F_\eps^{(h)}+t^{(h)}_\eps)+S_\eps^{(h)}(J)} \int P_\eps^{(\leq h)}(D\varphi^{(\leq h)}) e^{V_\eps^{(h)}(\varphi^{(\leq h)},J)}
\label{eq:2.26bis}
 \ee
 where $P_\epsilon^{(\leq h)}$ has a covariance $g^{(\leq h)}_\eps$ defined by  the same expression as $g_\eps^{(\leq -1)}$, 
 modulo replacing $\chi^{(1)}_{-1}$ by $\chi^{(1)}_{h}(q):=\chi^{(1)}_{0}(2^{-h}q)$ and $r_{-1}$, $\a_{-1}$, $\z_{-1}$ by $r_h:=\boldsymbol{r}_h(0)$,
 $\a_h:=\boldsymbol{\a}_h(0)$,
 $\z_h:=\boldsymbol{\z}_h(0)$, respectively, where $\boldsymbol{r}_h(q),\boldsymbol{\a}_h(q),\boldsymbol{\z}_h(q)$ are inductively defined below.
The constant $t^{(h)}_\eps$ is obtained by the iterative normalization of the Grassmann integration while the effective interaction is recursively defined by \begin{equation}
\hspace{-8pt}
     L^{2}\widetilde F_\eps^{(h)}+\widetilde S_\eps^{(h)}(J)+V_\eps^{(h)}(\varphi',J) = \sum_{n \geq 1} \frac{1}{n!}\mathcal{E}^T_{h+1} \left (\mathcal{V}_\eps^{(h+1)}(\varphi'+\cdot,J); n \right ) \label{eq:4.68}
\end{equation}
where $\widetilde F_\eps^{(h)}=F_\eps^{(h)}-F_\eps^{(h+1)}$,
$\widetilde S_\eps^{(h)}(J)=S^{(h)}_\eps(J)-S^{(h+1)}_\eps(J)$, $\mathcal{V}_\eps^{(h+1)}(\varphi,J):=2^{h}\nu_{h} N(\varphi)$ $+\sum_{s}Y_{h;\eps,s}F_{Y,s}(\varphi,J)$ $+\mathcal{R}_{1,\eps}V_\eps^{(h+1)}(\varphi,J)
$,  and satisfies  $S_\eps^{(h)}(0)=V_\eps^{(h)}(0,J)=0$; moreover, \be 2^h \nu_h:=\hat W^{(h+1);\io}_{2,0;0}(0), \qquad  Y_{h;\eps,s}:=\hat W^{(h+1);\io}_{2,1;\eps}(0,0),\label{eq:rcc1regime} \ee
and $\mathcal{R}_{1,\eps}V^{(h+1)}_\eps$ admits a representation as in \eqref{eq:3.3.30}-\eqref{eq:interpopsi2} with scale label $0$ replaced by $h+1$, where the operator $\mathcal{T}$, in analogy with \eqref{eq:operatorT}, is defined as follows:
\be \mathcal{T} ( \hat W_{2,0;\epsilon}^{(h+1);\io}(0)-\hat W_{2,0;0}^{(h+1);\io}(0)):= \sum_{k > h} r_k  \partial_{\textbf{r}_k}\hat W^{(0);\io}_{2,0;0}(0). \label{eq:operatorTscaleh}\ee 
 The truncated expectation $\mathcal{E}^T_{h+1}$ in the right side of \eqref{eq:4.68} is the one associated with the Gaussian integration $\tilde P_\epsilon^{(h+1)}$ with propagator    $\tilde g^{(h+1)}_\eps(y,y')=\int \tilde P^{(h+1)}(D\varphi) \varphi_y^-\varphi^+_{y'}$ given by
 \begin{equation}
 \tilde{g}_\eps^{(h+1)}(y,y')=\int dq \frac{e^{-iq(y-y')}f^{(1)}_{h+1}(q)}{\boldsymbol{r}_{h}-iq_1(1+ {\boldsymbol \z}_{h}) +2^{-1}q_2^2(1+{\boldsymbol \a}_{h})+\rho(q)}
 \label{eq:singlescalepropI}
\end{equation}
where $f^{(1)}_{h+1}=\chi^{(1)}_{h+1}-\chi_h^{(1)}$,  $\rho(q)$ is defined as in \eqref{eq:2.26biss}, while
\be  \begin{split}  & {\boldsymbol r_{h}(q)}=r_{h+1}-\chi^{(1)}_{h+1}(q)\sum_{k =h+1}^0r_k \partial_{\boldsymbol{r}_k} \hat W^{(h+1);\io}_{2,0;0}(0), \qquad \boldsymbol{r}_0:=:r_0:=r
\\ & {\boldsymbol \z}_{h}(q)=\z_{h+1}-i \chi^{(1)}_{h+1}(q) \partial_{1} \hat W^{(h+1);\io}_{2,0;0}(0), \qquad {\boldsymbol \a}_{h}(q)=\a_{h+1}-\chi^{(1)}_{h+1}(q) \partial^2_{2} \hat W^{(h+1);\io}_{2,0;0}(0). \end{split}\label{eq:2.32bis}\ee
 $\widetilde S^{(h)}_\eps(J)+V_\eps^{(h)}(\varphi,J)$ admits a representation analogous to \eqref{eq:3.3.49}-\eqref{eq:2.27biss} with scale label $-1$ replaced by $h$ and in Fourier representation the kernels $\hat W^{(h)}_{2n,m;\eps,\underline s}(\un q,\un p)$  satisfy the following symmetries (recall \eqref{eq:S1S2bis} and see App. \ref{app:symmetries} for a proof): 
\be   \begin{split}
        & \hat W^{(h)}_{2n,m;\epsilon, \un s}(\underline{q},\un p)=\overline{\hat W^{(h)}_{2n,m;\epsilon, \un s}(-\underline{q},-\un p)} , \\ & \left [\hat W^{(h)}_{2n,m;\epsilon, \un s}(\underline{q},\un p)\right ]_{\vartheta_0}=\left [\hat W^{(h)}_{2n,m;\epsilon,\un s^*}(\underline{\sigma_3 q},\underline{\sigma_3 p})\right ]_{\sigma_1 \vartheta_0}  \label{eq:2.33bis}
        \end{split}
\ee
where $\sigma_1,\sigma_3$ are the first and third Pauli matrices, respectively, and $\underline{\sigma_3 q}:=(\sigma_3 q_1,\dots, \sigma_3 q_{2n})$; $\underline{s}^*=(s^*_1,\dots,s^*_m)$, with $s^*_i=A$ if $s_i=A$, while $s^*_i=B$ if $s_i=C$ and viceversa; in the second line we specified the boundary condition $\vartheta_0$, because of the definition of the symmetry $\SS_2$, see \eqref{eq:S1S2bis}, and recall that the multiscale analysis for $\sigma_1 \vartheta_0=(0,1)$ is the same as that of $\vartheta_0$. Note that, since as $L\to \io$ the kernels are independent of the boundary condition $\vartheta_0$, then  \eqref{eq:2.33bis}   guarantees that ${\boldsymbol r}_h, {\boldsymbol \a}_h, {\boldsymbol \z}_h \in \mathbb{R}$ (and, therefore, $r_h,\a_h,\z_h$ are real as well) and that $\partial_2 W^{(h+1);\io}_{2,0:0}(0)=0.$

\begin{remark}
\label{rem:3} The effective constants
$\a_h,\z_{h},\nu_{h}$ of this paper  correspond to $\a_{h},z_{h},{\mu_{h}} _{\rvert{r=0}}$ in \cite[Sec. 2]{BM36}, respectively. The main differences between 
the construction of this paper and the one of \cite{BM36} are: (i) we keep $\nu_{h}$ in the interaction while the new constant $r_{h}$ is moved at each step in the Gaussian integration, thus contributing to 
the iterative dressing of the propagator, and in particular to the iterative shift of its singularities in momentum space; (ii) the presence of the external fields $J$, which in particular induces the presence of the local term proportional to $F_{Y,s}$, 
with coupling $Y_{h;\eps,s}$. Differently from \cite[eq.(98) and below]{BM36}, where a symmetry guarantees that the location of the transition from massless to massive phase is not modified by the interaction, in our case the transition takes place at $\varepsilon=\varepsilon_0(\lambda)\neq0$;
the criticality condition is equivalent to 
fixing appropriately the initial datum on $\nu$, $\nu=\nu_0^*(\lambda)$, with $\nu_0^*(\lambda)$ fixed uniquely by a fixed point argument: the choice of such an initial datum guarantees that the flow of 
$\nu_h$ remains bounded, uniformly in the scale index and in the choice of $\epsilon$. 
This is discussed in the next section.
\end{remark}

\begin{remark}\label{rem:kernelsdependences} As it follows from the construction, the kernels $W^{(h)}_{2n,m;\eps,\un s}$ of the effective potentials \eqref{eq:4.68} can be seen as functions of the running coupling `functions' at higher scales,  $\{\boldsymbol \a_{h'},\boldsymbol \z_{h'}, \boldsymbol r_{h'}\}_{h'> h}$, via the single scale propagators $\tilde g^{(h)}$ (see \eqref{eq:singlescalepropI}) and of the running coupling constants $\{\nu_{h'},Y_{h';\eps,s}\}_{h' > h}$ appearing in the interaction, see  \eqref{eq:4.68} and following line. 
Note in particular that  the dependence on $\eps$ is all mediated by the couplings $\{\boldsymbol{r}_{h'}\}_{h'= h+1}^0$ appearing in the single scale propagators \eqref{eq:singlescalepropI} and by the bare interaction $\tilde V_\eps$ (see item \ref{item2after} after \ref{inftvolprop}). In order to emphasize that the dependence of the kernels upon $\eps$ is mediated by 
$\un r_{h+1}:=\{\boldsymbol{r}_{h'}\}_{h'=h+1}^0$, we will sometimes write $\hat W^{(h+1);\io}_{2,0;\eps,\un s}(0)=\hat W^{(h+1);\io}_{2,0;\un s}(0;\un r_{h+1})$.
\end{remark}

  \paragraph{The flow of $r_h$ and the scale $h_\eps$.}\label{par:florrscalaheps}
By definition, the flow of $r_h=\boldsymbol{r}_h(0)$ is linear in the sequence $\{r_{h'}\}_{h' \geq h}$ (cf. \eqref{eq:2.32bis}) so we can write \[
r_h=r Q_h(\l),\] for some $Q_h$ independent of $\eps.$ We will show that for $|\l|$ small enough
 $|Q_h-1|\leq C|\l|$ for some $C>0$ independent of $h,\eps$ (see \eqref{eq:2.29bis} and App. \ref{app:A2}), which implies that \textit{$r_h$ remains close to the initial datum $r$ uniformly in $h$}; more precisely, $r_h=-\epsilon(1+O(\eps,\lambda))$. We will also prove that $|\alpha_h|,|\zeta_h|\le C|\lambda|$ uniformly in $h$.  
 For any $\eps\neq0$, the multiscale procedure described above stops at the largest negative scale such that, 
 on the support of $f_{h+1}^{(1)}$, the norm 
 of ${\boldsymbol{r}_{h}-iq_1(1+ {\boldsymbol \z}_{h}) +2^{-1}q_2^2(1+{\boldsymbol \a}_{h})+\rho(q)}$, which is nothing but the denominator of the Fourier transform of $\tilde g_\epsilon^{(h+1)}$, see \eqref{eq:singlescalepropI}, is of the same order as the norm of $iq_1+q_2^2/2$. Given the definition and properties of 
 ${\boldsymbol{r}}_{h}$, ${\boldsymbol{\z}}_{h}$, 
 ${\boldsymbol{\a}}_{h}$, $\rho(q)$, and $f_{h+1}^{(1)}$, we can choose such a scale to be
\begin{equation}\label{eq:4.69}
    h_\eps:=\inf\left \{h \in \mathbb{Z}: 2^{h}>|\eps| \right \},
\end{equation}
and we shall do so in the following. 
The reader can check that for $\eps>0$ such a definition guarantees that the support of $\chi^{(1)}_{h_\eps}$ contains the two singularities of the Fourier transform of $g^{(\le h_\eps)}_\eps$, and that these are at a distance $O(\sqrt\eps)$ from the complement of the support of $\chi^{(1)}_{h_\eps}$, as shown in the right part of Fig. \ref{fig:enter-label}. Note that $2^{h_\eps} \asymp |\eps|$. If $\epsilon=0$, then $h_\eps=-\infty$; in this case, the multiscale procedure described above continues until we reach the first scale below which the support of $\chi_h^{(1)}$ no longer contains points of $\DD'_{\vartheta_0,R}$, which are at mutual distance of order $L^{-1}$; we denote such a scale by 
$\tilde h_L$, which is such that $\tilde h_L
\asymp -\log L$. 

\paragraph{The transition to the second regime}

Let us consider the case $\eps\neq0$, distinguishing between the cases of $\eps$ positive or negative. 

\medskip 

\par{If  $\boldsymbol{\eps<0}$}: In this case the Fourier transform of the propagator $g^{(\leq h_\eps)}_\eps$ associated with $P_\eps^{(\leq h_\eps)}$ in  \eqref{eq:2.26bis} at $h=h_\eps$ is non-singular in momentum space. 
More precisely, $g^{(\leq h_\eps)}_\eps$
 satisfies the same bounds as those stated in Lemma \ref{lem:propag1} for $\tilde g^{(h_\eps+1)}_\eps$ \eqref{eq:singlescalepropI}. We thus integrate all the scale smaller than $h_\eps$ in one step and obtain \begin{equation}
    \mathcal{W}^{(  \vartheta_0)}_{\eps+\varepsilon_0,\l,L}(\AAA(J))=L^2(F_\eps^{(h_\eps-1)}+t_\eps^{(h_\eps)})+S^{(h_\eps-1)}_\eps(J)\label{eq:epsnegativo}
\end{equation}
where $L^{2}\widetilde F_\eps^{(h_\eps-1)}+\widetilde S_\eps^{(h_\eps-1)}(J) = \sum_{n \geq 1} \frac{1}{n!}\mathcal{E}^T_{\leq h_\eps} (V^{(h_\eps)}_\eps(\cdot,J); n )$ with $\widetilde F_\eps^{(h_\eps-1)}=F_\eps^{(h_\eps-1)}-F_\eps^{(h_\eps)}$,
$\widetilde S_\eps^{(h_\eps-1)}=S_\eps^{(h_\eps-1)}-S_\eps^{(h_\eps)}$, $\widetilde S^{(h_\eps-1)}_\eps(0)=0$ and $\mathcal E^T_{\le h_\eps}$ the truncated expectation w.r.t. $g_\eps^{(\leq h_\eps)}$. This concludes the multiscale integration procedure in the case of negative $\eps$. 

\medskip 

\par{If $\boldsymbol{\eps>0}$}: In this case, as mentioned above,
the Fourier transform of the propagator $g^{(\leq h_\eps)}_\eps$ has two singularities. For this reason, the integration procedure 
involving scales smaller than $h_\eps$ still needs to be performed in a multiscale fashion, although different from the one described above. In the second regime, the local terms produced by such a multiscale integration iteratively induce a shift of 
the two singularities. 
For technical convenience, in order to identify the location of the singularities of the `dressed theory' resulting from the integration of all the infrared degrees of freedom, 
we introduce a \emph{counterterm} $\mu_{h_\eps}$ such that $|\mu_{h_\eps}|\le C_\theta|\lambda|\eps^\theta$ for any 
$\theta\in[0,1/2)$ and some $C_\theta>0$, to be fixed 
appropriately below, see Remark \ref{rem:4} . We proceed as follows: our starting point is \eqref{eq:2.26bis} at $h=h_\eps$, which we rewrite as
\begin{equation}
    \eqref{eq:2.26bis}=e^{L^2 (F^{(h_\eps)}+ t^{(h_\eps-1)})+S^{(h_\eps)}(J)}\int \tilde P_\eps^{(\leq h_\eps)} (D\varphi)e^{V_\eps^{(h_\eps)}(\varphi,J)+\eps\mu_{h_\eps} N(\varphi)}
    \label{eq:2.34}
\end{equation}
where $\tilde P_\eps^{(\leq h_\eps)}$ has covariance $\tilde g_\eps^{(\leq h_\eps)}(y,0)$ obtained from \eqref{eq:4.64p} by replacing the scale label $-1$ with $h_\eps-1$ and by adding a term 
$\eps \mu_{h_\eps}\chi^{(1)}_{h_\eps}(q)$ to the denominator of the right hand side (the constant $ t^{(h_\eps-1)}$ accounts for the change in the normalization of the reference Gaussian Grassmann integration). The singularities of the Fourier transform of  $\tilde g_\eps^{(\leq h_\eps)}$ are now $\pm  \boldsymbol{q}_\eps=(0,\pm\tilde q_\eps)$, where $\tilde q_\eps$ is the positive solution to \be
r_{h_\eps-1}+\frac{\tilde q_\eps^2}{2}\a_{h_\eps-1}+1-\cos \tilde q_\eps+ \eps\mu_{h_\eps}=0. 
\label{eq:2.35}\ee
Next we split $\chi_{h_\eps}^{(1)}(q)=\tilde f_{h_\eps}^{(1)}(q)+\sum_{\o=\pm}\chi^{(2)}_{h_\eps-1}(q-\o\boldsymbol{q}_\eps)$, thus implicitly defining $\tilde f^{(1)}_{h_\eps}$,  where  $\chi_{h_\eps-1}^{(2)}(q):=\chi(\d2^{-h_\eps+1}\|q\|_{(2)})$, $\chi$ is the same as in Sect. \ref{ssec:uv}, $\|q\|^2_{(2)}:=q_1^2+v^2_\eps q_2^2$, with $v_\eps$ to be defined momentarily, and $\d$ is a large enough constant\footnote{We take $\d$ large enough that $\cap_\o S_\o = \emptyset$ and $\chi^{(1)}_{h_\eps}=1$ in $\cup_\o S_\o$ where $S_\o:=\{q:  \chi^{(2)}_{h_\eps-1}(q-\o \boldsymbol{q}_\eps)\neq 0\}$.}, independent of $\eps,\lambda$.
We correspondingly rewrite the propagator as: 
\begin{equation}
\tilde{g}_\eps^{(\leq h_\eps)}(y,z)=\sum_{\o=\pm}e^{-i\omega (y-z) \boldsymbol{q}_\eps}g^{(\leq h_\eps-1)}_{\o}(y,z)/Z_{h_\eps-1}+\tilde g^{(h_\eps)}_\eps(y,z),\label{eq:2.55zz}
\end{equation}
where $Z_{h_\eps-1}=\cos \tilde q_\eps+\zeta_{h_\eps-1}$ and, 
letting $D_{\o}(q):=-iq_1+\o q_2 v_\eps$ with $v_\eps=Z_{h_\eps-1}^{-1}(\tilde q_\eps \a_{h_\eps-1}+\sin \tilde q_\eps)$, 
\begin{equation}
g^{(\leq h_\eps-1)}_{\o}(y,z)=\int_\o dq\frac{e^{-iq(y-z)} \chi^{(2)}_{h_\eps-1}(q)}{D_{\o}(q)+\rho_{\o}(q)/Z_{h_\eps-1}}. 
\label{eq:2.37}
\end{equation}
Here $\int_\o dq :=L^{-2} \sum_{q \in \mathcal{D}'_{R,\vartheta_0,\o}}$
and, recalling the definition of $\rho$ in \eqref{eq:2.26biss} and  letting $\tilde D_\eps(q):=r_{h_\eps-1}-iq_1(1+\z_{h_\eps-1})+2^{-1}q_2^2(1+\a_{h_\eps-1})+\rho(q)+\eps \mu_{h_\eps}$, we defined
$\rho_{\o}(q):=\tilde D_\eps(q+\o \boldsymbol{q}_\eps)-\tilde D_\eps(\o \boldsymbol{q}_\eps)-q\cdot\partial \tilde D_\eps(\o \boldsymbol{q}_\eps)$. 
We now use the addition principle associated with the decomposition \be
    \varphi_y^{\pm,(\leq h_\eps)}=\sum_{\omega=\pm}e^{\pm i\omega y \boldsymbol{q}_\eps}\varphi^{\pm,(\leq h_\eps-1)}_{y,\omega}+{\varphi}^{\pm,(h_\eps)}_y, 
    \label{eq:qp}\ee 
where 
$\varphi^{\pm,(\leq h_\eps-1)}_{y,+}$, $\varphi^{\pm,(\leq h_\eps-1)}_{y,-}$, ${\varphi}^{\pm,(h_\eps)}_y$
are three independent Gaussian Grassmann fields with propagators 
$g^{(\leq h_\eps-1)}_{+}/Z_{h_\eps-1}$,
$g^{(\leq h_\eps-1)}_{-}/Z_{h_\eps-1}$, $\tilde g^{(h_\eps)}_\eps$, respectively, 
thus obtaining 
\begin{equation}
\eqref{eq:2.34}=e^{L^2 (F_\eps^{(h_\eps-1)}+t_\eps^{(h_\eps-1)})+S_\eps^{(h_\eps-1)}(J)}\int P_{Z_{h_\eps-1}}^{(\leq h_\eps-1)}(D\varphi) e^{ V_\eps^{(h_\eps-1)}(\varphi,J)}\label{eq:4.91}
\end{equation}
where the field $\varphi$ over which 
we are integrating in the right hand side is labelled by an index $\omega$, and 
$P_{Z_{h_\eps-1}}^{(\leq h_\eps-1)}$ 
is the Gaussian Grassmann integration with propagator 
$\int P_\eps^{(\leq h_\eps-1)}(D\varphi) \varphi^-_{y,\omega}\varphi^+_{z,\omega'}=\frac{\delta_{\o,\o'} }{Z_{h_\eps-1}}g^{(\leq h_\eps-1)}_{\o}(y,z)$. Moreover, 
$\widetilde F_\eps^{(h_\eps-1)}$,
$\widetilde S_\eps^{(h_\eps-1)}$,
$V_\eps^{(h_\eps-1)}$ satisfy 
the analogue of \eqref{eq:4.68} at $h=h_\eps-1$ and, as usual, are normalized in such a way that $S_\eps^{(h_\eps-1)}(0)=
V_\eps^{(h_\eps-1)}(0,J)=0$. Due to the normalization constant 
$Z_{h_\eps-1}$ appearing in the 
reference Grassmann integration, it is convenient to rescale the effective potential as follows: 
$V_\eps^{(h_\eps-1)}(\varphi,J)\equiv 
\mathcal V^{(h_\eps-1)}(\sqrt{Z_{h_\eps-1}}\varphi,J)$, 
so that 
\begin{equation}
\eqref{eq:4.91}=e^{L^2 (F_\eps^{(h_\eps-1)}+ t_\eps^{(h_\eps-1)})+S_\eps^{(h_\eps-1)}(J)}\int P_{Z_{h_\eps-1}}^{(\leq h_\eps-1)}(D\varphi) e^{\mathcal V^{(h_\eps-1)}(\sqrt{Z_{h_\eps-1}}\varphi,J)}
\label{eq:4.91bbis}
\end{equation}
The effective potential admits a representation analogous to 
\eqref{eq:3.3.49}-\eqref{eq:2.27biss}; more precisely, 
we shall write
$\widetilde S_\eps^{(h_\eps-1)}(J)+\mathcal V^{(h_\eps-1)}(\varphi,J)=\sum_{\substack{n,m \in \mathbb{N},n+m \geq 1}} V^{(h_\eps-1)}_{2n,m;\eps}(\varphi,J)$ with  \begin{equation}\begin{split}
          V_{2n,m;\eps}^{(h_\eps-1)}(\varphi,J)& :=\int d\un y d\un z d \un s d \un \o d \un D\,  \partial^{\un D}(e^{i\un \o {\boldsymbol q}_\eps \un y}  \varphi_{\un y,\un \o}) J_{\un z, \un s}W^{(h_\eps-1)}_{2n,m;\eps, \un s,\un D}(\un y, \un z) \\ & \equiv\int d\un y d\un z d \un s d \un \o d \un D\,  \partial^{\un D} \varphi_{\un y,\un \o} J_{\un z, \un s}W^{(h_\eps-1)}_{2n,m;\eps,\un s,\un \o, \un D}(\un y, \un z), 
        \end{split}
        \label{eq:2.39bis}
        \end{equation}
 where  $\partial^{\un D}(e^{i\un \o {\boldsymbol q}_\eps \un y}  \varphi_{\un y,\un \o}):=\prod_{j=1}^{2n}\bar \partial^{D_{j}}(e^{i\tilde \sigma_j \o_{j}\boldsymbol{q_\eps}y_{j}}\varphi^{\tilde \sigma_j}_{y_{j},\o{_{j}}})$, and the notation $\bar \partial^{D_j}$ is analogous to the one in \eqref{eq:3.3.49}; moreover $\int d \un \o$ denotes $\sum_{\un \o \in \{-1,1\}^{2n}}$, while the other conventions are as in \eqref{eq:3.3.49},\eqref{eFPotent(0)}.
 The kernels $W^{(h_\eps-1)}_{2n,m;\eps,\un s,\un \o, \un D}(\un y, \un z)$ can be written as $W^{(h_\eps-1)}_{2n,m;\eps, \un s,\un \o, \un D}(\un y, \un z)=e^{i\sum_{j=1}^{2n}\tilde \sigma_j( \o_j \boldsymbol{q}_\eps+q_F) y_j} \widetilde W^{(h_\eps-1)}_{2n,m;\eps, \un s,\un \o, \un D}(\un y, \un z)$ with $\widetilde W^{(h_\eps-1)}_{2n,m;\eps, \un s,\un \o, \un D}$ translation invariant, $\vartheta_0$-periodic and defined as
 \be 
 \begin{split}
 \hspace{-6pt}& 
 \widetilde W^{(h_\eps-1)}_{2n,m;\eps, \un s,\un \o, \un D}(\un y, \un z)=\\ &\int^\ast d \un D'  W^{(h_\eps-1)}_{2n,m;\eps,\un s, \un D'}(\un y+\un \t,\un z) \, \prod_{j=1}^{2n} \binom{D'_{j2}}{D'_{j2}} e^{2i\tilde \sigma_j \o_j \tilde q_\eps D_{j2}} e^{i\tilde\sigma_j \t_j(\o_j\boldsymbol{q}_\eps+q_F)} \Big (\frac{i\o_j \tilde\sigma_j\sin 2\tilde q_\eps}{2}\Big )^{D'_{j2}-D_{j2}} ,
 \label{eq:2.44bis}
 \end{split}
 \ee 
 where $^\ast$ denotes the constraint $D'_{j1}=D_{j1}, D'_{j2}\geq D_{j2}$, for all $j=1,\dots,2n$, while $\un \t=\un \t(\un D,\un D')=(\t_1,\dots,\t_{2n})$ with $\t_j=2(0,D'_{j2}-D_{j2})$; in passing from the first to the second line of \eqref{eq:2.39bis} with the definition in \eqref{eq:2.44bis} we used that $\boldsymbol{q}_\eps=(0,\tilde q_\eps)$, the Leibniz rule for $\bar \partial_2$ with $\bar \partial_2 e^{i \o \boldsymbol{q}_\eps y}=\frac{i \o \sin 2\tilde q_\eps}{2}\, e^{i \o \boldsymbol{q}_\eps y}$, and we exchanged the summation over the derivative labels\footnote{For any $f,g$ one has  $\bar \partial^n_2 (fg)_y=\sum_{m=0}^n \binom{n}{m} \bar \partial_2^m f_{y-(0,2(n-m))} \bar \partial_2^{n-m} g_{y+(0,2m)}$.}.
   If desired, in Fourier space one can equivalently rewrite
        \begin{equation}
             V^{(h_\eps-1)}_{2n,m;\eps}(\varphi,J)=\int_{\un \o} d \un s d \un q d \un p d \un \o\,  \hat{W}_{2n,m;\eps,\un s, \un \o}^{(h_\eps-1)}(\un q,\un p)\hat\varphi_{\un{q},\un \o}\hat J_{\un p,\un s}\delta ({\underline{q}+\un \o \boldsymbol{q}_\eps},\un p)\label{eq:2.27bisss}
        \end{equation}
 where the pedex ${\un \o}$ in the integration symbol means that $\un q=(q_1,\dots, q_{2n})$ is summed (with the appropriate normalization $L^{-2n}$) over the set $(\mathcal{D}'_{R,\vartheta_0,\o_i})^n$, 
 with $\mathcal{D}'_{R,\vartheta_0,\o_i}:=\mathcal{D}'_{R,\vartheta_0}-\omega_i\boldsymbol{ q}_\eps$; moreover letting $\f^\pm_{q,\o}:=\f^{\pm}_{q+\o \tilde q_\eps}$, $\hat \varphi_{\un q,\un \o}:=\prod_{i=1}^n \hat \varphi^+_{q_{2i-1},\o_{2i-1}}\hat \varphi^-_{q_{2i},\o_{2i}}$,  $\hat J_{\un p,\un s}=\prod_{i=1}^m J_{p_i,s_i}$, $\un q+\un \o \boldsymbol{ q}_\eps:=(q_1+\o_1\boldsymbol{  q}_\eps,\dots,q_{2n}+\o_{2n}\boldsymbol{  q}_\eps)$;   $\hat W^{(h_\eps-1)}_{2n,m;\eps,\un s, \un \o}(\un q,\un p):=\int d \un D \hat{\widetilde{W}}{}^{(h_\eps-1)}_{2n,m;\eps,\un s,\un \o,\un D}(\un q,\un p) \, \prod_{j=1}^{2n}\prod_{\ell=1}^2 [i\tilde\sigma_j 2^{-1}\sin 2 q_{j\ell}]^{D_{j\ell}}$ with $\hat{\widetilde{W}}^{(h_\eps-1)}_{2n,m;\eps,\un s,\un \o,\un D}(\un q,\un p)$ being the Fourier transform of $\widetilde W^{(h_\eps-1)}_{2n,m;\eps,\un s ,\un \o, \un D}(\un y,\un z)$ computed at momenta $(q_1+\o_1 \boldsymbol{q}_\eps+q_F,\dots,q_{2n}+\o_{2n} \boldsymbol{q}_\eps+q_F,p_1\dots,p_m)$ and we are using the convention $q_j=(q_{j1},q_{j2})$ and similarly for $D_j$, while all the other expressions are as in \eqref{eq:3.3.49},\eqref{eFPotent(0)}.

\begin{remark}
\label{rem:4}
The constants  $\a_{h_\eps-1},\z_{h_\eps-1},r_{h_\eps-1}$ appearing in \eqref{eq:2.35} and in the definitions of $Z_{h_\eps-1}$, $v_\eps$ and $\rho_\omega(q)$
are known functions, obtained via the iterative procedure described above (see \eqref{eq:2.32bis} at $q=0$) starting from their values at scale $h=0$, while $\tilde q_\eps$ and $\mu_{h_\eps}$ are a priori arbitrary, although related by \eqref{eq:2.35}. They will be fixed conveniently below: the counterterm $\mu_{h_\eps}$ will be uniquely determined by a fixed point equation, whose solution will be denoted by $\mu_{h_\eps}^*(\eps,\l,\tilde q_\eps)$, and will be shown to satisfy  $|\mu^*_{h_\eps}|\leq C_\theta \eps^\theta|\l|$ for any $\theta \in (0,1/2)$, see Proposition \ref{lem:kernbounds2}.
Fixing $\mu_{h_\eps}=\mu_{h_\eps}^*(\eps,\l,\tilde q_\eps)$ guarantees that the flow of the running constants of the second regime remains bounded uniformly in the scale index. Moreover, plugging this 
choice of $\mu_{h_\eps}$ in \eqref{eq:2.35}, we obtain an implicit equation for the dressed Fermi point $\tilde q_\eps$, which admits, for $|\lambda|$ small enough, a unique positive solution $\tilde q_\epsilon(\lambda,\epsilon)$ in the vicinity of the origin, satisfying $\tilde q_\eps(\lambda,\eps) \asymp \sqrt{\eps}$. By computing all the involved quantities at $\tilde q_\eps=\tilde q_\eps(\lambda,\eps)$ and $\mu_{h_\eps}=\mu_{h_\eps}^*(\l,\eps,\tilde q_\eps(\l,\eps))$, we find $Z_{h_\eps-1}=1+O(\l)$ and   $v_\eps\asymp \sqrt{\eps}$ for $\eps>0$ small; a posteriori, the resulting expansion for the observables of the theory will only depend upon $\l$ and $\eps$, as desired. See the proof of Item 3 of Lemma \ref{lem:derivateepsilon} in Appendix \ref{app:regularitydensityeps} for more details on how  $\mu_{h_\eps}$ and $\tilde q_\eps$ are fixed.
\end{remark}

\begin{remark} By applying the symmetries $\SS_1,\SS_2$ (cf. \eqref{eq:S1S2bis}) to \eqref{eq:qp} one finds that their action on the variables $\varphi^\pm_{y,\o}$ is given by $\mathcal{S}_1: \varphi^{\sigma}_{y,\o} \longrightarrow \varphi^{\sigma}_{y,-\o},$ together with $c \to \overline{c}$ and $\mathcal{S}_2: \varphi^{\sigma}_{y,\o} \longrightarrow \varphi^{\sigma}_{\sigma_3 y,-\o}$ together with $ c \to c$ and $\vartheta_0 \to \sigma_1\vartheta_0$, for any constant $c \in \mathbb{C}$ appearing in the effective action. In particular one can check (see App. \ref{app:symmetries}) that the symmetries \eqref{eq:2.33bis} imply the following symmetries  of the kernels in \eqref{eq:2.27bisss}: 
\begin{equation}
\begin{split}
   &  \hat W^{(h_\eps-1)}_{2n,m;\eps,\un s,\un \o}(\un q,\un p)=\overline{\hat W^{(h_\eps-1)}_{2n,m;\eps,\un s,-\un \o}(-\un q,-
    \un p)}, \\ &  \left [\hat W^{(h_\eps-1)}_{2n,m;\eps,\un s,\un \o}(\un q,\un p)\right ]_{\vartheta_0}=\left [\hat W^{(h_\eps-1)}_{2n,m;\eps,\un s,-\un \o}(\underline{\sigma_3 q},
    \underline{\sigma_3 p})\right ]_{\sigma_1\vartheta_0},
    \label{eq:2.43bis}
    \end{split}
\end{equation}
where in the second line we used the same conventions as in \eqref{eq:2.33bis}.\label{rem:remark11}
\end{remark}

\paragraph{Dimensional bounds: first regime}
In order to derive a bound on $W^{(h)}_{2n,m;\eps,\un s}$, for $h \geq h_\eps-1$, it is necessary to have an estimate on the propagators involved in the truncated expectations \eqref{eq:4.68} at each scale $h \geq h_\eps$, as well as a Gram representation thereof.  
Given $\un n=(n_0,n_1,n_2) \in \mathbb{N}^3$ define \begin{equation}\label{eq:tildegnm}\tilde{g}^{(h);\un n;\un m}_{\eps}(y,0):=y_1^{m_1}y_2^{m_2}\partial^{n_0}_{\boldsymbol{r}_{h-1}}\bar{\partial}^{n_1}_1\bar{\partial}^{n_2}_2 \tilde{g}_{\eps}^{(h)}(y,y'),
    \end{equation}
    and the norms $\lVert \tilde g_{\eps}^{(h);\un n; \un 0} \rVert_{\infty}:=\sup_{y \in \tilde \Lambda} \rvert \tilde g_{\eps}^{(h);\un n}(y,0) \rvert, \lVert\tilde g_{\eps}^{(h);\un n;\un m}\rVert_{1}:=\sum_{y \in \tilde \Lambda} \lvert \tilde g_{\eps}^{(h);\un n;\un m} (y,0) \rvert$; moreover, given $y=(y_1,y_2)\in\tilde \Lambda$, 
    we let $\|y_i\|=\min_{n\in\mathbb Z^2}|y_i+nL|$. Then, 
     we have the following:
\begin{lemma}\label{lem:propag1}Let $h\geq h_\eps-1$ and suppose that there exists $C>0$, independent of $L,\eps$, such that \be  \max \{\sup_{h \leq h' \leq -1 }|\a_{h'}|,\sup_{h \leq h' \leq -1 }|\z_{h'}|\} \leq C|\l|, \qquad  \sup_{h \leq h' \leq -1} |r_{h'}-r| \leq C|\l| |r|.\label{eq:2.42bis}\ee  
Then there exist positive constants $c_0=c_0(\un n)$ and $c$, independent of $L,\eps$, such that 
\begin{equation}
| \tilde{g}_{\eps}^{(h);\un n; \un 0}(y,0)| \leq c_0 2^{h\left (\frac{1+n_2}{2}+n_1-n_0 \right )}e^{-c\sqrt{ d^{(h)}(y) }}
\label{eq:prop1reg}    \end{equation}
    where $d^{(h)}(y):=2^h\|y_1\|+2^{h/2}\|y_2\|$. As a consequence, $\| \tilde{g}_{\eps}^{(h);\un n; \un m}\|_1\leq c_1 2^{h(-1+\frac{n_2-m_2}{2}+n_1-m_1-n_0)}$ for some $c_1=c_1(\un n,\un m)$; in particular, $\|\tilde g^{(h)}_{\eps}\|_\io \leq c_0 2^{h/2}$ while $\|\tilde g^{(h)}_{\eps} \|_1 \leq c_1 2^{-h}$. 
Moreover, $\tilde{g}_{\eps}^{(h);\un n; \un 0}$ admits a Gram representation, i.e., there exists a Hilbert space $\mathcal H$ and two vectors $\gamma^{(h);\un n}_y,\eta^{(h)}_{y'}\in\mathcal H$ such that 
$\tilde{g}_{\eps}^{(h);\un n; \un 0}(y,y')=(\gamma^{(h);\un n}_y,\eta^{(h)}_{y'})$ and 
$\|\gamma^{(h);\un n}_y\|^2\le c_0(\un n) 2^{h\left (\frac{1+n_2}{2}+n_1-n_0 \right )}$, $\|\eta^{(h)}_{y'}\|^2\le c_0(\un 0) 2^{h/2}$, where $\|\cdot\|$ is the norm induced by the scalar product in $\mathcal H$. 
\end{lemma}
This is the counterpart of    \cite[Lemma 2.1]{BM36}, and can be proved in the same way; the Gram representation is the analogue of \cite[Eqs.(77)-(78)]{BM36}; the improved stretched exponential decay is a consequence of the Gevray-2 cutoff we chose in \eqref{eq:5.39}, see \cite[App. A]{GMR} for more details. 
Using Lemma \ref{lem:propag1}, one can derive the desired bound for the kernels of the effective potential in the first regime. 

\begin{proposition}
\label{lem:bounds1}{\cite{BM36}}
   Let $h \geq h_\eps-1$. There exists $\l_0,\bar \eps>0$ and two functions $C,C_1:[0,1/2)\to (0,\infty)$ such that, if  \eqref{eq:2.42bis} holds  together with $\sup_{h' \geq h}  2^{-\theta h'}|\nu_{h'}| \leq C(\theta) |\l|$ for $\theta\in [0,1/2)$, then, for $|\l|\leq \l_0$ and $\eps\le \bar\eps$, \begin{equation}
        \|W^{(h)}_{2n,m;\eps,\un s, \un D}\|\leq  [C_1(\theta)]^{n+m}|\l|^{d_{n,m}}2^{h(\frac{3}{2}-\frac{n}{2}-m+\d_{m,0}\theta-\|\un D\|^{(1)})} \left (\sup_{\substack{h' \geq h }}\sup_s |Y_{h';\eps,s}| \right )^m,
\label{eq:2.29}    \end{equation}
    where $\|\un D\|^{(1)}:=\sum_{i=1}^{2n}\sum_{j=1}^2 |D_{ij}|j^{-1}$, while $d_{n,m}$ and $\|\cdot \|$ are the same as in \eqref{kernestim}.
\end{proposition}

This result is the analogue of \cite[Lemma 2.2]{BM36} in our setting. Apart from minor technical points, among which the presence of the external fields $J$ (discussed in App. \ref{app:A2}), the proof is essentially identical to \cite{BM36}, and is based on the Gallavotti-Nicol\`o tree expansion of the kernels of the effective potential; the details are left to the reader. 

Finally, by properly and uniquely tuning the parameter $\nu$ as a function of $\l$ one can control the flow of the running constants in \eqref{eq:rcc1regime},  \eqref{eq:2.42bis}  for all the scales $h \geq h_\eps-1$:

\begin{proposition}[Flow of the running constants]
\label{lem:flowRC}
     There exists positive constants $\l_0,\bar \eps$ and a function $C:[0,1/2)\to(0,+\infty)$ s.t. for $|\l|\leq \l_0$  there exists a unique analytic function $\l \mapsto \nu^*_0(\l)$ of order $\lambda$ as $\lambda\to0$,      
     such that if $\nu=\nu^*_0(\l)$,  then for $h_\eps-1 \leq h \leq -1$ and any $\theta\in[0,1/2)$
    \be
     \begin{split} \max\left \{|\nu_h|,|\z_h-\z_{h+1}|,|\a_h-\a_{h+1}|,e^{\eps \d_{s,A}}|Y_{h;\eps,s}-Y_{h+1;\eps,s}|\right \}& \leq C(\theta)2^{\theta h}|\l|, \\  \left |r_h r^{-1}_0-1\right |& \leq C(0)|\l|
     \label{eq:2.29bis}\end{split}\ee
     uniformly in $\eps \leq \bar \eps,$ where $\d_{\cdot,\cdot}$ is the Kronecker's delta and we let $Y_{0;\eps,A}:=e^{-\eps-\varepsilon_0(\l)}, Y_{0;\eps,B}=Y_{0;\eps,C}:=\frac{1}{2}$.
\end{proposition}

In Appendix \ref{app:A2} we prove this proposition, also showing how to constructively compute $\nu^*_0(\l)$.  As already commented in item 1 below  \eqref{inftvolprop}, an important difference with \cite{BM36} is that here $\nu^*_0(\l)\neq 0$, which corresponds to a non-trivial shift $\varepsilon_0(\l)=-\log(1-\nu_0^*(\l))$ of the liquid-frozen transition line, see \eqref{eq:5.71} and Theorem \ref{th:5}.
Note that as a consequence of \eqref{eq:2.29bis}, $Y_{-\io;0,s}=\lim_{h \to -\io} Y_{h;0,s}$ exists, and similarly 
for $\alpha_{-\infty},\zeta_{-\infty}$; they are such that 
$Y_{-\io;0,s}=Y_{0;0,s}+O(\l)$ and 
$\a_{-\io},\z_{-\io}=O(\l)$. As discussed in Section \ref{sec:2.3.1} below, see in particular \eqref{eq:2.81}, these are the functions entering the definition of $A(\l)$ in Theorem  \ref{th:5}.     
 Finally observe that, thanks to \eqref{eq:2.29bis}, the factor $\sup_{\substack{h' \geq h }}\sup_s |Y_{h';\eps,s}|$ in \eqref{eq:2.29} can be bounded uniformly in $h$ and, therefore, it can be reabsorbed in the definition of $C_1$.

\subsubsection{Infrared integration: second regime}
\label{sec:secondregime} We consider the case $\eps>0$, otherwise there is no second regime at all. In order to lighten the notation, we drop the label $\eps$ from the kernels and running coupling constants, keeping the dependence upon $\eps$ implicit. 
We follow \cite[Sec. 3]{BM36}, keeping track of the presence 
of the external field $J$, whose effect was not discussed 
in \cite{BM36}. 

\medskip 

We start by discussing the first step of integration. For this purpose, we introduce a different \textit{localization operation} w.r.t.  \eqref{eq:loc1reg}; next, we will iterate the integration procedure used in 
the first step, and we shall thus obtain a general inductive statement for the scales $h \leq h_\eps-2$.

We start from \eqref{eq:4.91bbis} and we split 
$\mathcal V^{(h_\eps-1)}=\LL_2  \mathcal V^{(h_\eps-1)}+\RR_2  \mathcal V^{(h_\eps-1)}$, where, recalling the representation \eqref{eq:2.27bisss}, we let 
\begin{equation}
\begin{split}
\LL_2\mathcal V^{(h_\eps-1)}(\varphi,J)&:=\sum_{\o=\pm} \int_{\o} dq
 \hat \varphi^+_{q,\o} [\hat W^{(h_\eps-1);\infty}_{2,0;(\o,\o)}(0)+q\cdot \partial\hat W^{(h_\eps-1);\infty}_{2,0;(\o,\o)}({0})\big]\hat \varphi^-_{q,\o}\\
& + 4 
 \int dy\,  \varphi^+_{y,+}\varphi^-_{y,+}\varphi^+_{y,-}\varphi^-_{y,-} \hat W^{(h_\eps-1);\infty}_{4,0;(+,+,-,-)}(\un 0)\\
& + \sum_{\un \o \in \{\pm\}^2}\sum_{s=A,B,C} \int dy \,  \varphi^+_{y,\o_1}\varphi^-_{y,\o_2} J_{y,s}e^{i\boldsymbol{q}_\eps y(\o_1- \o_2)}\hat W^{(h_\eps-1);\infty}_{2,1;s,(\o_1,\o_2)}(0,\boldsymbol{q}_\eps(\o_1-\o_2)). 
\end{split}
 \label{eq:loc2reg}
\end{equation}
Note that in the first line only the term $(\o,\o)$ appears because of the support properties of the cutoffs $\chi^{(2)}_{h_\eps-1}$, while in the second line we used anticommutation of the Grassmann variables and antisymmetry of the kernels. 
Now let $z_{h_\eps-1}:=-i\partial_1 \hat W^{(h_\eps-1);\io}_{2,0;(\o,\o)}(0)$, which is real and independent of $\omega$, thanks to the symmetries \eqref{eq:2.43bis}. Therefore, in \eqref{eq:loc2reg} we rewrite $q_1 \partial_1 \hat W^{(h_\eps-1)}_{2,0;(\o,\o)}=iq_1z_{h_\eps-1}$. We now add and subtract a term $z_{h_\eps-1}\sum_\omega\int_\omega \omega q_2 v_\eps \hat \varphi^+_{q,\omega}\hat\varphi^-_{q,\omega}$ from $\mathcal L_2 \mathcal V^{(h_\eps-1)}(\varphi,J)$, thus rewriting $\mathcal L_2\mathcal V(\varphi,J)=
-z_{h_\eps-1}\sum_\omega\int_\omega 
D_\omega(q)\hat \varphi^+_{q,\omega}
\hat\varphi^-_{q,\omega}+\mathcal L_2\widetilde{\mathcal V}^{(h_\eps-1)}(\varphi,J)$, where, 
if we define $F_\mu(\varphi):=\sum_\o\int_\o dq \hat \varphi ^+_{q,\o}\hat \varphi^-_{q,\o}$, $F_\d(\varphi):=\sum_\o \int_\o dq \hat \varphi^+_{q,\o} \hat \varphi^-_{q,\o} \o q_2$, $ F_\l(\varphi):=\int dy \varphi^+_{y,+}\varphi^-_{y,+}\varphi^+_{y,-}\varphi^-_{y,-}$, $ F_{Y,s,\un \o}(\varphi,J):=\int dy \varphi^+_{y,\o_1} \varphi^-_{y,\o_2}J_{y,s}e^{i\boldsymbol{q}_\eps (\o_1-\o_2)y}$, 
\begin{equation}
\mathcal L_2\widetilde{\mathcal V}^{(h_\eps-1)}(\varphi,J)=
2^{h_\eps-1} \tilde m_{h_\eps-1} F_\mu(\varphi)+\tilde l_{h_\eps-1}F_\l(\varphi)+\tilde d_{h_\eps-1}F_\d(\varphi)+\sum_{s,\un \o}\tilde Y_{h_\eps-1;s,\un \o} F_{Y,s, \un \o}(\varphi,J)\label{eq:2.54} 
\end{equation}
with $2^{h_\eps-1}\tilde m_{h_\eps-1}:=\hat W^{(h_\eps-1);\io}_{2,0;(\o,\o)}(0)$, 
$\tilde l_{h_\eps-1}:=4\hat W^{(h_\eps-1);\io}_{4,0;(+,+,-,-)}(\un 0)$, 
$\tilde d_{h_\eps-1}:=\omega\partial_2 \hat W^{(h_\eps-1);\io}_{2,0;(\o,\o)}(0)+v_\eps z_{h_\eps-1}$,
which are all real and independent of $\omega$, while 
$\tilde Y_{h_\eps-1;s,\un \o}:=\hat W^{(h_\eps-1);\io}_{2,1;s,(\o_1,\o_2)}(0,\boldsymbol{q}_\eps(\o_1-\o_2))$
is real and satisfies 
$\tilde Y_{h_\eps-1;s,\un \o}=
\tilde Y_{h_\eps-1;s,-\un \o}$, as it  follows from the symmetries in
\eqref{eq:2.43bis}, see Appendix \ref{app:symmetries}.

Given these definitions, 
and denoting $\widetilde{\mathcal V}^{(h_\eps-1)}(\varphi,J):=
\mathcal L_2\widetilde{\mathcal V}^{(h_\eps-1)}(\varphi,J)+\mathcal R_2\mathcal V^{(h_\eps-1)}(\varphi,J)
$, we rewrite: 
\begin{equation}
\begin{split}
   \eqref{eq:4.91bbis}&= e^{{L^2 (F^{(h_\eps-1)}+t^{(h_\eps-1)})+
  S^{(h_\eps-1)}(J)}}\int P^{(\leq h_\eps-1)}_{Z_{h_\eps-1}}(D\varphi)e^{-Z_{h_\eps-1} z_{h_\eps-1}\int _\o dq D_\o(q)\hat \varphi^+_{q,\o} \hat \varphi^-_{q,\o} +\widetilde{\mathcal{V}}^{(h_\eps-1)}\left (\sqrt{Z_{h_\eps-1}}\varphi,J\right )}
  \\ & = e^{L^{2}(F^{(h_\eps-1)}+t^{(h_\eps-2)})+S^{(h_\eps-1)}(J)}\int \tilde P^{(\leq h_\eps-1)}_{\tilde Z_{h_\eps-2}}(D\varphi) e^{\widehat{\mathcal{V}}^{(h_\eps-1)}\left (\sqrt{Z_{h_\eps-2}}\varphi,J\right )}
  \end{split}
  \label{eq:2.51bis}
\end{equation}
where, letting $ \tilde Z_{h_\eps-2}(q):=Z_{h_\eps-1}(1+\chi^{(2)}_{h_\eps-1}(q)z_{h_\eps-1})$ and $Z_{h_\eps-2}=\tilde Z_{h_\eps-2}(0)$, $\tilde P^{(\leq h_\eps-1)}_{\tilde Z_{h_\eps-2}}$ is the Gaussian Grassmann integration satisfying $\int \tilde P^{(\leq h_\eps-1)}_{\tilde Z_{h_\eps-2}}(D\varphi) \varphi^-_{y,\o}\varphi^+_{y',\o'}=\d_{\o,\o'}\tilde g_\omega^{(\leq h_\eps-1)}(y,y')Z^{-1}_{h_\eps-2}$ with propagator given by \be 
 \hspace{-10pt}
 \frac{\tilde g_\omega^{(\leq h_\eps-1)}(y,y')}{Z_{h_\eps-2}}=\int_\o dq \frac{e^{-iq(y-y')}\chi^{(2)}_{h_\eps-1}(q)}{\tilde Z_{h_\eps-2}(q)D_\o(q)+\rho_{\o}(q)},
  \label{eq:2.58bis}\ee
and the prefactor $\exp\{L^2 t^{(h_\eps-2)}\}$ in the second line of \eqref{eq:2.51bis} accounts for the change of normalization in the 
reference Gaussian Grassmann integration. Moreover, $\widehat{\mathcal V}^{(h_\eps-1)}$ 
is obtained from $\widetilde{\mathcal V}^{(h_\eps-1)}$
via a rescaling by $\sqrt{Z_{h_\eps-2}/Z_{h_\eps-1}}$ of its first argument, so that 
\begin{equation}
\begin{split}
\widehat{\mathcal V}^{(h_\eps-1)}(\varphi,J)&=
2^{h_\eps-1} \mu_{h_\eps-1} F_\mu(\varphi)+\lambda_{h_\eps-1}F_\l(\varphi)+\delta_{h_\eps-1}F_\d(\varphi)+\sum_{s,\un \o}Y_{h_\eps-1;s,\un \o}Z_{h_\eps-2}^{-1} F_{Y,s, \un \o}(\varphi,J)\\
&+\mathcal R_2\mathcal V^{(h_\eps-1)}\left (\sqrt{\tfrac{Z_{h_\eps-1}}{Z_{h_\eps-2}}}\varphi,J\right ),
\end{split}
\label{eq:2.540} 
\end{equation}
with $\mu_{h_\eps-1}=\frac{Z_{h_\eps-1}}{Z_{h_\eps-2}}\tilde m_{h_\eps-1}$,
$\lambda_{h_\eps-1}=(\frac{Z_{h_\eps-1}}{Z_{h_\eps-2}})^2\tilde l_{h_\eps-1}$,
$\delta_{h_\eps-1}=\frac{Z_{h_\eps-1}}{Z_{h_\eps-2}}\tilde d_{h_\eps-1}$,
and 
$Y_{h_\eps-1;s,\underline\omega}=Z_{h_\eps-1}\tilde Y_{h_\eps-1;s,\un \omega}$.

Finally, rescale the cutoff $\chi^{(2)}_{h_\eps-2}(q):=\chi^{(2)}_{h_\eps-1}(2q)$, split $\chi^{(2)}_{h_\eps-1}=\chi^{(2)}_{h_\eps-2}+f^{(2)}_{h_\eps-1}$,  so that  $\tilde g^{(\leq h_\eps-1)}_\o=g^{(\leq h_\eps-2)}_\o+\tilde g^{(h_\eps-1)}_\o$ where $\tilde g^{( h_\eps-1)}_\o$ is obtained by replacing $\chi^{(2)}_{h_\eps-1}$ with $f^{(2)}_{h_\eps-1}$ in the right hand side of \eqref{eq:2.58bis}. Correspondingly, decompose $\varphi_\omega^{(\leq h_\eps-1)}=\varphi_\omega^{(\leq h_\eps-2)}+
\varphi_\omega^{(h_\eps-1)}$, 
where $\varphi_\omega^{(\leq h_\eps-2)}$, 
$\varphi_\omega^{(h_\eps-1)}$ are two independent Gaussian Grassmann fields associated with the propagators $g^{(\leq h_\eps-2)}_\o/Z_{h_\eps-2}$, $\tilde g^{(h_\eps-1)}_\o/Z_{h_\eps-2}$, respectively. Integrating out $\varphi^{(h_\eps-1)}$, we find that
\eqref{eq:2.51bis} can be rewritten as 
\begin{equation}
\begin{split}
\eqref{eq:2.51bis}=e^{L^2(F^{(h_\eps-2)}+t^{(h_\eps-2)})+S^{(h_\eps-2)}(J)}\int P^{(\le h_\eps-2)}_{Z_{h_\eps-2}}(D\varphi)e^{\mathcal V^{(h_\eps-2)}(\sqrt{Z_{h_\eps-2}}\varphi,J)}
\end{split} \nn 
\end{equation}
where $F^{(h_\eps-2)}=
F^{(h_\eps-1)}+\widetilde F^{(h_\eps-2)}$, $S^{(h_\eps-2)}
=S^{(h_\eps-1)}+\widetilde S^{(h_\eps-2)}$ and $\mathcal V^{(h_\eps-2)}$ are defined in such a way that $L^{2}\widetilde F^{(h_\eps-2)}+\widetilde S^{(h_\eps-2)}(J)+\mathcal V^{(h_\eps-2)}(\varphi,J) =$ $\sum_{n \geq 1} \frac{1}{n!}\mathcal{E}^T_{h_\eps-1} \left (\mathcal{V}^{(h_\eps-1)}(\varphi+\sqrt{Z_{h_\eps-2}}\cdot,J); n \right )
$ and $S^{(h_\eps-2)}(0)=\mathcal V^{(h_\eps-2)}(0,J)=0$, where 
the truncated expectation is the one relative to the integration $\tilde P_{Z_{h_\eps-2}}^{(h_\eps-1)}$
with propagator $\tilde g_\omega^{(h_\eps-1)}/Z_{h_\eps-2}.$ 
These definitions imply that 
 $\widetilde S^{(h_\eps-2)}+\mathcal V^{(h_\eps-2)}$ admits an expression analogous to  \eqref{eq:2.39bis} and preceding line, with scale label replaced by $h_\eps-2$; since $\tilde g_\omega^{(h_\eps-1)}$ is symmetric under $\SS_1,\SS_2$, see Remark \ref{rem:remark11}, then the kernels of $\widetilde S^{(h_\eps-2)}+\mathcal V^{(h_\eps-2)}$ satisfy the same symmetries as \eqref{eq:2.43bis}, see App. \ref{app:symmetries} for more details.

\paragraph{The inductive statement} By iterating the procedure described in the previous subsection, we find the following inductive representation of the generating function.  Let $\chi_h^{(2)}(q):=\chi^{(2)}_{h_\eps-1}(2^{-h+h_\eps-1}q)$, $S^\o_h:=\{q \in \DD'_{\vartheta_0}: \chi^{(2)}_h(q-\o \boldsymbol{q}_\eps)\neq 0\}$. Then for any $h_L < h \leq h_\eps-1$ with $h_L:=\sup\{h \leq  h_\eps-1: S^+_h \cap \DD'_{\vartheta_0,R} = \emptyset\}$,  we can express \eqref{eq:4.91bbis} as
\begin{equation}
    \eqref{eq:4.91}=e^{L^2(F^{(h-1)}+t^{(h-1)})+S^{(h-1)}(J)}\int P^{(\leq h-1)}_{Z_{h-1}}(D\varphi)e^{\mathcal V^{(h-1)}(\sqrt{Z_{h-1}}\varphi,J)}
    \label{eq:2.39}
\end{equation}
where $P^{(\leq h-1)}_{Z_h-1}$ 
has propagator of the same form as 
\eqref{eq:2.37} with $h_\eps$ replaced by $h$, and $Z_{h-1}:=Z_{h}(1+z_{h})$, with $z_{h} \in \mathbb{R}$ defined as in  \eqref{eq:2.43biss} below. For 
$h-1=h_L$, the integral in the r.h.s. of \eqref{eq:2.39} must be interpreted as being equal to $1$. 

The constant $t^{(h-1)}$ accounts for the iterative change of Grassmann integration while the effective interaction satisfies $S^{(h-1)}(0)=\mathcal V^{(h-1)}(0,J)=0$ and can be written  as $\widetilde S^{(h-1)}(J)+\mathcal V^{(h-1)}(\varphi,J)=\sum_{n,m \in  \mathbb{N}, n+m \geq 1}V^{(h-1)}_{2n,m;\eps}(\varphi,J)$ with \begin{equation}
    \begin{split}
         V_{2n,m}^{(h-1)}(\varphi,J)& = \int d\un y d\un z d \un s d \un \o d \un D \,   \partial^{\un D} \varphi_{\un y,\un \o} J_{\un z, \un s}W^{(h-1)}_{2n,m;\un s, \un \o,\un D}(\un y,\un z) \\ &= \int_{\un \o} d \un s d \un q d \un p d \un \o\,\hat\varphi_{\un{q},\un \o}\hat J_{\un p,\un s}  \hat{W}_{2n,m;\un s}^{(h-1)}(\un q,\un p)\delta ({\underline{q}+\un \o \boldsymbol{q}_\eps},\un p)
    \end{split}\label{eq:2.51biss}
        \end{equation} where $\partial ^{\un D}\varphi_{\un y,\un\o}:=\prod_{j=1}^{2n} \bar \partial^{D_{j}} \varphi^{\tilde \sigma_j}_{y_{j,\o_{j}}}$  and the conventions are the same as in \eqref{eq:2.39bis},\eqref{eq:2.27bisss}; in particular the kernels satisfy the following symmetries, see Appendix \ref{app:symmetries},
  \begin{equation}
      \begin{split}
    \label{eq:upps}
   & \hat  W^{(h-1)}_{2n,m;-\underline\o,\underline s}(\un q,\un p)=     \overline{\hat W^{(h-1)}_{2n,m;\underline\o,\underline s}(-\un q,-\un p)},  \\ & \left [\hat  W^{(h-1)}_{2n,m;-\underline\o,\underline s}(\un q,\un p)\right ]_{\vartheta_0}=   \left [ \hat W^{(h-1)}_{2n,m;\underline\o,\underline s^*}(\underline{\sigma_3 q},\underline{\sigma_3 p})\right ]_{\sigma_1\vartheta_0},
  \end{split}
  \end{equation}
  which are the analogue of \eqref{eq:2.33bis} and \eqref{eq:2.43bis}.  Moreover $F^{(h-1)}=F^{(h)}+\widetilde F^{(h-1)}$,
  $S^{(h-1)}=S^{(h)}+\widetilde S^{(h-1)}$, $\mathcal V^{(h-1)}$ are inductively defined by \begin{equation}
\hspace{-8pt}
     L^{2}\widetilde F^{(h-1)}+\widetilde S^{(h-1)}(J)+\VV^{(h-1)}(\varphi,J) = \sum_{n \geq 1} \frac{1}{n!}\mathcal{E}^T_{h} \left (\hat{\mathcal{V}}^{(h)}(\varphi+\sqrt{Z_{h-1}}\cdot,J); n \right )\label{eq:2.52bis}
\end{equation}
where $\mathcal{E}^T_{h}$ is the truncated expectation associated 
with the Gaussian Grassmann integration $\tilde P^{(h)}_{Z_{h-1}}$ satisfying $\int \tilde P^{(h)}_{\tilde Z_{h-1}}(D\varphi) \varphi^-_{y,\o}\varphi^+_{z,\o'}=\d_{\o,\o'}\tilde g_\omega^{(h)}(y,z)Z^{-1}_{h-1}$ with propagator given by 
\begin{equation}
    \frac{\tilde g^{(h)}_\o(y,z)}{Z_{h-1}}=\int_\o dq \frac{e^{-iq(y-z)}f^{(2)}_h(q)}{\tilde Z_{h-1}(q)D_\o(q)+\rho_{\o}(q)},\qquad \tilde Z_{h-1}(q):=Z_h(1+\chi^{(2)}_h(q)z_h).
\label{eq:singlescaleprop2}
\end{equation}
where $f^{(2)}_h(q)=f_{h_\eps-1}^{(2)}(2^{-h+h_\eps-1}q)$, cf. below \eqref{eq:2.540}. In the r.h.s. of \eqref{eq:2.52bis}, 
\[\hat{\mathcal{V}}^{(h)}(\varphi,J)=2^h \mu_h F_\mu(\varphi)+\l_hF_\l(\varphi)+\d_hF_\d(\varphi)+\sum_{s,\un \o}Y_{h;s,\un \o}Z^{-1}_{h-1}F_{Y,s, \un \o}(\varphi,J)+\mathcal{R}_2 V^{(h)}\left (  \sqrt{\frac{Z_h}{Z_{h-1}}}\varphi,J\right )\] with $F_\mu,F_\l,F_\d,F_{Y,s,\o}$ as in \eqref{eq:2.54} and preceding lines, while
\begin{equation}
\begin{split}
& 2^h\mu_h:=\frac{Z_h}{Z_{h-1}}\hat W^{(h);\io}_{2,0;(+,+)}(0), \quad \delta_h:=\frac{Z_h}{Z_{h-1}}\big(\omega\partial_2 \hat W^{(h);\io}_{2,0;(+,+)}(0)+v_\eps z_h\big),   \\ &  z_h:=-i\partial_1W^{(h);\io}_{2,0;(+,+)}(0), \quad \l_h:=4\left (\frac{Z_h}{Z_{h-1}}\right )^2 \hat W^{(h);\io}_{4,0;(+,+,-,-)}(\un 0),   \\ &Y_{h;s,\un \o}:=Z_h \hat W^{(h);\io}_{2,1;s,(\o_1,\o_2)}(0,\boldsymbol{q}_\eps(\o_1-\o_2)).
\end{split}
\label{eq:2.43biss}
\end{equation}
The constants in \eqref{eq:2.43biss} are all real, and, moreover, $Y_{h;s,\un \o}=Y_{h;s,-\un \o}$,    as a consequence of the $L\to \io$ counterpart of the symmetries in  \eqref{eq:upps}, so we can further define
 \begin{equation}
     Y^{(1)}_{h;s}:=Y_{h;s,(    \omega,\omega)}, \qquad Y^{(2)}_{h;s}:=Y_{h;s,(\omega,-\omega)},
     \label{eq:2.63bis}
 \end{equation}
which are independent of $\omega$. 
 
\paragraph{Dimensional bounds: second regime}

We first state the analogue 
of Lemma \ref{lem:propag1}, 
summarizing the key properties of 
the propagator $\tilde g^{(h)}_{\o}$ 
of the second regime. 
We write $\tilde g^{(h)}_{\o}=\tilde g^{(h)}_{R,\o}+\tilde g^{(h)}_{S,\o}$ where $\tilde g^{(h)}_{R,\o}$ is defined by 
replacing $\rho_\omega(q)$ with zero 
in the definition of $\tilde g^{(h)}_\omega$ in \eqref{eq:singlescaleprop2}. In analogy with \eqref{eq:tildegnm}, we denote $\tilde{g}^{(h);\un n;\un m}_{\o}(y,0):=y_1^{m_1} y_2^{m_2}\partial^{n_0}_{v_\eps}\bar{\partial}^{n_1}_1\bar{\partial}^{n_2}_2 \tilde{g}_{\o,\io}^{(h)}(y)$, 
where $v_\eps$ was defined right 
below \eqref{eq:2.55zz}, 
and similarly for $\tilde g^{(h)}_{R,\o},\tilde g^{(h)}_{S,\o}$.
\begin{lemma}
\label{lem:3bis}
    Suppose that there exists $C>0$, independent of $L,\eps$, such that $\sup_{h \leq h' < h_\eps} |z_{h'}| \leq C|\l|$, then there exists $L,\eps$ independent constants $c_0(\un n),c>0$ such that for $h < h_\eps-1$
    \begin{equation}
    |\tilde g^{(h);\un n;\un 0}_{R,\o}(y,0)|\leq c_0(\un n) v_\eps^{-1-n_2-n_0}2^{h(1+n_1+n_2)}e^{-c \sqrt{2^h(\|y_1\|+v_\eps^{-1}\|y_2\|)}},
    \end{equation}
and $\tilde g^{(h)}_{S,\o}$ satisfies the same estimates times an extra $2^{h-h_\eps}$ factor. 
As a consequence, $\|\tilde g^{(h);\un n;\un m}_{R,\o}\|_1\leq c_1 v_\eps^{-1-n_2-n_0+m_2}2^{h(-1+n_1+n_2-m_2-m_1)}$
for some $c_1=c_1(\un n,\un m)$; in particular,
$\|\tilde g^{(h)}_{\o}\|_\io \leq c_0 v_\eps^{-1}2^h$, while $
\|\tilde g^{(h)}_{\o}\|_1 \leq c_1 2^{-h}$. Moreover, $\tilde{g}_{R,\omega}^{(h);\un n; \un 0}$ admits a Gram representation, i.e., there exists a Hilbert space $\mathcal H$ and two vectors $\gamma^{(h);\un n}_{R,\omega;y},\eta^{(h)}_{R,\omega;y'}\in\mathcal H$ such that 
$\tilde{g}_{R,\omega}^{(h);\un n; \un 0}(y,y')=(\gamma^{(h);\un n}_{R,\omega;y},\eta^{(h)}_{R,\omega;y'})$ and 
$\|\gamma^{(h);\un n}_{R,\omega;y}\|^2\le 
c_0(\un n) v_\eps^{-1-n_2-n_0}2^{h(1+n_1+n_2)}$, $\|\eta^{(h)}_{R,\omega;y'}\|^2\le 
c_0(\un 0) v_\eps^{-1}2^{h}$, where $\|\cdot\|$ is the norm induced by the scalar product in $\mathcal H$. 
\end{lemma}

Lemma \ref{lem:3bis} is a key ingredient in the derivation of 
the following upper bound on 
the $L_1$ norm of the kernels of the effective potentials in the second regime, provided the running constants in \eqref{eq:2.43biss}
remain small uniformly in the scale index. 

\begin{proposition}
\label{lem:4biss}
    Suppose that Proposition \ref{lem:bounds1} holds together with the hypothesis of Lemma \ref{lem:3bis} and that there exists some $L,\eps$-independent constant $C>0$ such that
   \begin{equation}\sup_{h \leq h'\leq h_\eps} |\mu_{h'}|  \leq C|\l|, \qquad \max \left \{\sup_{h \leq h'<h_\eps} |\l_{h'}|,\sup_{h \leq h'<h_\eps} |\d_{h'}|\right \} \leq C\eps^{1/2} |\l|. \label{eq:4.140}
 \end{equation}   There exists $C_2>0$ independent of $L,\epsilon$, such that if $|\l|\leq \l_0$ then for $h<h_\eps-1$
 \begin{equation}
\lVert W^{(h)}_{2n,m;\un s,\un \o,\un D} \rVert \leq C_2^{n+m}|\l|^{d_{n,m}}2^{h_\eps(-\frac{1}{2}+\frac{n}{2})}2^{h(2-n-m-|\un D|)}\left (\sup_{h<h'\leq h_\eps-1}\sup_{s,\un \o}\frac{|Y_{h';s,\un \o}|}{Z_{h'-1}}\right )^m
\label{eq:2.61bis}
\end{equation}
uniformly in $\eps \leq \bar \eps$, 
where $|\un D|=\sum_{i=1}^{2n}\sum_{j=1}^2 D_{ij}$ and $d_{n,m}$ as in Proposition \ref{lem:bounds1}.
\end{proposition}
As it was the case for Proposition \ref{lem:bounds1}, the proof 
of this result is a straightforward 
adaptation of the proof of 
\cite[Lemma 3.2]{BM36}, whose details are left to the reader. 

\medskip

In view of these results, we see that the key point that needs to be controlled is the flow of the running coupling constants: once that is proved, the bounds on the 
kernels of the effective potential (and, a posteriori, of the correlation functions) follow from the propositions above. 
As commented in Remark \ref{rem:4}, while $\mu_{h_\eps}$ still needs to be fixed in a suitable way (see Proposition \ref{lem:kernbounds2} below),  $z_{h_\eps-1},\lambda_{h_\eps-1}, \d_{h_\eps-1}$ are fixed and completely determined by the output of the first regime of integration, and they satisfy the following estimates:
 \begin{lemma}\label{lem:costantiascalaheps}
    Under the hypothesis of Prop. \ref{lem:bounds1}, there exists $C:[0,1/2)\to(0,+\infty)$ such that, for $|\l| \leq \l_0$, $|z_{h_\eps-1}|\leq C(\theta) |\l| \eps^\theta$ and $\max\{|\l_{h_\eps-1}|,|\d_{h_\eps-1}|\} \leq C(\theta) |\l| \eps^{1/2+\theta}$.
\end{lemma}
The proof is given in App. \ref{app:scalaheps}. The following result shows that, by properly fixing $\mu_{h_\eps}$, the 
running coupling constants are all uniformly bounded in the second regime. 

\begin{proposition}
\label{lem:kernbounds2}
    Under the validity of Proposition \ref{lem:bounds1}, there exists $C:[0,1/2)\to(0,+\infty)$ independent of $L,\eps$, an analytic function $(-\l_0,\l_0) \ni \l \mapsto \mu_{h_\eps}(\l,\eps)$ satisfying $|\mu_{h_\eps}(\l,\eps)|\leq C(\theta)\eps^{\theta}|\l|$, such that if $\mu_{h_\eps}=\mu_{h_\eps}(\l,\eps)$ then for any $h \leq h_\eps-1$
\begin{equation} 
\begin{split}
& \max\left \{2^{-\theta h}|\mu_h|,\eps^{-1/2}2^{-\theta h}|\d_h-\d_{h-1}|,\left|Y_{h-1;s,\un \o}(Y_{h;s,\un \o})^{-1}-1\right |\right \}\leq C(\theta)|\l|, \\   & \max\{2^{-\theta h}|z_h-z_{h-1}|,\eps^{-1/2}§2^{-\theta h}|\l_h-\l_{h-1}|\}\leq C(\theta)|\l|^2.
\end{split}\label{eq:2.7}\end{equation}
\end{proposition}

In the case $J=0$, the proof is an adaptation of that of \cite[Lemma 3.2]{BM36}.  The function $\mu_{h_\eps}(\l,\eps)$ corresponds to the function $\nu_{h_\eps}$ of \cite{BM36}, see \cite[eq.(145)]{BM36}, which is determined by a fixed point argument analogous to the one  used for $\nu^*_0(\l)$ appearing in Proposition \ref{lem:flowRC} (see Remark \ref{rem:4} and  App.\eqref{app:A2}). See also Section 3.1 of Appendix \ref{app:regularitydensityeps} for more details on how to fix $\mu_{h_\eps}(\l,\eps)$. 

Note that \eqref{eq:2.7} clearly implies the hypothesis of Lemma \ref{lem:3bis}, as well as the bounds \eqref{eq:4.140} (recall that $v_\eps \asymp \sqrt{\eps}\asymp2^{h_\eps/2}$, see Remark \ref{rem:4}); it actually implies more: if we denote $\BB^\l_h:=\l_{h-1}-\l_{h}$, we can write $    \l_{h}=\l_{h_\eps-1}+\sum_{j=h+1}^{h_\eps-1} \BB^\l_j$ (and similarly for $z,\d,\mu$) so that combining \eqref{eq:2.7} with Lemma \ref{lem:costantiascalaheps}, we find: 
\begin{equation}
    |\d_h|, |\l_h|\leq C_\theta|\l| \eps^{1/2+\theta}, \quad |z_h|\leq C_\theta|\l|\eps^\theta,
    \label{eq:2.70biss}
\end{equation}
for any $\theta \in [0,1/2)$ and some $C_\theta>0$.
The extra factors $\eps^\theta$ will play a role in the proof of Theorem \ref{th:5}, see Proposition \ref{prop:7}.

 When $J \neq 0$, since the local quartic term in $\mathcal V^{(h)}(\varphi,0)$ is no longer vanishing (see e.g. \eqref{eq:loc2reg},  differently from what happens in the first regime, it turns out that the running coupling  
 $Y_{h;s,\un \o}$ is not bounded uniformly in the scale label: rather, it acquires a \emph{critical exponent}:
 \be Y_{h;s,\un \o}=2^{-\eta_{\un \o}(h-h_\eps)}Y_{h_\eps-1;s,\un \o}(1+O_\theta(|\l|\eps^\theta)) \label{eq:2.71bis}\ee where 
 $\eta_{\un \o}$ is analytic in $\l \in (-\l_0,\l_0)$, and depends only on $\l_{-\io}(\l,\eps):=\lim_{h \to -\io} \l_h$ and on the product $\o_1\o_2.$ A similar behavior is found for the renormalization constant $Z_{h}$ (more details about \eqref{eq:2.71bis} are discussed in Lemma \ref{YZ-lemma} below).
 \begin{remark}
Note that as a consequence of \eqref{eq:2.70biss} one has  $|\l_{-\io}(\l,\eps)|\leq C_\theta|\l|\eps^{1/2+\theta}$ which in turn implies that the critical exponents $\eta_{\un \o}$ are vanishing for $\eps \to 0^+$. These properties are at the core of the non-renormalization (i.e.  universality) of the exponent $1/2$ in Theorem \ref{th:5}, as it can be seen in the next section,  Proposition \ref{prop:7} and its proof.
 \end{remark}
 
\subsection{RG output and proof of Theorem \ref{th:5}}
\label{sec:2.3}

In this section we first state the representation of the density $\rho_{\eps+\varepsilon_0(\lambda),\lambda}$ 
following  {from \eqref{eq:mlimit}} and from the multiscale integration procedure described above, which is used to prove Theorem \ref{th:5} with the sharp asymptotics of the density as $\epsilon\to0^+$ and, by integration in $\epsilon$, the analogue for the free energy.

Let  $\|\cdot \|_\io=\sup_{y \in R^T\mathbb{Z}^2} |\cdot(y)|, \|\cdot \|_1=\sum_{y \in R^T \mathbb{Z}^2} |\cdot(y)|$ (recall \eqref{eq:transf}); for every function $\eps \mapsto f_{\eps}$ we denote  $\mathfrak{d}_\eps f_\cdot :=f_\eps-f_0$ and similarly for $\mathfrak{d}_\l.$ Also, regarding the dependence on $\eps$, we use the same convention as in the previous section, that is, 
we specify it by explicitly adding a 
label $\eps$ to the kernels 
of the effective potential on scale $h$ only 
for the scale labels in the first regime, i.e., for $h \geq h_\eps-1$; if $\eps>0$ and $h<h_\eps-1$, then we drop the label $\eps$ and 
keep the dependence implicit. 

\begin{proposition}
\label{prop:7}    Let $\l_0, \bar \eps>0$ and $(-\lambda_0,\l_0) \ni\l \mapsto \nu^*_0(\l), \mu_{h_\eps}(\l,\eps)$  as in Propositions \ref{lem:flowRC}-\ref{lem:kernbounds2}   and fix $\nu=\nu^*_0(\l)$ (cf. \eqref{eq:5.71})  and $\mu_{h_\eps}=\mu_{h_\eps}(\l,\eps)$ (cf. \eqref{eq:2.35}). {The limit  $\rho_{\eps+\varepsilon_0(\l),\l}:=\lim_{L \to \io}\rho^{(\vartheta_0)}_{\eps+\varepsilon_0(\l),\l,L,0}$ exists, is reached uniformly in $|\eps|\leq \bar \eps$,}  is analytic in $\l \in (-\l_0,\l_0)$ uniformly in 
$-\bar\eps\le \eps\le 0$ and $C^\infty$ in 
$\l \in (-\l_0,\l_0)$ uniformly in $0<\eps\le \bar\eps$. It can be expressed as
\begin{equation}
 \rho_{\eps+\varepsilon_0(\l),\l}=\sum_{h= h_\eps-1}^0 W^{(h);\io}_{0,1;\eps,B}(0)+\mathbbm{1}_{\{\eps>0\}}\sum_{h <h_\eps-1}W^{(h);\io}_{0,1;B}(0)\label{eq:2.8}
\end{equation}
where $(W^{(h);\io}_{0,1;\eps,B}(0))_{h= h_{\eps}-1}^0$ are the limits as $L \to \infty$ of the kernels of $\widetilde S_\eps^{(h)}$
defined in \eqref{eq:4.68} for $h\ge h_\eps$ and, for $h=h_\eps-1$,  after \eqref{eq:4.91} if $\eps>0$ and in \eqref{eq:epsnegativo} if $\eps<0$ (with the understanding that $h_\eps\big|_{\eps=0}=-\infty$), while $(W^{(h);\io}_{0,1;B}(0))_{h<h_\eps-1}$ are those of
the kernels of $\widetilde S^{(h)}$
defined in \eqref{eq:2.52bis}. 
They can be further rewritten as
\begin{equation}
     \begin{split}
     & W^{(h);\io}_{0,1;\eps,B}(0)=  -\tilde g^{(h+1)}_{\eps,\io}(0)Y_{h;\eps,B}+\Theta^{(h)}_{\rho,\eps}  \qquad \qquad  \quad h_\eps-1 \leq   h  \leq 0 \\ & W^{(h);\io}_{0,1;B}(0)= \sum_{\o=\pm}\left (-\tilde g^{(h+1)}_{\o,\io}(0)Z^{-1}_h Y^{(1)}_{h;B}+\Theta^{(h)}_{\rho,\o} \right ) \qquad  h< h_\eps-1
\end{split}\label{eq:2.9}
\end{equation}
 where, in the case $\eps=0$, 
only the first line is present. Here $Y_{h;\eps,B}$, $Y^{(1)}_{h;B}$ 
are defined as in \eqref{eq:rcc1regime}, \eqref{eq:2.63bis}, respectively, and satisfy \eqref{eq:2.29bis}, \eqref{eq:2.7}; moreover,   $\tilde g^{(h)}_{\eps,\io}$, $\tilde g^{(h)}_{\o,\io}$ are the limits as $L \to \infty$ of the single scale propagators \eqref{eq:singlescalepropI}, \eqref{eq:singlescaleprop2}; finally, $\Theta^{(h)}_{\rho,\eps}$,
$\Theta^{(h)}_{\rho,\o}$ satisfy the following estimates:
\begin{equation}
    \begin{split}
    &   \left | \Theta^{(h)}_{\rho,\eps} \right |\leq C_\theta 2^{(\frac{1}{2}+\theta)h}, \qquad  \quad  \left |\Theta^{(h)}_{\rho,\o}\right |\leq C_\theta 2^{h-h_\eps}\eps^{\frac{1}{2}+\theta},
    \end{split}\label{eq:2.10bis}
\end{equation}
for any $\theta\in[0,\frac12)$
and a suitable $C_\theta>0$. 
Moreover, if $h \geq h_\eps-1,$
letting $\mathcal B^Y_{h;\eps,B}:=
Y_{h-1;\eps,B}-Y_{h;\eps,B}$, 
the action of $\mathfrak{d}_\eps$  
on $\tilde g^{(h)}_{\eps,\infty}$, 
$\mathcal B^Y_{h;\eps,B}$, 
$\Theta^{(h)}_{\rho,\eps}$, for $0<|\eps|\leq \bar \eps$,
can be bounded as follows:
\begin{equation}
    \|\mathfrak{d}_\eps \tilde g^{(h)}_{\cdot,\infty}\|_\io\leq C_0|\eps| 2^{-\frac{h}{2}},    \qquad |\mathfrak{d}_\eps \BB^Y_{h;\cdot,B}|\leq C_\theta|\eps|\,|\l|2^{h(-1+\theta)}, \qquad  |\mathfrak{d}_\eps \Theta^{(h)}_{\rho,\cdot}|\leq C_\theta|\eps| 2^{(-\frac{1}{2}+\theta)h},\label{eq:2.73bis}
\end{equation}
that is, $\mathfrak d_\eps$ has a dimensional effect of producing an additional factor $|\eps|2^{-h}$ as compared to the bounds on the functions it acts on: 
compare with the bound on $\|\tilde g^{(h)}_\eps\|_\io$ in Lemma \ref{lem:propag1}, the one on $Y_{h-1;\eps,B}-Y_{h;\eps,B}$ in Proposition \ref{lem:flowRC} and the first bound for $\Theta_{\rho,\eps}$
in  \eqref{eq:2.10bis}. 
\end{proposition}

 The proof of this proposition is postponed to Appendix \ref{app:A3} and it is based on the so called \textit{Gallavotti-Nicol\`o tree expansion}, which is briefly recalled in App.\ref{app:A}. The rationale behind the result is the following: the renormalized expansion of Sections \ref{sec:firstregime}-\ref{sec:secondregime} implies a multiscale decomposition of the interacting dimer density \eqref{eq:2.8}. Using the scaling properties of the propagator and the dimensional bounds on the kernels of the effective potential summarized in Propositions \ref{lem:bounds1} and \ref{lem:4biss}, we can isolate from each single-scale contribution $W^{(h)}_{0,1;\eps,B}$ 
an explicit dominant term, corresponding to the first addend in the right hand side of  \eqref{eq:2.9}; the subdominant terms, denoted by $\Theta_{\rho,\eps}$ and $\Theta_{\rho,\omega}$ admit an improved dimensional bound as compared to the dominant ones, which 
allow us to show that their contribution to the density is subdominant as $\eps\to0^+$. In the next section, we discuss how to extract from the dominant terms the 
sharp asymptotics of the density 
as $\eps\to0^+$.

\subsubsection{Proof of \eqref{eq:mainthrm} {and \eqref{eq:PTbehavior}}}
\label{sec:2.3.1}
In this section we use Proposition \ref{prop:7} in order to first prove  \eqref{eq:mainthrm} and then \eqref{eq:PTbehavior}, which is a simple consequence of \eqref{eq:mainthrm}. The proof of the other claims stated in Theorem \ref{th:5}, concerning the regularity of $\rho_{\eps+\varepsilon_0(\l),\l}$ in $\eps$ for $\eps \neq 0$, will be proved in App. \ref{app:regularitydensityeps},  since it requires the same technical tools necessary to prove Proposition \ref{prop:7} (whose proof is also deferred to App.\ref{app:A}).

\\ $\bullet$ \textit{Case $\eps> 0$.} We start by considering the case of positive $\epsilon$.  We will prove below (in the part of proof dedicated to the case $\eps\le 0$) that $\rho_{\varepsilon_0(\l),\l}=0$, so that 
\begin{equation}
 \rho_{\eps+\varepsilon_0(\l),\l}=\mathfrak{d}_\eps   \rho_{\cdot+\varepsilon_0(\l),\l}=\overbrace{\sum_{h =h_\eps-1}^0\mathfrak{d}_\eps W^{(h)}_{0,1;\cdot;B}}^{:=I}+
\overbrace{\sum_{h <h_\eps-1}\left (  W^{(h)}_{0,1;B}-W^{(h)}_{0,1;0,B} \right)}^{:=J}\label{eq:2.66}
\end{equation}
where we used \eqref{eq:2.8}, the definition of $\mathfrak{d}_\eps$ (cf. the beginning of Section \ref{sec:2.3}) and the fact that $\mathfrak{d}_\eps(fg)_\cdot=(\mathfrak{d}_\eps f_\cdot)g_0+f_\eps \mathfrak{d}_\eps g_\cdot$. Using \eqref{eq:2.9} and recalling that 
$Y_{h-1;\eps,B}=Y_{h;\eps,B}+\mathcal B^Y_{h;\eps,B}$, that $Y_{0;\eps,B}=1/2$ (see Proposition \ref{lem:flowRC}) and $h_\eps\big|_{\eps=0}=-\infty$, 
we can rewrite
\begin{equation}
    \begin{split}
    I=&\sum_{h= h_\eps}^1\left [-\left (\mathfrak{d}_\eps \tilde g_{\cdot,\io}^{(h)}(0)\right )\left (Y_{-\io;0,B}-\sum_{k< h} \BB^Y_{k;0,B}\right )-\tilde g^{(h)}_{\eps,\infty}(0)\sum_{k=  h}^0 \mathfrak{d}_\eps \BB^Y_{k;\cdot,B}+\mathfrak{d}_\eps \Theta_{\rho,\cdot}^{(h-1)} \right ] \\ 
    J=&\sum_{\substack{h < h_\eps \\ \o=\pm}} \left [ -\tilde g^{(h)}_{\o,\io}(0)\frac{Y^{(1)}_{h-1;B}}{Z_{h-1}}+\Theta^{(h-1)}_{\rho,\o}\right ]- \sum_{h < h_\eps}\left [-\tilde g^{(h)}_{0,\io}(0)\left (Y_{-\io;0,s}-\sum_{k <h} \BB^Y_{k;0;B} \right )+\Theta^{(h-1)}_{\rho,0} \right ].
    \end{split} \label{eq:2.922222}\nn
\end{equation} 
Proposition \ref{lem:flowRC} implies that 
$|\sum_{k < h} \BB^Y_{k;0;B}|\leq C_\theta |\l|2^{\theta h}$ for any $\theta\in[0,1/2)$ and some $C_\theta>0$, so that, in particular, $|Y_{-\io;0,B}-1/2|\leq C|\l|$ for some $C>0$. 
The dominant behavior in $\epsilon$ as $\epsilon\to0^+$ of the right hand side of 
the first line of \eqref{eq:2.922222} comes from the term $-Y_{-\io;0,B}\sum_{h=h_\eps}^1\mathfrak{d}_\eps \tilde g_{\cdot,\io}^{(h)}(0)$, which is of order $\sqrt\eps$, as it follows from Lemma \ref{lem:propag1}; by the same lemma, it follows that $\mathfrak{d}_\eps \tilde g_{\cdot,\io}^{(1)}(0)=O(\eps)$ so, if desired, we can drop the 
term with $h=1$ 
from the sum over $h$ of $\mathfrak{d}_\eps \tilde g_{\cdot,\io}^{(h)}(0)$ without changing the value of the sum at dominant order in $\eps$
(it is actually convenient to do so, in order to compute exactly the asymptotic behavior of such a dominant term as $\eps\to0^+$, see below). 
Therefore, using \eqref{eq:2.73bis} and Lemma \ref{lem:propag1}, we can rewrite 
\begin{equation}
\begin{split} \left |\frac{I}{Y_{-\io;0,B}}+\sum_{h= h_\eps}^0\mathfrak{d}_\eps \tilde g^{(h)}_{\io,\cdot}(0)\right | \leq & C_\theta \eps\left (1+ \sum_{h= h_\eps}^02^{h(-\frac{1}{2}+\theta)}\right ) \leq  C_\theta' \eps ^{\frac{1}{2}+\theta},
\end{split}\label{eq:2.20}
\end{equation}
valid for any $\theta\in[0,1/2)$ and some $C_\theta,C_\theta'>0$. In order to extract the dominant term in $\eps$ from the right hand side
of the second line of \eqref{eq:2.922222}, we use the following remarkable identity, proved in Lemma
\ref{YZ-lemma} below: 
$ Y^{(1)}_{h-1;B}Z^{-1}_{h-1}=Y_{-\io;0,B}\left (1+\z_{-\io}\right )^{-1}(1+O(\eps)+O_\theta(|\l|\eps^\theta))$ (recall that $\zeta_{-\infty}$ is an $\eps$-independent constant of order $\lambda$, see Proposition \ref{lem:flowRC} and the discussion right after its statement), which is valid uniformly in $h$, for $h<h_\eps$. By using also 
\eqref{eq:2.10bis} and Lemma \ref{lem:3bis}, we find: 
\begin{equation}
J=Y_{-\io;0,B}\sum_{\substack{h < h_\eps}}\left [ 
  -\frac{\sum_{\o=\pm} \tilde g^{(h)}_{\o,\io}(0)}{1+\z_{-\io}} + \tilde g^{(h)}_{0,\io}(0) \right ]+O_\theta(\eps^{\frac{1}{2}+\theta}). 
\label{eq:2.21}
\end{equation}   
We now intend to compute explicitly the leading term in $\eps$ as $\eps\to0^+$ in the expressions for $I$ and $J$ derived above. 
We rewrite $\tilde g^{(h)}_{\eps,\io}(0)=G^{(h)}_\eps+R_{h,\eps}$ for $h \geq h_\eps$, 
where, letting $\xi_1:=1+\z_{-\io}$
and $\xi_2:=1+\a_{-\io}$, 
\begin{equation}\label{eq:Gheps}
G^{(h)}_\eps:=\int_{\mathbb{R}^2} \frac{dq}{2\pi^2} \frac{f^{(1)}_h(q)}{-\frac{\tilde q_\eps^2}{2}\xi_2
-iq_1\xi_1+\frac{q_2^2}{2}\xi_2},\qquad \forall h>h_\eps;\end{equation}
for $h=h_\eps$, $G^{(h)}_\eps$ is defined by a similar formula, with $\tilde f^{(1)}_h$ replacing $f^{(1)}_h$ in the right hand side of \eqref{eq:Gheps} (see the line following \eqref{eq:2.35} for the definition of $\tilde f^{(1)}_h$). Similarly, for  
$h <h_\eps$, we rewrite 
$\tilde g^{(h)}_{\o,\io}(0)=G^{(h)}_{\o}+R_{h,\o}$, with 
\begin{equation}G^{(h)}_{\o}:=\int_{\mathbb{R}^2}   \frac{dq}{2\pi^2} \frac{f^{(2)}_h(q)}{-iq_1+\xi_1^{-1}\xi_2 (\o \tilde q_\eps q_2+\frac{q_2^2}{2})}, \qquad \forall h< h_\eps. 
\end{equation}
The constants $G^{(h)}_\eps, G^{(h)}_{\o}$ satisfy the same bounds as $\tilde g^{(h)}_{\eps,\io}(0),\tilde g^{(h)}_{\o,\io}(0)$, respectively,
namely, 
$|G^{(h)}_\eps|\le C2^{h/2}$, $|G^{(h)}_{\o}|\le C\eps^{-1/2}2^{h}$ 
(see Lemma \ref{lem:propag1} and Lemma \ref{lem:3bis}), while the remainder terms  $R_{h,\eps},R_{h,\o}$ satisfy 
\be\begin{split} |R_{h,\eps}|&\leq C_\theta 2^{h(\frac{1}{2}+\theta)}, \quad |\mathfrak{d}_\eps R_{h,\cdot}|\leq C_\theta\eps 2^{h(-\frac12+\theta)},\\ |R_{h,\o}|&\leq 
C 2^h \eps^{-1/2}\Big(\eps+C_\theta |\lambda| \eps^{\theta}+2^{h-h_\eps}\min\{\eta|h-h_\eps|,1\}\Big), \end{split} \label{eq:Rbounds}\ee 
for any $\theta\in[0,1/2)$ and some $C,C_\theta>0$;
in the last term in the second line, 
$\eta=\eta_{+,+}$ is the same critical exponent as 
in \eqref{eq:2.71bis}, which can be bounded as $0\le \eta\le C_\theta|\lambda|^2\eps^{2\theta}$
for any $\theta\in[0,1/2)$ and some $C_\theta>0$ independent of $\eps,\lambda$ (cf. Lemma \ref{YZ-lemma}).  
The first bound in  \eqref{eq:Rbounds} follows from: (1) the bounds in \eqref{eq:2.29bis}, which imply in particular that $|\a_h-\a_{-\io}|,|\z_h-\z_{-\io}|\leq C_\theta|\l|2^{\theta h}$
and that, on the support of $f^{(1)}_h$, 
$|\boldsymbol{\alpha}_h(q)-\alpha_h|, |\boldsymbol{\zeta}_h(q)-\zeta_h|
\leq C_\theta|\l|2^{\theta h}$; (2) the fact that, on the support of $f^{(1)}_h$, 
$\rho(q)=O(2^{2h})$ (see \eqref{eq:2.26biss}); (3) the fact that $|r_h-r_{h_\eps-1}|\le C_\theta \eps|\lambda|2^{\theta h}$, 
that $|r_{h_\eps-1}+\xi_2\frac{\tilde q_\eps^2}{2}|\leq C\eps^{1+\theta}$ (to see this, use the very definition of $\tilde q_\eps$
as the positive solution to \eqref{eq:2.35}, recalling that $\mu_{h_\eps}=O_\theta(\lambda\eps^\theta)$)
and that, on the support of $f^{(1)}_h$, 
$|\boldsymbol{r}_h(q)-r_h|\leq C_\theta\eps|\l|2^{\theta h}$ (see 
\eqref{eq:4.71} and following lines).
The proof of the second bound in \eqref{eq:Rbounds} is analogous to that of the first of \eqref{eq:2.73bis}. 
The last bound in \eqref{eq:Rbounds} follows from the results summarized in Lemma \ref{YZ-lemma} below, in particular from the fact that, for any $\theta\in[0,1/2)$: (1) 
on the support of $f^{(2)}_h$, 
$|\tilde Z_{h-1}(q)/Z_{h-1}-1|\leq C_\theta|\l|^2\eps^\theta$
and $\rho_\omega(q)-\xi_2\frac{q_2^2}{2}=O(2^{2h})+O_\theta(2^{2h}\lambda \eps^{-1+\theta})$; 
(2) $v_\eps=\xi_1^{-1}\xi_2 \tilde q_\eps(1+O(\eps)+O_\theta(|\lambda|\eps^{\theta}))$; (3) $Z_{h-1}=\xi_1 2^{-\eta(h-h_\eps)}(1+O(\eps)+O_\theta(\lambda \eps^\theta))$.

The bounds \eqref{eq:2.20}, \eqref{eq:2.21}, \eqref{eq:Rbounds}
readily imply that 
\begin{equation}
\begin{split}
    \frac{I+J}{Y_{-\io;0,B}}=& - \sum_{h=h_\eps}^0 \mathfrak{d}_\eps G_\cdot^{(h)}-\sum_{h <h_\eps}\left ( \sum_{\o=\pm} \xi_1^{-1}G^{(h)}_{\o}-G^{(h)}_0 \right )+O_\theta(\eps^{1/2+\theta}) \\ = & -\sum_{h \geq h_\eps} \mathfrak{d}_\eps G_\cdot^{(h)}-\sum_{h <h_\eps}\left ( \sum_{\o=\pm} \xi_1^{-1}G^{(h)}_{\o}-G^{(h)}_0 \right ) +O_\theta(\eps^{1/2+\theta}) 
        \label{eq:2.71}
    \end{split}
\end{equation}
where in passing from the first to the second line we added and subtracted $\sum_{h \geq 1} \mathfrak{d}_\eps G^{(h)}_\cdot$, 
which is of order $\eps$. We now further manipulate the 
last line of \eqref{eq:2.71} as follows: first of all, in the integral expression defining $G^{(h)}_\omega$ we perform the change of variables $q \mapsto q-\o\tilde{\boldsymbol q}_\eps$, which gives
\[\sum_{h <h_\eps}\xi_1^{-1}G^{(h)}_{\o}=\int_{\mathbb{R}^2} \frac{dq}{2\pi^2}\frac{\sum_{h <h _\eps} f^{(2)}_h(q-\o {\boldsymbol q}_\eps)}{-\xi_2\frac{\tilde q_\eps^2}2 -iq_1\xi_1+\xi_2q_2^2/2}.\] 
We now use the fact that 
$\sum_{h > h_\eps} f^{(1)}_{h}(q)+\tilde f^{(1)}_{h_\eps}+\sum_{\o=\pm}\sum_{h < h_\eps} f^{(2)}_h(q-\o \tilde q_\eps)\equiv1$, which implies 
\be \sum_{h \geq h_\eps} \mathfrak{d}_\eps G^{(h)}_\cdot+\sum_{h < h_\eps}\left ( \sum_{\o=\pm}\xi_1^{-1} G^{(h)}_{\o}-G^{(h)}_0\right )=\int_{\mathbb{R}^2} \frac{dq}{2\pi^2} \mathfrak{d}_\eps\Big(-\xi_2\frac{\tilde q_\eps^2}{2}-iq_1 \xi_1+\frac{q_2^2}{2}\xi_2\Big)^{-1}=-\frac{2\tilde q_\eps}{\pi\xi_1},\ee 
where the last identity requires an explicit computation of
the integral via residue's theorem. Plugging this into 
\eqref{eq:2.71} and then into \eqref{eq:2.66} gives: 
\be  \rho_{\eps+\varepsilon_0(\lambda),\l}=Y_{-\io;0,B}\frac{2\tilde q_\eps}{\pi\xi_1}+O_\theta(\eps^{1/2+\theta})\equiv \frac{\sqrt{2\eps}}{\pi}A(\l)+O_\theta(\eps^{1/2+\theta}), 
\label{eq:2.81}\ee
where $A(\lambda)$ is an analytic function of $\lambda$, such that $A(0)=1$, defined as $ A(\l):=\frac{2Y_{-\io;0,B}}{1+\z_{-\io}}\lim_{\eps\to0^+}\frac{\tilde q_\eps}{\sqrt{2\eps}}$. For a proof that $A(\lambda)-1=O(\lambda^2)$, see App. \ref{app:firstorder}. This concludes the proof of  \eqref{eq:mainthrm} in the case $\eps>0$. 

\medskip

$\bullet$ \textit{Case $\eps \le 0$.} We first consider the case of $\eps$ negative, in which case we will prove that the density is identically zero. 
By continuity we will then infer that the density vanishes at $\eps=0$ as well.
Thanks to Proposition \ref{prop:7} and \eqref{eq:mlimit}, we know that {$\rho_{\epsilon+\varepsilon_0(\lambda),\lambda}$} is analytic in $\lambda$ uniformly in {$\epsilon<0$. 
} This implies that the power series expansion
\begin{equation}
{\rho_{\epsilon+\varepsilon_0(\lambda),\lambda}}=\sum_{n\ge0}\frac{\lambda^n}{n!}c_n(\epsilon)\label{eq:analyticexpan}
\end{equation}
is convergent for $|\lambda|$ sufficiently small. Since analyticity holds uniformly in $L$ and $m$ (as long as $\eps<0$), we can exchange sums, limits and derivatives below. Therefore, from \eqref{eq:dimercorr} and \eqref{eq:3.2.1} {at $m=0$}, it follows that
\begin{equation}
c_n(\epsilon)=\lim_{L\to\infty}\partial_{A_{\bar e}}\partial^n_\lambda\log\sum_{M\in\Omega_L}w_{\epsilon,0}(M)\,e^{B_\lambda(M)+\sum_{e\in E_L}A_e\mathbbm 1_e(M)}\Big|_{\underline A=0,\,\lambda=0},\label{eq:rhoepsser}
\end{equation}
where
 $\bar e$ is an arbitrary edge of type $B$ or $C$, say the edge of type $B$ whose white vertex 
has coordinate $x=(0,0)$, and 
$B_\lambda:=\l V+\varepsilon_0(\l)(N_B+N_C)$. In order to evaluate $c_n$, we can first expand 
the right hand side of \eqref{eq:rhoepsser} in $\lambda$ and $\varepsilon_0$ separately, and then use the fact that $\varepsilon_0(\lambda)$ is analytic in $\lambda$. Namely, $c_n$ is the coefficient of the series obtained by composing $\sum_{s,t \geq 0} \frac{\l^s \varepsilon_0^t(\l)}{s!t!}d_{s,t}(\epsilon)$ with  $\varepsilon_0(\l)=\sum_{k \geq 1} \frac{\l^k}{k!} \varepsilon^{(k)}_0(0)$, where  
\begin{equation}d_{s,t}(\epsilon)=\lim_{L \to \infty} \mathbb{E}_{\eps;0,L}\Big(\overbrace{V;\cdots;V}^{s {\rm\,  times}}\, ;\overbrace{N_B+N_C; \cdots; N_B+N_C}^{t {\rm \, times}}\, ;\mathbbm{1}_{\bar e}\Big)\label{eq:dsteps}\end{equation}
is the joint cumulant of its $s+t+1$ arguments
with respect to the expectation $\mathbb E_{\eps;0,L}$ 
associated with the probability distribution $\mathbb P_{\eps;0,L}$ on $\Omega_L$ whose probability
weight is $w_{\eps,0}(M)/\sum_{M'\in\Omega_L}w_{\eps,0}(M')$, $\forall M\in\Omega_L$. We now show that $d_{s,t}(\epsilon)=0$, 
which implies $c_n(\epsilon)=0$. The results is a simple corollary of the following lemma. 
\begin{lemma}
\label{lem:3.5.0.1}
For any $\eps<0$ and any non-negative integer $m$,  
there exists $\gamma(\eps)$, $L_0(\eps)$, $c(m)>0$ such that, given any $m$-ple of sets $S_1,\ldots,S_m\subset E_L$, and letting 
$\mathbbm{1}_{S_i}:=\prod_{e \in S_i} \mathbbm{1}_e$, then, for $L\ge L_0(\eps)$,
\begin{equation}\label{eq:3.14}
    \Big |\mathbb{E}_{\epsilon;0,L}\Big(\mathbbm{1}_{S_1};\cdots;\mathbbm{1}_{S_m};\mathbbm 1_{\bar e}\Big) \Big | \leq c(m) e^{-\g(\eps) L}. 
\end{equation}
\end{lemma}
\begin{proof}[Proof of Lemma \ref{lem:3.5.0.1}]
We let $S_{m+1}:=\{\bar e\}$ and expand the truncated into simple expectations, thus finding  
\begin{equation}\label{eq:3.15}
\mathbb{E}_{\epsilon; 0,L}\Big( \mathbbm{1}_{S_1};\cdots;\mathbbm{1}_{S_{m+1}}\Big)=\sum_{\pi} (-1)^{|\pi|-1}(|\pi|-1)!\prod_{Y \in \pi} \mathbb{E}_{\epsilon;0,L}\big(\mathbbm{1}(Y)\big) 
\end{equation}
where the
sum runs over partitions $\pi$ of $\{1,\dots,m+1\}$ and $\mathbbm{1}(Y)=\prod_{i \in Y} \mathbbm{1}_{S_i}$. We now brutally bound the absolute value of the right hand side 
of \eqref{eq:3.15} by $\big(\sum_\pi (|\pi|-1)!) \mathbb{E}_{\epsilon;0,L}(\mathbbm{1}_{\bar e})$, where, letting 
$\bar b,\bar w$ be the black and white vertices of $\bar e$ (we shall write $\bar e=(\bar b,\bar w)$), and
using Kasteleyn's solution, see e.g. \cite[eq.(2.38)]{notedimeri},
we can explicitly write: 
\be \mathbb{E}_{\eps;0,L}(\mathbbm{1}_{\bar e})=\frac{\sum_{\vartheta}\frac{c_\vartheta}{2}  K_\vartheta(\bar b, \bar w)  K_{\vartheta}^{-1}(\bar w, \bar b)D_\vartheta}{\sum_{\vartheta}\frac{c_\vartheta}{2}  D_\vartheta},  \label{eq:blablablabla}\ee where: $c_\vartheta$ was defined 
right after \eqref{eq:Zinteract}; letting $K(b,w)=K_{s(b,w)}$, with $K_s$ as defined after \eqref{eq:3.2.1} and $s(b,w)\in\{A,B,C\}$ the type 
of the edge $e=(b,w)$, we denoted by 
$K_\vartheta$ the $|\Lambda|\times|\Lambda|$ matrix of elements $K_\vartheta(b,w):=(-1)^{\vartheta_1}K(b,w)$ if 
$x_1(w)=L$ and $x_1(b)=1$, $K_\vartheta(b,w)=(-1)^{\vartheta_2}K(b,w)$ if $x_2(w)=L$ and $x_2(b)=1$, and $K_\vartheta(b,w)=K(b,w)$ otherwise; we defined 
$D_\vartheta:=\prod_{k \in \DD_\vartheta} \mu(k)$. Recall that, for $\eps<0$, $|\mu(k)|
=|1+\frac{e^\eps}2(e^{ik_1}+e^{ik_2})|\ge 
1-e^\eps>0$, uniformly in $k\in[-\pi,\pi]^2$; therefore, using Poisson's summation formula, 
$D_\vartheta=\exp\{L^2f_\infty+O(e^{-\gamma(\eps) L})\}$, where $f_\infty:=\int_{[-\pi,\pi]^2}\frac{dk}{(2\pi)^2}\log |\mu(k)|$
and the remainder $O(e^{-\gamma(\eps) L})$
is uniform in $\vartheta$, for any $L\ge L_0(\eps)$. Using also the fact that $\sum_{\vartheta}c_\vartheta/2=1$, we readily find that 
\begin{equation}
\mathbb{E}_{\epsilon;0,L}(\mathbbm 1_{\bar e}) 
\le C \sup_\vartheta |K^{-1}_{\vartheta} (\bar w, \bar b)|
\end{equation}
for some $C>0$. In view of this bound, 
the desired exponential decay, \eqref{eq:3.14}, 
follows from Lemma \ref{lem:13} and \eqref{eq:A.16}, because the  difference of coordinates of $\bar w $ and $\bar b$ for $\bar e$ of type $B$ is $(-1,0)$, so that $K^{-1}((-1,0))=0$ for any $\eps<0$. 
\end{proof}

Let us go back to the proof that the cumulant $d_{s,t}(\eps)$ in \eqref{eq:dsteps} vanishes. 
Recall that 
$V$ is the sum over faces $f$ of $\mathbbm{1}_{f;1}\mathbbm{1}_{f;4}+\mathbbm{1}_{f;2}\mathbbm{1}_{f;5}+
\mathbbm{1}_{f;3}\mathbbm{1}_{f;6}$, see
\eqref{defintermodel}; therefore, by the multi-linearity of the cumulant, $d_{s,t}(\eps)$ can be written as a linear combination of terms of the same form as the left hand side of \eqref{eq:3.15}. The number of terms in this linear combination is $O(L^{2(t+s)}).$  In particular, by Lemma \ref{lem:3.5.0.1} the cumulant can be bounded by a constant times $L^{2(t+s)}e^{-\g(\eps) L}$, which vanishes 
in the limit $L\to\infty$.

\noindent This concludes the proof that $\rho_{\epsilon+\varepsilon_0(\lambda),\lambda}=0$ for any $-\bar\eps\le\epsilon<0$. Recall that $\eps\mapsto F_{\eps+\varepsilon_0(\l),\l} $ is convex (hence continuous); its derivative  $\eps\mapsto\rho_{\eps+\varepsilon_0(\l),\l} $ is non-negative and non-decreasing. By monotonicity of the density, we immediately infer that 
$\rho_{\eps+\varepsilon_o(\lambda),\lambda}$ is identically zero for all $\eps<-\bar\eps$ as well. 
In order to conclude that the density vanishes also for $\eps=0$ we use continuity of $\lim_{L\to\infty}\rho^{(\vartheta_0)}_{\eps+\varepsilon_0(\lambda),\lambda,L,0}$, which follows from 
the uniform convergence in $\eps$ of $\rho^{(\vartheta_0)}_{\eps+\varepsilon_0(\lambda),\lambda,L,0}$ to its $L\to\infty$ limit stated in Proposition \ref{prop:7}, see Appendix \ref{app:A} for the proof and additional details. 

\paragraph{Proof of \eqref{eq:PTbehavior}.}
\label{app:corollary}
Given \eqref{eq:mainthrm},  \eqref{eq:PTbehavior} is an immediate consequence. For $\eps\le 0$ the limit density of $B,C$ dimers is $0$, hence the limit density of $A$ dimers is one. The model is in a frozen phase, with vanishing entropy and the energy per unit volume is exactly $\lambda$. This implies  \eqref{eq:PTbehavior} for $\eps\le 0$. For $\eps>0$,  \eqref{eq:PTbehavior} follows 
from the facts that $\eps\mapsto F_{\eps+\varepsilon_0(\l),\l,L,m}$ is convex, that it tends to $\lambda$ (as $L\to \infty,m\to 0^+$) at $\eps=0$ and that its derivative tends to a continuous limit given by the r.h.s. of \eqref{eq:mainthrm}. Integrating the latter from $0$ to $\eps$ gives \eqref{eq:PTbehavior}.

\subsubsection{{Proof of \eqref{eq:mlimit}}}\label{sec:2.3.2}

Letting in shorthand notation $\rho_{L,m}:=\rho_{\epsilon+\varepsilon_0(\lambda),\lambda,L,m}$ and $F_{L,m}:=F_{\epsilon+\varepsilon_0(\lambda),\lambda,L,m}$, we rewrite: 
\begin{equation}\label{eq:sectors}
e^{L^2F_{L,m}}=\frac12\sum_\vartheta c_\vartheta\,\Xi^{(\vartheta)}_{L,m},\qquad
\rho_{L,m}=\frac{\sum_\vartheta c_\vartheta\,\Xi^{(\vartheta)}_{L,m}\,\rho^{(\vartheta)}_{L,m}}{\sum_\vartheta c_\vartheta\,\Xi^{(\vartheta)}_{L,m}},
\end{equation}
where for $\vartheta\in\{0,1\}^2$ and $m\ge0$,
$\Xi^{(\vartheta)}_{L,m}:=e^{\mathcal W^{(\vartheta)}_{\epsilon+\varepsilon_0(\lambda),\lambda,L,m}(0)},\rho^{(\vartheta)}_{L,m}:=\partial_{A_{(0,B)}}\mathcal W^{(\vartheta)}_{\epsilon+\varepsilon_0(\lambda),\lambda,L,m}(0).$

The desired result follows from the following proposition, proved in Appendix \ref{app:proofprop7}. 

\begin{proposition}\label{prop:rgmassa}   Let $|\eps| \leq \bar \eps, \eps \neq 0$, $0<m<c_0\sqrt{|\eps|}$, with $c_0$ a small constant. Let $\vartheta_0$ be equal either to $(1,0)$ or to $(0,1)$. \begin{itemize}
    \item[(i)] There exists $C,c,L_0>0$, independent of $m$ if $\eps<0$, such that for $L \geq L_0$ and any $\vartheta \in \{0,1\}^2$ \[\left |\frac{\Xi^{(\vartheta)}_{L,m}}{\Xi^{(\vartheta_0)}_{L,m}}-s_\vartheta\right |\leq C e^{-c\sqrt{L}}, \qquad |\rho^{(\vartheta)}_{L,m}-\rho^{(\vartheta_0)}_{L,m}|\leq Ce^{-c\sqrt{L}}, \] where $s_\vartheta=
    1$ for $\vartheta \neq (0,0)$ and $s_{(0,0)}=-\sgn \eps.$
    \item[(ii)] The limits $m\to 0^+$ and $L\to\infty$ of $\rho^{(\vartheta_0)}_{L,m}$, commute, i.e.,     
    $\lim_{m \to 0^+} \lim_{L \to \io} \rho^{(\vartheta_0)}_{L,m}=\lim_{L \to \io} \rho^{(\vartheta_0)}_{L,0}$ and the result is independent of the choice of $\vartheta_0$; the same is true for $L^{-2}\WW^{(\vartheta_0)}_{\eps+\varepsilon_0(\l),\l,L,m}(0)$.
\end{itemize}
\end{proposition}Note that from item (i), the effect of the regularization  for $m>0$, is to select, in the thermodynamic limit, the boundary condition $\vartheta_0$; if $\eps<0$ this holds uniformly in $m\to 0^+$, hence also at $m=0$ by continuity at fixed $L$. Now let us deduce \eqref{eq:mlimit} by combining (i) and (ii).

\begin{proof}[Proof of \eqref{eq:mlimit}]
We first consider the case $\epsilon\neq 0$, so that we can apply Proposition \ref{prop:rgmassa}.
Let $a_{L,m}^{(\vartheta)}:=\Xi_{L,m}^{(\vartheta)}/\Xi_{L,m}^{(\vartheta_0)}$ and $\Sigma_{L,m}:=\sum_{\vartheta}c_\vartheta a_{L,m}^{(\vartheta)}/2.$ We can rewrite \eqref{eq:sectors} as 
\begin{equation}
 \rho_{L,m}=\rho_{L,m}^{(\vartheta_0)}+\frac{1}{\Sigma_{L,m}}\sum _\vartheta \frac{c_\vartheta}{2} a^{(\vartheta)}_{L,m} (\rho_{L,m}^{(\vartheta)}-\rho_{L,m}^{(\vartheta_0)}), \quad F_{L,m}=L^{-2}( \WW^{(\vartheta_0)}_{\eps+\varepsilon_0(\l),\l,L,m}(0)+\log \Sigma_{L,m}).
\end{equation}
From the definition of $c_\vartheta$ in \eqref{eq:Zinteract} and item (i) of Proposition \ref{prop:rgmassa}, since $\sum_{\vartheta}c_\vartheta s_\vartheta/2=(3+\sgn \eps)/2$,  it follows that $|\Sigma_{L,m}-(3+\sgn \eps)/2|\leq Ce^{-c\sqrt{L}}$ and, in turn, that \begin{equation}
        |\rho_{L,m}-\rho^{(\vartheta_0)}_{L,m}|\leq C' e^{-c\sqrt{L}}, \qquad |F_{L,m}-L^{-2}\WW^{(\vartheta_0)}_{\eps+\varepsilon_0(\l),\l,L,m}(0)|\leq C'L^{-2}.
    \end{equation}
    Letting first $L\to\infty$ and then $m\to0^+$,  item (ii) of Proposition \ref{prop:rgmassa} gives
\[
\rho_{\epsilon+\varepsilon_0(\lambda),\lambda}=\lim_{m\to0^+}\lim_{L\to\infty}\rho_{L,m}=\lim_{m\to0^+}\lim_{L\to\infty}\rho^{(\vartheta_0)}_{L,m}=\lim_{L\to\infty}\rho^{(\vartheta_0)}_{L,0},
\]
and similarly $F_{\epsilon+\varepsilon_0(\lambda),\lambda}=\bar F_{\epsilon+\varepsilon_0(\lambda),\lambda}$, that is the first of \eqref{eq:mlimit}. \\ 

Finally, we consider the case $\epsilon=0.$ Consider the second equation in \eqref{eq:mlimit}. The left and right sides coincide for $\epsilon\ne0,$ as we have just proven, and they equal zero for $\epsilon<0$, as proven in Section \ref{sec:2.3.1} The right hand side is continuous in $\epsilon$, by Proposition \ref{prop:7}. Since the left-hand side is non-negative and non-decreasing in $\epsilon$, it follows that it also equals zero at $\epsilon=0$, hence the second equality in \eqref{eq:mlimit} holds for $\epsilon=0$ as well. This readily implies the first equality in \eqref{eq:mlimit} as well.
\end{proof}

\appendix
\section{Symmetries of the effective theory}\label{app:symmetries}

In this appendix we sketch the proof of the  symmetries of the massless effective interaction, i.e. at $m=0$, at scales $h \leq 0.$ We drop the label $\epsilon$ from the kernels and 
running coupling constants, since their dependence upon $\epsilon$ is not relevant for the incoming discussion. On the other hand, since the symmetries act non-trivially on the boundary condition $\vartheta$, cf. \eqref{eq:S1S2}, we restore the dependence upon $\vartheta$ wherever relevant, by introducing an extra label: e.g., we will denote the kernels in momentum space by $\hat W^{(h)}_{2n,m;\un s,\vartheta}(\underline{q},\un p)$.

\medskip 

We use the following general fact: if the Gaussian Grassmann integration $P(d\varphi)$ and the potential $V(\varphi,J)$ are symmetric\footnote{By saying that a Gaussian Grassmann integration $P(d\varphi)\propto D\varphi e^{Q(\varphi)}$ is symmetric we mean that 
the quadratic action $Q$ is, that is $\mathcal{S}[Q(\psi)]=Q(\psi)$.} under a transformation $\mathcal{S}$ acting on the Grassmann algebra $\{\varphi^\pm_y\}_{y \in \tilde \L}$ and on the external fields $\{J_{z,s}\}_{z \in \tilde \Lambda}^{s=A,B,C}$,  then the effective potential $V'$ defined by
\begin{equation}
    e^{V'(\varphi',J)}=\int P(d\varphi)e^{V(\varphi+\varphi',J)}\label{eq:B.4.3}
\end{equation}
is symmetric under $\mathcal{S}$ as well; note that the definition of $V'$ in \eqref{eq:B.4.3} is equivalent to $V'(\varphi',J)=\sum_{n \geq 1} \frac{1}{n!}\mathcal{E}^T(V(\varphi'+\cdot,J);n)$, to be compared with \eqref{eq:4.68},\eqref{eq:2.52bis}. As a consequence of this fact we find that the symmetries \eqref{eq:S1S2bis} satisfied by the bare action $\tilde S_\vartheta(\varphi)+\nu N_\vartheta(\varphi)+\tilde V_\vartheta(\varphi,\AA(J))$ in \eqref{eq:5.71}  are preserved at each scale of the multiscale construction, in the following sense. Let $I, II$ denote the two regimes of scales $h \geq h_\eps,h <h_\eps$, respectively.
\begin{lemma}\label{lem:symmetries}
    The kernels of the effective potential $V^{(h)}(\varphi,J)$, see \eqref{eq:4.68}, the lines below \eqref{eq:2.32bis}, \eqref{eq:2.39bis} and following lines, and \eqref{eq:2.51biss} and following lines,  satisfy the symmetries stated in \eqref{eq:2.33bis},\eqref{eq:2.43bis},\eqref{eq:upps}, i.e. 
\begin{equation}
\begin{split}
& \hspace{5pt} I:  \hat W^{(h)}_{2n,m;\un s,\vartheta}(\underline{q},\un p)=\overline{\hat W^{(h)}_{2n,m;\un s,\vartheta}(-\underline{q},-\un p)}, \qquad \hspace{12pt} \hat W^{(h)}_{2n,m;\un s, \vartheta}({\underline{\sigma_3q}},{\underline{ \sigma_3 p}})=\hat W^{(h)}_{2n,m;\un s^*,{\sigma_1 \vartheta}}(\underline{q},\un p)
        \\ & II: \overline{\hat{W}^{(h)}_{2n,m;\un s,\underline{\o},\vartheta}(\underline{q},\underline{p})}=\hat{W}^{(h)}_{2n,m;\un s,-\underline{\o},\vartheta}(-\underline{q},-\un p), \qquad  \hat{W}^{(h)}_{2n,m;\un s,\underline{\o},\vartheta}({\underline{\sigma_3q}},{\underline{ \sigma_3 p}})=\hat{W}^{(h)}_{2n,m;\un s^\ast,-\underline{\o},{\sigma_1 \vartheta}}(\underline{q},\un p) 
        \label{eq:B.4.5}
             \end{split}
\end{equation}
where $\sigma_1,\sigma_3$ denote the first and third Pauli matrices, $\underline{\sigma_3 q}:=(\sigma_3 q_1,\dots,\sigma_3q_{2n})$ and similarly for $\underline{\sigma_3 p}$; $\underline{s}^*:=(s_1^*,\dots,s_m^*)$ where $A^*:=A,B^*:=C,C^*:=B.$ As a consequence, taking the $L \to \io$ limit of \eqref{eq:B.4.5}, we have that the running constants in \eqref{eq:rcc1regime}, those in \eqref{eq:2.32bis} evaluated at $q=0$, and those in \eqref{eq:2.43biss}, satisfy \begin{equation}
\begin{split}
    \nu_h, r_h, \a_h, \z_h, z_h, \l_h, \mu_h, Y_{h;s},Y_{h;s,\un \o} \in \mathbb{R}, \qquad  Y_{h;s,(+,+)}=Y_{h;s,(-,-)}, \quad Y_{h;s,(+,-)}=Y_{h;s,(-,+)} 
\end{split}
\end{equation}
\end{lemma}
\begin{proof}[Proof of Lemma]
Assume inductively the statement valid up to  scale $h$, with $h>h_\eps$. Using the multiscale decomposition and the observation in \eqref{eq:B.4.3} we only need to check that the Guassian measure $\tilde P^{(h)}$ with propagator \eqref{eq:singlescalepropI} is symmetric; we can write $\tilde P^{(h)}(D\varphi) \propto D\varphi\, e^{S_{\vartheta,h}(\varphi)}$  where 
\begin{equation}
\begin{split}
  S_{\vartheta,h}(\varphi)=-L^{-2} \sum_{\substack{q \in \mathcal{D}'_{\vartheta,R} \\ f^{(1)}_h(q) \neq 0}} \hat\varphi^{+}_{q+q_F} (f^{(1)}_h(q))^{-1}\tilde{\mu}_h(q) \hat\varphi^{-}_{q+q_F}
\end{split}
\end{equation}
with $D'_{\vartheta,R}$ as in \eqref{eq:5.71}, and $\tilde \mu_h(q):={\boldsymbol r}_{h-1}(q)-iq_1(1+{\boldsymbol \z}_{h-1})+2^{-1}q_2^2(1+\boldsymbol{\a}_{h-1})+\rho(q)$, where $\rho(q)$ was defined in \eqref{eq:2.26biss}, $f^{(1)}_h$ right after \eqref{eq:singlescalepropI}, and $q_F=(\pi,0)$. Note that ${\boldsymbol r}_{h-1},{\boldsymbol \a}_{h-1},\boldsymbol{\z}_{h-1} \in \mathbb{{R}}$, as it follows from the inductive hypothesis using the definition in  \eqref{eq:2.32bis} and the relations in \eqref{eq:B.4.5}
in the limit $L \to \io$ (recall that the thermodynamic limit of the kernels is independent of $\vartheta$, see the lines after  \eqref{inftvolprop}). Using that ${\mathcal{D}'_{\vartheta,R}}=-\mathcal{D}'_{\vartheta,R}$ and  $\sigma _3{\mathcal{D}'_{\vartheta,R}}=\mathcal{D}'_{\sigma_1 \vartheta,R}$ mod $(2\pi,2\pi)$\footnote{This follows from the observations that $q_F=-q_F$, $\DD_{\vartheta,R}=-\DD_{\vartheta,R}$, and  $\sigma_3q_F=q_F, \sigma_3D_{\vartheta,R}=D_{\sigma_1\vartheta,R}$ mod $(2\pi,2\pi)$, which are just restatements 
of the claims done after \eqref{Steta} in the rotated lattice.} together with $f^{(1)}_h(q)=f^{(1)}_h(-q)=f^{(1)}_h(\sigma_3q) \in \mathbb{R}$,  $\tilde \mu_h(-q)=\overline{\tilde \mu(q)}, \tilde \mu(\sigma_3q)=\tilde \mu(q)$,  we find that  $S_{\vartheta,h}$ is symmetric under $\SS_1,\SS_2$.
Thanks to \eqref{eq:B.4.3}, the interaction $V^{(h-1)}$ (cf.  \eqref{eq:4.68}) is symmetric under $\SS_1,\SS_2$ as well and, since the symmetries preserve the degree of Grassmann monomials, the invariance is valid order by order, for any fixed value of $n$ and $m$, i.e., 
$\mathcal S_1 V^{(h)}_{2n,m;\underline{s},\vartheta}(\varphi,J)=
V^{(h)}_{2n,m;\underline{s},\vartheta}(\varphi,J)$ and 
$\mathcal S_2 V^{(h)}_{2n,m;\underline{s},\vartheta}(\varphi,J)=
V^{(h)}_{2n,m;\underline{s}^*,\sigma_1\vartheta}(\varphi,J)$. 
By using the explicit representation 
of $V^{(h)}_{2n,m;\underline{s},\vartheta}$ in terms of its kernel, the symmetry property \eqref{eq:B.4.5}, i.e. \eqref{eq:2.33bis}, readily follows (see \eqref{eq:concrete}
and following lines for an illustration of how concretely the proof of this conclusion goes in the case $n=1$, $m=0$). 

Once we reach scale $h_\eps $, we rewrite
\eqref{eq:qp}
\begin{equation}
    \f_{y,s}^{\sigma,(\leq h_\eps )}=\f^{\sigma,(\leq h_\eps -1)}_y+{\f}^{\sigma,(h_\eps )}_y, \qquad \f^{\sigma,(\leq h_\eps -1)}_y=\sum_{\omega=\pm}e^{-i\sigma\omega y \boldsymbol{q}_\eps}\f^{\sigma,(\leq h_\eps -1)}_{y,\omega}
\end{equation}
which implies that
the symmetries $\mathcal{S}_i$, $i=1,2$, 
act on the quasi-particle fields as follows: 
\begin{equation}
\begin{split}
\mathcal{S}_1:\hat\varphi^{\sigma}_{q,\o} \longrightarrow \hat\varphi^{\sigma}_{-q,-\o}, \quad c \to \overline{c}, \quad \vartheta \to \vartheta; \qquad   \qquad 
   \mathcal{S}_2:\hat\varphi^{\sigma}_{q,\o} \longrightarrow \hat\varphi^{\sigma}_{\sigma_3 q,-\o}, \quad  c \to c, \quad \vartheta \to \sigma_1\vartheta,  \label{eq:B.4.10}
   \end{split}
\end{equation}
where $\hat \varphi^\pm_{q,\o}:=\hat \varphi^\pm_{q+q_F+\o{\boldsymbol{q}_\eps}}$, and $\boldsymbol{q}_\eps$ is defined as in \eqref{eq:2.35}. 
Note that in \eqref{eq:2.34}  the term $\eps\mu_\eps$ is real by hypothesis, so it preserves both  the symmetry of the Gaussian Grassmann integration and that of the interaction at scale $h_\eps -1$. An iterative application of \eqref{eq:B.4.3}, together with the fact that $\tilde P^{(h)}_{Z_{h-1}}$ in \eqref{eq:singlescaleprop2} is symmetric under \eqref{eq:B.4.10}, implies that the symmetries \eqref{eq:B.4.10} are preserved at each scale $h <h_\eps.$  Using again the fact that the symmetries preserve the degree of every Grassmann monomial, by proceeding as described above the statement in the second line of  \eqref{eq:B.4.5} readily follows.

For illustration purposes, let us explicitly show, 
using the Fourier representation of the effective potentials (see \eqref{eq:upps} and lines above), 
how the symmetries of the kernels 
are proved, for $n=1, m=0$ and $h<h_\eps$; the other cases follow from a simple adaptation of the same argument, and are left to the reader. Let us  consider the action of $\mathcal S_2$ on 
the terms with $n=1$ and $m=0$ of the effective potential 
at scales $h<h_\epsilon$: 
\begin{equation}\label{eq:concrete}
\begin{split}
    & L^{-2}\sum_{q \in \mathcal{D}'_{\vartheta,R,\o}} \hat \varphi^+_{q,\o} \hat W^{(h)}_{2,0;\o,\vartheta}(q) \hat \varphi^-_{q,\o} \stackrel{\mathcal{S}_2}{=}L^{-2}\sum_{q \in \mathcal{D}'_{\sigma_1\vartheta,R,\o}}\hat\varphi^+_{\sigma_3 q,-\o}W^{(h)}_{2,0;\o,\sigma_1\vartheta}(q)\hat\varphi^-_{\sigma_3q,-\o}= \\ &=L^{-2}\sum_{q \in \mathcal{D}'_{\vartheta,R,\o}} \hat \varphi^+_{q,\o} \hat W^{(h)}_{2,0;-\o,\sigma_1\vartheta}(\sigma_3q) \hat \varphi^-_{q,\o}
    \end{split}
\end{equation}
where we used the definition of the kernels \eqref{eq:2.27bisss}, of $\hat \varphi^\pm_{q,\o}$ in \eqref{eq:B.4.10}, and the fact that $\sigma_3{\mathcal{D}_{\vartheta,R,\o}}=\mathcal{D}_{\sigma_1 \vartheta,R,-\o}$ mod $(2\pi,2\pi)$.
In particular, taking the infinite volume limit, and recalling that the limiting kernels are independent of the boundary conditions $\vartheta$, we find that $\hat W^{(h);\infty}_{2,0;\o}(\sigma_3q)=\hat W^{(h);\infty}_{2,0;-\o}(q)$. Repeating the same argument for $\SS_1$ implies similarly that 
$\hat W^{(h);\infty}_{2,0;\o}(q)=\hat W^{(h);\infty}_{2,0;-\o}(-q)$, so that $\hat W^{(h);\infty}_{2,0;\o}(0) \in \mathbb{R}, \hat W^{(h);\infty}_{2,0;\o}(0)=\hat W^{(h);\infty}_{2,0;-\o}(0)$, $ \partial_1\hat W^{(h);\infty}_{2,0;\o}(0) \in i\mathbb{R}$, $ \partial_1\hat W^{(h);\infty}_{2,0;\o}(0)=\partial_1\hat W^{(h);\infty}_{2,0;-\o}(0), \partial_2\hat W^{(h);\infty}_{2,0;\o}(0) \in \mathbb{R}, \partial_2\hat W^{(h);\infty}_{2,0;\o}(0)=-\partial_2\hat W^{(h);\infty}_{2,0;-\o}(0)$, which proves that $\mu_h,z_h,\d_h\in\mathbb R$. A similar procedure proves the reality of the other 
running coupling constants, as well as the symmetries of $Y_{h:s,(\omega,\omega')}$
and of the kernels with general $n$ and $m$, both in the first and second regimes of scales. 
\end{proof}

\section{Tree expansion}
\label{app:A}
In this appendix we recall the \textit{Gallavotti-Nicol\`o} tree expansion and explain how to adapt it to the present context, in order to show, in particular: 
\medskip

\par{1.} how to control the flow of the constants of the first regime (Proposition \ref{lem:flowRC}), and specifically how to determine the function $\varepsilon_0(\l)$ in Theorem \ref{th:5}; 

\medskip

\par{2.} how to prove Proposition \ref{prop:7}, and specifically how to obtain \eqref{eq:2.9}-\eqref{eq:2.10bis}.
\medskip

Our main references for this section will be \cite[Sec. 2.1 \& 3.1]{BM36} and references therein; for additional details, the reader can also look at \cite[Sec. 5.1]{GM01} or \cite[Sec. 5.2.1]{GMT17}.

\paragraph{The tree expansion for the effective potentials}\label{sec:treexp}
The detailed structure of $S^{(h)}(J),V^{(h)}(\varphi,J)$ arising from the iterative construction of Section \ref{sec:firstregime}-\ref{sec:secondregime}  can be naturally
   represented in terms of a  tree expansion. This is because the effective potential at scale $h$ is recursively defined by a relation of the following form: 
   \begin{equation}
   \hspace{-10pt}        L^{2}\widetilde{\mathbb{F}}^{(h)}+\widetilde{\mathbb{S}}^{(h)}(J)+\mathbb{V}^{(h)}(\varphi',J)= \sum_{n \geq 1} \frac{1}{n!}\mathcal{E}^T_{h+1} \left (\hat{\mathbb{V}}^{(h+1)}(\varphi'+\sqrt{\Xi_h}(\, \cdot\,),J); n\right ) 
\label{eq:4.147}
   \end{equation}
where  $\widetilde{\mathbb{F}}^{(h)}:=\mathbb{F}^{(h)}-\mathbb{F}^{(h+1)}$, $\widetilde{\mathbb{S}}^{(h)}(J):=\mathbb{S}^{(h)}(J)-\mathbb{S}^{(h+1)}$, $\mathbb{V}^{(h)}, \hat{\mathbb{V}}^{(h+1)}$  satisfy $\mathbb{S}^{(h)}(0)=\hat{\mathbb{V}}^{(h+1)}(0,J)={\mathbb{V}}^{(h)}(0,J)=0$; the specific definition of these quantities, as well as 
of $\Xi_h$, depends upon 
the range of scales $h$ belongs to: 
\begin{enumerate}
\item 
For $h \geq h_\eps-1$, we set $\mathbb{F}^{(h)}:=F_\eps^{(h)}, \mathbb{S}^{(h)}(J):=S_\eps^{(h)}(J), \mathbb{V}^{(h)}(\varphi,J)=V_\eps^{(h)}(\varphi,J)$, $\Xi_h:=1$ and $\hat{\mathbb{V}}^{(h+1)}(\varphi,J)=\VV^{(h+1)}(\varphi,J)$, where we recall that 
\begin{equation}
\VV^{(h+1)}(\varphi,J)=(2^{h}\nu_{h}+\eps\mu_{h_\eps}\mathds{1}_{\{h=h_\eps-1\}}) N(\varphi) +\sum_{s}Y_{h;\eps,s}F_{Y,s}(\varphi,J) +\mathcal{R}_{1,\eps}V_\eps^{(h+1)}(\varphi,J),\label{eq:onestepprimo}
\end{equation}
see \eqref{eq:4.68} and following lines (for the case $h\ge h_\epsilon$), as well as \eqref{eq:2.34}, \eqref{eq:4.91} and following lines (for the case $h=h_\epsilon-1$);
\item
For $h<h_\eps-1$, we set  $\mathbb{F}^{(h)}:=F^{(h)}, \mathbb{S}^{(h)}(J):=S^{(h)}(J), \mathbb{V}^{(h)}(\varphi,J)=\VV^{(h)}(\varphi,J)$, $\Xi_h:=Z_h$ and $\hat{\mathbb{V}}^{(h+1)}(\varphi,J)=\hat \VV^{(h+1)}(\varphi,J)$, where we recall  
that
\begin{equation}
\hat \VV^{(h)}(\varphi,J)=2^{h} \mu_{h} F_\mu(\varphi)+\l_{h}F_\l(\varphi)+\d_{h}F_\d(\varphi)+\sum_{s,\un \o}\frac{Y_{h;s,\un \o}}{Z_{h-1}}F_{Y,s, \un \o}(\varphi)+\mathcal{R}_2 \VV^{(h)}\Big(\sqrt{\frac{Z_h}{Z_{h-1}}}\varphi,J\Big ), \label{eq:onestepsecondo} 
\end{equation}
see \eqref{eq:2.52bis}.
\end{enumerate}

\begin{figure}[ht]
       \centering
       \includegraphics[scale=0.8]{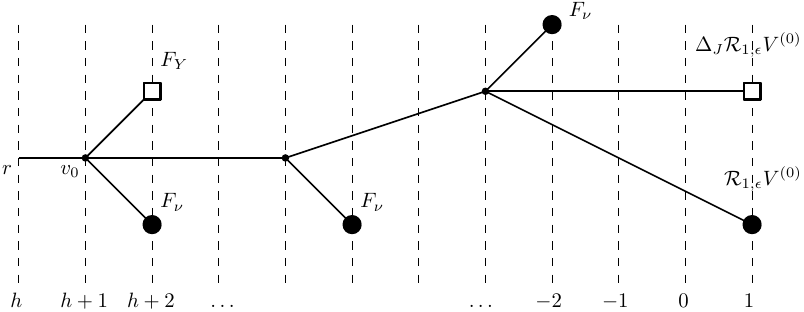}
       \caption{A tree of the first regime. Normal endpoints are represented by large black dots, while special endpoints are represented by white squares. The branching points (called non-trivial vertices) are represented by small black dots; trivial vertices (i.e., all the other nodes of the tree located at the intersection between the tree branches and the vertical dotted lines at scales $h\le h'\le 1$) are not explicitly drawn; similarly, the label $\mathcal R$ indicating the action of the operator  $\mathcal{R}_{1,\eps}$ at all trivial and non-trivial vertices $v$ different from the endpoints and from $v_0$ (the only immediate successor of the root, at scale $h+1$) is dropped, in order not to overwhelm the picture. A similar graphic representation holds for the second regime, up to a change in the allowed range of scale labels and in the types of allowed endpoints.}       \label{fig:alberello}
\end{figure}

The above single-step procedure can be iterated: given $h \geq h_\eps-1$ (resp. $h<h_\eps-1$), using the multi-linearity of the truncated expectation and the explicit expression of $\hat{\mathbb V}^{(h+1)}$ in \eqref{eq:onestepprimo} (resp. in \eqref{eq:onestepsecondo}), we can rewrite the $n$-th term in the right hand side of \eqref{eq:4.147}
as $1/n!$ times a sum of terms of the form 
$\mathcal E^T_{h+1}(A_1;\ldots;A_n)$, where every 
$A_i=A_i(\varphi'+\sqrt{\Xi_h}(\cdot),J)$ corresponds to one of the terms in the right hand side of \eqref{eq:onestepprimo} (resp. \eqref{eq:onestepsecondo}); whenever $A_i$ equals 
$(\mathcal{R}_{1,\eps}V_\eps^{(h')} )_{h' \geq h+1}$ (resp. $ (\RR_2V^{(h')} )_{h'\geq h+1}$), 
we re-express it via the analogue of \eqref{eq:4.147} with $h$ replaced by $h+1$; 
we do so until all the arguments in the 
truncated expectations obtained via this recursive procedure either correspond to one of the terms in the right hand side of \eqref{eq:onestepprimo} (resp. \eqref{eq:onestepsecondo}) proportional to a running coupling constant, or to the term $\mathcal{R}_{1,\eps}V_\eps^{(0)}$ (resp. $\mathcal{R}_2V^{(h_\eps-1)}$). 

The result is a series expansion of nested truncated expectations (similar to \cite[(5.22)]{GM01}) in which each term can be graphically 
represented by a labelled tree $\t$, defined via a procedure analogous to the one described in 
\cite[Sect.8.4]{GM01}, and pictorially represented as in Figure \ref{fig:alberello}. The resulting set of labels and the definition of value of a labelled tree is in our case very similar to those described in \cite[Sec. 2.1]{BM36}, which we refer to for additional details about notations and definitions. The main features of the trees appearing in our expansion
are the following: 

   \paragraph{The first regime: $h \geq h_\eps-1$.}\label{sec:treexpfirst}
 A tree $\t$ contributing to $V_\eps^{(h)}$, $\tilde S_\eps^{(h)}(J):=S_\eps^{(h)}(J)-S_\eps^{(h+1)}(J)$, or to $\tilde F_\eps^{(h)}=F_\eps^{(h)}-F_\eps^{(h+1)}$  
has root $r$ on scale $h$, it has exactly one immediate successor of the root on scale $h+1$, denoted by $v_0$, 
and can have endpoints, the set of which we denote by $E(\t)=E_n(\tau)\sqcup E_s(\t)$ (either normal \say{$n$} or special \say{$s$}, represented as black dots or white
squares, respectively, cf. Fig. \ref{fig:alberello}), \textit{on all possible scales}\footnote{In \cite{BM36}, the trees of the first regime have endpoints that are all normal (because no external fields are considered there) and that all correspond to $\mathcal R_{1,\eps}V^{(0)}(\f,0)$ (because the interaction does not include local terms proportional to $\nu_h$: in \cite{BM36} the analogue of those local quadratic terms are step by step included in the reference Gaussian integration), see the definitions of GN trees in \cite[Sect.2.1]{BM36}.} between $h+2$ and $1$. The family of GN trees with root on scale $h$, $N_n$ normal endpoints and $N_s$ special endpoints 
is denoted by $\mathcal T^{(h)}_{N_n,N_s}$. Its elements have the following features. 

\begin{enumerate}
    \item  A normal endpoint $v$ on scale $h_v\le 0$ is associated with $2^{h_v-1}\nu_{h_v-1} F_\nu$, and called of type $\nu$; on scale $h_v=1$, it can either be associated with $\nu_{0} F_\nu$ (in which case it is called of type $\nu$), or with     
    (one of the monomials contributing to) $\mathcal R_{1,\eps}V^{(0)}(\f,0)$. The   predecessor $v'$ at scale $h_{v'}=h_v-1$ of an endpoint $v$ of type $\nu$ at scale $h_v$ is necessarily a branching point, unless $v'=v_0$.  

 \item A special endpoint $v$ on scale $h_v\le 0$ is associated with $\sum_{s=A,B,C}Y_{h_{v}-1,s;\eps}F_{Y,s}$, and called of type $Y$; 
 on scale $h_v=1$, it can either be associated with $\sum_{s=A,B,C}Y_{0,s;\eps}F_{Y,s}$ (in which case it is called of type $Y$), or with (one of the monomials contributing to) $\Delta_J\RR_{1,\eps}V_\eps^{(0)}:=\RR_{1,\eps}V_\eps^{(0)}(\f,J)-\RR_{1,\eps}V_\eps^{(0)}(\f,0)$.
The predecessor $v'$ at scale $h_{v'}=h_v-1$ of an endpoint $v$ of type $Y$ at scale $h_v$ is necessarily a branching point, unless $v'=v_0$.  
\end{enumerate}

\paragraph{The second regime: $h<h_\eps-1$.}\label{ssec:treexpsecond}
 In this case the only difference with \cite{BM36} comes from the presence of the external source $J$. A tree $\t$ contributing to $V^{(h)}$, $\tilde S^{(h)}(J)$, or to $\tilde F^{(h)}$ has root $r$ on scale $h$, a single 
immediate successor of the root on scale $h+1$, denoted by $v_0$, and can have endpoints $v \in E(\tau)$ (either normal or special as before) on all the scale between $h+2$ and $h_\eps$; as before, we denote the family of GN trees with $N_n$ normal endpoints and $N_s$ special endpoints by $\mathcal T^{(h)}_{N_n,N_s}$; its elements havethe following features. 
\begin{enumerate}
\item[3.] A normal endpoint $v$ on scale $h_v \leq h_\eps-1$ is associated either with $2^{h_v-1}\mu_{h_v-1} F_\mu$, or with $\d_{h_v-1} F_\d$, or with $\l_{h_v-1} F_\l$, in which case it is called of type $\mu$, or $\delta$, or $\lambda$, respectively; on scale 
$h_v = h_\eps$, it is 
associated either with $2^{h_\eps-1}\mu_{h_\eps-1} F_\mu$, or with $\d_{h_\eps-1} F_\d$, or with $\l_{h_\eps-1} F_\l$ (in which case it is called of type $\mu$, or $\delta$, or $\lambda$, respectively), or with (one of the monomials contributing to) $\RR_2 \VV^{(h_\eps-1)}\big(\sqrt{\frac{Z_{h_\eps-1}}{Z_{h_\eps-2}}}\varphi,0\big)$. The predecessor $v'$ at scale $h_{v'}=h_v-1$ of an endpoint $v$ of type $\mu$ or $\delta$ or $\lambda$ at scale $h_v$ is necessarily a branching point, unless $v'=v_0$. 

\item[4.] A special endpoint $v$ on scale 
on scale $h_v \leq h_\eps-1$ is associated
with $\sum_{s=A,B,C}\sum_{\un \o} Y_{h_{v}-1,s,\un \o}F_{Y;s,\un \o}(\f,J)$, and called of type $Y$; 
on scale $h_v=h_\eps$ it can either be associated with $\sum_{s=A,B,C}\sum_{\un \o} Y_{h_{\eps}-1,s,\un \o}F_{Y;s,\un \o}(\f,J)$
(in which case it is called of type $Y$), or with (one of the monomials contributing to) $\Delta_J\RR_2 V^{(h_\eps-1)}:=\RR_2\big [ V^{(h_\eps-1)}\big (\sqrt{\frac{Z_{h_\eps-1}}{Z_{h_\eps-2}}}\varphi,J\big )- V^{(h_\eps-1)}\big (\sqrt{\frac{Z_{h_\eps-1}}{Z_{h_\eps-2}}}\varphi,0\big )\big ]$.
The predecessor $v'$ at scale $h_{v'}=h_v-1$ of an endpoint $v$ of type $Y$ is necessarily a branching point, unless $v'=v_0$. 
\end{enumerate}

In both cases, each vertex of the tree that is not an endpoint and that is not the special vertex $v_0$ is associated with the action of an $\RR_{1,\eps}$ (if $h \geq h_\eps-1$) or $\RR_2$ (if $h < h_\eps-1)$ operator; for brief, we will simply say that an $\mathcal R$ operator acts on all such vertices; 
sometimes this action is recalled by adding a label $\mathcal R$ to all vertices $v>v_0$ that are not endpoints (see e.g. \cite[Chapt.8]{BG_RGbook}; however, we prefer to drop such a label, in order not to overwhelm the resulting figures, see Fig.\ref{fig:alberello} and its caption.

The GN tree expansion sketched above allows us to re-express the effective potentials as a sum over trees of the corresponding tree values, which leads to the following representation formula: 
if, e.g., $h \geq h_\eps-1$, then $ L^2 \tilde F_\eps^{(h)}+\tilde S_\eps^{(h)}(J)+V_\eps^{(h)}(\varphi,J)=\sum_{\substack{N_n,N_s\ge 0:\\ N_n+N_s\ge 1}}\sum_{\t\in \mathcal T^{(h)}_{N_n,N_s}} V_\eps^{(h)}(\t,\varphi,J)$ where
\begin{equation} 
\begin{split}
& V_\eps^{(h)}(\t,\varphi,{J})=\sum_{
{\bf P}\in\PP_\t}\sum_{ \bf D\in\mathcal D_\tau}\sum_{T \in \boldsymbol{T}_\tau}\sum_{ \xx_{v_0}}W^{(h)}_{\t,{\bf P},{T}, {\bf D},\eps}
(\xx_{v_0}) \varphi_{\bf D}({P_{v_0}^\varphi})\, J({P_{v_0}^J}),
\;\label{eq:4.150}
\end{split}
\end{equation}
and 
$P_{v_0}^\varphi$ and $P_{v_0}^J$ are two sets of indices labelling the Grassmann external fields and the external fields of type $J$ at the vertex $v_0$, respectively \cite[(72)]{BM36}, in terms of which 
\be J({P_{v_0}^J})=\prod_{f\in P_{v_0}^J}J_{x(f),s(f)}, \qquad  \varphi_{\bf D}({P_{v_0}^\varphi})=\prod_{f\in P_{v_0}^\varphi} \bar \partial^{D(f)}\varphi_{y(f)}^{\tilde \s(f)}, \label{eq:A.4bis}\ee
where, in the second of \eqref{eq:A.4bis}, we denoted $D(f)=(D_1(f),D_2(f))$, so that $\bar \partial^{D(f)}:=\bar \partial^{D_1(f)}_1\bar \partial^{D_2(f)}_2.$ Eq.\eqref{eq:4.150} above  is the analogue of \cite[(80)]{BM36} or \cite[(6.64)]{GMT17}, 
which we refer to for the definitions of the set of indices $\mathcal P_\tau$, $\mathcal D_\tau$, ${\boldsymbol{T}}_\tau$ (modulo minor notational changes; e.g., in our formula, the indices $\bf D\in\mathcal D_\tau$ replace the 
multi-indices $\b\in B_T$ of \cite[(6.64)]{GMT17}), as well as for the recursive definition 
of the kernels associated with a tree $\tau$ and indices $\mathcal P,\mathcal D,T$ follows from the recursive construction sketched above. In the simplest case where all the $\mathcal R$ operators associated with the vertices of $\tau$ 
act like the identity (which is the case when ${\bf D}=\{D_v\}_{v\in \tau}={\bf 0}$ and the indices ${\bf P}=\{(P_v^\varphi,P_v^J)\}_{v\in\tau}$ are such that either $|P_v^J|=0$ and $|P_v^\varphi|>6$ or $|P_v^J|+|P_v^\varphi|>3$ for all vertices $v>v_0$ of $\tau$ that are not endpoints) 
then $W^{(h)}_{\tau,{\bf P},T,{\bf 0},\eps}$ admits the following explicit expression, resulting from the iterative application of the so called \emph{Battle-Brydges-Federbush-Kennedy} (BBFK) determinant formula (see \cite[(75)]{BM36}, \cite[Lemma 3, (6.63)]{GMT17})  to the truncated expectation $\mathcal{E}^T_{h_v}$ at each $v \notin E(\t)$ (a similar, more involved expression, is valid in the general case, where one needs to recursively take into account the action of the $\mathcal R$ operators, 
see \cite[Sec. 3.10]{BM4} for details):
\begin{equation}
\hspace{-10pt}
\begin{split}
    & W^{(h)}_{\t,{\bf P},{T},{\bf 0},\eps}(\xx_{v_0})=\alpha_T
\left(\prod_{v \in E(\tau)} K^{(h_v)}_{v}(\xx_{v})\right)
\prod_{v \notin E(\t)}\frac1{s_v!} \int
dP_{T_v}({\bf t}_v) \det (G^{h_v,T_v}({\bf t}_v))
\prod_{\ell\in T_v} g^{(h_v)}(x_\ell,y_\ell)\end{split}\label{eq:kernoren}
\end{equation}
where $\alpha_T$ is a sign, $K_v^{(h_v)}$ is the kernel associated with the endpoint $v$, and 
$dP_{T_v}$ and $G^{h_v,T_v}$ are defined as in \cite[eq.(75)]{BM36} and following lines (in particular $G^h_v,T_v$ is a Gram matrix, whose elements are interpolated versions of the propagators $g^{(h_v)}$).  

\medskip 

A similar representation holds for the second regime of scales $h<h_\eps-1$, with the difference that the kernel $W_{\tau,{\bf P},T,{\bf D}}$ has an extra argument $\boldsymbol{\o}_{v_0}$, coming from the decomposition \eqref{eq:qp}; correspondingly, $\varphi_{\boldsymbol{D}}(P_{v_0}^\f)=\prod_{f \in P_{v_0}^\f} \bar \partial^{D(f)} \f^{\tilde \sigma(f)}_{y(f),\o(f)}$. Moreover, $W^{(h)}_{\t,{\bf P},{T},{\bf D}}(\xx_{v_0},\boldsymbol{\o}_{v_0*})$
is given by an expression analogous to \eqref{eq:kernoren} (or, better, to the generalization of \eqref{eq:kernoren} to the case where the operators $\mathcal R$ may act non-trivially on some vertex), modulo the presence of an extra factor $\prod_{v \notin E(\t)} (Z_{h_v}/{Z_{h_v-1}})^{|P_v^\varphi|/2}$, keeping track of the rescaling of the Grassmann fields at each step of the renormalization procedure (see, e.g., the analogous expression in \cite[Eq.(6.63)]{GMT17}). 

\medskip 

Clearly, the kernels of the effective potentials (see lines below \eqref{eq:2.32bis} and \eqref{eq:2.51biss}) are obtained by summing $W_{\t,{\bf P},{T},{\bf D},\eps}
(\xx_{v_0})$ if $h \geq h_\eps-1$, or $W_{\t,{\bf P},{T},{\bf D}}
(\xx_{v_0},\boldsymbol{\o}_{v_0})$, if $h<h_\eps-1$, over $\t\in \mathcal T^{(h)}_{N_n,N_s}$ and over $N_n,N_s$, 
under the constraint that the number of external fields of type $\f$ and $J$ is equal to $2n$ and $m$, respectively, that the elements of ${\bf D}$ are the same as $\un D$, etc.

\medskip

Let us now focus on the tree expansion of the first regime, in order to study the flow of the running coupling constants, as anticipated in Item 1 at the beginning of Section \ref{app:A}.

\subsection{The first regime: bounds on the kernels of the effective potential}
\label{app:technaspectsfirstregime}

Using the tree expansion, we obtain the following bound for $W^{(h)}_{\t,\boldsymbol{P},T,\boldsymbol{D},\eps}$.

\begin{lemma}\label{prop:an} There exists $L$-independent constants $\d_0,C,c,\kappa >0, \theta \in (0,1/2)$ such that, if $h \geq h_\eps$ and
\be \sup_{h'>h}\{|\l|,2^{-\theta h'}|\n_{h'}|,|\a_{h'}|,|\z_{h'}|,|r_{h'}/r-1|\}\le \d_0,\label{ggg0}\ee
and $\tau\in\mathcal T^{(h)}_{N_n,N_s}$, then 
\bea
&& \|W^{(h)}_{\t,{\bf P},T,{\bf D},\eps}\|\le C^{N_s}
\,(C\d_0)^{\max\{N_n,c|I_{v_0}^\psi|\}}\,
2^{h\left (\frac{3}{2}-\frac14|P_{v_0}^\varphi|-
|P_{v_0}^J|-|{\bf D}|^{(1)})\right )} \nn \\
 && \times \left( \prod_{v \in E_n(\t)}2^{\theta h_v}\right )\left(\prod_{v \in E_s(\t)}\sup_{s=A,B,C}
\left |Y_{h_v-1,s;\eps}\right|\right)\prod_{v \notin E(\t)}\hskip-.02truecm\frac{C^{|Q_{v}|}}{s_v!} \;2^{(h_v-h_{v'})\left(\frac32-\frac14|P_v^\varphi|-|P_v^J|-z_1(P_v)\right)}\;, \hspace{30pt} \label{eq:A.6bis}
\eea
where: $|I_{v_0}^\psi|=\sum_{v\ {\rm e.p.}}|P_{v}^\psi|$ is the total number of Grassmann fields associated with the endpoints of the tree; $|Q_v|=\sum_{i=1}^{s_v} |P_{v_i}|-|P_v|$ is the number contracted fields at the vertex $v$; $v'$ is the immediate predecessor of $v$. 
Moreover $z_1(P_v)=2,\frac{3}{2},1,\frac{1}{2}$ if $(|P_v^\varphi|,|P_v^J|)=(6,0),(2,0),(4,0),(2,1)$, respectively, and $z_1(P_v)=0$ otherwise; $|{\bf D}|^{(1)}=\sum_{f\in P_{v_0}^\varphi}\sum_{j=1,2}D_j(f)2^{-\mathbbm{1}_{j=2}}$.
\end{lemma}
The proof of this lemma is essentially the same as that of \cite[Lemma 2.2]{BM36}, which we refer the reader to for details. Here we just comment 
about the minor differences between our case and theirs. 
 
\begin{enumerate}
    \item The fact that in our setting the trees can have normal endpoints of type $\nu$ on all scales $\le 0$, whose coupling constants 
    $\nu_h$ are assumed to satisfy the hypothesis \eqref{ggg0}, leads to the first factor in the second line \eqref{eq:A.6bis}. This term is crucial to obtain the factor $2^{\theta h}$ appearing in the right hand side of \eqref{eq:2.29} of Proposition \ref{lem:bounds1} for $m=0$, see also the discussion around \eqref{eq:4.158} below.

    \item The gain factor $2^{-(h_v-h_{v'})z_1(P_v)}$ is due to the action of  $\mathcal{R}_{1,\eps}$. When $(|P^\psi_v|,|P_v^J|)=(4,0)$, $\RR_{1,\eps}$ acts as in \eqref{eq:3.3.30}, with its renormalization effect discussed lines below it, leading to $z_1(4,0)=1$; the cases $P_v=(6,0)$ and $(2,1)$ are completely analogous and lead to $z_1(6,0)=2$ and $z_1(2,1)=1/2$. 
When $(|P_v^\varphi|,|P_v^J|)=(2,0)$ instead, 
denoting by $\hat W^{(h_v)}_{\tau_v,\eps}(q)$ the 
value of the kernel associated with the subtree $\tau_v$ rooted in $v'$, the action of the renormalization operator on such a kernel 
can be written as $\RR_{1,\eps}\hat W^{(h_v)}_{\tau_v,\eps}(q)=\widetilde \RR_{1,\eps}\hat W^{(h_v)}_{\tau_v,\eps}(q)+\RR_{1,\eps}^{\mathcal{T}}\hat W^{(h_v)}_{\tau_v,\eps}(q)$, see \eqref{eq:2.33biss}-\eqref{eq:interpopsi2} and preceding lines; the rewriting \eqref{eq:interpopsi2} allows one to 
bound the effect of $\widetilde \RR_{1,\eps}$ in the same way as for $\RR_{1,\eps}$ in the cases $P_v=(4,0),(6,0),(2,1)$, and to show that its action can be dimensionally bounded by a gain factor $2^{-(h_v-h_{v'})z_1(2,0)}$, with $z_1(2,0)=3/2$. We are left with bounding the effect of the operator $\mathcal{R}^{\mathcal{T}}_{1,\eps}$. 
Neglecting finite volume effects, we rewrite \begin{equation}\label{eq:splitting}
\begin{split}
\mathcal{R}^{\mathcal{T}}_{1,\eps}
\hat W^{(h_v)}_{\tau_v,\eps}(q)&=
[(1-\mathcal{T})(\hat W^{(h_v)}_{\tau_v,\eps}(0)-\hat W^{(h_v)}_{\tau_v,0}(0))]+
[\hat W^{(h_v)}_{\tau_v,\eps}(q)-\hat W^{(h_v)}_{\tau_v,0}(q)-(\hat W^{(h_v)}_{\tau_v,\eps}(0)-\hat W^{(h_v)}_{\tau_v,0}(0))]\\
&\equiv [I]+[II].\end{split}\end{equation}
We start with the first term. Recalling that the 
dependence upon $\eps$ of the kernels can all be encoded in $r$ 
(see Item 2 below \eqref{inftvolprop}) and, therefore, in view of the multiscale construction, in the sequence of running couplings $\boldsymbol{r}_{h}$, we rewrite $\hat W^{(h_v)}_{\tau_v,\eps}(q)\equiv \hat W^{(h_v)}_{\tau_v}(q;\underline r_{h_v})$, 
in order to make such a dependence explicit (here 
$\underline{r}_{h_v}$ is the sequence of
the couplings $\boldsymbol{r}_\ell$ at scales $\ell\ge h_v$). Using the formula for the Taylor remainder we can rewrite the first term in 
the right hand side of 
\eqref{eq:splitting} as: 
\begin{equation}
[I]=\int_{0}^1 dt \sum_{\ell,m \geq h_v} r_\ell r_m \partial^2_{\boldsymbol{r}_\ell,\boldsymbol{r}_m}  \hat W^{(h_v)}_{\tau_v}(0;t\un r_{h_v}).\label{eq:4.150bis}
\end{equation}
Rewriting 
$\hat W^{(h_v)}_{\tau_v}(0;t\un r_{h_v})$ as in 
\eqref{eq:4.150}, we see that the derivative with respect $\boldsymbol{r}_\ell$ either act on one of the propagators in $G^{h_v,T_v}$ (if $h_v=\ell$) or on one of the propagators  in the rightmost product in the 
same formula (if $h_v=\ell$ or on one of the kernels associated with the endpoints on scale $1$ (if $\ell=0$). From Lemma \ref{lem:propag1}
and the explicit expression of the kernels of 
the endpoints on scale $1$, it turns out that any such derivative is dimensionally bounded by $2^{-\ell}$; on the other hand, the assumption \eqref{ggg0} implies that $\sup_{j > h}|r_j|\leq |r|(1+\d_0) \leq |r|C'$; moreover, $|r| \leq C''/C'2^{h_\eps}$ as it follows from the definition \eqref{eq:4.69}. All in all, the action of $r_\ell\partial_{\boldsymbol{r}_\ell}$
on $\hat W^{(h_v)}_{\tau_v}$ produces a dimensional factor proportional to $2^{h_\eps-h_v}$; in combination with the factor produced by the action of $r_m\partial_{\boldsymbol{r}_m}$
we obtain $2^{2(h_\eps-h_v)}$, 
which is even better than what is needed, 
since $2^{2(h_\eps-h_v)}\leq2^{2(h_{v'}-h_v)}\leq 2^{(h_{v'}-h_v)z_1(P_v)}$ with $P_v=(2,0)$. Note that the sum over $\ell,m$ also produces a combinatorial factor that can be bounded proportionally to  $(\sum_{v\in\tau}|Q_v|)^2+N_s+N_s$, 
with $|Q_v|=\sum_{i=1}^{s_v}|P_{v_i}|-|P_v|$ (here $v_1,\ldots,v_{s_v}$ are the successors of $v$ on $\tau$), which can be re-absorbed in the factor $C^{N_s+N_n}$ in the right hand side of 
\eqref{eq:A.6bis}, possibly at the cost of increasing $C$. A similar discussion allows one to bound the second term in the right hand side of \eqref{eq:splitting}, by Taylor-expanding 
$[II]$ at first order both in $q$ around $q=0$ and in $\underline r_{h_v}$. The effect of Taylor expanding in $\underline r_{h_v}$
can be bounded along the same lines above and leads to a dimensional factor $2^{h_\eps-h_v}$, 
while that of expanding in $q$ can be bounded as discussed after \eqref{eq:3.3.30}, and leads to a dimensional factor $2^{-\frac12(h_v-h_{v'})}$. Putting things together, we see that the action of $\mathcal R_{1,\eps}$ on kernels with $P_v=(2,0)$ leads to the desired gain factor $2^{-\frac32(h_v-h_{v'})}$
as stated in the lemma. 
\end{enumerate}

As a corollary of Lemma \ref{prop:an}, we  obtain the bound in Proposition \ref{lem:bounds1} by summing \eqref{eq:A.6bis} over $N_s,N_n\ge 0$, over $\t\in \mathcal T^{(h)}_{N_n,N_s}$, over ${\bf P}\in\PP_\t$, ${\bf D}\in\mathcal D_\tau$ and over $T\in{\bf T}$, under the constraint that $|P_{v_0}^\f|=n$ and $|P_{v_0}^J|=m$: this is possible because all the renormalized scaling dimensions \[d^{(1)}_{v,z}:=\frac32-\frac14|P^\f_v|-|P^J_v|-z_1(P_v)\] are negative (and actually $d^{(1)}_{v,z}\leq-1/2)$  thanks to the definition of $z_1(P_v)$; see \cite[App. 6.1]{GM01} for details on how to sum over the labels. \medskip 

\textbf{The short memory property.}\, The previous discussion implies that when $J=0$ (i.e. $m=0$) a generic $\tau$ has at least one normal endpoint $w$ which carries a factor $2^{\theta h_w}$. Denoting by $\Gamma_\tau(w):=(v_0,v_1,\dots,w)$, the ordered collection of vertices from $v_0$ to $w$ ($v_i > v_{i-1}$ with respect to the partial order induced by the tree), we can rewrite part of the factor in the bound \eqref{eq:A.6bis} as
\begin{equation}
  2^{\theta h_w}\prod_{v \notin E(\tau)} 2^{(h_v-h_{v'})d^{(1)}_{v,z}}=2^{\theta h}\prod_{v \notin E(\t)} 2^{(h_v-h_{v'})\left(d^{(1)}_{v,z}+\theta\mathbbm{1}_{v \in \Gamma_\tau}\right)}\label{eq:4.158}.
\end{equation}
Since $d_{v,z}^{(1)}\leq -1/2$, this implies that for $\theta\in[0,1/2)$ one can still perform the summation over the trees and its labels (since $d^{(1)}_{v,z}+\theta'\mathbbm{1}_{v \in \Gamma_\tau}<0$), thus obtaining a qualitatively 
analogous bound up to an extra dimensional factor $2^{\theta h}$ w.r.t \eqref{eq:A.6bis}; in this way one obtains the bound of Proposition \ref{lem:bounds1} when $m=0$, with an overall constant $C(\theta)$ which is not uniformly bounded as $\theta \to 1/2^-$, because in this limit the exponent $d_{v,z}^{(1)}+\theta\mathds 1_{v\in\Gamma_\tau}$ in \eqref{eq:4.158} may tend to zero for some vertices $v$ (those for which $d_{v,z}^{(1)}=-1/2$ and $P_v^J=\emptyset$, i.e., those with $P_v=(2,0),(4,0)$). 

\medskip 
The previous argument does not apply to the case $m \neq 0$, i.e., to the case of trees with $N_s\ge 1$ special endpoints. In fact, the value of a special endpoint does not carry any exponentially small factors (because the running couplings $Y_{h;0,s}$ do not tend to zero as $h\to-\infty$). By proceeding as above, we can ``just" 
extract a factor $2^{\theta (h-k)}$ from the value
of trees with root at scale $h$ and whose special endpoints are all at scales $\ge k$. On the other hand, as we will see in the discussion of the flow of $Y_{h;\eps,s}$ (App.  \ref{app:A2}) and of the proof of Proposition \ref{prop:7} (App. \ref{app:A3}), 
even in the case of contributions to kernels or running couplings with $m\neq0$, we can typically isolate from the tree expansion a class of dominant term, whose sum provides the leading asymptotics of the quantity of interest in the infrared limit, plus a remainder, which can be proved to be exponentially small thanks to an argument similar to the one spelled out after \eqref{eq:4.158}. 

\subsection{The flow of the first regime}
\label{app:A2}
In this section we prove Proposition \ref{lem:flowRC}. This is done by re-interpreting the problem of constructing and controlling the flow of the running constants as a fixed point problem in a Banach space of sequences satisfying the bounds given in Lemma \ref{prop:an}, 
in the same spirit of \cite[Sec. 6.4]{GMT20}. 
In virtue of the definition of the running constants, we first study the case $\eps=J=0$, in which case it turns that $\nu$ must be fixed uniquely as $\nu=\nu^*(\l)$, in order to control and bound the flow of $\n_h,\z_h,\a_h$. Then, under the condition $\nu=\nu^*(\l)$, the statements of Proposition \ref{lem:flowRC} concerning $r_h$ and $Y_{h;\eps,s}$ at $\eps,J \neq 0$ follow by induction, using Lemma \ref{prop:an}.

\paragraph{Critical $\eps=0$ theory: flow of $\nu_h$ and the value of $\varepsilon_0(\l)$}
 The aim of this section is to prove Proposition \ref{lem:flowRC} at $\eps=J=0$, by showing how to determine $\nu^*(\l)=a_0\l+O(\l^2)$ with $a_0\neq0$; an explicit first order calculation, performed in App.\ref{app:firstorder} below, see Lemma \ref{lem:firstordercalc}, shows that $a_0=2$, i.e., $\nu^*(\lambda)=2\lambda+O(\lambda^2)$. 

 We start by rewriting the flow of $\nu_h,\a_h,\z_h$ (cf.\eqref{eq:rcc1regime},\eqref{eq:2.32bis} and \eqref{eq:2.2bis},\eqref{eq:2.26biss}) at $\eps=J=0$, for $h \leq -1$, as 
 \begin{equation}
     \nu_h=2\nu_{h+1}+\BB^\nu_{h+1}, \quad \a_h=\a_{h+1}+\BB^\a_{h+1}, \quad \z_h=\z_{h+1}+\BB^\z_{h+1}, 
     \label{eq:beta2}
 \end{equation}
 where $\nu_0:=\nu, \a_0:=0,\z_0:=0$ and $\BB^\nu_h,\BB^\a_h,\BB^\z_h$ depend on $(\nu_k,\a_k,\z_k)_{h \leq   k \leq 0}$ and $\l$. Let $\boldsymbol{\mathcal{S}}:=\mathbb{R}^{\mathbb{Z}_{\leq 0}}$ and denote one of its elements by $\un s =(s_h)_{h \leq 0}$. Let  $\t:\boldsymbol{\mathcal{S}} \to \boldsymbol{\mathcal{S}}$ and $\rVert \cdot \rVert_\theta: \boldsymbol{\mathcal{S}} \to \mathbb{R}$ be given by
         \be (\t \un s)_h:=s_{h-1}; \qquad \quad   \rVert \un s \rVert_\theta:=\sup_{h \leq 0}2^{-\theta h}|s_h|, 
         \ee  
with $\theta \in [0,1/2)$. Note that $\|\cdot \|_0$ is the usual $\ell_\io$ norm. 
Next, we consider $\un u_0=(\un \nu,\un \a,\un \z) \in \boldsymbol{\mathcal{S}}^3$,  where  $\un \nu=(\nu_h)_{h \leq 0}$, $\un \a=(\a_h)_{h \leq 0}$, $\un \z=(\z_h)_{h \leq 0} \in \boldsymbol{\mathcal{S}}$ are arbitrary sequences: for the moment they do not need to be the solutions to the beta function equations in \eqref{eq:beta2}. Rather, we note that these solutions can be rephrased as the solutions to a suitable fixed point equation: in fact, assuming that $\|\nu\|_0 < \io$ and using that $\nu_{h}=2^{k-h}\n_{k}-\sum_{k<j\leq h}2^{j-h}\BB^\n_{j}(\l,\un u_0)$ (with the understanding that $\BB^\n_{j}(\lambda,\underline u_0)$ only depends upon the restriction of 
the sequences $\un \nu,\un \a,\un \z$ in $\underline u_0$ to scale indices larger or equal to $j$), together with $\a_0=\z_0=0$, we can rewrite \eqref{eq:beta2} as
\begin{equation}
    \un u_0=\un T_\lambda(\un u_0)\label{eq:bfixp}
\end{equation}
for a map $\un T_\lambda=(T^\nu_\lambda,T^a_\lambda,T^\z_\lambda)$ such that $T^\#_\l(\un u_0)=(T_{\l;h}^{\#}(\un u_0))_{h \leq 0}$, with $T_{\l;h}^{\nu}=-\sum_{j\leq h}2^{j-h}\BB^\n_{j}(\l,{\un u}_0)$, $T^\a_{\l;h}=\sum_{j \geq h} \BB^\a_h(\l,\un u_0), T^\z_{\l;h}=\sum_{j \geq h} \BB^\z_h(\l,\un u_0)$.
Given $\un u_0 \in \boldsymbol{\mathcal{S}}^3$,
we let
\begin{equation}
\begin{split}
 &\rVert \un u_0 \rVert^{(0)}:=\max\left \{\rVert \un \nu \rVert_\theta,\kappa_\theta\rVert (\t-1)\un \a  \rVert_\theta,\kappa_\theta\rVert (\t-1) \un \z \rVert_\theta \right \}, \qquad \kappa_\theta:=1-2^{-\theta}.
\end{split}\label{eq:3.4.34}
\end{equation}
Note that $\mathbb{X}^{(0)}:=(\boldsymbol{\mathcal{S}}^3,\rVert \cdot \rVert^{(0)})$ is a Banach space of convergent sequences in which $\nu_h \to 0$, $\a_h \to \a_{-\infty} \in \mathbb{R}$, $\z_h \to \z_{-\infty} \in \mathbb{R}$ exponentially fast with rate $\theta$ (thus $\|\a\|_0,\|\z\|_0,\|\nu\|_0 < \io$).

 Given $\delta>0$, let $\mathbb{X}^{(0)}_\d:=\{\un u_0:\, \rVert \un u_0 \rVert^{(0)} \leq \delta\}\subset \mathbb{X}^{(0)}$ and let $d:(\mathbb{X}^{(0)}_\d)^2\to\mathbb{R}^+$ be the distance induced by the norm $
    \rVert \un u_0 \rVert:=\max\{\rVert \un \nu \rVert_\theta,\rVert \un \a \rVert_0,\rVert \un \z \rVert_0\}$. We want to show that if $\delta,|\l|$ are small enough then\begin{enumerate}\item  ${\un T}_\lambda(\mathbb{X}^{(0)}_\delta)\subset (\mathbb{X}^{(0)}_\delta)$  \item  ${\un T}_\lambda$ is a contraction on the metric space $(\mathbb{X}^{(0)}_\delta,d)$, 
\begin{equation}
    \rVert {\un T}_\l(\un u_0)-{\un T}_\l(\un u'_0)\ \rVert \leq 2^{-1} \rVert \un u_0- \un u'_0 \Vert, \quad \forall \un u_0,\un u_0' \in \mathbb{X}_\d^{(0)}.\label{eq:contractionprop}
\end{equation} \end{enumerate}
The Fixed Point Theorem then implies that there exists a unique solution of \eqref{eq:bfixp} in $\mathbb{X}^{(0)}$: otherwise stated, there is a unique way to fix $\nu$, the only free parameter, i.e. $\nu=\nu^*$, such that the iterative construction of Section \ref{sec:firstregime}, produces a flow $\{\nu^*_h,\a^*_h,\z^*_h\} \in \mathbb{X}^{(0)}_\d$ satisfying  \eqref{eq:2.29bis}.

We start by discussing the immersion property in item 1. We limit ourselves to the component $\nu$, since the others follow from a similar reasoning, which we leave to the reader. 

\paragraph{1: The bound on $T^\nu_\l$:}  We want to  prove that, if $\underline u_0\in\mathbb X_\delta^{(0)}$, then $|T^\nu_{\lambda;h}(\underline u_0)|\leq 2^{\theta h}\d$.  By using the tree expansion for $W^{(j);\io}_{2,0}$ (recall \eqref{eq:beta2},\eqref{eq:rcc1regime},\eqref{eq:kernoren}), we can rewrite 
 \begin{equation} \BB^\n_{j}(\l,\underline u_0)=C_j^\nu \l+\sum_{N_n\ge 1}\sum^\ast_{\t\in \mathcal T^{(j)}_{N_n,0}} \sum_{{\bf P}\in\PP_\t}\sum_{T\in{\bf T}}\BB^\n_{j}(\l,\underline u_0;\t,{\bf P},T),\label{6.75-0}
 \end{equation}
where $*$ means that the summation is restricted  to trees that either have one endpoint of type $\mathcal{R}_{1,0}V_0^{(0)}(\varphi,0)$ of order $\l^2$ or $\l\d$, or at least two endpoints: recall indeed that the endpoints are either of type $\RR_{1,0} V_0^{(0)}$ or $\nu_kF_\nu$, and the first order in $(\nu_{h})_{h>j}$ does not contribute to $\BB_j^\nu$ by definition; also, because of the presence of $\RR_{1,0}$ in front of $V^{(0)}_0$ and the definition in \eqref{effectivepot}, one has that the first order in $\nu$ cannot contribute to $\RR_{1,0} V_0^{(0)}$.  The term $C_j^\nu$ in \eqref{6.75-0}, corresponding to a tree with one endpoint of type $\mathcal{R}_1V^{(0)}$, when we select the linear contribution in $\l$, can be bounded by $|C_j^\nu|\leq \bar C |\l|2^{\theta j}$ using  Lemma \ref{prop:an} and  the short memory property (see \eqref{eq:4.158}), for some explicit constant $\bar C>0$: note that the factor $2^{h}$ of Lemma \ref{prop:an} is compensated by the $2^{-h}$ in the definition of $\BB^\nu_h$ (cf. \eqref{eq:beta2}). Proceeding similarly for the second term in \eqref{6.75-0}, and recalling that $\un u_0 \in \mathbb{X}_\d$,
we find that, for $|\lambda|$ small enough
\begin{equation}
    |\BB_j^\nu(\lambda,\underline u_0)|\leq 2^{\theta j}\left(\bar C |\l|+C\max\{\l^2,\d|\l|,\d^2\}\right), \label{eq:4.171}
\end{equation}
where $C>0$ is the same of Proposition \ref{lem:bounds1}; this implies that, if $|\l| \leq \frac{\d}{4}\min\{1,\bar C^{-1}\}$, then $
 |T^\nu_{\lambda;h}(\underline u_0)| \leq \sum_{j \leq h}2^{j-h}|B_j^\nu(\l,\un u_0)|\leq  2^{\theta h-1} \d (1+C\d) \leq \d 2^{\theta h}$, provided that $\d$ is small enough, e.g. $\d < \min\{1,C^{-1}\}$. 

\paragraph{2: The contraction on  the $\nu$ component}

By definition  we can write $
    \left (T_{\lambda;h}^\n(\un u_0)-T_{\l;h}^\n(\un u'_0)\right )=\sum_{j \leq h}2^{j-h}\left(\BB^\nu_j(\l,\un u_0)-\BB^\nu_j(\l,\un u'_0)\right)$ 
and the tree expansion for $\BB^\nu_j$ allows us to rewrite
\begin{equation}
\BB^\nu_j(\l,\un u_0)-\BB^\nu_j(\l,\un u'_0)=\int_0^1 dt\sum_{N_n \ge 1}\sum_{\t\in \mathcal T^{(j)}_{N_n,0}}^* \sum_{{\bf P}\in\PP_\t}\sum^*_{T\in{\bf T}}\frac{d}{dt}\BB^\n_{j}(\l,\underline u_0(t);\t,{\bf P},T)
\end{equation}
where $\un u_0(t)=t\un u_0+(1-t)\un u'_0 \in \mathbb{X}_\d^{(0)}$. 
Note that $\BB^\n_{j}(\l,\underline u_0(t);\t,{\bf P},T)$ satisfies a bound analogous to the one in \eqref{eq:4.171}, uniformly in $t \in [0,1]$. The effect of the derivative $d/dt$ can be of two types:
(a) either it acts on an endpoint $v$ of type $\nu$ on scale $h<k\leq 1$, in which case it replaces $\nu_k(t)$ by $\nu_{k}-\nu_{k}'$, 
(b) or it act on a propagator $\tilde g^{(k)}_0$,
in which case its effect is, dimensionally, to multiply it by a factor $\a_k-\a'_k$ or $\z_k-\z'_k$.\footnote{The  derivatives $\partial_{\a_k},\partial_{\z_k}$ applied to $\tilde g^{(k)}_0$ are ``dimensionless", i.e. 
they qualitatively satisfy the same dimensional 
bounds as $\tilde g^{(k)}_0$ itself. This can be seen from \eqref{eq:singlescalepropI} and the support properties of the cutoff $f^{(1)}_k$.}
Then, by proceeding as in the proof of \eqref{eq:4.171}, we obtain $
    |\BB_j^\nu(\l,\un u_0)-\BB_j^\nu(\l,\un u'_0)| \leq C 2^{\theta j} \rVert \un u_0-\un u'_0 \rVert (|\l|+\d)$ which indeed implies  the bound in \eqref{eq:contractionprop} when summed over $j \leq h$, for $|\l|,\d$ small enough. 

    \paragraph{The value of $\nu^*_0(\l)$.}
Thanks to the convergence of the tree expansion for the components of the beta function, the components of $\un u_0^*$ and, in particular, 
those at $h=0$, are all real analytic functions
$\lambda$ in a sufficiently small complex ball centered at the origin. In order to compute the coefficients of the Taylor expansion of $\nu^*_0(\lambda)$, one can use the fact that, thanks to \eqref{eq:bfixp}, 
\begin{equation}
    \nu^*_0=-\sum_{h \leq 0}2^{h}\BB_h^\nu(\l,\un u_0^*).
    \label{eq:2.48bis}
\end{equation}
In particular, an explicit calculation shows  that $\nu^*_0(\l)=2\l+O(\l^2)$ (see Lemma \ref{lem:firstordercalc}) and thus
\begin{equation}
\varepsilon_0(\l)=-\log(1-\nu^*_0(\l))=2\l+O(\l^2).\label{eq:eps0nu0}
\end{equation} 

\paragraph{The flow of $r_h$}
\label{par:flowrh}
Now let $\eps>0$, $J=0$ and fix $\un u_0=\un u_0^*$, i.e. $\nu=\nu^*_0(\l).$ We want to prove the statement \eqref{eq:2.29bis} for $r_h$ by induction, assuming that there exists a constant $C'>0$ such that $\sup_{h'>h}|r_{h'}-r| \leq C'|r\l|$. For the induction base, i.e.  $h=-1$, the statement holds by using the definition $r_{-1}=\boldsymbol{r}_{-1}(0)$ together with \eqref{eq:2.26bis}, and  \eqref{kernestim} for $n=1,m=0$ and $C'>C$ therein. For $h<-1$,  from $r_h=\boldsymbol{r}_h(0)$ and \eqref{eq:2.32bis} we can bound \begin{equation}\label{eq:4.71}
    |r_h-r_{h+1}| \leq |r|(1+C'|\l|)\sum_{k \geq h+1} |\partial_{\boldsymbol{r}_k}\hat{W}^{\infty;(h+1)}_{2,0;0}(0)| \leq \tilde{C}(r,\l)\sum_{k \geq h+1}2^{(1+\theta)h-k}
\end{equation}
for $\tilde{C}(r,\l)=C_1|r \l|(1+C'|\l|)$, $C_1$ as in Proposition \ref{lem:bounds1}, which implies that $r_h$ is a Cauchy sequence. In the last inequality of \eqref{eq:4.71} we used that the action of $\partial_{\boldsymbol{r}_k}$ produces an extra dimensional factor $2^{-k}$ w.r.t. to the upper bound of $|\hat W^{\io;(h+1)}_{2,0;0}(0)|$ provided by Proposition \ref{lem:bounds1}, i.e. $|\partial_{\boldsymbol{r}_k}\hat{W}_{2,0;0}^{\infty;(h+1)}(0)|\leq C_1|\l| 2^{\theta h}2^{h-k}$, see also the discussion after \eqref{eq:4.150bis}. Thus using \eqref{eq:4.71} we can bound
$|r_h-r|\leq \tilde{C}(r,\l)\sum_{j=h}^{0}2^{\theta j}\leq C'|r||\l|$
where we used that for any positive $\theta$
in $(0,1/2)$ and $C'$ larger (say) than $2C_12^\theta/(2^\theta-1)$, then $(1+C'|\l|)C_1\leq C'(2^\theta-1)$, provided that $|\l|$ is small enough. 

Lastly note that by iteration of \eqref{eq:2.32bis} we can rewrite for any $h_\eps -1\leq h \leq 0$,  $r_h=Q_h r$ where, for $h=0$ we set $Q_0:=1$, while for $h \leq -1$, 
\be Q_h=\sum_{s=1}^{|h|} \sum_{\un \ell}^* \prod_{j=0}^{s-1} \t^{\ell_{j}+1}_{\ell_{j+1}}, \qquad \t^h_k:=\d_{h,k}-\partial_{\boldsymbol{r}_k}W^{(h);\io}_{2,0;\eps}\rvert_{\eps=0}
\label{eq:A.23bis}
\ee  where the second sum runs over ordered indices $h=\ell_0<\ell_1<\cdots<\ell_s=0$. In particular from \eqref{eq:4.71} and lines below one has that $Q_h$ is a Cauchy sequence and that $|Q_h-1|\leq C'|\l|.$ This implies that \be r_{h_\eps-1}=Q_{h_\eps-1}r=-\xi_0\eps+O(\eps^{1+\theta}), \qquad \xi_0:=e^{-\e_0}Q_{-\io}, \qquad  Q_{-\io}:=\lim_{\eps \to 0^+} Q_{h_\eps-1},
\label{eq:A.24bis}\ee where we used \eqref{eq:2.3} and $|Q_{h_\eps-1}-Q_{-\io}|\leq C'|\l|\eps^{\theta}.$ 

\paragraph{The flow of $Y_{h;\eps,s}$}
\label{par:flowYprimoregime}

Now let also $J \neq 0$, and set $\nu=\nu^*_0(\l)$, and recall that $\nu^*_0(\l)=O(\l)$. We prove by induction that, letting $\BB^Y_{h;\eps,s}:=Y_{h-1;\eps,s}-Y_{h;\eps,s}$, then  \be |\BB^Y_{h;\eps,s}|\leq C{e^{-\eps \d_{s,A}}}|\l| 2^{\theta h}, \label{eq:goalbetaY}\ee for any $h_\eps\leq h\leq  0$ (i.e. \eqref{eq:2.29bis}) and for some constant $C=C(\theta)>0$. 

\medskip

For $h=0$, the fact that $\BB^Y_{h;\eps,s}$ is $O(\l)$ follows from the definition of $Y_{-1;\eps,s}$ in \eqref{eq:Yfirstreg} and the relation \eqref{effectivepot}, which guarantee that the only contribution to $Y_{-1;\eps,s}$ at $\l=0$ is exactly $Y_{0;\eps,s}$ (defined in Proposition \ref{lem:flowRC}): to see this,  note that the only $O(J)$ term at $\l=0$ in the r.h.s. of  \eqref{effectivepot} can be obtained by retaining the $n=1$ term in the first sum by picking from $\tilde V_\eps(\varphi,J)$ the only $\l=0$ term of first order in $J_{z,s}$. In formulae, $\sum_{s=A,B,C}F_{Y,s}(\varphi,J)Y_{-1;\eps,s}\rvert_{\l=0}=\LL_{1,\eps}(\sum_{e  \in E_L} E_{e,\vartheta}(\psi(\varphi)) J_{e})=\sum_{s=A,B,C} \int dy \,   \varphi^+_{y}\varphi^-_{y} J_{y,s} \tilde K_s$ where $\tilde K_s=e^{-\eps-\varepsilon_0(\l)}K_s$  (cf. below \eqref{eq:Scoordinates} and recall the rescaling below \eqref{eq:2.2bis}). This clearly implies that $Y_{-1;\eps,s}\rvert_{\l=0}=Y_{0;\eps,s}$ as defined in Proposition \ref{lem:flowRC}. To conclude, the bound \eqref{eq:goalbetaY} at $h=0$ follows from \eqref{kernestim}. 

\medskip 

For $h\leq-1$, we inductively assume \eqref{eq:goalbetaY} to be valid up to scale $h+1$. Using the tree expansion for $Y_{h-1;\eps,s}$ and isolating the contribution $Y_{h;\eps,s}$ (which correspond to the only tree with exactly one endpoint, special, on scale $h+1$), we can rewrite $\BB^Y_{h;\eps,s}=\sum_{k =h}^{0}Y_{k;\eps,s}\BB^Y_{h,k;\eps,s}
+Y_{k;\eps,s}\tilde\BB^Y_{h,k;\eps,s}$, where \be 
\BB^Y_{h,k;\eps,s}=\sum_{N_n \geq 1} \sum^*_{\t \in \TT^{(h)}_{N_n,1}}\sum_{\boldsymbol{P} \in \PP_\t} \sum_{T \in \boldsymbol{T}} \sum_{\boldsymbol{x}_{v_0}} \tilde W^{(h)}_{\t,\boldsymbol{P},T}(\boldsymbol{x}_{v_0})
\label{eq:A.23}
\ee
 and $^*$ means that each $\t$ has the special endpoint on scale $h<k+1\leq 0$ and such endpoint is of type $Y$; a similar expression holds for $\tilde \BB^Y_{h,0;\eps,s}$ with the difference that $N_n \geq 0$ and the constraint 
 on the sum over $\tau$ is replaced by a new one, 
indicating that the special endpoint is of type 
 $\RR_{1,\eps}V_\eps^{(0)}(\f^{(\leq 0)},J)-\RR_{1,\eps}V_\eps^{(0)}(\f^{(\leq 0)},0)$. Note that we used linearity of $\BB^Y_{h;\eps,s}$ in the sequence $(Y_{h';\eps,s})_{h \leq h'<0}$ which implies in particular that $\tilde W^{(h)}_{\t,\boldsymbol{P},T}$ (thus $\BB^Y_{h,k;\eps,s}$) is independent of $Y_{h';\eps,s}$ for $h\leq h'\leq -1$.

  Using Lemma \ref{prop:an} together with the short memory property \eqref{eq:4.158}, applied to a normal endpoint in \eqref{eq:A.23} and to the special endpoint at scale $0$ in the expansion for $\tilde\BB^Y_{h,0;\eps,s}$, we obtain $|\BB^Y_{h,k;\eps,s}|,|\tilde \BB^Y_{h,0;\eps,s}|\leq \tilde C(\theta')|\l|2^{\theta' h}$ for any $\theta' \in (0,1/2)$ with $\tilde C(\theta)>0$ independent of the inductive constant $C(\theta)$; since we are free to choose $\theta<\theta'<1/2$, we fix $\theta'=2^{-1}(2^{-1}+\theta)$. 
On the other hand, using the inductive hypothesis one finds that, for any $k=h,\dots,-1$,  $|Y_{k;\eps,s}|\leq |Y_{0;\eps,s}|+\sum_{j=k+1}^{0} |\BB^Y_{j;\eps,s}|\leq \bar C $, for some $\bar C $ independent of $C$ (e.g. $\bar C=3/4$ for $|\l|$ small, and we used the definition of $Y_{0;\eps,s}$ in Proposition \ref{lem:flowRC}). This implies that 
\begin{equation}
    |\BB^Y_{h;\eps,s}|\leq |\l| \tilde C 2^{\theta' h}(\bar  C|h|+1)\leq C |\l|2^{\theta h}
    \label{eq:A.24}
\end{equation}
for $C=C(\theta)$ large enough w.r.t. $\tilde C(\theta)$, e.g. $C\geq 2 \tilde C \bar C\max\{1,\frac{4}{e(1-2\theta)\log 2}\}.$

\subsection{Proof of Proposition \ref{prop:7}}
\label{app:A3}
Here $C>0$ denotes a constant independent of $\eps,h,L$, possibly changing from line to line.\medskip 

\par\textbf{Proof of \eqref{eq:2.8}}. 
 By definition (see \eqref{eq:mlimit} and following line) $\rho^{(\vartheta_0)}_{\eps+\varepsilon_0(\l),\l,L,0}=\partial_{J_{0,B}}\WW^{(\vartheta_0)}_{\eps+\varepsilon_0(\l),\l,L,0}(\AA(J))|_{J=0}$, with $J_{z,s}=e^{A_e}-1$ as stated after  \eqref{eq:4.20}. The multiscale integration of Sections \ref{ssec:uv}-\ref{sec:secondregime} gives, see \eqref{eq:epsnegativo} and \eqref{eq:2.39},
\begin{equation}\label{eq:B.29}
\WW^{(\vartheta_0)}_{\eps+\varepsilon_0(\l),\l,L,0}(\AA(J))-\WW^{(\vartheta_0)}_{\eps+\varepsilon_0(\l),\l,L,0}(0)=\begin{cases}\sum_{h=(h_\eps-1)\vee \tilde h_L}^0\tilde S^{(h)}_\eps(J), & \eps\leq0 \\ \sum_{h=h_L}^{h_\eps-2}\tilde S^{(h)}(J)+\sum_{h=h_\eps-1}^0\tilde S^{(h)}_\eps(J), & \eps>0 \nonumber
\end{cases}\end{equation}
where: $(h_\eps-1) \vee \tilde h_L:=\max\{h_\eps-1,\tilde h_L\}$ and $\tilde S^{(0)}_\eps:=S^{(0)}_\eps$, see \eqref{eq:4.20} and the following lines, while for $h<0$ the single scale contributions $\tilde S^{(h)}_\eps$ (for $h\ge h_\eps-1$) and $\tilde S^{(h)}$ (for $h<h_\eps-1$)\footnote{Concerning the label $\eps$, we are using the same convention stated at the beginning of Section \ref{sec:2.3.1}.} are defined in \eqref{eq:4.68bis}, \eqref{eq:4.68}, \eqref{eq:epsnegativo}, \eqref{eq:4.91} and \eqref{eq:2.52bis}; finally, if $\eps\ge0$, $h_L,\tilde h_L\asymp-\log_2L$ are the last scales of the integration, see the lines preceding \eqref{eq:2.39} for $\eps>0$ and those after \eqref{eq:4.69} for $\eps=0$. Differentiating the above relation with respect to $J_{0,B}$ at $J=0$ gives
\begin{equation}\label{eq:Brho}
\rho^{(\vartheta_0)}_{\eps+\varepsilon_0(\l),\l,L,0}=
\begin{cases}\sum_{h=(h_\eps-1)\vee \tilde h_L}^0W^{(h)}_{0,1;\eps,B}(0)&\eps\leq0,\\ \sum_{h=h_L}^{h_\eps-2}W^{(h)}_{0,1;B}(0)+\sum_{h=h_\eps-1}^0W^{(h)}_{0,1;\eps,B}(0)&\eps>0.
\end{cases}
\end{equation} By Proposition \ref{lem:bounds1}, 
for $h>h_\eps$ we have $|W^{(h)}_{0,1;\eps,B}(0)|\le C2^{\frac h2}$; by Proposition \ref{lem:4biss}, for $h\le h_\eps$
we have $|W^{(h)}_{0,1;B}(0)|\le C2^{h-\frac{h_\eps}2}\le C2^{\frac h2}$; in both cases the bounds are 
uniform in $L$ and $\eps$; moreover, for fixed $h$, the kernels converge as $L\to\io$, see the discussion after \eqref{inftvolprop}, and thus by dominated convergence $\lim_{L\to\io}\rho^{(\vartheta_0)}_{\eps+\varepsilon_0(\l),\l,L,0}$ exists and is given by \eqref{eq:2.8}. Furthermore, estimating the finite size corrections of the kernels on each scale by the Poisson summation formula, Lemma \ref{lem:10Poisson}, as in \cite[App. B]{GMT20}, one can show that $\rho^{(\vartheta_0)}_{\eps+\varepsilon_0(\l),\l,L,0}$ converges to its limit uniformly in $\eps$ in a neighborhood of $0$: for a similar discussion see the proof of Proposition \ref{prop:rgmassa} after Lemma \ref{lem:ultimosperiamo}.

\medskip

\par
\begin{figure}[ht]
    \centering
    \includegraphics[width=0.85\linewidth]{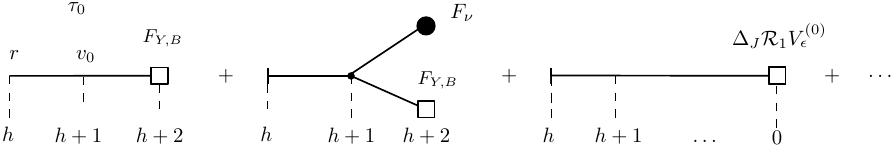}
    \caption{Left: the tree contributing to the first term in the r.h.s. of the first line of  \eqref{eq:2.9}. Mid-Right: two trees contributing to $\Theta^{(h)}_{\rho,\eps}.$}
    \label{fig:alberoGN}
\end{figure}

 \textbf{Proof of \eqref{eq:2.9}-\eqref{eq:2.10bis}}. 
If $h\geq h_\eps-1$, we can write a tree expansion for $W^{(h);\io}_{0,1;\eps,B}(0)$ analogous to \eqref{eq:A.23} where $N_n \geq 0$, $P_{v_0}^\varphi=  \emptyset, P_{v_0}^J=\{f\}$ with $x(f)=(0,0)$ and $s(f)=B$. In particular, the contribution to $W^{(h);\io}_{0,1;\eps,B}(0)$,  coming from the tree $\tau_0$ with $N_s=1,N_n=0$ with the special endpoint on scale $h+2$ (see Fig. \ref{fig:alberoGN}) is given by
\begin{equation}\partial_{J_{0,B}}\sum_{s=A,B,C}\mathcal{E}^T_{h+1}\left ( Y_{h+1;\eps,s} F_{Y,s}(\varphi,J)\right )=-Y_{h+1;B;\eps} \tilde g^{(h+1)}_{\eps}(0)  \label{eq:B.34bis}
\end{equation}
as it follows from the definitions \eqref{eq:Yfirstreg},\eqref{eq:4.68},\eqref{eq:singlescalepropI}. Let  $\Theta^{(h)}_{\rho,\eps}$ be the sum of all the other terms, by construction associated to trees $\t \in \TT^{(h)}_{N_n,1}$ either with $N_n \geq 1$  or with $N_n=0$ and having the special endpoint on scale $0$: in both cases one can use \eqref{eq:A.6bis} for $n=0,m=1$ and the \emph{short memory property} (see the discussion after \eqref{eq:4.158}) to obtain that  $|\Theta^{(h)}_{\rho,\eps}|\leq C2^{(1/2+\theta)h}$.\footnote{Note that $\Theta_{\rho,\eps}^{(h)}$ is not $O(\l)$ because of the contribution coming from the tree with only one endpoint, special, at scale $h=0$ (see Fig. \ref{fig:alberoGN}).}

\medskip

In the case $h<h_\eps-1$ the idea is similar: the contribution to $W^{(h);\io}_{0,1;\eps,B}(0)$ coming from the tree $\t_0 \in \TT^{(h)}_{0,1}$, having the special endpoint at scale $h+2$,  is now given by $\partial_{J_{0,B}} \sum_{s=A,B,C} \sum_{\un \o}\mathcal{E}^T_{h+1}(Y_{h+1;s,\un \o} F_{Y,s,\un \o} )=-\sum_{\o} Z_h^{-1}Y^{(1)}_{h;B} \tilde  g^{(h+1)}_{\o}(0)$ where we used definition of $F_{Y,s,\un \o}$ in \eqref{eq:2.54}, the fact that the integration \eqref{eq:singlescaleprop2} is diagonal in $\un \o$ and the definition of $Y_{h;B}^{(1)}$ in \eqref{eq:2.63bis}.
By definition, $\Theta^{(h)}_{\rho,\o}$ is again the sum of the remaining contributions, associated to trees $\t \in \TT_{N_n,1}^{(h)}$ having either $N_n \geq 1$  or $N_n=0$ and the only endpoint, special, being at scale $h_\eps-1$ (item $3,4$ above \eqref{eq:4.150}). We decompose \be \Theta^{(h)}_{\rho,\o}=\Theta_1+\Theta_2+\Theta_3\label{eq:decomptheta} \ee with $\Theta_1,\Theta_2,\Theta_3$ being the contributions in which the special endpoint is associated to $Y^{(1)}_{h;B},Y^{(2)}_{h;B}$ or $\RR_2 (V^{(h_\eps-1)}(\varphi,J)- V^{(h_\eps-1)}(\varphi,0))$, respectively.
\medskip 

In the tree expansion for $\Theta_1$, the normal endpoints all come with an `extra' factor $\eps^\theta$;
more precisely, we refer here to: (1) the factor $\eps^\theta$ in the right hand side of the 
upper bound on 
$\l_h$ and $\d_h$ in the first of \eqref{eq:2.70biss}; (2)) the short memory factor 
$2^{\theta h}$ associated with $\mu_h$ 
(see \eqref{eq:2.7}) and with the endpoints of type $\mathcal{R}_2 V^{(h_\eps-1)}(\psi,0)$ (see  \eqref{eq:2.29}), which can be bounded as
$2^{\theta h}\le 2^{\theta h_\eps}\asymp \eps^\theta$ for all $h\le h_\eps$, cf. \eqref{eq:4.69}. 
Thus from $\eqref{eq:2.61bis},$ \[|\Theta_1|\leq C|\l|2^{h-h_\eps/2}\eps^\theta \sup_{h<h'<h_\eps} |Z_{h'}^{-1}Y^{(1)}_{h';B}|\leq C2^{(h-h_\eps)}\eps^{1/2+\theta}, \] where in the last step we used that $\sup_{h'<h_\eps}|Z_{h'}^{-1}Y^{(1)}_{h';B}|\leq C$: this last property is based on the fact that the critical exponents of $Y^{(1)}_{h;B}$ (cf. \eqref{eq:2.71} for $\o_1=\o_2$) and $Z_h$ are the same, see Lemma \ref{YZ-lemma} below.
\medskip 

In the tree expansion for $\Theta_2$, there must be at least a normal endpoint $\bar v$ of type $\RR_2 V^{(h_\eps-1)}(\varphi,0)$ with kernel $K_{\bar v}^{(h_\eps-1)}(\xx_{\bar v})$ satisfying $\sum_{f \in P^\psi_{\bar v}} \sigma(f) \o(f) \neq 0$ (recall notations \eqref{eq:4.150}-\eqref{eq:kernoren}). This is because of the definition of $Y^{(2)}_{h;B}$ in \eqref{eq:2.63bis}, the fact that every contraction of Grassmann variables, occurring at vertices in the tree which are not endpoints, is diagonal in the $\o$ index (see below \eqref{eq:2.51bis}) and the fact that all the Grassmann variables must be contracted at some scale larger than $h$ (recall that we are discussing the contribution to $W^{(h)}_{0,1}$ which has no external Grassmann fields). These considerations, together with \eqref{eq:2.61bis}, imply \be |\Theta_2|\leq C|\l| 2^{h-h_\eps/2}2^{(h-h_\eps)\theta'}\eps^\theta \sup_{h < h' <h_\eps}|Z_{h'}^{-1}Y^{(2)}_{h'}|\leq C|\l|2^{(h-h_\eps)(1+\theta'-(\eta_1-\eta))} \eps^{1/2+\theta} \ee
where, in the first inequality the factor $\eps^\theta$ has the same origin as discussed for $\Theta_1$, while the factor $2^{\theta'(h-h_\eps)}$, with $\theta' \in (0,1)$, comes from the short memory property\footnote{Now $\theta'\in (0,1)$ because the renormalized scaling dimensions $d^{(2)}_{v,z}$ of the second regime, being the analogue of $d^{(1)}_{v,z}$ (cf. \eqref{eq:4.158}), satisfies $d^{(2)}_{v,z}\geq 1$.}  applied to the branch of the tree connecting the root to $\bar v$ (cf. \eqref{eq:4.158}). In the second inequality we used Lemma \ref{YZ-lemma}: note that $|\eta_1-\eta|\leq C|\l| \eps^{\theta}$, with $\theta \in (0,1/2)$, which is negligible w.r.t. to $1+\theta'$ (see \eqref{eq:eta1eta}). 

\medskip

Concerning $\Theta_3$, recall that we have the special endpoint of type $\RR_2(V^{(h_\eps-1)}(\varphi,J)-V^{(h_\eps-1)}(\varphi,0))$.  We can use the tree expansion of the first regime for $V^{(h_\eps-1)}(\varphi,J)-V^{(h_\eps-1)}(\varphi,0)$, the kernel of which has a similar expression to the r.h.s. of \eqref{eq:A.23}, with $N_n \geq 0$. The key observation is that because of the presence of the operator $\RR_2=1-\LL_2$ (cf. \eqref{eq:loc2reg}), the contribution to $\RR_2( V^{(h_\eps-1)}(\varphi,J)-V^{(h_\eps-1)}(\varphi,0))$, coming from the tree $\t_0$ with a unique endpoint, special, on scale $h_\eps+1$ (associated to $F_{Y,B}Y_{h_\eps;\eps,B}$, pictorially shown in the first of Fig. \ref{fig:alberoGN}), vanishes.\footnote{In fact, the contribution to $V^{(h_\eps-1)}(\varphi,J)-V^{(h_\eps-1)}(\varphi,0)$ coming from $\t_0$ in the tree expansion, corresponds to the relation in \eqref{eq:4.68} for $h=h_\eps-1$, when in the r.h.s. we pick $n=1$, the term $Y_{h_\eps;\eps,B}F_{Y,B}(\varphi,J)$ in $\VV^{(h_\eps-1)}$, without contracting any fields in the truncated expectation (recall that $V^{(h_\eps-1)}(0,J)=0$): since $Y_{h_\eps;\eps,B}F_{Y,B}(\varphi,J)$ is local (cf. \eqref{eq:Yfirstreg}) one has that $\LL_2$ acts as the identity, from which the claim for $\RR_2(V^{(h_\eps-1)}(\varphi,J)-V^{(h_\eps)}(\varphi,0))$ follows.}  Again, this implies that in the tree expansion of $\RR_2( V^{(h_\eps-1)}(\varphi,J)-V^{(h_\eps-1)}(\varphi,0))$ there is at least one normal endpoint; therefore, by the same line of reasoning discussed after \eqref{eq:A.23} and the short memory property, we find that the kernels of $\RR_2( V^{(h_\eps-1)}(\varphi,J)-V^{(h_\eps-1)}(\varphi,0))$ can be bounded 
in the same way as those of $V^{(h_\eps-1)}(\varphi,0))$ times an extra $2^{h_\eps \theta}\asymp \eps^\theta$. These considerations, together with \eqref{eq:2.61bis}, imply  $|\Theta_3|\leq C 2^{h-h_\eps/2}2^{\theta'(h-h_\eps)}\eps^\theta\leq C2^{(1+\theta')(h-h_\eps)}\eps^{1/2+\theta}$.

\medskip 

All in all we find 
\begin{equation}
    |\Theta^{(h)}_{\rho,\o}|\leq \sum_{i=1}^3
|\Theta_i|\leq C 2^{h-h_\eps}\eps^{1/2+\theta},\end{equation}
for some $C=C(\theta).$
\medskip

  \par \textbf{Proof of \eqref{eq:2.73bis}: effect of $\mathfrak{d}_\eps$ on the kernels.}
  
  We start with the first of \eqref{eq:2.73bis}. Recalling that the only $\eps$ dependent quantity in $\tilde g^{(h)}_{\io,\eps}$ is $\boldsymbol{r}_{h-1}\equiv\boldsymbol{r}_{h-1}(q;\eps)$ (cf. \eqref{eq:singlescalepropI},\eqref{eq:5.71}), we rewrite \be \mathfrak{d}_\eps \tilde g^{(h)}_{\io,\cdot}(y,0)=\eps \int_0^1 ds\int_{\mathbb{R}^2}dq\, e^{-iqy}f^{(1)}_{h}(q)  [\partial_{\boldsymbol{r}_{h-1}}\hat g_{\tilde \eps}^{(h)}(q) \partial_{\tilde \eps} \boldsymbol{r}_{h-1}(q;\tilde \eps)] \rvert_{\tilde \eps=s\eps} \label{eq:A.29bis}\ee where $\hat g^{(h)}_{\eps}(q):=(\boldsymbol{r}_{h-1}(q;\eps)-iq_1(1+ {\boldsymbol \z}_{h-1}) +2^{-1}q_2^2(1+{\boldsymbol \a}_{h-1})+\rho(q))^{-1}$. 
  Secondly, using that $r_h(\eps)=Q_h r(\eps)$ with $Q_h=1+O(\l)$ independent of $\eps$ (cf. \eqref{eq:A.23bis} and lines below) we find that \be |\partial_{\tilde \eps} \boldsymbol{r}_h(q;\tilde \eps)|\leq |\partial_{\tilde \eps}r(\tilde \eps)(Q_h-\chi^{(1)}_h\sum_{k\geq h}Q_k\partial_{\boldsymbol{r}_k}\hat W^{(h);\io}_{2,0;0}(0)|\leq C \label{eq:parepsboldsymbr}\ee uniformly in $q \in \mathbb{R}^2$, $|\tilde \eps| \leq \bar \eps$, for $|\l|$ small enough, where we used the definition of $r$ in \eqref{eq:5.71} and proceeded as in \eqref{eq:4.71} and the lines below it. Then we find $|\mathfrak{d}_\eps \tilde g^{(h)}_{\io,\cdot}(y,0)|\leq C|\eps| \sup_{0\leq \tilde \eps \leq \eps} \int_{\mathbb{R}^2} dq f^{(1)}_h(q) |\partial_{\boldsymbol{r}_{h-1}} \hat g^{(h)}_{\tilde \eps} (q)|\leq CC'|\eps|2^{-h/2},$ where in the last step we used that the integral can be bounded exactly in the same way as $\|\tilde g^{(h);(1,0,0);\un 0}_{\tilde \eps,\io}\|_\io\leq C'2^{-h/2}$, uniformly in $|\tilde \eps|\leq \bar \eps$, see Lemma \ref{lem:propag1}. This proves the first of \eqref{eq:2.73bis}. 
Note also that since $\boldsymbol{r}_h(q;\eps)$ satisfies \eqref{eq:parepsboldsymbr} and depends on $q$ only via $\chi_h^{(1)}(q)$,  one can prove that $\mathfrak{d}_\eps\tilde g_\cdot^{(h);\un n;\un 0}(y)$ admits the same bound as in \eqref{eq:prop1reg} times an additional factor $\eps2^{-h}.$ 
  
  \medskip 
  
  Next, we prove by induction the second of \eqref{eq:2.73bis}. Recall the definition of $B^Y_{h;\eps,B}=Y_{h-1;\eps,B}-Y_{h;\eps,B}$ (see \eqref{eq:goalbetaY}). 
  
  \medskip

  \emph{Inductive base.}  When $h=0$, one has $\mathfrak{d}_\eps B^Y_{0;\eps,B}=\mathfrak{d}_\eps Y_{-1;\eps,B}$ since $Y_{0;\eps,B}=1/2$. From the definition of $Y_{-1;\eps,s}$ in \eqref{eq:Yfirstreg} and the relation \eqref{effectivepot}, one sees that the operator $\mathfrak{d}_\eps$ either acts on one of the propagator $\tilde g^{(1)}_{\io,\eps}$ involved in the truncated expectation $\EE^T_1$ or it acts on one of the coefficients of $\tilde V_\eps(\psi,J)$,  which are nothing but the products of the Kasteleyn weights $K(b,w)$ in \eqref{eq:Zinteract} rescaled by $e^{-\eps-\varepsilon_0}$ (cf. \eqref{eq:2.2bis} and below).
 In both cases the action of $\mathfrak{d}_\eps$ produces an extra $\eps$ factor w.r.t. the bound in \eqref{kernestim} for $W_{2,1;\eps:B}^{(h);\io}$. 
 
 \medskip 
 
 \emph{Inductive step. } For $h<-1$ we write a tree
 expansion for $\mathfrak{d}_\eps \BB^Y_{h;\cdot,B}$ as in the r.h.s. of \eqref{eq:A.23} without the constraint $^*$ and the kernel replaced by $\mathfrak{d}_\eps W_{\t,\boldsymbol{P},T;\cdot}(\boldsymbol{x}_{v_0})$. Then $\mathfrak{d}_\eps$ can act on the kernel of an endpoint $v$  or on a propagator $\tilde g_\eps^{(k)}$ involved in some truncated expectation at scale $k\geq h$. In the former case, if the endpoint has scale $h_v=1$ one can repeat the same discussion as we did for the base of the induction; if, otherwise, $h_v<1$, we use that the endpoints of type $\nu$ do not depend on $\eps$ and that $\mathfrak{d}_\eps Y_{h_v-1;\cdot,B}=
 \sum_{k=  h_v}^{0}\mathfrak{d}_\eps \BB^Y_{k;\cdot,s}$, which, by the inductive hypothesis, can be bounded as $|\mathfrak{d}_\eps Y_{h_v-1,s;\cdot}|\leq C_\theta \eps 2^{-h_v}$. If $\mathfrak{d}_{\eps}$ acts on some propagator, we can repeat the same reasoning as discussed after \eqref{eq:A.29bis} to obtain an extra $\eps 2^{-k}$ factor. By performing the summation over the labels of the tree expansion and over the scale at which $\mathfrak{d}_\eps$ acts, one finds that $|\mathfrak{d}_\eps \BB^Y_{h;\cdot,B}|\leq C_\theta |\l| 2^{\theta h}|\eps| 2^{-h}$, to be compared with \eqref{eq:A.24}.

 \medskip
 
 Once we have a bound for $\mathfrak{d}_\eps \BB^Y_{h;\cdot,s}$, thus for $\mathfrak{d}_\eps Y_{h;\cdot,s}$\footnote{For general $s=A,B,C$ the previous proof gives $|\mathfrak{d}_\eps \BB^Y_{h;\cdot,s}|\leq C_\theta |\l|2^{\theta h}e^{-\eps \d_{s,A}}\eps 2^{-h}$, which has to be compared with \eqref{eq:A.24}. Note that the condition $|\eps|\leq \bar \eps$ stated before  \eqref{eq:2.73bis} comes from the fact that $\mathfrak{d}_\eps Y_{0;\cdot,A}=e^{-\eps-\varepsilon_0}-e^{-\varepsilon_0}\equiv r$ can be bounded proportionally to $|\eps|$ only for $|\eps|$ small.}, we can use the same strategy used above by tree expanding $W^{(h)}_{2n,m;\eps,\un s}$ and distinguishing whether $\mathfrak{d}_\eps$ acts on a propagator or on some endpoint (necessarily of type $Y_{h;\eps,s}$ if $h<1$) to obtain that $\|\mathfrak{d}_\eps W^{(h)}_{2n,m;\cdot,\un s}\|\leq C\eps 2^{-h} \|W^{(h)}_{2n,m;\eps,\un s}\|$: in particular, for $(2n,m)=(0,1)$ and $s=B$, this implies the statement for $\mathfrak{d}_\eps\Theta^{(h)}_{\rho,\cdot}$, i.e. the last of \eqref{eq:2.73bis}.  

\subsection{Proof of Lemma \ref{lem:costantiascalaheps}}\label{app:scalaheps} 
Let us first consider $\l_{h_\eps-1},$ which was defined in \eqref{eq:2.54}-\eqref{eq:2.540}. We can write 
\be \l_{h_\eps-1}=(1+z_{h_\eps-1})^{-1} \hat W^{(h_\eps-1);\io}_{4,0;(+,+,-,-)}(\un 0),\label{eq:lambdaheps-1}\ee with $z_{h_\eps-1}$ defined after \eqref{eq:loc2reg}. As we will shortly prove, $|z_{h_\eps-1}|\leq C|\l|\eps^{\theta}$ for $\theta \in (0,1/2)$, so in particular one has $|1+z_{h_\eps-1}|^{-1}\leq 2$ for $|\l|,\eps$ small.  Comparing with \eqref{eq:2.39bis},\eqref{eq:2.27bisss} (and lines below), we find that 
\be \begin{split}
\hat W^{(h_\eps-1)}_{4,0;(+,+,-,-)}(q_1,q_2,q_3,q_4)& =\int \hspace{-3pt}d \un D \hspace{-2pt} \int^\ast \hspace{-7pt}d\un D' \hat{\widetilde{W}}{}^{(h_\eps-1)}_{4,0;\eps,(+,+,-,-),\un D'}(q_2,q_3,q_4)\prod_{j=1}^4 \binom{D'_{j2}}{D_{j2}}e^{i\tilde\sigma_j (2\o_j \tilde q_\eps- \t_jq_j)} \times \\ &   \times \left ( \frac{i \tilde\sigma_j \o_j \sin 2\tilde q_\eps}{2} \right )^{D'_{j2}-D_{j2}}\prod_{\ell=1}^2 \left (\frac{i \tilde\sigma_j \sin 2 q_{j\ell}}{2}\right )^{D_{j\ell}},
\end{split}\label{formulalambdaeps}\ee see \eqref{eq:3.3.49} and following lines for the notations and conventions used in this formula. Eq.\eqref{formulalambdaeps} shows that the only non-vanishing contribution to $\l_{h_\eps-1}$ coming from the sum $\int d\un D$ is the one with $\un D=\un 0$; therefore, $\l_{h_\eps-1}$ can be bounded by \begin{equation}
\begin{split}
    |\l_{h_\eps-1}|& \leq 2 \int^\ast \hspace{-6pt} d\un D' \tilde q_\eps^{\sum_{i=1}^4 D'_{i2}} \int_{(R^T\mathbb{Z}^2)^3} dy_1 dy_2 dy_3 |W^{(h_\eps-1); \io}_{4,0;\eps,\un D'}(0,y_1,y_2,y_3)|\leq \\ & \leq  C |\l| \int^\ast \hspace{-6pt}d\un D' 2^{\frac{h_\eps}{2}\sum_{j=1}^4 D'_{j2}}2^{h_\eps(\frac12+\theta-\frac12\sum_{j=1}^4 D'_{j2})}\leq C |\l| \eps^{1/2+\theta}, \label{eq:A.30bis}
    \end{split}
\end{equation}
where in the second inequality we used that $q_\eps \asymp \sqrt{\eps} \asymp 2^{h_\eps/2}$ (cf. \eqref{eq:4.69} and Remark \ref{rem:4}) and Proposition \ref{lem:bounds1} together with the fact that $\un D'=(D'_1,\dots,D'_4)$ with $D'_{j1}=0, D'_{j2} \in \{0,1,2\}$ (recall the constraint $^\ast$ in \eqref{eq:2.44bis} and the notation defined after \eqref{eq:3.3.49}).

\medskip 

In order to compute and estimate $z_{h_\eps-1}=-i\partial_1 \hat W^{(h_\eps-1);\io}_{2,0;(+,+)}(0)$, using the definitions \eqref{eq:2.39bis}-\eqref{eq:2.27bisss} (and lines below) we write \begin{equation}
\begin{split}
    & \hspace{-8
    pt}\hat W^{(h_\eps-1); \io}_{2,0;(+,+)}(q)= \\ & \hspace{-10pt}\int \hspace{-4pt} d \un D\int^\ast \hspace{-5pt} \un d D'\prod_{j=1}^2\binom{D'_{j2}}{D_{j2}} e^{i\tilde\sigma_j \o_j (2\tilde q_\eps D_{j2}-\t_j q)}  \Big (\frac{i \o_j \tilde \sigma_j \sin 2\tilde q_\eps}{2}\Big )^{D'_{j2}-D_{j2}}\prod_{\ell=1}^2 \Big (\frac{i\tilde\sigma_j \sin 2q_{\ell}}{2}\Big )^{D_{j\ell}} \hat W^{(h_\eps-1); \io}_{2,0;\eps,\un D'}(q).\label{eq:A.31bis}
    \end{split}
\end{equation}
where the symbol $\ast$ in the summation recall the constraint $\mathds{1}_{\{D'_{j1}=D_{j1}, D'_{j2}\geq D_{j2}\}}.$ Because of the action of $\partial_1$ in the definition of $z_{h_\eps-1}$, 
 the only non-zero terms contributing to $z_{h_\eps-1}$ in \eqref{eq:A.31bis} have $\sum_{i=1,2}D_{i1}\leq 1$ and $\sum_{i=1,2}D_{i2}=0$. Now, applying $\partial_{1}$ and setting $q=0$ in \eqref{eq:A.31bis}, we can decompose $z_{h_\eps-1}=z^{(0)}_{h_\eps-1}+z^{(1)}_{h_\eps-1}$ where $z^{(p)}_{h_\eps-1}$ is associated to the constraint $\sum_{i=1,2}D_{i1}=p$, so that 
 \begin{equation}
 \begin{split}
& |z^{(0)}_{h_\eps-1}|\leq \int d \un D' \mathds{1}_{\{\sum_{i=1}^2 D_{i1}=0\}} \tilde q_\eps^{\, D'_{12}+D'_{22}} \| W^{(h_\eps-1)}_{2,0;\eps,\un D'}\|\leq C|\l|\eps^\theta \\ & |z^{(1)}_{h_\eps-1}|\leq \int d \un D' \mathds{1}_{\{\sum_{i=1}^2 D_{i1}=1\}} \tilde q_\eps^{\, D'_{12}+D'_{22}} \|(\cdot)_1 W^{(h_\eps-1)}_{2,0;\eps,\un D'}\|\leq C|\l| \eps^\theta
\end{split}
 \end{equation}
 where in the first line we used Proposition \ref{lem:bounds1} and the fact that again $\tilde q_\eps \asymp 2^{h_\eps/2}\asymp{\sqrt{\eps}}$ while in the second line we also used the fact that $|\partial_{q_1}\hat W^{(h_\eps-1); \io}_{2,0;\eps, \un D'}(0)|\leq \|(\cdot)_1W^{(h_\eps-1); \io}_{2,0;\eps, \un D'}\|:=\int_{R^T\mathbb{Z}^2} dy |y_1W^{(h_\eps-1); \io}_{2,0;\eps, \un D'}(y,0)| \leq C|\l| 2^{h_\eps(3/2-1/2+\theta-1-\|\un D'\|_{(1)})}=C|\l| 2^{h_\eps(\theta-\|\un D'\|_{(1)})}$ which follows by observing that the extra factor $y_1$
 in the norm gives an extra dimensional factor $2^{-h}$ w.r.t. the bound of $\|W^{(h)}_{2n,m;\eps,\un D}\|$ provided by Proposition \ref{lem:bounds1}. In fact, using the tree expansion \eqref{eq:4.150}, for the sake of concreteness under the assumption $^*$ in \eqref{eq:kernoren} (and lines above)  and $\un D'=0$, we have
 \begin{equation}
W^{(h);\io}_{2,0;\eps,\emptyset}(y,0)=\sum_{\substack{N_n\ge 1}}\sum_{\t\in \mathcal T^{(h)}_{N_n,0}}\sum_{
{\bf P}\in\PP_\t}^*\sum_{T \in \boldsymbol{T}}\sum_{ \xx_{v_0}}^*W^{(h),*;\io}_{\t,{\bf P},{T},\emptyset}
(\xx_{v_0})\label{eq:A.33bis}
 \end{equation}
 where the constraints $^*$ in the sums means that $P_{v_0}=\{f_1,f_2\}$ with $x(f_1)=y,x(f_2)=0$. As explained right after \cite[Eq.(80)]{BM36}, $T$ is a tree graph on the set of vertices associated to the coordinates in $\boldsymbol{x}_{v_0}$, among which are the coordinates $y,0$; therefore, there exists a path in $T$, $\g^T_{y \to 0}=\{y^{(0)}=0,y^{(1)},\dots,y^{(|\g|)}=y\}$ such that $y=\sum_{i=0}^{|\g|-1} (y^{(i)}-y^{(i+1)})$. Then in bounding $\sum_{\boldsymbol{x}_{v_0}\setminus\{0\}}(y^{(i)}-y^{(i+1)})_1W^{(h),*;\io}_{\t,{\bf P},{T},\emptyset}
(\xx_{v_0})$ we proceed as in \cite[Eq. (85),(86)]{BM36}: for the branch of the tree $T$ connecting $y^{(i)},y^{(i+1)}$, we replace the standard bound $\|\tilde g_\eps^{(h_{\bar v})}\|_1\leq C2^{-h_{\bar v}}$ by $\int dY |Y_1 \tilde g^{(h_{\bar v})}_\eps(Y,0)|\leq C2^{-2h_{\bar v}}$, which follows from \eqref{eq:prop1reg} and differs from the standard bound by an extra factor $2^{-h_{\bar v}}$, which can be bounded from above by $2^{-h}$ (since $h_{\bar v} > h$ for all the terms involved in the summation, and 
and $h_{\bar v}=h+1$ for at least one term). 
Repeating the same discussion leading to the bound in Proposition \ref{lem:bounds1}, we  conclude that $|z_{h_\eps-1}| \leq C |\l| \eps^\theta$, as desired. An analogous proof 
can be repeated for $\d_{h_\eps-1}$, details left to the reader. 

\subsection{Completion of the proof of Theorem \ref{th:5}: regularity in $\eps$}\label{app:regularitydensityeps}

In this appendix we show that 
\be \eps \mapsto \rho_{\eps+\varepsilon_0(\l),\l}\in C^\infty((-\bar \eps,0)\cup(0,\bar \eps)).\label{eq:cinftyeps} \ee 
The statement is trivial for $\eps<0$, since $\rho_{\eps+\varepsilon_0(\l),\l}=0$ there. 
We thus focus on $\eps \in (0,\bar \eps)$, where we rely on the multiscale analysis.

Recall that $h_\eps$, defined in \eqref{eq:4.69}, is a right-continuous step function with 
jump set $\mathcal{J}_2=\{2^h:h \in \mathbb{Z}\}$. We first restrict to 
$\eps \in \EE_2:=(0,\bar \eps) \cap \mathcal{J}^c_2$ where $h_\eps$ is locally constant. The proof of \eqref{eq:cinftyeps} 
then consists of the following two steps.

\medskip

\textbf{I.} \emph{(Differentiability away from the jump set.)} For every integer $k \geq 1$ 
there exists a constant $C(k)>0$ such that 
\be |\partial^k_\eps \rho_{\eps+\varepsilon_0(\l),\l}|\leq C(k) \eps^{1/2-k}, 
\qquad \forall \eps \in \EE_2, \label{eq:epskrho}\ee
and consequently $\eps\mapsto\rho_{\eps+\varepsilon_0(\l),\l}\in C^\infty(\EE_2)$. This is 
the content of Lemma \ref{lem:derivateepsilon} below, and its proof occupies the rest of 
this appendix. Here and in the following the dyadic choice of the scale parameter is made 
only for notational convenience: for any $\gamma>1$, the same construction and the same 
proof, with $2^h$ replaced by $\gamma^h$ throughout, give \eqref{eq:epskrho} on 
$\EE_\gamma:=(0,\bar\eps)\cap\JJ_\gamma^c$, where $\JJ_\gamma:=\{\gamma^h:h\in\mathbb Z\}$.

\medskip

\textbf{II.} \emph{(Removing the restriction $\eps\in\EE_2$.)} Observe that 
$\rho_{\eps+\varepsilon_0(\lambda),\lambda}$ is independent of the particular multiscale 
decomposition used in its construction. Indeed, changing the scale parameter only modifies 
the decomposition of the propagator into single-scale components, and therefore the 
associated tree expansion, while the resulting absolutely convergent series still represents 
the same quantity. More precisely, let $\gamma>1$ and replace the dyadic decomposition by a 
$\gamma$-adic one, using cutoff functions $\chi_h^{(j),\gamma}$, $j=1,2$, satisfying
$\chi_h^{(j),\gamma}(q)=\chi_{h+1}^{(j),\gamma}(\gamma^{-1}q)$ with the same ultraviolet 
cutoff $\chi^{(1),\gamma}_0=\chi^{(1)}_0$, and define the corresponding transition scale by 
$h_\eps^{(\gamma)}:=\inf\{h\in\mathbb Z:\gamma^h>\eps\}$. By Item \textbf{I} one obtains
\[
\eps\longmapsto\rho_{\eps+\varepsilon_0(\lambda),\lambda}\in C^\infty(\EE_\gamma).
\]
We now choose $\gamma>1$ such that $\JJ_2\cap\JJ_\gamma\cap(0,\bar\eps)=\emptyset$, which 
implies $(0,\bar\eps)=\EE_2\cup\EE_\gamma$. Therefore, around every $\eps\in(0,\bar\eps)$ 
there exists a multiscale decomposition for which the transition scale is locally constant, 
so that Item \textbf{I} applies. Since both constructions represent the same function 
$\rho_{\eps+\varepsilon_0(\lambda),\lambda}$, we conclude
\begin{equation}\label{dopostepII}
\eps\longmapsto\rho_{\eps+\varepsilon_0(\lambda),\lambda}\in C^\infty((0,\bar\eps)).
\end{equation}

\medskip

So, in order to completes the proof of \eqref{eq:cinftyeps}, we need to prove Step \textbf{I}, i.e.\ of the 
bound \eqref{eq:epskrho}, whose proof occupies the rest of this appendix. We begin by recalling how the multiscale representation of 
$\rho_{\eps+\varepsilon_0(\l),\l}$ depends on $\eps$.

\medskip

Recall that $\rho_{\epsilon+\varepsilon_0(\lambda),\lambda}$ can be written as a multiscale sum of the kernels
$W^{(h);\infty}_{0,1;\epsilon,B}$ and $W^{(h);\infty}_{0,1;B}$ (cf. \eqref{eq:2.8}), which is absolutely convergent uniformly in $|\eps|\leq \bar \eps$ (and therefore throughout $\EE_2$) under the assumptions of Proposition \ref{prop:7}, in particular under the requirement that $\mu_{h_\eps}(\lambda,\eps)$ has been fixed 
as in Proposition \ref{lem:kernbounds2}; recall also (see Remark \ref{rem:4}) that once $\mu_{h_\eps}$ is fixed in this way, $\tilde q_\eps$ is fixed uniquely as a function of $(\l,\eps)$ as the unique positive solution to 
the implicit equation $\eqref{eq:2.35}$ in some small neighborhood of $(0,\eps)$ with $\eps \in (0,\bar \eps)$.

Using \eqref{eq:2.8}, it is then necessary to distinguish how the dependence on $\eps$ appears in the single $h$-scale contribution: 
 as a consequence of the construction of Sections \ref{sec:firstregime}-\ref{sec:secondregime}  
 and of the tree expansion described above, we have that for a generic kernel:

\begin{itemize}
    \item[(1)] \label{itemi1desc}if $h \geq h_\eps-1$, the dependence on $\eps$ of $W^{(h)}_{2n,m;\eps,\un s}$ is given either via the sequence $\boldsymbol{r}_\ell$ in the single scale propagator $\tilde g_\eps^{(\ell+1)}$ (cf. \eqref{eq:singlescalepropI},\eqref{eq:2.32bis}) for some $\ell=h,\dots,0$, or via an endpoint, either of type $\mathcal{R}_{1,\eps} V^{(0)}(\varphi,J)$ at scale $1$, or special of type $Y_{\ell,B;\eps}F_Y$ for $\ell= h+1,\dots,0$; recall instead that $\nu_h$ is independent of $\eps$.  Note also that $\boldsymbol{r}_h$ is analytic in $\eps$ since  $r_h=Q_h r$ with $Q_h$ independent of $\eps$ and $r$ explicit and analytic in $\eps$ (cf. above \eqref{eq:A.23bis} and \eqref{eq:5.71}).
    
\item[(2)] \label{itemi2desc} if $h < h_\eps-1$ the dependence on $\eps$ of $W^{(h)}_{2n,m;\un s}$ is partly mediated by $\tilde q_\eps(\l,\eps)$, once implicitly determined from \eqref{eq:2.35}, and partly mediated by the fact that the output of the 
tree expansion of the first regime, which enters 
the definition of trees of the second regime, depends on $\eps$ explicitly, as discussed in the previous item. To be more explicit about how to distinguish these different dependencies, 
let us recall that the $\tilde q_\eps$ appears in the definition of the functions $v_\eps, Z_{h_\eps-1},\rho_{\o}(q)$ in the single scale propagators $\tilde g^{(\ell)}_{\o}$, $\ell=h+1,\dots,h_\eps-1$, see  \eqref{eq:2.37},\eqref{eq:singlescaleprop2},
and actually these formulas show that the whole dependence of these propagators upon $\eps$ is all mediated by $\tilde q_\eps$ itself. 
In the tree expansion of the second regime, 
$\tilde q_\eps$ also appears in the oscillating factors associated with the kernels of the endpoints $\mathcal R_2V^{(h_\eps-1)}$ (cf. \eqref{eq:2.39bis},\eqref{eq:2.44bis}); note, however, that these kernels also depend on $\eps$ 
directly, since they are obtained via 
\eqref{eq:2.44bis} from the kernels of the first regime at scale $h_\eps-1$, i.e., the output of the tree expansion of the first regime, which depend upon $\eps$ as discussed in the first item. In addition to these `direct' dependencies, let us recall that 
the kernels of the second regime, computed as sums over trees $\tau$ of the corresponding tree values, also depend upon the running couplings $\l_{h_v-1}$, $\d_{h_v-1}$, $\mu_{h_v-1}$, $Z^{-1}_{h_v-2}Y_{{h_v-1},s,\un \o}$ 
associated with the endpoints $v\in E(\tau)$ of type $\lambda$, etc., and upon the rescaling factors $(Z_{h_v}/Z_{h_v-1})^{|P_v^\f|/2}$ associated with the vertices $v \notin E(\t)$ (see below \eqref{eq:kernoren}); once we choose these couplings and re-scalings to be the solutions 
to the beta function equation, these inherit the same dependence upon $\tilde q_\eps$ and $\eps$ 
of the beta function itself; correspondingly, we shall write $\lambda_h=\lambda_h(\tilde q_\eps,\eps)$, etc., thus distinguishing the 
direct dependence upon $\eps$ from the one mediated by $\tilde q_\eps$. Similarly, once these values of running couplings  
are plugged in the formula for the tree values, 
we shall think of the kernels as functions 
of $\tilde q_\eps$ and $\eps$ separately, and 
write $W^{(h)}_{2n,m;\un s, \un D,\un \o}=W^{(h)}_{2n,m;\un s, \un D,\un \o}(\tilde q_\eps,\eps)$,  
\end{itemize}

\paragraph{Proof of \eqref{eq:epskrho}}  
In view of the two items above, in order to 
prove \eqref{eq:epskrho} from  \eqref{eq:2.8}, 
we first need to quantitatively discuss: 
\begin{itemize} 
\item[a)] the regularity of $\eps \mapsto \tilde q_\eps(\l,\eps)$ and, for $h \geq h_\eps$, of  $\eps \mapsto W^{(h);\io}_{0,1;\eps,B}$;
\item[b)] the regularity of $\eps \mapsto W^{(h);\io}_{0,1;B}(\tilde q_\eps,\eps),\tilde q_\eps \mapsto W^{(h);\io}_{0,1;B}(\tilde q_\eps,\eps)$, for $h <h_\eps$, when $\tilde q_\eps,\eps$ are thought as independent variables ranging in $I_{\k_0}:=\{(\tilde q_\eps,\eps): \eps \in \EE_2, \, \k_0^{-1}\sqrt{\eps}\leq \tilde q_\eps \leq \k_0 \sqrt{\eps}\}$ for some large enough positive constant $\k_0$.\footnote{Recall that even before fixing the parameter $\mu_{h_\eps}=\mu_{h_\eps}(\tilde q_\eps,\eps,\l)$, as long as one has $|\mu_{h_\eps}|\leq C |\l|$, then $\tilde 	q_\eps \asymp \sqrt{\eps}$ for any given $\eps \in \EE_2$ (cf. Remark \ref{rem:4}), i.e. $(\tilde q_\eps,\eps)\in I_{\k_0}$ for some $\k_0$. Moreover, for every fixed $h$, the bounds for $W^{(h);\io}_{0,1;B}(\tilde q_\eps,\eps)$ in \eqref{eq:2.61bis} do not require $\tilde q_\eps$ to be fixed but only that $(\tilde q_\eps,\eps)\in I_{\k_0}$.}
\end{itemize}

The desired bounds on the regularity of these 
functions with respect to $\eps$ are summarized 
in the following lemma, proven below.

\begin{lemma}\label{lem:derivateepsilon} 
\begin{enumerate}
    \item Under the hypotheses of Proposition \ref{lem:bounds1}, for any integer $k \geq 1$ there exists $C(k)>0$ such that   $|\partial_\eps^k W^{(h);\io}_{0,1;\eps,B}|\leq C(k) 2^{h(1/2-k)}$ for $h=h_\eps,\dots, 0$;
    \item Under the hypotheses of Proposition \ref{lem:kernbounds2}, for any pair of integers $k_1,k_2 \geq 0$ there exists $C(k_1,k_2)>0$ such that \be |\partial^{k_1}_{\tilde q_\eps}\partial^{k_2}_\eps W^{(h);\io}_{0,1;B}(\tilde q_\eps,\eps)|\leq C(k_1,k_2)2^{h-h_\eps/2}\eps^{-k_1/2-k_2}, \label{eq:item2lemma7}\ee for any $h< h_\eps$; 
    \item Fix $\mu_{h_\eps}=\mu_{h_\eps}(\lambda, \tilde q_\eps,\eps)$ as in Proposition \ref{lem:kernbounds2} and under such choice let $\tilde q_\eps(\l,\eps)$ be the unique  solution of \eqref{eq:2.35} for $|\l|\leq \l_0$,  $\eps \in \EE_2$. Then for any integer $k \geq 1$, there exists $C(k)>0$ such that
\begin{equation}
|\partial_\eps^k \tilde q_\eps(\l,\eps)|\leq C(k) \eps^{1/2-k}. \label{eq:B.47bis}
\end{equation}
\end{enumerate}
\end{lemma}

Given Lemma \ref{lem:derivateepsilon}, the bound \eqref{eq:epskrho} easily follows. In fact, using 
\eqref{eq:2.8}, the chain rule, 
and the bounds stated in Lemma \ref{lem:derivateepsilon}, we can bound 
$\partial^k_\eps\rho_{\eps+\varepsilon_0(\lambda),\lambda}$ for $k\ge 1$ as follows: 
\begin{equation}
\begin{split}
|\partial^k_\eps \rho_{\eps+\varepsilon_0(\l),\l}|&\leq  
\sum_{h=h_\eps}^0 |\partial_\eps^k W^{(h);\io}_{0,1;\eps,B}|+ C(k)
\sum_{h <h_\eps} \max_{\substack{k_1, k_2\ge 0,\\ j_1,\ldots,j_{k_1}\ge 1 :\\ 
j_1+\cdots+j_{k_1}+k_2=k}}
|\partial^{k_1}_{\tilde q_\eps}\partial^{k_2}_\eps W^{(h);\io}_{0,1;B}(\tilde q_\eps,\eps)|\prod_{i=1}^{k_1} |\partial_\eps^{j_i} \tilde q_\eps(\l,\eps)| \\ & \le C(k)\Big( \sum_{h=h_\eps}^0 2^{h/2-kh}
+\sum_{h<h_\eps} \max_{\substack{k_1, k_2\ge 0,\\ j_1,\ldots,j_{k_1}\ge 1 :\\ 
j_1+\cdots+j_{k_1}+k_2=k}} 2^{h-h_\eps/2}\eps^{-k_1/2-k_2}\prod_{i=1}^{k_1}\eps^{1/2-j_i}
\Big )\\ &\leq C(k)\eps^{1/2-k}, 
\end{split}\label{eq:B.48bis}
\end{equation}
for some $C(k)>0$, possibly changing from line to line. \hfill $\blacksquare$ 

\medskip 

Let us now discuss the proof of Lemma \ref{lem:derivateepsilon}. 

\begin{proof}[Proof of Lemma \ref{lem:derivateepsilon}]

As usual, we express the kernels of interest, 
$W^{(h);\infty}_{0,1;\eps,B}$ for $h\ge h_\eps-1$, and $W^{(h);\infty}_{0,1;B}(\tilde q_\eps,\eps)$, for $h< h_\eps-1$, as 
sums over GN trees. When acting on the tree value associated with a tree $\tau$, 
the derivatives with respect to $\eps$ or $\tilde q_\eps$ can either act on a propagator associated with a vertex of $\tau$ that is not an endpoint, or on the kernel associated with an endpoint of $\tau$; 
the effect of these actions on the dimensional bounds of the kernels is discussed below. 
Before we do that, let us preliminarily note two important facts:\\
(i) The trees contributing to the kernels of interest have exactly one special endpoint, whose kernel is linear in $J$.\\
(ii) With no loss of generality, the external field $J$ can be chosen to be supported on momentum zero, i.e., to be constant in real space. 

\medskip

Let us now prove the three items separately. 

\medskip

\textbf{Item 1.} Recall from Item (1) after \eqref{dopostepII} that the dependence upon $\eps$ in a tree value contributing to $W^{(h);\infty}_{0,1;\eps,B}$
is either carried by a propagator $\tilde g_\eps^{(\ell)}$, $\ell=h+1,\dots,1$, or by an 
endpoint of type $\mathcal{R}_{1,\eps}V^{(0)}(\varphi,J)$ at scale $1$, or by an endpoint of 
type $Y$ at scale $h_v\in\{h+2,\dots,1\}$.

If $\partial_\eps$ acts on a propagator on scale $k$ we can write $\partial_\eps g^{(k)}_{\cdot}(y,0)=\int dq\, e^{-iqy}f^{(1)}_{k}(q)  (\partial_{\boldsymbol{r}_{k-1}}\hat g_{\eps}^{(k)}(q) \partial_{\eps} \boldsymbol{r}_{k-1}(q; \eps))$ that, if combined with the definition of $\boldsymbol{r}_k$ and the property \eqref{eq:parepsboldsymbr}, readily implies that the $L^1,L^\io$ norms of $\partial_\eps g^{(k)}_\cdot$ have an extra factor $2^{-k}$ as compared to the corresponding bounds in Lemma \ref{lem:propag1}. Since $k>h$, 
we can bound this extra factor by $2^{-h}$. 
Similar considerations allow us to bound in a similar way the actions of $\partial_\eps$ on 
the kernels associated with the endpoint, but we do not belabor the corresponding details here, since they are similar to those spelled several times above. 

\medskip 

\textbf{Item 2.} Recall from Item (2) after \eqref{dopostepII} that the dependencies upon 
$\tilde q_\eps$ and $\eps$ in a tree value 
contributing to $W^{(h);\infty}_{0,1;B}(\tilde q_\eps,\eps)$ are either `direct', or `mediated'
by the dependencies of the running coupling constants $\lambda_{k}=\lambda_{k}(\tilde q_\eps,\eps)$, etc., or by the rescaling factors 
$Z_{k}=Z_{k}(\tilde q_\eps,\eps)$. For simplicity, here we limit ourselves to estimate the effect of the derivatives on the direct dependencies; those mediated by the running coupling constants or by the rescaling factors 
can be proved to lead to the same qualitative bounds, via an inductive argument based once again on the use of the tree expansion and of its absolute convergence. 

The direct dependence upon $\eps$ comes from the 
$\eps$-dependence of the kernels of the endpoints 
$\mathcal R_2 V^{(h_\eps-1)}$ on scale $h_\eps$; these are nothing but the output kernels at the end of the first regime and, therefore, the action of each derivative $\partial_\eps$ on them is bounded dimensionally as discussed in the previous item, 
leading to a factor $2^{-h_\eps}\propto \eps^{-1}$ per derivative with respect to $\eps$, as desired. 

The direct dependence upon $\tilde q_\eps$ either comes from the propagators, or from the oscillating factors associated with the endpoints; note (see below for a more detailed discussion of this point) that the $\tilde q_\eps$-dependent oscillating factors are necessarily associated with the endpoints of type $\mathcal R_2 V^{(h_\eps-1)}$ on scale $h_\eps$ or of type $Y^{(2)}$ on scale $k\le h_\eps$. 
Let us discuss separately the action of $\partial_{\tilde q_\eps}$ on these two kinds of contributions. 

\textbf{$\bullet$ } \textit{Action of $\partial_{\tilde q_\eps}$ on the single scale propagators.} In this case, using 
\eqref{eq:singlescaleprop2} and the fact that $\chi_{h_\eps-1}^{(2)}(q)=\chi(\d2^{-h_\eps+1}\|q\|_{(2)})$ (cf. below \eqref{eq:2.35}), for any $k\le h_\eps-1$ we can write \[\partial_{\tilde q_\eps} \tilde g^{(k)}_{\o}= \sum_{j=0}^4\tilde G_{\o,j}^{(k)}, \quad \qquad \tilde G_{\o,j}^{(k)}(y)=\int_{\mathbb{R}^2}\frac{e^{-iqy}f_k^{(2)}(q) S_j^{(k)}(q)}{\frac{\tilde Z_{k-1}(q)}{Z_{k-1}}D_\o(q)+\frac{\rho_\o(q)}{Z_{k-1}}}\]  
where $S_0^{(k)}(q)=\d2^{-k} \|q\|_{(2)}^{-1}v_\eps \partial_{\tilde q_\eps} v_\eps q_2^2$ and, for $j\ge 1$, $S_j^{(k)}(q)=-N_j^{(k)}(q)(\tilde Z_{k-1}(q) D_\o(q)+\rho_{\o}(q))^{-1}Z_{k-1}$ with: 
\begin{equation}\begin{split} 
& N_1^{(k)}(q)=
\frac{\tilde Z_{k-1}(q)}{Z_{k-1}}\o q_2 \partial_{\tilde q_\eps}v_\eps, \qquad
N_2^{(k)}(q)= \frac{\partial_{\tilde q_\eps}\rho_{\o}(q)}{Z_{k-1}}, \\
& N_3^{(k)}(q)=D_\o(q)\partial_{\tilde q_\eps} \frac{\tilde Z_{k-1}(q)}{Z_{k-1}},\qquad 
N_4^{(k)}(q)=-\frac{\rho_{\o}(q)}{Z_{k-1}^2}\partial_{\tilde q_\eps}Z_{k-1}.\end{split}
\end{equation}

We want to show that on the support of $f_k^{(2)}(q)$, in which $\|q\|_{(2)}\asymp 2^k$ (and thus $|q_1|, v_\eps |q_2| \asymp 2^k$, cf. \eqref{eq:singlescaleprop2} and below \eqref{eq:2.35}), the factor $S_j^{(k)}(q)$ can be bounded from above by $\eps^{-1/2}$ in absolute value. This implies that the bound 
on $\partial_{\tilde q_\eps} \tilde g_\omega^{(k)}$ is qualitatively the same as that on $\tilde g_\omega^{(k)}$, up to an additional factor $\eps^{-1/2}$, and similarly for higher order derivatives, as desired.

In order to bound $S_0^{(k)}$, we start by recalling that $|v_\eps| \asymp \eps^{1/2}$; from the identity  $\partial_{\tilde q_\eps} v_\eps=Z_{h_\eps-1}^{-1} (\a_{h_\eps-1}+\cos \tilde q_\eps+v_\eps \sin \tilde q_\eps)$ and the fact that $\a_{h_\eps-1}=O(\l), Z_{h_\eps-1}-1=O(\l)+O(\eps)$, we find that $|\partial_{\tilde q_\eps} v_\eps| \asymp 1$. This implies that on the support of $f_k^{(2)}$, $|S_0^{(k)}| \le C 2^{-2k} \eps^{1/2} 2^{2k} \eps^{-1} \asymp \eps^{-1/2}$ for some $C>0$, as desired.   

The bound on $S_1^{(k)}$ is very similar. We choose $q$ in the support of $f_k^{(2)}(q)$. Recalling the fact that $\tilde Z_{k-1}(q)/Z_{k-1}=1+(\chi_k^{(2)}-1)z_k/(1+z_k)$, with $z_k$ satisfying the bounds of Lemma \ref{lem:costantiascalaheps} and Proposition \ref{lem:kernbounds2}, and using again that,
for the $q$'s of interest, $v_\eps |q_2| \le C 2^k$, $|v_\eps| \asymp \eps^{1/2}$ and $|\partial_{\tilde q_\eps} v_\eps| \asymp 1$, 
we find that, on the support of $f_k^{(2)}(q)$,  $|N_{1}^{(k)}(q)|\le C 2^k\eps^{-1/2}$. Plugging this into the formula for $S^{(k)}_1$, and 
using the fact that for the $q$'s of interest $|D_\o(q)|, |(D_\o(q)\tilde Z_{h-1}(q)+\rho_\o(q))/Z_{h-1}|\asymp 2^{h}$, we find that, on the support of $f_k^{(2)}$, $|S_1^{(k)}(q)|\le C\eps^{-1/2}$, as desired. 

Let us now focus on the case $j=2$. 
Note that, from its definition, stated right after \eqref{eq:2.37},
we can rewrite $\rho_\o$ as follows: 
\be
\rho_\o(q)
=\frac{q_2^2}{2}\a_{h_\eps-1}
 +\cos\widetilde q_\eps
  \bigl(1+iq_1-e^{iq_1}\cos q_2\bigr)
 +\o\sin\widetilde q_\eps
  \bigl(e^{iq_1}\sin q_2-q_2\bigr),
\label{eq:rhoidentityinappB}\ee
so that $\partial_{\tilde q_\eps} \rho_\o(q)=-\sin \tilde q_\eps(1+iq_1-e^{iq_1}\cos q_2)+\o\cos \tilde q_\eps(e^{iq_1}\sin q_2-q_2)$. 
Therefore, $|\partial_{\tilde q_\eps} \rho_\o(q)|\asymp 2^{2k}\eps^{-1/2}$ in the support of $f_k^{(2)}$. Plugging this bound in the expression for $N_2^{(k)}$ leads once again to the desired estimate for $S_j^{(k)}$, for $j=2$. 

The cases $j=3,4$ can be bounded similarly, taking however into account that the derivative with respect to $\tilde q_\eps$ now acts on 
$z_k$ (recall in fact that $\tilde Z_{k-1}(q)/Z_{k-1}=1+(\chi_k^{(2)}-1)z_k/(1+z_k)$ and that $Z_{k-1}/Z_k=1+z_k$). An inductive argument on the scale label, in combination with the 
iterative definition of $z_k$, allows one to prove that also in this case the action of $\partial_{\tilde q_\eps}$ can be dimensionally bounded by $\eps^{-1/2}$. We do not belabor the details here, since they are similar to 
other discussions made above, and we leave them to the reader. 

\medskip

\textbf{$\bullet$ } \textit{Action of
of $\partial_{\tilde q_\eps}$ on the oscillating factors associated with the endpoints.} Recall that the endpoints associated with a $\tilde q_\eps$ oscillating factor can be either of type 
$Y^{(2)}$ on scale $k\le h_\eps$ or of type $\RR_2 V^{(h_\eps-1)}$. In the former case, 
the oscillating factor attached to the local kernel associated with $F_{Y,s,(\omega,-\omega)}(\varphi,J)$ is
$e^{2i\omega\tilde q_\eps y_2}$, where $y$ is 
the location of the fields, see the line before \eqref{eq:2.54}; 
in the latter case, the oscillating factor 
attached to the kernel $\widetilde W^{(h_\eps-1)}_{2n,m;\eps,\un s, \un D}(\underline y,\underline z)$ is $e^{i\sum_{j=1}^{2n}\tilde\sigma_j(\omega_j\boldsymbol{q}_\eps+q_F)y_j}$, see \eqref{eq:2.44bis} and preceding lines. Suppose that the 
derivative with respect to $\tilde q_\eps$ acts on one of these oscillating factors, associated with an endpoint $\bar v$ of the tree $\tau$ we are evaluating the value of; its effect is that of multiplying the kernel 
associated with $\bar v$ by a
factor $\sum_{f \in P_{\bar v}^\f} \tilde\sigma(f) \o(f) y_2(f)$ (see above for the conventions and notations about the field labels, see in particular the two paragraphs after \eqref{eq:kernoren}). In order to evaluate the effect of this term we define: 
\be U_v:= \sum_{f  \in P^+_v} \o(f)-\sum_{f\in P_v^-} \o(f)=\sum_{f \in P_v^\varphi}\tilde \sigma(f)\o(f) \ee where we decomposed $P_v^\f=P_v^+ \cup P_v^-$ with $P_v^{\tilde \sigma}=\{f \in P_v^\varphi: \tilde \sigma(f)=\tilde \sigma\}$ (note that  $|P_v^+|=|P_v^-|$; if $P_v^{\varphi}=\emptyset$, we set by convention $U_v=0$). By construction, for $v=\bar v$, the combination 
$U_{\bar v}:=\sum_{f \in P_{\bar v}^\f} \tilde\sigma(f) \o(f)$ is equal to $\pm 2$ in the case that $\bar v$ is of type $Y^{(2)}$, while 
$U_{\bar v}\in\{0,\pm2, \ldots, \pm|P_{\bar v}^{\varphi}|\}$ in the case that $\bar v$ is of type $\mathcal R_2 V^{(h_\eps-1)}$. 

\medskip

If $U_{\bar v}=0$, then $\bar v$ is necessarily of type $\mathcal R V^{(h_\eps-1)}$ and the effect of the factor
$\sum_{f \in P_{\bar v}^\f} \tilde\sigma(f) \o(f) y_2(f)$ on the bounds is easy to evaluate: in fact, in this case, we can rewrite such factor as 
a linear combination of differences, of the form
$\sum_{i=1}^{|P_{\bar v}^\varphi|/2}\big(y_2(f_{2i-1})-y_2(f_{2i})\big)$, for a suitable 
relabeling of the field variables, such that $(f_1,\ldots,f_{|P_{\bar v}^\varphi|})$ is a permutation of
$P_{\bar v}^\varphi$. In turn each coordinate difference can be bounded as discussed after \eqref{eq:A.33bis}, with the difference that now we replace 
the standard bound $\|\tilde g_\omega^{(h_{\bar v})}\|_1\le C 2^{-h_{\bar v}}$ 
by $\int dy |y_2 \tilde g^{(h_{\bar v})}_\omega(y,0)|\leq C 2^{-2h_{\bar v}+h_\eps/2}$ (see Lemma \ref{lem:3bis}), which differs from the standard bound by an extra factor proportional to $2^{-h_{\bar v}+h_\eps/2}$. Recalling the fact that $h_{\bar v}=h_\eps$, we find that such an extra factor is 
proportional to $2^{-h_\eps/2}\asymp \eps^{-1/2}$, as desired.
There are $|P_{\bar v}^\varphi|/2$ such factors, and the resulting additional combinatorial factor $|P_{\bar v}^\varphi|/2$ can be reabsorbed in the small factor
$|\lambda|^{d_{n,m}}$ (with $n=|P_{\bar v}^\varphi|$
and $m=|P_{\bar v}^J|$) which the kernel associated with $\bar v$
is proportional to, see \eqref{eq:2.29}, possibly at the cost of 
decreasing the constant $c$ in the definition of $d_{n,m}$, see the line after \eqref{kernestim}.

\medskip 

Let us now consider the case that $U_{\bar v}\neq0$, in which case $\bar v$ can either be of type $Y^{(2)}$ or of type $\mathcal R_2 V^{(h_\eps-1)}$. In this case the factor $\sum_{f \in P_{\bar v}^\f} \tilde\sigma(f) \o(f) y_2(f)$ should not be bounded `by itself' but, rather, in combination with analogous factors stemming from other endpoints, as we now explain. Note the following:
\medskip 

(a) if $v$ is an endpoint of type $\l, \d, \mu, Y^{(1)}$, then $U_v=0;$ 
\medskip 

(b) since for the kernels under consideration 
$P^{\varphi}_{v_0}=\emptyset$, then $U_{v_0}=0$;

\medskip 

(c) 
for every $v$ of $\tau$ that is not an endpoint, $U_v=\sum_{w \geq v: w \in E(\t)} U_w$ (recall notations below Fig. \ref{fig:alberello}) since every two contracted fields $\f^+_{y(f_1),\o(f_1)}$,  $\f^-_{y(f_2),\o(f_2)}$ satisfy $\o(f_1)=\o(f_2)$, see the lines after \eqref{eq:2.52bis}; 

\medskip

(d) as a corollary of the previous two items, letting $\bar v_1, \ldots, \bar v_k$ the endpoints of $\tau$ such that $U_{\bar v_i}\neq 0$, we have  $\sum_{i=1}^k U_{\bar v_i}=0$; we denote by $v^*$ the rightmost common ancestor 
of $\bar v_1,\ldots,\bar v_k$, with $k\ge 2$; note that  $U_{v^*}=0$. 

\medskip

For simplicity, we only consider the case where 
the number of endpoints mentioned in item (d) above is $k=2$, in which case the rearrangement of the terms coming from the derivative with respect to $\tilde q_\eps$ is a bit simpler, leaving the general case to the reader. In fact,
the action of the derivative with respect to $\tilde q_\eps$ on the oscillating factors
associated with the two endpoints $\bar v_1, \bar v_2$, produces an overall factor $\sum_{i=1}^2
\sum_{f \in P_{\bar v_i}^\f} \tilde\sigma(f) \o(f) y_2(f)$. Recalling from item (d) above that 
$U_{\bar v_1}=-U_{\bar v_2}$, we can rearrange this factor in the form of a linear combination of differences of coordinates, similarly to the case $U_{\bar v}=0$, thus rewriting it as 
\begin{equation}\sum_{i=1}^{N_1-M}\big(y_2(f_{2i-1})-y_2(f_{2i})\big)+\sum_{i=1}^{N_2-M}\big(y_2(f'_{2i-1})-y_2(f'_{2i})\big)+\sum_{i=1}^{2M}\big(y_2(\tilde f_{i})-y_2(\tilde f'_{i})\big)\label{eq:relab}\end{equation}
where $2N_1=|P_{\bar v_1}^\varphi|$, $2N_2=|P_{\bar v_2}^\varphi|$, $2M=|U_{\bar v_1}|=|U_{\bar v_2}|$, and the field variables in \eqref{eq:relab} are relabelings of those in 
$P_{\bar v_1}^\varphi$, $P_{\bar v_2}^\varphi$, such that 
$(f_1,\ldots, f_{2N_1-2M}, \tilde f_1,\ldots,\tilde f_{2M})$ 
is a permutation of $P_{\bar v_1}$ and 
$(f_1',\ldots, f_{2N_2-2M}', \tilde f_1',\ldots,\tilde f'_{2M})$ 
is a permutation of $P_{\bar v_2}$. Now, proceeding as in the case $U_{\bar v}=0$, we see that 
the terms in the first two sums can be bounded dimensionally by 
an overall factor proportional to $(N_1+N_2-2M)2^{-h_{\eps}+h_\eps/2}\asymp (N_1+N_2-2M)\eps^{-1/2}$, as desired
(the combinatorial factor $N_1+N_2-2M$ can be reabsorbed in the definition of the exponent $d_{n,m}$, as discussed above for $U_{\bar v}=0$). The terms in the third sum can instead be bounded proportionally to 
$M2^{-h_{v^*}+h_\eps/2}$, where we recall that $v^*$ is the rightmost common ancestor of $\bar v_1$ and $\bar v_2$. The latter factor is potentially dangerous: in order to compensate it, we note that whenever $h_\eps-h_{v^*}$ is big, we can extract 
an additional small factor from the tree value, as follows. 
Let $\tilde v_1$ be the first child of $v^*$ on the path along the tree $\tau$ from $v^*$ to $\bar v_1$, and similarly for 
$\tilde v_2$. By item (c) above, we know that $U_{\tilde v_1}=U_{\bar v_1}$ and $U_{\tilde v_2}=U_{\bar v_2}=-U_{\bar v_1}$. Denoting by $q(f)$ with $f\in P_{v}^\varphi$ the quasi-momentum associated
with the Grassmann field $\hat \varphi^{\tilde\sigma(f)}_{q(f),\omega(f)}$, by conservation of momentum and the support properties of the Grassmann fields 
we have: 
\begin{equation}M\tilde q_\eps=|\sum_{f\in P_{\tilde v_1}^\varphi}q(f)|\le C|P_{\tilde v_1}^\varphi|2^{-h_{v^*}+h_\eps/2},
\label{boundonM}\end{equation}
for some $C>0$, and similarly for $\tilde v_2$. Moreover, the same proof leading to the factor $|\lambda|^{d_{n,m}}$ in the right hand side of \eqref{eq:2.61bis}, with $d_{n,m}$ defined right after \eqref{kernestim}, allows us to extract from the tree value an additional factor $|\lambda|^{c'(|P_{\tilde v_1}|+|P_{\tilde v_2}|-2}$, with a constant $c'$ smaller than the constant $c$ in the definition of $d_{n,m}$ (e.g., half of it). Using \eqref{boundonM}, we see that such a small factor is smaller than $|\lambda|^{c''M2^{-h_{v^*}+h_\eps}}$ for some $c''>0$. 
Putting this together with the (potentially dangerous) factor 
$M2^{-h_{v^*}+h_\eps/2}$ we find an overall extra factor proportional to 
\begin{equation}\label{maximi}M2^{-h_{v^*}+h_\eps/2}|\lambda|^{c''M2^{-h_{v^*}+h_\eps}}. \end{equation}
Maximizing this over $h_{v^*}\le h_\eps$ we find that \eqref{maximi} is smaller than (const.)$2^{-h_\eps/2}\asymp \eps^{-1/2}$, as desired. 

\medskip 

\textbf{Item 3.} In order to discuss the effect of the derivatives with respect to $\eps$ on $\tilde q_\eps(\lambda,\eps)$ we preliminarily need to explain how $\tilde q_\eps$ is defined and fixed. The general strategy was explained in Remark \ref{rem:4}. More in detail, note that in order to control the flow of the second regime, $\mu_{h_\eps}$ must be fixed via the implicit equation \be \mu_{h_\eps}=-\sum_{h \leq h_\eps} 2^{h-h_\eps-1}\BB^\mu_h, \quad \Longleftrightarrow \quad \Phi_1(\mu_{h_\eps},\tilde q_\eps,\eps,\l)=0 \label{eq:Phi1}\ee  where $\BB^\mu_h:=\mu_{h-1}-2\mu_h$ depends on $\mu_{h_\eps},\tilde q_\eps, \eps,\l$ (cf. with \cite[eq. (140),(145)]{BM36} and \cite[Eq. (6.87)]{GMT20}), and satisfies $|\BB^\mu_h|\leq C|\l| \eps^{\theta}$ as a consequence of Proposition \ref{lem:kernbounds2}.\footnote{The tree expansion for $2^{h-1}\BB^\mu_h$ is the same as the one for $2^{h-1}\mu_{h-1}=\hat W^{(h);\io}_{2,0;(\o,\o)}(0)$, up to the contribution from the tree with exactly one endpoint, of type $\mu$, at scale $h$. The extra factor $\eps^\theta$ in the stated bound for $\mathcal B^\mu_h$, as compared to the bound of Proposition \ref{lem:kernbounds2}, follows from a discussion analogous to the one made for $\Theta_1$ after \eqref{eq:decomptheta}; see also \eqref{eq:2.70biss}.}
This is the analogue of the fixed point equation for $\nu^\ast_0$ in App. \ref{app:A2}; in particular proceeding with a strategy similar to the one described after \eqref{eq:2.48bis}, we find that $|\partial_{\mu_{h_\eps}} \BB_h^\mu|\leq C |\l| \eps^{\theta}$ which implies that $\partial_{\mu_{h_\eps}} \Phi_1(\mu_{h_\eps},\tilde q_\eps,\eps,\l)\neq0$, for $|\l|\leq \l_0$ small, and thus the existence of a unique solution $I_{\k_0}\times (-\l_0,\l_0)\to \mu_{h_\eps}(\tilde q_\eps,\eps,\l)$. Plugging this solution into \eqref{eq:2.35} we can rewrite it as an implicit equation for $\tilde q_\eps$ as a function of $(\eps,\l)$:
\be Q_{h_\eps-1}r+\frac{\tilde q_\eps^2}{2}\a_{h_\eps-1}+1-\cos \tilde q_\eps+ \eps\mu_{h_\eps}(\tilde q_\eps,\eps,\l)=0 \quad \Longleftrightarrow \quad \Phi_2(\tilde q_\eps,\eps,\l)=0,\label{eq:Phi2}\ee
where we used \eqref{eq:A.24bis}. 

Using the fact that $\a_{h_\eps-1},(Q_{h_\eps-1}-1)=O(\l)$ and temporarily \textit{assuming} that $|\partial_{\tilde q_\eps} \mu_{h_\eps}|\leq C|\l|\eps^{\theta-1/2}$, we find that $\partial_{\tilde q_\eps}\Phi_2(\tilde q_\eps,\eps,0)=-\sin \tilde q_\eps \neq 0$ for $(\tilde q_\eps,\eps) \in I_{\k_0}.$ The unique solution $ \EE_2\times (-\l_0,\l_0) \mapsto \tilde q_\eps(\eps,\l)$ of \eqref{eq:Phi2} has then the same regularity of $\Phi$, thus the same of $\mu_{h_\eps}(\tilde q_\eps(\eps,\l),\eps,\l)$. In particular, from the bounds of Proposition \ref{lem:bounds1} and Proposition \ref{lem:kernbounds2} for $\a_h,\BB_h^\mu$ and $Q_h$, we find that $(-\l_0,\l_0) \ni \l \mapsto \tilde q_\eps(\eps,\l)$ is analytic, while differentiability of $\eps \mapsto \tilde q_\eps(\eps,\l)$ follows from the differentiability of $I_{\k_0} \ni (\tilde q_\eps,\eps) \mapsto \mu_{h_\eps}(\tilde q_\eps,\eps,\l)$, which follows from the same proof of Item 2, now applied to the beta function of $\mu$, see \eqref{eq:Phi1}, 
which admits the same kind of GN tree expansion as the kernels studied in Item 2. More specifically, by repeating step by step the detailed analysis of Item 2, we find 
 \begin{equation}
     |\partial^{s}_{\eps}\mu_{h_\eps}(\tilde q_\eps,\eps,\l)|\leq C(s)|\l|\eps^{\theta-s}, \qquad |\partial^{k}_{\tilde q_\eps}\mu_{h_\eps}(\tilde q_\eps,\eps,\l)|\leq C(k)|\l|\eps^{\theta-k/2}\qquad \forall (\tilde q_\eps,\eps) \in I_{\k_0}.
 \label{eq:B.52bis}\end{equation}
The details of the proof of these bounds are left to the reader. 

\end{proof}

\section{First order calculations}\label{app:firstorder}

 In this appendix, we prove that $\varepsilon_0(\lambda)=2\lambda+O(\lambda^2)$
 and $A(\lambda)=1+O(\lambda^2)$, as stated in Theorem \ref{th:5}. 
 
 In order to compute $\varepsilon_0$, we use its expression in terms of $\nu_0^*$, see \eqref{eq:eps0nu0}, 
 and notice that its expression at first order in $\lambda$ is the same as that of $\nu_0^*$ at the same precision. 
 
 In order to compute $A(\lambda)$, we recall its expression: $ A(\l):=\frac{2Y_{-\io;0,B}}{1+\z_{-\io}}\lim_{\eps\to0^+}\frac{\tilde q_\eps}{\sqrt{2\eps}}$, see \eqref{eq:2.81} and following line. In this formula, $\tilde q_\eps=\sqrt{(-2r_{h_\eps-1}+O(\eps^{1+\theta)})/\xi_2}$
with $\xi_2=1+\alpha_{-\infty}$ (see a few lines after \eqref{eq:Rbounds} and the line preceding \eqref{eq:Gheps}); moreover, $r_{h_\eps-1}=-\xi_0\eps+O(\eps^{1+\theta})$, see \eqref{eq:A.24bis}. In conclusion: 
\begin{equation}\label{eq:C.1}
A(\l):=\frac{2Y_{-\io;0,B}}{1+\z_{-\io}}\sqrt{\frac{\xi_0}{1+\alpha_{-\infty}}}. 
\end{equation}
Therefore, in order to prove that $\varepsilon_0(\lambda)=2\lambda+O(\lambda^2)$ and $A(\lambda)=1+O(\lambda^2)$, 
we need to compute at first order in $\lambda$ the functions $\nu_0^*,\a_{-\io},\z_{-\io},Y_{-\io;0,B}, \xi_0$. For this purpose, it is useful to write down the explicit expression of the bare potential 
$\tilde V_\eps(\varphi,\mathcal A)$ appearing in \eqref{eq:5.71} and \eqref{eq:2.2bis} at first order in $\lambda$; that is, we are interested in: 
$\tilde V_{\eps,\vartheta}'(\varphi(\psi),\AA(J)):=\partial_{\l} \tilde V_{\eps,\vartheta}(\varphi(\psi),\AA(J))\rvert_{\l=0}$, where $\varphi(\psi)$ is the map in \eqref{eq:3.29}, and 
$\AA(J)$ the inverse of the map relating $J$ and $A$ defined two lines after \eqref{eq:4.20}. A straightforward computation shows:  
 \begin{equation}
 \begin{split}\hspace{-8pt}
     &\tilde V'_{\eps,\vartheta}(\varphi(\psi),\AA(J))=\\ &= \varepsilon_0'(0)\sum_{x \in \Lambda} J_{x,A} e^{-\eps} E_{x,A}(\psi)+\sum_{x \in \Lambda}\sum_{s=A,B,C} \prod_{j=0}^1e^{-\eps} E_{x+u_s+j w_s,s}(\psi)(1+J_{x+u_s+j w_s,s})
     \end{split}
     \label{eq:B.5}
 \end{equation}
 where $E_{x,s}(\psi)$ stands for $E_{\vartheta,e}(\psi)$ (see \eqref{eq:Scoordinates} and following lines) with $e$ the edge of type $s$ with white vertex of coordinates $x$; 
 here $e^{-\eps}=e^{-\eps-\varepsilon_0(\l)}\rvert_{\l=0}$
 comes from the rescaling in \eqref{eq:5.71}; also, recalling $\hat e_B=(1,0), \hat e_C=(0,1)$, then \[u_A:=u_C:=0,\qquad u_B:=w_C:=-\hat e_B, \qquad w_A:=\hat e_C-\hat e_B, \qquad  w_B:=\hat e_C.\]
 Note also that at $J=0$, using the Fourier representation in  \eqref{fourierfieldsk}, we can write
 \begin{equation}\begin{split}
     &\tilde V_{\eps,\vartheta}'(\varphi(\psi),0)=\int dp  d\ell dm\,  \hpsp_{p-\ell+m}\hpsm_p \hpsp_\ell \hpsm_m \, \hat v(p,\ell,m) \\  & \, \hat v(p,\ell,m)=e^{-2\eps} e^{i(\ell-m)(\hat e_C-\hat e_B)}+ \frac14e^{ip \hat e_B+i\ell \hat e_C-im(\hat e_C-\hat e_B)}+\frac14e^{ip \hat e_C-i\ell\hat e_B+im(\hat e_C+\hat e_B)},
     \end{split}
     \label{eq:B.6}
 \end{equation}
 where $\int dp d\ell dm$ stands for $L^{-6} \sum_{p,\ell,m \in \DD_\vartheta}$ etc.
\begin{lemma}
\label{lem:firstordercalc}
    Under the hypothesis of Prop. \ref{lem:flowRC} and Prop. \ref{lem:kernbounds2} we have: $\nu_0^*(\l)=2\l+O(\l^2)$, and  $\a_{-\io},\z_{-\io},$ $Y_{-\io;0,B}-1/2,\xi_0-1=O(\l^2)$.   
\end{lemma}

Using \eqref{eq:C.1} and the fact that $\a_{-\io},\z_{-\io}=O(\lambda^2)$, $Y_{-\io;0,B}=1/2+O(\lambda^2)$ and $\xi_0=1+O(\lambda^2)$, it readily follows that $A(\lambda)=1+O(\lambda^2)$, as stated in Theorem \ref{th:5}. 

\begin{proof}

\par{\textbf{1. $\nu^*_0(\l)$ at first order.}} Recall that $\nu^*_0(\l)$ satisfies  the implicit equation in \eqref{eq:2.48bis} and that $2^h \BB^\nu_h$ is the same as $\hat W^{(h+1);\io}_{2,0;0}(q)\rvert_{q=0}-2^{h+1}\nu_{h+1}$ (here $q$ is the relative momentum to $q_F=(\pi,0)$ in the coordinates \eqref{eq:transf}; moreover, we used the definition of beta function in \eqref{eq:beta2}). Therefore, $2^h\BB^\nu_h$ has the same tree expansion as $\hat W^{(h+1);\io}_{2,0;0}(q)\rvert_{q=0}$, up to the `trivial' tree with just one normal endpoint of type $\nu$, which corresponds to $2^{h+1}\nu_{h+1}$. 
In view of this, the first order contribution to $2^h\BB^\nu_h$ comes from the tree with just one endpoint of type $\RR_{1,0} V_0^{(0)}(\varphi,0)$ with $V_0^{(0)}(\varphi,0)$ replaced by its first order approximation (which is nothing but $\lambda\tilde V'_{0,\vartheta}(\varphi,0)$), when we contract two of the four Grassmann variables in the right hand side of (the first line of) \eqref{eq:B.6} at scale $h+2$\footnote{If the two Grassmann variables are contracted at higher scales, then the action of the localization operator makes the corresponding contribution vanish, since by construction $\LL_{1,0} \RR_{1,0}=0.$}: in formulas, recalling that $\mathcal R_{1,0}$ acts as the identity operator on 
$V^{(0)}_0$ (see the discussion around \eqref{eq:3.3.30}), we get:
\begin{equation}
\begin{split}
    \partial_{\l} 2^{h}\BB^\nu_h \rvert_{\l=0}=&\lim_{L \to \io}L^2 \left [\partial_{ \hpsm_k}\partial_{\hpsp_k} \mathcal{E}^T_{h+2}\left (  \tilde V'_{0,\vartheta}(\varphi(\psi+\, \cdot\, ),0)\right )  \right ]\rvert_{k=k_F,\eps=\l=0}=\\ & \int_{\mathbb{T}^2} \frac{d\ell}{(2\pi)^2} (\hat g^{(h+2)}_{0}(R^{-1}(\ell-k_F))\rvert_{\l=0} \left( -2\hat v(k,\ell,\ell)+\hat v(\ell,\ell,k)+\hat v(k,k,\ell)  \right )\rvert_{k=k_F,\eps=0}
    \end{split}\label{eq:B.7}
\end{equation}
where $k_F=Rq_F=(\pi,\pi)$, we used that $\mathcal{E}^T_{h}(\hpsm_\ell\hpsp_{\ell'})=\int \tilde P_\eps^{(h)}(D\psi)\hpsm_\ell\hpsp_{\ell'}=\d_{\ell,\ell'}L^{2}\hat g^{(h)}_\eps(R^{-1}(\ell-k_F))$, and $\hat g^{(h+2)}_0$ is the Fourier transform of $\tilde g^{(h+2)}_\eps|_{\eps=0}$ in \eqref{eq:singlescalepropI}, which still depends on $\l$ via  $\boldsymbol{\a},\boldsymbol{\z}.$ Recall that, from the definitions (see \eqref{eq:2.3}-\eqref{eq:5.39}, as well as the discussion after \eqref{eq:3.29}), $\hat g_\eps^{(h)}(R^{-1}(\ell-k_F))\big|_{\lambda=0}=\frac1{\mu(\ell)}f^{(1)}_h(R^{-1}(\ell-k_F))$, where $f^{(1)}_h$ is a function with compact support satisfying $\sum_{h \leq 1}f^{(1)}_{h}(R^{-1}(\ell-k_F))\equiv 1$. 
By inspection, from \eqref{eq:B.6} we can rewrite \be (-2\hat v(k_F,\ell,\ell)+\hat v(\ell,\ell,k_F)+\hat v(k_F,k_F,\ell))\rvert_{\eps=0}=-2+P(\ell) \label{eq:polyrewriting}\ee  with  $P(\ell)=
-\frac12 e^{i(k_F+\ell)\hat e_B}-\frac12e^{i(k_F+\ell)\hat e_C}+2\cos((k_F-\ell)(\hat e_B-\hat e_C))+\frac12 e^{i\ell(\hat e_B+\hat e_C)}\cos(k_F(\hat e_B-\hat e_C))
+\frac12 e^{i k_F(\hat e_B+\hat e_C)}
\cos(\ell(\hat e_B-\hat e_C))$. For the computation of interest, more than the specific expression of $P(\ell)$ it is important that we can write 
$P(\ell)=\sum_{j=1}^8c_j e^{-i\ell x^{(j)}}$
for suitable coefficients $c_j \in \mathbb{R}$ and suitable coordinates $x^{(j)}=(x^{(j)}_1,x^{(j)}_2)$ satisfying $\min\{x^{(j)}_1,x^{(j)}_2\}<0$. Using \eqref{eq:2.48bis},\eqref{eq:B.7} and the properties discussed sfter \eqref{eq:B.7} we find: 
\begin{equation} \partial_\l \nu^*_0\rvert_{\l=0}=-\sum_{h \leq -1}  \partial_{\l} 2^h \BB_h^\nu \rvert_{\l=0}=\big [2\mathfrak{g}(0)+\sum_{j=1}^8 c_j \mathfrak{g}(x^{(j)})\big ]\rvert_{\varepsilon=0}, \label{eq:B.8bis}
\end{equation} where $\mathfrak{g}(x)$ is defined below \eqref{eq:2comptsprop}, and is nothing but the inverse Kasteleyn matrix in the infinite volume limit, computed at $(x,0)$. For $\varepsilon=0$, we get $\mathfrak g(x)\rvert_{\varepsilon=0}=K^{-1}(x)\rvert_{A=1,B=C=1/2}$ with $K^{-1}(x)$ as in \eqref{eq:A.16} of Lemma \ref{lem:14} below;
in particular, using that formula, we conclude that 
$\mathfrak g(0)=1$ and $\sum_{j=1}^8 c_j \mathfrak{g}(x^{(j)})=0$, which implies $\partial_\l \nu^*_0\rvert_{\l=0}=2$, as desired. 

\par{\textbf{2. $\z_{-\io}(\l),\a_{-\io}(\l)$ at first order}.} Let us discuss the first order computation for $\zeta_{-\infty}$, 
the one for $\alpha_{-\infty}$ being completely analogous. 
From \eqref{eq:beta2} we have that \[\z_{-\io}=\sum_{h \leq 0} \BB^\z_h, \qquad \BB^\z_h=i\partial_{q_1}\hat W^{(h+1);\io}_{2,0;0}(q)\rvert_{q=0}\] where $q$ is the quasi-momentum relative to $q_F$. Passing to the variables $k=Rq$, 
see \eqref{eq:transf}, the first order contribution to $\mathcal B^\zeta_h$ is equal to $\partial_\l (\partial_{k_1}+\partial_{k_2})\hat W^{(h+1);\io}_{2,0;0}(R^{-1}(k-k_F))\rvert_{k=0,\l=0}.$ Similarly to the case discussed above for $\mathcal B^\nu_h$, such a first order contribution 
comes from the tree with just one endpoint of type $\partial_\l \RR_{1,0} V_0^{(0)}(\varphi,0)\rvert_{\l=0}$ when two of the four Grassmann variables associated with the endpoint 
are contracted on scale $h+2$.\footnote{The contribution associated with the tree with just one endpoint of type $\nu$ vanishes thanks to the presence of $\partial_{k_i}$, since the kernel of the endpoint, $2^{h+1}\nu_{h+1}$, is independent of the momentum $k$.} Proceeding as in \eqref{eq:B.7}, and taking into account the presence of the operator $\sum_{i=1,2}\partial_{k_i}$ acting now on the second factor of the second line of \eqref{eq:B.7}, we find:  
\begin{equation}
    \partial_{\l}\z_{-\io}\rvert_{\l=0}=\sum_{h \leq 0} \partial_\l \BB^\z_h \rvert_{\l=0}=\sum_{j=1}^8 i \tilde c_jK^{-1}(x^{(j)})\rvert_{A=1,B=C=\frac12},
    \label{eq:B.8}
\end{equation}
for suitable real coefficients $\tilde c_j$ and the same coordinates $x^{(j)}$ obtained above, see the discussion after \eqref{eq:polyrewriting}. Finally, the right hand side of \eqref{eq:B.8} vanishes, for the same reason discussed after 
\eqref{eq:B.8bis}, which implies that $\partial_{\l}\z_{-\io}\rvert_{\l=0}=0$, as desired.
A similar discussion holds for ${\a}_{-\io}$, details left to the reader. 
\medskip

\par{\textbf{3. $Y_{-\io;0,B}$ at first order.}} Recall that $Y_{-\io;0,B}=\sum_{h \leq 0} \BB_{h;0,B}^Y+Y_{0,;0,B}$, with $\BB^Y_{h;0,B}$ satisfying the bound in \eqref{eq:goalbetaY},  and $Y_{0;0,B}=1/2$. Using the tree expansion for $Y_{h-1;0,B}$, we see once again that the first order contribution to $\BB^Y_{h;0,B}$ is obtained from the tree with just one endpoint of type $\RR_{1,0} V^{(0)}_{0}$ when  two of the four Grassmann variables of $\partial_{J_{x,B}} \tilde V'_{0,\vartheta}(\f(\psi),\AA(J))\rvert_{J=0}=E_{x,B}(E_{x+\hat e_C,B}+E_{x-\hat e_C,B})$ in \eqref{eq:B.5}
are contracted  exactly at scale $h+1$. In formulas,
\begin{equation}\label{eq:splittedY}
\begin{split}
 \partial_{\l} \BB^Y_{h;0,B}\rvert_{\l=0}&=\lim_{L \to \io} \Big [ \partial_{\psi^-_x}\partial_{\psi^+_x} \LL_{1,0}\, \mathcal{E}^T_{h+1}\Big (\partial_{J_{x,B}} \tilde V'_{0,\vartheta}(\f(\psi+ \,\cdot\, ),\AA(J))\rvert_{J=0}\Big ) \Big ]_{\psi=0, \l=0} \\ & = 2^{-2}\big [-2e^{ik_F\hat e_B}  \tilde g_0^{(h+1)}(-R^T\hat e_B)+e^{-ik_F(\hat e_C-\hat e_B)}  \tilde g_0^{(h+1)}(R^T(\hat e_C-\hat e_B))+ \\ & + e^{ik_F(\hat e_C+\hat e_B)}  \tilde g_0^{(h+1)}(-R^T(\hat e_C+\hat e_B))\big]\rvert_{\l=0}.
    \end{split}
\end{equation} where we used that $\EE^T_{h+1}(\psi^-_x \psi^+_{x'})=\int P_0^{(h+1)}(d\varphi) \psi^-_x \psi^+_{x'}=e^{-ik_F(x-x')}\tilde g^{(h+1)}_0(R^T(x-x'))$. Plugging \eqref{eq:splittedY} into the formula for $Y_{-\infty;0,B}$, 
we find: $Y_{-\io;0,B}=\frac12+\lambda 2^{-2}(-2K^{-1}(-\hat e_B)+K^{-1}(\hat e_C-\hat e_B)+K^{-1}(-\hat e_C-\hat e_B))+O(\lambda^2)$. Once again, the linear term in $\lambda$ vanishes, because it is a linear combination of 
terms of the form $K^{-1}(\tilde x^{(i)})$, where $\tilde x^{(i)}=(\tilde x^{(i)}_1,\tilde x^{(i)}_2)$ with 
min$\{\tilde x^{(i)}_1,\tilde x^{(i)}_2\}<0$ (we used 
again \eqref{eq:A.16} of Lemma \ref{lem:14} below). 
\medskip

\par{\textbf{4. $\xi_0$ at first order.}} Recall the definitions in \eqref{eq:A.23bis}-\eqref{eq:A.24bis}, in particular  $\xi_0=e^{-\varepsilon_0}Q_{-\io}$. We already proved the fact that $\varepsilon_0(\lambda)=2\lambda+O(\lambda^2)$. We will now prove that $Q_{-\io}=1+2\lambda+O(\lambda^2)$, which 
implies $\xi_0=1+O(\lambda^2),$ as desired. In order to compute 
$Q_{-\infty}$ at first order precision, we use the definition 
\eqref{eq:A.23bis} and we observe that \begin{equation} \t^h_k=\begin{cases}1+O(\l) & k=h \\ O(\l) & k=0 \\ O(\l^2) & k \neq 0,h. \label{eq:C.7bisbis}\end{cases}\end{equation} The first two identities readily follow from the definition of $\t^h_k$ in \eqref{eq:A.23bis}. In  order to prove the last identity, we note that $\partial_{\l} \t^h_k \rvert_{\l=0}$ can be 
written essentially in the same way as the second line of \eqref{eq:B.7}, 
up to the following differences: (i) the overall sign is different, (ii) the scale label $h+2$ should be replaced by $h+1$, (iii) before setting $\eps=0$, we should act with the operator $\partial_{\boldsymbol{r}_k}$ on the integrand; in formulas: 
\begin{equation}\label{eq:dlambdatauhk}\hspace{-5pt}\partial_{\l} \t^{h}_k\rvert_{\l=0}=-\int_{\mathbb{T}^2}\, d\ell \,\partial_{\boldsymbol{r}_k}\Big[(\hat g^{(h+1)}_{\eps}(R^{-1}(\ell-k_F)) \left( -2\hat v(k_F,\ell,\ell)+\hat v(\ell,\ell,k_F)+\hat v(k_F,k_F,\ell)  \right )\Big]\Big|_{\eps=\l=0}.
\end{equation}
Note that the 
expression in square brackets in the right hand side has no other dependence upon $\{\boldsymbol{r}_k\}_{h\le k\le 0}$ besides the one upon $\boldsymbol{r}_h$ induced by $\hat g_\eps^{(h+1)}$ (see \eqref{eq:singlescalepropI})
and the one upon $r_0=r$ due to the coefficient  $e^{-2\eps}=(1+r)^2$ of $\hat v(k_F,\ell,\ell)$. Therefore, because of the action of 
$\partial_{\boldsymbol{r}_k}$ under the integral sign, we conclude that \eqref{eq:dlambdatauhk}
is zero unless $k=h$ or $k=0$, as stated above.  From \eqref{eq:C.7bisbis} and \eqref{eq:A.23bis} we thus find that, for any $h<-1$: 
\be Q_h=\sum_{s=1}^{|h|} \left (\prod_{j=1}^{s-1} \t^{h+j}_{h+j}\right )\t^{h+s}_0+O(\l^2) \quad  \Longrightarrow \quad  \partial_{\lambda}Q_h \rvert_{\l=0}=\sum_{k=h+1}^{0} \partial_{\l} \t^{k}_k\rvert_{\l=0} +\sum_{k=h+1}^{-1} \partial_{\l}\t_0^k\rvert_{\l=0}, \label{eq:exprQ} \ee  
where the product over $j$ in the first equation
should be interpreted as being equal to 1 if $s=1$. Note that the contribution from the term $\partial_{\l} \t^0_0 \rvert_{\l=0}$ is special because, using \eqref{eq:dlambdatauhk}, we see that its value is a linear combination of two terms, one coming from the action of $\partial_{r}$ on $\hat g_\eps^{(1)}$, the other from the action of $\partial_{r}$ on $\hat v(k_F,\ell,\ell)$. Proceeding as in \eqref{eq:B.8bis}, \eqref{eq:B.8} and \eqref{eq:splittedY}, recalling that $Q_{-\io}=\lim_{\eps \to 0^+} Q_{h_\eps-1}$, and noting the fact that $\partial_{r} e^{-2\eps}\rvert_{\eps=0}=\partial_r(1+r)^2\rvert_{r=0}=2$, we find: 
\begin{equation}
    \partial_\l Q_{-\io}
    \rvert_{\l=0}=\left ( 2\partial_AK^{-1}(0)+\sum_{j=1}^8 c'_j\partial_AK^{-1}(x^{(j)})+4K^{-1}(0)\right )\bigg\rvert_{\substack{A=1 \\ B=C=\frac12}}, \nn
\end{equation}
where $K^{-1}(x)$ is the same as in \eqref{eq:6.12}, $c'_j$ are suitable real constants and $x^{(j)}$ are the same of \eqref{eq:B.8bis}, satisfying $\min\{x^{(j)}_1,x^{(j)}_2\}<0$. 
Using once again \eqref{eq:A.16} of 
Lemma \ref{lem:14} (which implies, in particular, that $\partial_A K^{-1}(0)=-1$), we finally obtain 
$\partial_\l Q_{-\io}
    \rvert_{\l=0}=2$, as desired. \end{proof}

\section{Proof of Proposition \ref{prop:rgmassa}}\label{app:proofprop7}
In this appendix, we discuss the multiscale integration of $\WW_{\eps+\varepsilon_0(\l),\l,L,m}^{(\vartheta)}(\AA)$ (cf. \eqref{eq:5.71}) for $m>0$. We first summarize its output 
in Lemma \ref{lem:ultimosperiamo} and explain how to deduce from it the validity of Proposition \ref{prop:rgmassa}. Next, we prove Lemma \ref{lem:ultimosperiamo} by adapting the multiscale analysis described in Sections \ref{ssec:uv}-\ref{sec:secondregime} to the case $m>0.$ Throughout this appendix, we assume that the parameter $\varepsilon_0(\l),\mu_{h_\eps}(\l,\eps)$ are fixed as in Proposition \ref{prop:7}, that $|\l|\leq \l_0$ and that $0 <|\eps| \leq \bar \eps$; moreover, without loss of generality, since $m\to 0^+$ at fixed $\eps\neq 0$, we assume that $m\leq c_0\sqrt{|\eps|}$, for some sufficiently small $c_0>0$.

\paragraph{The multiscale integration for $m>0$: setup and output.}
We consider the Grassmann integral in \eqref{eq:5.71} and normalize it as follows: letting $\tilde{\mathcal{E}}_m(q):=|\tilde \mu(q)|^2+4\tilde m^2 \sin^2(2q_2)$ with $\tilde \mu(q)=1+r-e^{iq_1}\cos q_2$ and $\tilde m:=e^{-\eps-\varepsilon_0(\l)}m$, we rewrite \eqref{eq:5.71} as  \be e^{\WW^{(\vartheta)}_{\eps+\varepsilon_0(\l),\l,L,m}(\AA)}=s_\vartheta e^{L^2t^{(0)}_{\vartheta,m}}\int P_{\vartheta,m}(D\varphi)e^{\nu N(\varphi)+\tilde V_{\eps,\vartheta}(\varphi,\AA)} \label{eq:primaeqappE}\ee 
where $t^{(0)}_{\vartheta,m}:=\eps+\varepsilon_0(\l)+2^{-1}L^{-2} \sum_{q \in \DD'_{\vartheta,R}}\log \tilde{\mathcal{E}}_m(q)$, and $s_\vartheta$ is a sign, equal to $1$ for $\vartheta \neq (0,0)$, while $s_{(0,0)}=-\sgn \eps$;  in order to normalize the integration we used that, for $\eps\neq 0$,  $\tilde{\mathcal{E}}_m(q)\neq 0$ for all $q \in \DD_{\vartheta,R}'$, all $\vartheta \in \{0,1\}^2$ and all $m>0$\footnote{These properties follow from those of $\mathcal{E}_m$ defined in \eqref{eq:proppsi} and the relation $\tilde{\mathcal{E}}_m(q)=(1+r)^2 
\mathcal{E}^\ast_{m'}(R(q+q_F))$  where here $ \mathcal{E}^\ast_{m'}$ is defined as in  \eqref{eq:proppsi} with $m'=\tilde m/(1+r)$ in place of $m$ and $\varepsilon':=-\log(1+r)$ in place of $\varepsilon$: together with the bounds on $\EE_m$ stated after \eqref{eq:2comptsprop}, this implies that $\tilde{\mathcal E}_m\ge r^2$ if $\eps<0$ and $\tilde{\mathcal E}_m\ge c\,m^2\eps$ for some $c>0$ if $\eps>0$. \label{footn:tildecalE}}.
Letting $\media{\varphi^\a_y \varphi^\b_{y'}}=\int P_{\vartheta,m}(D\varphi)\varphi^\a_y \varphi^\b_{y'}$ the propagator $\mathbf{g}_{\eps,m}$ associated with $P_{\vartheta,m}(D\varphi)$ takes the form
\begin{equation}\label{eq:Eprop}
\begin{split}
& \mathbf g_{\eps,m}(y,y'):=\begin{pmatrix}\langle\varphi^-_y\varphi^+_{y'}\rangle&\langle\varphi^-_y\varphi^-_{y'}\rangle\\[2pt] \langle\varphi^+_y\varphi^+_{y'}\rangle&\langle\varphi^+_y\varphi^-_{y'}\rangle\end{pmatrix}=(-1)^{y_1-y'_1}L^{-2}\!\sum_{q \in \DD_{\vartheta,R}'}\!e^{-iq(y-y')}M(q)^{-1}, \\ & M(q)^{-1}=\frac1{\tilde{\mathcal E}_m(q)}\begin{pmatrix}\overline{\tilde\mu(q)}&2i\tilde m\sin(2q_2)\\-2i\tilde m\sin(2q_2)&-\tilde\mu(q)\end{pmatrix}.
\end{split}
\end{equation}
We call \emph{normal} the diagonal entries of $M$ and of $\mathbf g$, which at $m=0$ reduce to the propagator \eqref{eq:2.3}, and \emph{anomalous} the off-diagonal ones. 

The Grassmann integral in \eqref{eq:Eprop} is computed via the multiscale procedure of Sections \ref{ssec:uv}-\ref{sec:secondregime}, modified by the presence of $m>0$ as described in App.\ref{app:ultimolemma} below. Its output is summarized in the following lemma, where $h_\eps$ is the same integer as in \eqref{eq:4.69}, so that $2^{h_\eps}\asymp |\eps|$. 
We denote the kernels of the effective potentials by the same symbols as in \eqref{eq:4.68},\eqref{eq:2.33bis} and \eqref{eq:2.51biss}, modulo the presence of extra labels $\vartheta,m$; in order to avoid confusion with the regularization parameter $m$, the number of external fields $J$ of a kernel is denoted by $\ell$ (rather than by $m$, as in the main text): e,g, $W^{(h)}_{0,1;B,\vartheta,m}(0)$ is the coefficient of $\sum_{y} J_{y,B}$ in the effective potential on scale $h$, and $W^{(h);\io}_{0,1;B,m}(0)$ its $L \to \io$ limit, which is independent of $\vartheta$. 

\begin{lemma}\label{lem:ultimosperiamo} Assume the hypotheses of Proposition \ref{prop:7} and suppose that, additionally, $\eps\neq0$ and 
$0<m<c_0\sqrt{|\eps|}$ for a suitable, sufficiently small, $c_0>0$. Then there exists an integer $h^*\leq h_\eps-1$, depending on $\eps,m,\l$ but not on $L,\vartheta$, satisfying, for $\eps>0$, 
\begin{equation}
    c(m\eps^{-1/2})^{1+C|\l|}\leq 2^{h^*-h_\eps} \leq C (m \eps^{-1/2})^{1-C|\l|},\label{eq:hstar}
\end{equation}
for some $C,c>0$, while $h^*:=h_\eps-1$ for $\eps<0$, such that \eqref{eq:primaeqappE} can be written as
\begin{equation}
    e^{\mathcal{W}^{(\vartheta)}_{\eps+\varepsilon_0(\l),\l,L,m}(\AA(J)))}=s_\vartheta \exp \big [L^2t^{(0)}_{L,m}+ \sum_{h=h^*}^0(L^2 \tilde F^{(h)}_{\vartheta,m}+\tilde S_{\vartheta,m}^{(h)}(J))\big ]\label{eq:reprfunzionalemassa}
\end{equation}
where the single scale  contributions to the free energy $\tilde F^{(h)}_{\vartheta,m}$, as well the kernels of $\tilde S_{\vartheta,m}^{(h)}(J)$ associated to external fields attached to the edges of type $A,B,C$, satisfy the same bounds as in the case $m=0$, i.e. those of Proposition \ref{lem:bounds1} and Proposition \ref{lem:4biss}. In particular for $(2n,\ell)=(0,0),(0,1)$ these read
\begin{equation}\label{eq:Edim}
\begin{aligned}
&|\tilde F^{(h)}_{\vartheta,m}|\le C_\delta|\lambda|\,2^{(\frac32+\delta)h},&\qquad&|W^{(h)}_{0,1;B,\vartheta,m}(0)|\le C\,2^{\frac h2}&\qquad&\text{if }h\ge h_\epsilon-1,\\
&|\tilde F^{(h)}_{\vartheta,m}|\le C|\lambda|\,2^{2h-\frac{h_\epsilon}2},&&|W^{(h)}_{0,1;B,\vartheta,m}(0)|\le C\,2^{h-\frac{h_\epsilon}2}&&\text{if }h^*\leq h<h_\epsilon-1,
\end{aligned}
\end{equation}
for any $\delta \in [0,1/2)$ and suitable positive constants $C_\delta,C,c>0$, independent of $L,\vartheta,\epsilon,m,\lambda$.
\end{lemma}

\noindent\emph{Comparison with the case $m=0$.} The representation \eqref{eq:reprfunzionalemassa} is the analogue of the one obtained at $m=0$ by iterating \eqref{eq:2.39}. Besides the minor fact that notationally-wise here $\tilde F^{(h)}_{\vartheta,m}$ also includes the change of the Grassmann integration on scale $h$ (which at $m=0$ is accounted for by the constants $t^{(h)}$ in \eqref{eq:2.39}) the main difference is the infrared cutoff: recall that at $m=0$ and $\epsilon>0$, the integration proceeds down to the scale $h_L$ defined right before \eqref{eq:2.39};  
at $m>0$, it stops at $h^*$, which is independent of $L$: after having integrated the degrees of freedom on scale $h^*$, 
the effective propagator we are left with is non-singular and satisfies the same qualitative bounds as the single scale propagator on scale $h^*$, so we can compute the whole 
functional integral we are let with in a single integration step, whose output is the term $h=h^*$ in \eqref{eq:reprfunzionalemassa}; if $\epsilon<0$, the integration stops at $h^*=h_\epsilon-1$ as in the case $m=0$, see \eqref{eq:epsnegativo}. 

\medskip

Let us now deduce Proposition \ref{prop:rgmassa} from Lemma \ref{lem:ultimosperiamo}, whose proof is postponed to App.\ref{app:ultimolemma}.

\begin{proof}[Proof of Proposition \ref{prop:rgmassa}]\emph{Item (i).}  Since $\AA(0)=0$, $\tilde S^{(h)}_{\vartheta,m}(0)=0$ and $\partial_{A(0,B)}=\partial_{J_{0,B}}$ at $J=0$, \eqref{eq:reprfunzionalemassa} together with the definitions in \eqref{eq:sectors} give
\begin{equation}\label{eq:EX}
\frac{\Xi^{(\vartheta)}_{L,m}}{\Xi^{(\vartheta_0)}_{L,m}}=s_\vartheta\,e^{X^{(\vartheta)}_{L,m}},\qquad\rho^{(\vartheta)}_{L,m}=\sum_{h=h^*}^0W^{(h)}_{0,1;B,\vartheta,m}(0),
\end{equation}
where $X^{(\vartheta)}_{L,m}:=L^2\big(t^{(0)}_{\vartheta,m}-t^{(0)}_{\vartheta_0,m}\big)+L^2\sum_{h=h^*}^0\big(\tilde F^{(h)}_{\vartheta,m}-\tilde F^{(h)}_{\vartheta_0,m}\big)$ and we used $s_{\vartheta_0}=1$, cf. \eqref{eq:primaeqappE}. By proceeding as in the case $m=0$, cf. \cite[App. B]{GMT20}, 
using the Poisson summation formula (see Lemma \ref{lem:10Poisson} below), we find that $\tilde F^{(h)}_{\vartheta,m}-\tilde F^{(h);\io}_m$ and $W^{(h)}_{0,1;B,\vartheta,m}(0)-W^{(h);\io}_{0,1;B,m}(0)$ satisfy the same bounds as in \eqref{eq:Edim} times an additional factor $e^{-c\sqrt{2^hL}}$. Concerning the normalization factor, note that $\log\tilde{\mathcal E}_m$ is analytic and bounded in a complex strip of width $c\min\{|\eps|,1\}$ if $\eps<0$ or of width $cm\sqrt\eps$ if $\eps>0$ (see footnote \ref{footn:tildecalE}). Then the same argument used after \eqref{eq:blablablabla} for $D_\vartheta$,  gives $|t^{(0)}_{\vartheta,L,m}-t^{(0)}_{\io,m}|\le C_{\eps,m}e^{-c_{\eps,m}L}$ where    $t^{(0)}_{\io,m}:=\eps+\varepsilon_0(\l)+(2\pi)^{-2}\int_{R^{-1}[0,2\pi]^2}dq\,\log\tilde{\mathcal E}_m(q)$ with $C_{\eps,m},c_{\eps,m}>0$  independent of $L,\vartheta$, and of $m$ if $\eps<0$.
Comparing the quantities with labels $\vartheta$ and $\vartheta_0$ with their common limits, we find
\[
|X^{(\vartheta)}_{L,m}|\le2C_{\eps,m}L^2e^{-c_{\eps,m}L}+C|\l|L^2\sum_{h=h^*}^02^{\frac32h}e^{-c\sqrt{2^hL}}\le C'e^{-c'\sqrt L},\qquad L\ge L_0,
\]
where $C',c'>0$ depend on $\eps,m$, and only on $\eps$ if $\eps<0$, while $L_0$ depends on $h^*$. Here we used that the right sides of \eqref{eq:Edim} for $\tilde F^{(h)}_{\vartheta,m}$ are both bounded by $C|\l|2^{\frac32h}$, and that $L^22^{\frac32h}e^{-c\sqrt{2^hL}}=\sqrt L\,y_h^3e^{-cy_h}\le C\sqrt L\,e^{-cy_h/2}$, with $y_h:=\sqrt{2^hL}$; since $y_h=2^{(h-h^*)/2}y_{h^*}$ grows geometrically in $h$, the sum is dominated by its first term, $h=h^*$. Since $|e^X-1|\le2|X|$ for $|X|\le1$, the first identity in \eqref{eq:EX} gives the first bound in item (i). The second identity of \eqref{eq:EX} gives the second one of item (i) in the same way, since the right sides of \eqref{eq:Edim} for $W^{(h)}_{0,1;B,\vartheta,m}(0)$ are bounded by $C2^{h/2}$.

\emph{Item (ii).} By the second identity in \eqref{eq:EX} we have $
\lim_{L\to\io}\rho^{(\vartheta_0)}_{L,m}=\sum_{h=h^*}^0W^{(h);\io}_{0,1;B,m}(0)$, 
simply because $h^*$ is independent of $L$ and each term admits a limit. Note that the bounds \eqref{eq:Edim} hold for $W^{(h);\io}_{0,1;B,m}(0)$ as well, uniformly in $m$, and that $\lim_{m \to 0^+}W^{(h);\io}_{0,1;B,m}(0)$ exists and equals the corresponding kernel in \eqref{eq:2.8}; recalling that, for 
$\eps>0$, $h^*\to-\io$ as $m\to0^+$ (cf.  \eqref{eq:hstar}) while $h^*$ is independent of $m$ fo $\eps<0$, we find that, by dominated convergence, $\lim_{m\to0^+}\lim_{L\to\io}\rho^{(\vartheta_0)}_{L,m}$ is equal to the right side of \eqref{eq:2.8}, that is, to $\lim_{L\to\io}\rho^{(\vartheta_0)}_{L,0}$ (see Proposition \ref{prop:7}).
\end{proof}

\begin{subsection}{Proof of Lemma \ref{lem:ultimosperiamo}
}\label{app:ultimolemma} 
Since we are interested in differentiating the generating functional with respect to $A_e$ with $e$ of type $A,B,C$, without loss of generality we set $J_e=0$ if $e$ is an edge of type $m$. In the following we describe the 
modifications of the multiscale analysis due to the presence of the additional contributions to the kernels and propagators coming from the mass term $m$, in 
particular those induced by the anomalous terms in \eqref{eq:Eprop}. The claim of the lemma follows 
from the convergence of the multiscale expansion, 
uniformly in $L$ for any $\eps,m\neq0$, and from the observation that, by construction, it turns out that the limit as $m\to 0^+$ of the infinite volume contributions to the 
generating function, as produced by the GN tree expansion, 
are the same as those obtained via the multiscale construction described in the bulk of the paper, i.e., 
the one for $\vartheta_0\in\{(1,0),(0,1)\}$ and $m=0$, 
in the limit $L\to\infty$. Many ideas are similar to those
discussed in greater detail in \cite{GMT17}, where a mass term 
was present, too, and its effect under iterations of the multiscale expansion was traced systematically; 
therefore, here
we limit ourselves to discuss the main structure of the new expansion and the main differences with respect to the case $\vartheta=\vartheta_0$ and $m=0$, without belaboring the fine details, which are left to the reader. 

\medskip \emph{Charge and symmetries.} We call \emph{charge} of a Grassmann monomial the number of its $\varphi^+$ fields minus that of its $\f^-$ fields. The action $\nu N+\tilde V_{\eps,\vartheta}(\f,\AA(J))$ has zero charge while the anomalous terms of $\tilde S_{\eps,\vartheta,m}$ have charge $\pm2.$ Since the normal entries of the propagator preserve the charge while the anomalous ones change it by $\pm2$, the effective potentials of scale $h\le 0$ contain 
terms of any even charge $Q$. Note that, for $m>0$, 
in addition to being symmetric under the action of 
$\SS_1,\SS_2$ in \eqref{eq:S1S2}, the effective potential 
is invariant also under the following two transformations:
\begin{equation}
    \SS_3: \hat\f^\pm_q \to \pm i \hat \f^\pm_q, \quad m \to -m; \qquad \quad  \SS_4: \hat \f^+_q \to \hat \f^-_q, \quad \hat \f^-_q \to -\hat \f^+_q,\quad  \AA \to \SS_4 \AA
\end{equation}
where $(S_4\AA)_e=A_{S_4e}$ and $S_4e$ is the inversion of the edge $e$, i.e., with the notation of \eqref{eq:S1S2psiA}, $\SS_4(x,s)=(-x-\hat e_s,s)$, $s=A,B,C$. The kernels of the effective potential at all scales inherit these symmetries; in particular, using the invariance under $\SS_3$, we find that the contributions 
of charge $Q$ are even or odd in $m$, depending on whether $Q/2$ is even or odd. The terms of vanishing charge and independent of $m$ have the same structure and scaling dimensions as those at $m=0$ discussed in the bulk of the paper. On the other hand, the terms that are linear, quadratic, or higher order in $m$, are better behaved; every 
extra power of $m$ corresponds to a gain of $1/2$ or $1$
in the scaling dimension of the kernel, depending on 
whether we are considering the first or second regime. 
This motivates
a modified localization procedure where the terms that 
are dimensionally relevant or marginal are further expanded in Taylor series in $m$; the terms of order 0 or 1 or $2$ in $m$ may contribute to the relevant or marginal coupling constants (in different ways in the first and second regimes), while the remainders of higher order in $m$ are irrelevant in a RG sense. As already anticipated above, we proceed in a way similar to \cite[Sects.6.1.2 and 6.1.4]{GMT17}, which we refer to for additional details. Here we limit ourselves to discuss the main modifications in the localization procedure, in the structure of the local 
terms, and in the control of the flow of the new running 
coupling constants, separately for the first and second 
regimes.   

\medskip\emph{First regime.} In the scale range $h \geq h_\eps,$ the modified localization procedure based on further expanding the relevant and marginal terms in 
Taylor series in $m$ (and/or in its dressed counterparts), in view 
also of the symmetries of the model, produces two extra marginal terms, 
independent of $J$ and 
quadratic in $\f$, which are re-absorbed step by step into the reference 
Gaussian Grassmann measure. After the integration of the scales greater than $h$, the dressed propagator has the same structure of \eqref{eq:Eprop} with $M^{-1}(q)$ replaced by $\chi^{(1)}_h(q)M^{-1}_h(q)$ with $\chi_h^{(1)}$ as in \eqref{eq:2.26bis} and
\begin{equation}
    M_h(q):=\begin{pmatrix}
        \tilde \mu_h(q)+\sigma_h & im_h q_2+i\rho_1(q_2) \\ -i m_hq_2-i\rho_1(q_2) & -\overline{\tilde \mu_h(q)}-\sigma_h
    \end{pmatrix},
    \nn
\end{equation}
where $\tilde \mu_h(q):=r_h-iq_1(1+\z_h)+2^{-1}q_2^2(1+\a_h)+\rho(q)$ being the same denominator in \eqref{eq:2.26bis}, so that $\tilde \mu_0:=\tilde \mu$, while $m_h,\sigma_h$ are new running coupling constants such that $m_0:=4\tilde m, \sigma_0=0$ and $\rho_1(q_2)=2\tilde m (\sin (2q_2)-2q_2)$ is a remainder. Note that all the running coupling constants but the new `anomalous' ones, denoted by $m_h$ and $\sigma_h$
are exactly the same as those computed in the bulk of the paper, because
they are defined by setting $m=0$ and sending $L\to\infty$, as we did in 
Section \ref{sec:multiscale}. On the other hand, $m_h$ and $\sigma_h$ are obtained by expanding at linear and quadratic order in $m$ the local parts of the kernels quadratic in $\f$,
of charge $\pm2$ and $0$, and they satisfy the following recursion relations: 
\begin{equation}\label{eq:Eflow}
\begin{split}
&m_h=m_{h+1}+2i\sum_{k=h+1}^0m_k\partial_{m_k}\partial_{q_2}\hat W^{(h+1);\io;(+,+)}_{2,0;0,0}(0),\\
&\sigma_h=\sigma_{h+1}-\frac12\sum_{k,k'=h+1}^0m_km_{k'}\,\partial_{m_k}\partial_{m_{k'}}\hat W^{(h+1);\io;(+,-)}_{2,0;0,0}(0)-\sum_{k=h+1}^0\sigma_k\,\partial_{\sigma_k}\hat W^{(h+1);\io;(+,-)}_{2,0;0,0}(0),
\end{split}
\end{equation}
where the derivatives computed at $\eps=m=0$ i.e. $r_k=m_k=\sigma_k=0$ for all $k \geq h+1$. The structure of the flow of the anomalous constants is due to symmetry constraints and to the different scaling of the entries of $m_h:$ recall that on the support of $f^{(1)}_h(q)$, $|q_1|\leq C 2^h$ and $|q_2|\leq C2^{h/2}$, so that the anomalous entries of $M^{-1}_h$ are smaller than the normal ones by a factor $C(\tilde m+m_h)2^{-h/2}$. Hence, as already anticipated above, each $m_k$ improves the scaling dimension by $1/2$, and each $\sigma_k$ by $1$.\footnote{A derivative $\partial_{\sigma_k}$ acting on a single scale propagator on scale $k$ produces an extra dimensional factor $2^{-k}$, exactly as for the case of $r_k$, see the discussion after \eqref{eq:4.150bis}, while $\partial_{m_k}$ gives a factor $2^{-k/2}$. Note also that the local parts $q_2\partial_{q_2} \hat W^{(h+1);\io;(+,+)}_{2,0;\eps,m}(0)$ and $\hat W^{(h+1);\io;(+,-)}_{2,0;\eps,m}(0)$ have scaling dimension $1/2$ and $1$, respectively; thanks to the $\SS_3$ symmetry, the first term is odd and the second is even in the anomalous constants: hence it is sufficient to expand the first term at first order in $\{m_k\}_{k>h}$, and the second at second order in $\{m_k\}_{k>h}$ and at first order in $\{\sigma_k\}_{k>h},$ both at $\eps=0:$ remainders in the Taylor series expansion in $\eps$ and/or in $m$ are dimensionally irrelevant.} The 
flow equation for $m_h$ is decaoupled from that of $\sigma_h$: therefore, we first solve the first equation in \eqref{eq:Eflow}, via the same strategy used for $r_h$ in Proposition \ref{lem:flowRC}; next, we plug the solution for $m_h$ in the second of \eqref{eq:Eflow}, and we solve it;  thanks to the short memory property, for any $\theta \in [0,1/2)$ and $h_\eps-1 \leq h\leq -1$ we find
\begin{equation}
    |m_h-m_{h+1}|\leq C(\theta) 2^{\theta h} |\l|\tilde m, \qquad |\sigma_h-\sigma_{h+1}|\leq C(\theta)2^{\theta h}|\l| \tilde m^2,
\end{equation}
so that, on all the scales, $|m_h-m_0|\le C|\l|\tilde m$ and $|\sigma_h|\le C|\l|\tilde m^2$. Since $2^{h}\ge2^{h_\eps-1}\asymp|\eps|$, this implies $(\tilde m+|m_h|)2^{-h/2}\le Cm|\eps|^{-1/2}\le Cc_0$ and $|\sigma_h|2^{-h}\le C|\l|m^2|\eps|^{-1}\le C|\l|c_0^2$: on the support of $f^{(1)}_h$, where $|\tilde\mu_h(q)|\ge c2^h$, $\sigma_h$ is a small, subleading, correction to the normal entries, and the anomalous entries are smaller than the normal ones by $Cm2^{-h/2}$. Hence, the single scale propagator satisfies the same dimensional bounds as the one in Section \ref{sec:multiscale}; moreover, it still admits a Gram representation analogous to the one in Lemma \ref{lem:propag1}; therefore, by adapting the tree expansion of App.\ref{app:A} to the presence of the anomalous terms, we obtain from it the dimensional bounds in the first line of  \eqref{eq:Edim}. If $\eps<0$, the integration procedure ends by integrating the scales $\leq h_\eps$ in a single step, as in the case $m=0$, and there is nothing else to compute or estimate. If $\eps>0$, instead, 
we integrate the remaining infrared degrees via a multiscale procedure analogous to the one described in Section \ref{sec:secondregime}, and briefly discussed next. 

\medskip\emph{Second regime.} 
In analogy with what we did in Section \ref{sec:multiscale}, 
see \eqref{eq:2.34} and following discussion, once we reach scale $h_\eps$ we first move the local part of $V^{(h_\eps)}$ into the Gaussian integration, thus defining $M_{h_\eps-1}$ in the same way as we did in thefirst regime, and then we add and subtract the counterterm $\eps\mu_{h_\eps}N(\f)$, i.e., we replace $\tilde\mu_{h_\eps-1}$ by $\tilde\mu_{h_\eps-1}+\eps\mu_{h_\eps}$ in $M_{h_\eps-1}$: the propagator on scales $\le h_\eps$ is thus equal to $\chi^{(1)}_{h_\eps}(q)M_{h_\eps-1}(q)^{-1}$. Then we decompose the field around the singularities $\pm q^\eps$, which are the same as at $m=0$, see \eqref{eq:2.55zz}-\eqref{eq:4.91}. Since the anomalous entries of $M_{h_\eps-1}$ do not vanish at $\pm q^\eps$, they couple the quasi-particle fields with opposite chiralities: in the basis $(\hat\f^-_{q,+},\hat\f^+_{-q,-})$, with $\hat\f^\pm_{q,\omega}$ the Fourier transform of the quasi-particle fields in \eqref{eq:qp}, the Gaussian integration on scales $\le h_\eps-1$ has propagator $\chi^{(2)}_{h_\eps-1}(q)  M^{-1}_{h_\eps-1}(q+\tilde q_\eps)/Z_{h_\eps-1}$, cf. \eqref{eq:2.55zz}-\eqref{eq:2.37}, where
\begin{equation}
    \tilde M_{h_\eps-1}(q+\tilde q_\eps)=\begin{pmatrix}
        D_+(q)+(\rho_+(q)+\sigma_{h_\eps-1})/Z_{h_\eps-1} & i(\Delta_{h_\eps-1}+\eta(q)/Z_{h_\eps-1}) \\ -i(\Delta_{h_\eps-1}+\eta(q)/Z_{h_\eps-1})  & -D_-(-q)-(\rho_-(-q)+\sigma_{h_\eps-1})/Z_{h_\eps-1}
    \end{pmatrix},
\label{eq:D.10}\end{equation}
where $\Delta_{h_\eps-1}$ and $\eta$ are defined by the conditions  $Z_{h_\eps-1}\Delta_{h_\eps-1}+\eta(q):=m_{h_\eps-1}(\tilde q_\eps+q_2)+\rho_1(q_2+\tilde q_\eps)$ and $\eta(0)=0,$ so that $\Delta_{h_\eps-1}\asymp m\sqrt{\eps}$ and $\eta(q)=O(mq_2).$ Eq.\eqref{eq:D.10} has the same structure as \cite[(6.3)-(6.4)]{GMT17},
modulo the presence of the terms $\rho_\pm,\sigma_{h_\eps-1},\eta$, which are subleading corrections to the dominant terms, namely to $D_\pm(q)$ (which plays the same role as $-i\sin q_1\pm \sin q_2$ in \cite[(6.4)]{GMT17})
and to $\Delta_{h_\eps-1}$ (which plays the same role as the mass $m_h$ in 
\cite[(6.4)]{GMT17}).  The scales $h<h_\eps-1$ are integrated as in \cite[Sec. 6]{GMT17}, which we refer to for more details, the symmetries $\SS_1$ to $\SS_4$ playing the same role as those in \cite[Claim 1]{GMT17}. In particular, the local terms of charge $0$ are the same as those at $m=0$, and the only new non-irrelevant term is marginal and
given by $\hat W^{(h);\io;(+,+)}_{2,0;(+,-);m}(0)$, the local part of the kernel of $\hat \f^+_{q,+}\hat \f^{+}_{-q,-}$. After the integration of the scales larger than $h$, the propagator is $\chi_h^{(2)}(q) M_{h}^{-1}(q)/Z_h$ where $M_h(q)$ is obtained from the previous equation by replacing $Z_{h_\eps-1}$ and $\Delta_{h_\eps-1}$ with $Z_h$ (cf. \eqref{eq:2.39}-\eqref{eq:2.43biss}) and $\Delta_h$, respectively, the latter playing the role of the mass $m_h(0)$ of \cite{GMT17}. The discussion for $\Delta_h$  is the same as the one for $Z_h$, cf. Lemma \ref{YZ-lemma}, and one finds that for any $\theta \in [0,1/2)$ \begin{equation}
    \Delta_{h-1}=2^{\eta_\Delta|h-h_\eps|}\Delta_{h_\eps-1}(1+O_\theta(|\l|\eps^\theta), \qquad |\eta_\Delta|\leq C |\l|\eps^{\theta}.
\end{equation}
The multiscale integration continues in the same fashion as long as the anomalous entries of $M_h$, computed at momenta in the support of $f^{(2)}(q)$, are smaller than the diagonal ones, that is, until 
$|\Delta_h|<2^h$; therefore, it stops at the scale $h^*$ such that
\begin{equation}
    h^*:=\sup\{h<h_\eps-1: |\Delta_h|>2^h\}.
\end{equation}
On the support of $\chi_{h^*}^{(2)}(q)$ one has $|\det M_{h^*}|\geq c 2^{2h^*}$, uniformly in $\vartheta,L$ so that the scales below $h^*$ are integrated in a single step as in \cite[Sec. 6.1.3]{GMT17}. Moreover, since $\Delta_{h_\eps-1} \asymp m\sqrt{\eps}$ and $2^{h_\eps}\asymp \eps$, from the definition of $h^*$ one finds $2^{h^*-h_\eps} \asymp (m/\sqrt{\eps})^{1/(1+\eta_\Delta)}$, which implies \eqref{eq:hstar}.
In conclusion, for all the scales $h^*<h<h_\eps-1$, the multiscale construction has the same structure as the one discussed in Section \ref{sec:secondregime}, modulo the presence of the modified propagator $f^{(2)}_h(q)M^{-1}_h(q)/Z_h$, which still admits a Gram representation; its normal entries satisfy the same bounds stated in Lemma \ref{lem:3bis}, while its anomalous ones are smaller by a factor $\Delta_h2^{-h}$. Therefore, the tree expansion of App. \ref{app:A} adapted to this setting (similarly to what was done in \cite[Sec. 6]{GMT17}) implies the bounds of Proposition \ref{lem:4biss} stated in the second line of \eqref{eq:Edim} for $h^*\leq h<h_\eps-1$.
\end{subsection}

\section{The last few more technicalities}

In this appendix we prove a few additional technical results about the model, both in its
non-interacting and in its interacting version, which are used throughout the paper.

Appendix \ref{A:6.3} concerns the non-interacting model in the frozen phase: in particular, we show that in the frozen phase the inverse Kasteleyn matrix $K^{-1}_\vartheta$ differs from its infinite volume counterpart $K^{-1}$ by an
exponentially small error in $L$ (Lemma \ref{lem:13}); we also compute $K^{-1}$
explicitly for $A\ge B+C$ (Lemma \ref{lem:14}). These results are used in the proof of Lemma
\ref{lem:3.5.0.1}, hence in the proof that the density vanishes for $\eps<0$ (Section
\ref{sec:2.3.1}), and in the first-order computations of Lemma \ref{lem:firstordercalc}.

Appendix \ref{app:interacting} concerns the interacting model in the second regime: Lemma
\ref{YZ-lemma} gives the anomalous scaling of $Z_h$ and $Y_{h;s,\un\o}$ announced in
\eqref{eq:2.71bis}, the identity $\eta=\eta_{(+,+)}$ and the bounds \eqref{eq:eta1eta}
(used in Section \ref{sec:2.3.1} for \eqref{eq:2.21} and \eqref{eq:Rbounds}, and in
Appendix \ref{app:A3} for \eqref{eq:decomptheta}), together with the asymptotics of
$v_\eps$, $\widetilde Z_{h-1}(q)/Z_{h-1}$ and $\rho_\o(q)$ on the support of $f^{(2)}_h$
(used in \eqref{eq:Rbounds} and in the proof of Lemma \ref{lem:derivateepsilon}).
\subsection{Frozen phase for the integrable model}\label{A:6.3}
We start by recalling two standard results. 
\begin{lemma}\label{lem:11}
Let $f:\mathbb{C}^n \longrightarrow \mathbb{C}$ be analytic in each argument in the strip $S_\eta:=\{z: |\Im z| \leq \eta, \Re z \in [0,2\pi) \}$ and be $2\pi$-periodic in the real direction in each of its arguments. Then,
\begin{equation}
	\Big |\frac{1}{(2\pi)^n}\int_{[0,2\pi)^n} d\phi_1\dots d\phi_n f(\phi)e^{-ix\cdot\phi}\Big | \leq e^{-\eta |x|_1}\sup_{\phi\in S_\eta\times\dots\times S_\eta}|f(\phi)|
\end{equation}
with $x=(x_1,\dots,x_n)\in \mathbb R^n$ and $|x|_1=|x_1|+\dots+|x_n|$.
\end{lemma}

\begin{lemma}[Poisson summation]\label{lem:10Poisson}
	If $\hat F:[0,2\pi)^2 \to \mathbb{C}$ is $C^\infty$, then
	\begin{equation}
	 L^{-2} \sum_{k \in \mathcal D_\vartheta} \hat F(k)=\sum_{n \in \mathbb{Z}^2} F(n_1 L,n_2 L)(-1)^{\vartheta_1 n_1+\vartheta_2 n_2} \label{eq:poissonsumm}
	\end{equation}
	where $F(x)=\frac{1}{(2\pi)^2}\int_{[0,2\pi)^2}dke^{-ik\cdot x} \hat F(k)$ and $\DD_\vartheta$ is the same as in \eqref{fourierfieldsk}.
\end{lemma}

If $A=1$, $B=C=e^\epsilon/2$ with $\epsilon<0$, we have that $f(z_1,z_2):=\mu^{-1}(z_1,z_2)=(1+\frac{e^\epsilon}{2}(e^{iz_1}+e^{iz_2}))^{-1}$
is analytic in each of its arguments in the strip $S_\frac{\epsilon}{2}$, thus by Lemma \ref{lem:11} we have  $|K_\eps^{-1}(x)|\leq e^{-\frac{\epsilon}{2}|x|_1}\sup \{|f(z)|: z\in S_\frac{\varepsilon}{2}\times S_\frac{\varepsilon}{2}\}$  where $|x|_1=|x_1|+|x_2|$ and $K_\eps^{-1}$ stands for $K^{-1}$ when $A=1,B=C=e^\eps/2$, with \begin{equation}\label{eq:6.12}
	\begin{split}
	& K^{-1}(x):=\int_{[0,2\pi)^2} \frac{dk}{(2\pi)^2} \frac{e^{-ikx}}{A+Be^{ik_1}+Ce^{ik_2}}.
	\end{split}
\end{equation}
Let $K^{-1}_{\vartheta}$ be obtained from the RHS of  \eqref{eq:6.12} by replacing $\int_{[-\pi,\pi]^2}dk$ with $L^{-2} \sum_{k \in \DD_\vartheta}$, and let $K^{-1}_{\vartheta,\eps}$ be equal to $K^{-1}_\vartheta$ computed at $A=1,B=C=e^\eps/2$.  We have the following.
\begin{lemma}\label{lem:13} Let $\eps<0$. Then there exist $c(\eps),C(\eps)>0$ such that 
$|K_\eps^{-1}(x)|\leq c(\eps) e^{-|x|_1/c(\eps)}$ for
 every $x\in \mathbb Z^2$, and 
\begin{equation}
	\sup_{\substack{x,y \in \Lambda:\\ |x_1-y_1|,|x_2-y_2|\le L/2}} |K^{-1}_{\vartheta,\eps}(x,y)-K_\eps^{-1}(x-y)|\leq C(\eps)e^{- L/C(\eps)}.
\end{equation}
\end{lemma}

\begin{proof} By definition, $K^{-1}_{\vartheta,\eps}(x,y)=L^{-2}\sum_{k\in\mathcal D_\vartheta} \hat F_{x-y}(k_1,k_2)$, where 
$\hat F_x(k_1,k_2):=\frac{e^{-ikx}}{1+\frac{e^\varepsilon}{2}(e^{ik_1}+e^{ik_2})}$. Using Poisson summation formula and the definition of $K_\eps^{-1}$ 
(see \eqref{eq:6.12} with $A=1$ and $B=C=e^\eps/2$), we find: 
	\begin{equation}\label{eq:sumexpsmallPoisson}
		K_{\vartheta,\eps}^{-1}(x,y)-K^{-1}(x-y)=\sum_{n \in \mathbb{Z}^2\setminus\{(0,0)\}} K_\eps^{-1}(x-y+nL)(-1)^{n_1\vartheta_1+n_2\vartheta_2}.
	\end{equation}
Recalling that, for $\eps<0$, $K_\eps^{-1}(x)$ decays exponentially in $|x|_1$ (see the discussion right after the statement of Lemma \ref{lem:10Poisson}), and noting that, if $|x_1-y_1|,|x_2-y_2|\le L/2$, then $|x-y+nL|_1\ge L/2$, we easily find that the sum 
in the right hand side of \eqref{eq:sumexpsmallPoisson} is exponentially small in $L$, as desired. 
\end{proof}

In the frozen phase, the inverse Kasteleyn operator admits a simple explicit expression, summarized in the following lemma. 

\begin{lemma}\label{lem:14}
Let $A,B,C$ such that $A\geq B+C$, then the infinite volume inverse Kasteleyn operator $K^{-1}$ in \eqref{eq:6.12} 
takes the following explicit form: 
\begin{equation}\label{eq:A.16}
K^{-1}(x)=\mathbbm{1}_{\{x_1,x_2 \geq 0\}} \frac{(-1)^{x_1+x_2}}{A} \left (\frac{B}{A}\right )^{x_1}\left (\frac{C}{A}\right )^{x_2}\binom{x_1+x_2}{x_1}.
    \end{equation}
    In particular $
|K^{-1}(x)|\le\mathbbm{1}_{\{x_1,x_2 \geq 0\}} \frac1A \left(\frac{B+C}{A}\right)^{x_1+x_2}$ and $\partial_AK^{-1}(x)=-A^{-1}(x_1+x_2+1)K^{-1}(x).$
\end{lemma}
\begin{proof}
Consider $A>B+C$ (the case $A=B+C$ follows from the continuity of $K^{-1}$ in the edge weights). We can write: 
\begin{equation}
    K^{-1}(x)=\frac{1}{(2\pi i)^2}\int_{|z|=1}dz \int_{|w|=1}dw\frac{z^{-x_1-1}w^{-x_2-1}}{A+Bz+Cw}.
\end{equation}
We perform the integral over $z$ first. Since $A>B+C$, the only pole inside the circle $|z|=1$ is at $z=0$ (if $x_1\ge0$, otherwise there is no pole inside and the integral is zero). The residue theorem then gives
    \begin{equation}\label{eq:D.8}
K^{-1}(x)= \mathbbm 1_{x_1\ge 0} (-B)^{x_1}\frac1{2\pi i}\int_{|w|=1}dw \frac{w^{-x_2-1}}{(A+C w)^{x_1+1}}.
    \end{equation}
    The integral over $w$ is performed similarly: for $x_2<0$ the integral in \eqref{eq:D.8} is zero because the integrand has no pole inside $|w|=1$, and for $x_2\ge0$ the residue theorem leads to \eqref{eq:A.16}.
 As for the upper bound on $|K^{-1}(x)|$ stated in the line after \eqref{eq:A.16}, it is enough to note that
    $ \Big(\frac{B}{B+C}\Big)^{x_1}\left (\frac{C}{B+C}\right )^{x_2}\binom{x_1+x_2}{x_1}\le1$.
\end{proof}

\subsection{The interacting model}\label{app:interacting}

\begin{lemma}\label{YZ-lemma}
Let $\l_0>0$ be fixed as in Theorem \ref{th:5}. For $|\l|\leq\l_0$ and
$0<\eps\leq \bar \eps$ with $\bar \eps$ sufficiently small, the following statements hold uniformly
in $h<h_\eps$.  
\begin{enumerate}
   \item There exist
$\eta=\eta(\l,\eps)$
$\eta_{\un\o}=\eta_{\un\o}(\l,\eps)$ analytic in $\l$ for fixed
$\eps$ and depending on $(\l,\eps)$ only through
$\l_{-\io}(\l,\eps)$, such that, for every
$\theta\in[0,1/2)$,
\begin{equation}
\begin{split}
    Z_{h-1}
    &=2^{-\eta(h-h_\eps)}Z_{h_\eps-1}
      \bigl(1+O_\theta(|\l|^2\eps^\theta)\bigr),\\
    Y_{h-1;s,\un\o}
    &=2^{-\eta_{\un\o}(h-h_\eps)}
      Y_{h_\eps-1;s,\un\o}
      \bigl(1+O_\theta(|\l|\eps^\theta)\bigr),
    \label{eq:A.10}
\end{split}
\end{equation} with $Z_{h_\eps-1},Y_{h_\eps-1;s,\un\o}$ as in \eqref{eq:2.55zz},\eqref{eq:2.540}.
The exponents $\eta_{\un \o}$ depend on $\un\o$ only through the product
$\o_1\o_2$ and satisfy $\eta_{\un\o}=\eta_{-\un\o}$; 
in addition, $\eta=\eta_{(+,+)}$
and,  setting $\eta_1:=\eta_{(+,-)}$, the following bounds hold:
\begin{equation}
    0\leq\eta\leq C_\theta|\l|^2\eps^{2\theta},
    \qquad
    |\eta_1-\eta|\leq C_\theta|\l|\eps^\theta.
    \label{eq:eta1eta}
\end{equation}
Recalling the definitions of $Y^{(j)},\xi_j$, for $j=1,2$, 
in \eqref{eq:2.63bis} and \eqref{eq:Gheps}, we can further rewrite \eqref{eq:A.10} as
\begin{equation}
\begin{split}
    Z_{h-1}
    &=\xi_1\,2^{-\eta(h-h_\eps)}
      \bigl(1+O(\eps)+O_\theta(|\l|\eps^\theta)\bigr),\\
    Y^{(1)}_{h-1;B}Z^{-1}_{h-1}
    &=\frac{Y_{-\io;0,B}}{\xi_1}
      \bigl(1+O(\eps)+O_\theta(|\l|\eps^\theta)\bigr),\\
    Y^{(2)}_{h-1;B}Z^{-1}_{h-1}
    &=2^{-(\eta_1-\eta)(h-h_\eps)}
      \frac{Y_{-\io;0,B}}{\xi_1}
      \bigl(1+O(\eps)+O_\theta(|\l|\eps^\theta)\bigr).
    \label{eq:YZ-ratios}
\end{split}
\end{equation}

    \item For every $\theta\in[0,1/2)$, the effective velocity
    satisfies $
        v_\eps=\xi_1^{-1}\xi_2\widetilde q_\eps
        \bigl(1+O(\eps)+O_\theta(|\l|\eps^\theta)\bigr).$ Moreover, for 
    $q\in\operatorname{supp}f^{(2)}_h$, we have: \be \left|Z^{-1}_{h-1}\widetilde Z_{h-1}(q)-1\right|
        \leq C_\theta|\l|^2\eps^\theta, \qquad \quad   \rho_\o(q)-\xi_2\frac{q_2^2}{2}
        =O(2^{2h})
        +O_\theta\bigl(|\l|2^{2h}\eps^{-1+\theta}\bigr).\label{eq:item3YZ}\ee 
\end{enumerate}
\end{lemma}

\begin{proof}
\textbf{Item 1.}
By definition,
$Z_{h-1}=Z_h(1+z_h)=Z_{h_\eps-1}\prod_{k=h}^{h_\eps-1}(1+z_k)$.
Since $\{z_k\}_{k<h_\eps}$ is a Cauchy sequence (see Proposition \ref{lem:kernbounds2}), we let
$z_{-\io}:=\lim_{k\to-\io}z_k$ and
$\eta:=\log_2(1+z_{-\io})$, so that
\begin{equation}
\begin{split}
    Z_{h-1}=Z_{h_\eps-1}2^{-\eta(h-h_\eps)}A_h,
    \qquad
    A_h:=\prod_{k=h}^{h_\eps-1}
    \left(1+\frac{z_k-z_{-\io}}{1+z_{-\io}}\right).
    \label{eq:YZ-Ah}
\end{split}
\end{equation}
Using $|z_k-z_{-\io}|\leq C_\theta|\l|^2 2^{\theta k}$
(cf. Proposition \ref{lem:kernbounds2}), together with
$2^{h_\eps}\asymp\eps$, we find:
\begin{equation}
    A_h=1+O_\theta(|\l|^2\eps^\theta),
    \qquad h<h_\eps.
    \label{eq:YZ-Ah-bound}
\end{equation}
The fact that $z_{-\io}$, and thus $\eta$, depends only on
$\l_{-\io}(\l,\eps)$ follows from \cite[Remark 13]{GMT20}. This proves the first of  \eqref{eq:A.10}.

Similarly, defining
$y_{k;s,\un\o}:=Y_{k-1;s,\un\o}/Y_{k;s,\un\o}-1$, the strategy used in
\cite[Sec. 6.4.9]{GMT20} gives the Cauchy property $|y_{k;s,\un \o}-y_{k-1;s,\un\o}|\leq C |\l| 2^{\theta k}$ and thus the existence of
$y_{-\io;s,\un\o}:=\lim_{k\to-\io}y_{k;s,\un\o}$, which depends only
on $\l_{-\io}(\l,\eps)$ and on $\o_1\o_2$. Repeating the same
argument leading to \eqref{eq:YZ-Ah}, setting
$\eta_{\un\o}:=\log_2(1+y_{-\io;s,\un\o})$ and using  $|y_{k;s,\un\o}-y_{-\io;s,\un\o}|
    \leq C_\theta|\l|2^{\theta k}$, we obtain the second identity in \eqref{eq:A.10}. The symmetries below
\eqref{eq:2.43bis} imply $\eta_{\un\o}=\eta_{-\un\o}$, while the
identity $\eta=\eta_{(+,+)}$ between the critical exponents of $Z_h$ and $Y_h^{(1)}$
(recall \eqref{eq:2.63bis}) is a very non-trivial fact, proved in 
\cite[Th. 3]{BM02}.

Using the GN tree expansions for $z_{-\infty}$ and $y_{-\infty;s,\underline\omega}$ (that is, the limiting beta functions for $Z_h$ and $Y_{h;s,\underline\omega}$ in the infrared limit $h\to-\infty$), it follows that these quantities, and, therefore, 
the critical exponents $\eta$ and $\eta_{\underline\omega}$, are real analytic functions of $\lambda_{-\infty}$, i.e., they are convergent power 
series in this parameter; see
\cite[Secs. 6.4.6-6.4.9]{GMT20}.
By direct inspection and the short memory property, it follows that the limiting beta functions $z_{-\infty}$ and $y_{-\infty;s,\underline\omega}$ can be expressed as sums of limiting tree values, where all the propagators are replaced by relativistic propagators of the following form: 
\[
    \widehat g^{(h)}_{R,\o}(q)
    =\frac{\widetilde f_h^{(2)}(q)}{-iq_1+\o v_\eps q_2},
    \qquad
    \widetilde f_h^{(2)}(q)
    :=f_h^{(2)}(q)\frac{Z_{h-1}}{\widetilde Z_{h-1}(q)}.
\]
Recall that the velocity satisfies $v_\eps\asymp\sqrt\eps$. By rescaling $q_2\to q'_2=v_\eps q_2$, we can reduce to the case of 
unit velocity, as in \cite{GMT20}. Under this rescaling, each quartic endpoint $\l_j$ is replaced by $\l_j/v_\eps$. Therefore, the
limiting beta functions of $z_h$ and
$y_{h;s,\un\o}$ (that, in order to make contact with the notation of \cite{BM36}, we also denote by 
$\overline\beta_{\sharp,v_\eps}$, with $\sharp=z,y_{\underline\omega}$), satisfy:
\begin{equation}
    \overline\beta_{\sharp,v_\eps}
       (\un\l)
    =\overline\beta_{\sharp,1}
       (\un\l/v_\eps),
    \qquad \sharp=z,y,
    \label{eq:YZ-beta-velocity-rescaling}
\end{equation}
where the dependence upon the sequence of ratios of wave-function renormalizations is
left implicit. This is the scaling property proved in
\cite[Sec. 3.1, Eq. (152), and Eqs. (156)-(158)]{BM36}. Therefore, in our case the critical exponents $\eta$, $\eta_1$ 
are the same as those of \cite{GMT20}, up to the replacement of $\lambda_{-\infty}$ by 
$\l_{-\io}/v_\eps$; in particular, a lowest order computation shows that
\begin{equation}
\begin{split}
    \eta
    &=a\left(\frac{\l_{-\io}}{v_\eps}\right)^2
      +O\left(\left|\frac{\l_{-\io}}{v_\eps}\right|^3\right),
      \qquad
    \eta_1=b\frac{\l_{-\io}}{v_\eps}
      +O\left(\left|\frac{\l_{-\io}}{v_\eps}\right|^2\right),
\end{split}
\label{eq:YZ-exponent-expansions}
\end{equation}
for $a>0$ and $b \neq 0$. 
Since $v_\eps\asymp\sqrt\eps$ and
$|\l_{-\io}|\leq C_\theta|\l|\eps^{1/2+\theta}$ by
\eqref{eq:2.70biss}, we find $|\l_{-\io}/v_\eps|
    \leq C_\theta|\l|\eps^\theta$ that, if plugged in \eqref{eq:YZ-exponent-expansions}, implies
\eqref{eq:eta1eta}.

We are left with \eqref{eq:YZ-ratios}, which follows by expanding the constants $Z_{h_\eps-1},Y_{h_\eps-1;s,\un\o}$ in \eqref{eq:A.10}.
Note that $Y_{h_\eps-1;B,\un\o}$ is obtained matching its value with the outcome of the first regime of integration, as follows: 
\begin{equation}
    Y_{h_\eps-1;B,\un\o}
    =Y_{h_\eps-1;\eps,B}+\BB^Y_{h_\eps-1;\eps,B},
    \qquad
    |\BB^Y_{h_\eps-1;\eps,B}|
    \leq C_\theta|\l|\eps^\theta,
    \label{eq:YZ-Y-transition}
\end{equation}
where the bound on $\BB^Y_{h_\eps-1;\eps,B}$ follows from \eqref{eq:A.24}.  Moreover, $Y_{0;\eps,B}=1/2$
is independent of $\eps$; hence, by the flow equation in the first-regime  and
the definition of $\mathfrak d_\eps$:
\begin{equation}
    Y_{h-1;\eps,B}
    =Y_{-\io;0,B}
     -\sum_{\ell<h}\BB^Y_{\ell;0,B}
     +\sum_{\ell=h}^{0}
       \mathfrak d_\eps\BB^Y_{\ell;\mathord\cdot,B}.
    \label{eq:YZ-Y-first-regime}
\end{equation}
The bounds \eqref{eq:2.29bis} and the second of
\eqref{eq:2.73bis}, evaluated at $h=h_\eps$, give
    $\left|\sum_{\ell<h_\eps}\BB^Y_{\ell;0,B}\right|
    \leq C_\theta|\l|2^{\theta h_\eps}$, $\left|\sum_{\ell=h_\eps}^{0}
      \mathfrak d_\eps\BB^Y_{\ell;\mathord\cdot,B}\right|
    \leq C_\theta|\l|\eps
      \sum_{\ell=h_\eps}^{0}2^{(\theta-1)\ell}
    \leq C_\theta|\l|\eps^\theta.$
Since $2^{h_\eps}\asymp\eps$ and
$Y_{-\io;0,B}=1/2+O(\l)$, equations
\eqref{eq:YZ-Y-transition}-\eqref{eq:YZ-Y-first-regime}  imply: 
\begin{equation}
    Y_{h_\eps-1;B, \un \o}
    =Y_{-\io;0,B}
      \bigl(1+O_\theta(|\l|\eps^\theta)\bigr)
    \label{eq:YZ-matching-Y}
\end{equation}

Concerning $Z_{h_\eps-1}$, its definition and the estimate
$|\z_h-\z_{-\io}|\leq C_\theta|\l|2^{\theta h}$ on the flow in the first-regime yield
\begin{equation}
\begin{split}
    Z_{h_\eps-1}
    &=\z_{h_\eps-1}+\cos\widetilde q_\eps=1+\z_{-\io}+O(\eps)+O_\theta(|\l|\eps^\theta)
      =\xi_1\bigl(1+O(\eps)+O_\theta(|\l|\eps^\theta)\bigr).
    \label{eq:YZ-matching-Z}
\end{split}
\end{equation}
Combining \eqref{eq:A.10}, \eqref{eq:YZ-Ah-bound},
\eqref{eq:YZ-matching-Z}, and \eqref{eq:YZ-matching-Y} we finally obtain 
\eqref{eq:YZ-ratios}, as desired.

\medskip 

\textbf{Item 3.} Inserting the identities
$\a_{h_\eps-1}=\a_{-\io}+O_\theta(|\l|\eps^\theta), 
    \z_{h_\eps-1}=\z_{-\io}+O_\theta(|\l|\eps^\theta)$, obtained telescopically from \eqref{eq:2.29bis}, 
into the definition of $v_\eps$ in \eqref{eq:2.37}, we obtain 
\be
    v_\eps
    =\frac{1+\a_{-\io}}{1+\z_{-\io}}\widetilde q_\eps
      \bigl(1+O(\eps)+O_\theta(|\l|\eps^\theta)\bigr),
\label{eq:velboundproof}\ee
where we used
$\widetilde q_\eps\asymp\sqrt\eps$ and thus
$\sin\widetilde q_\eps/\widetilde q_\eps=1+O(\eps)$.

\medskip

By \eqref{eq:singlescaleprop2},
\begin{equation}
    \frac{\widetilde Z_{h-1}(q)}{Z_{h-1}}-1
    =\bigl(\chi^{(2)}_{h}(q)-1\bigr)\frac{z_h}{1+z_h}.
    \label{eq:YZ-Ztilde-identity}
\end{equation}
Using \eqref{eq:2.70biss}, we obtain the first of \eqref{eq:item3YZ}. 
In order to prove the second of \eqref{eq:item3YZ}, recall the definition of $\rho_\o$, \eqref{eq:2.37}, as well as its rewriting in \eqref{eq:rhoidentityinappB}. 
Recall also that, on $\operatorname{supp}f_h^{(2)}$ we have $|q_1|\leq C2^h$, $
|q_2|\leq C2^h\eps^{-1/2}$, $2^h\leq C\eps$ and so, using the expansion of
$\a_{h_\eps-1}$ discussed right above \eqref{eq:velboundproof}, $\tilde q_\eps \asymp \eps^{-1/2}$,
$\xi_2:=1+\a_{-\io}$ and Taylor expanding \eqref{eq:rhoidentityinappB}, we find the second of \eqref{eq:item3YZ}, as desired.
\end{proof}

\section*{AI use statement}
The authors take full responsibility for the strategy and proofs of the results and for the writing of the manuscript. No AI was used at any stage in the production of this paper, with the exception of the right part of Fig.1, whose tikz code was written with AI support. Any headaches caused by reading this article, as well as any typos, mistakes, and inconsistencies, are entirely human in origin.

\section*{Acknowledgments}
A.G., V.M. and B.R. gratefully acknowledge financial support from the MUR, PRIN 2022 project MaIQuFi cod. 20223J85K3. F.T. was
supported by the Austrian Science Fund (FWF) through grant P35428-N. This work has been carried out under the auspices of the GNFM-INdAM Gruppo Nazionale per la Fisica Matematica.

\end{document}